\documentclass[12pt,twoside]{book}
\usepackage[a4paper,width=150mm,top=30mm,bottom=28mm,bindingoffset=6mm]{geometry}
\usepackage{amsmath,amssymb,amsthm}
\usepackage{mathrsfs}
\usepackage{bm}
\usepackage{floatrow}
\usepackage{amsmath}
\usepackage{amssymb}
\usepackage{amsthm}
\usepackage{setspace}
\usepackage{graphicx}
\usepackage{braket}
\usepackage{mathrsfs}
\usepackage{lscape,graphicx}
\usepackage{amsfonts}
\usepackage{longtable}
\usepackage{epsfig}
\usepackage{float}
\usepackage{rotating}
\usepackage[T1]{fontenc}
\usepackage{fancyhdr}
\usepackage[titletoc]{appendix}
\usepackage[english]{babel}
\usepackage{mathptmx}
\usepackage{parskip}
\usepackage{cite}
\usepackage{setspace}
\usepackage{caption}
\usepackage[hidelinks]{hyperref}
\usepackage{braket}
\usepackage{romannum}
\usepackage{bigints}
\usepackage{appendix}
\usepackage{longtable}
\usepackage[noabbrev]{cleveref}
\usepackage[acronym]{glossaries}
\usepackage{enumitem}
\usepackage{booktabs}
\usepackage{lastpage}

\usepackage{newtxtext}

\usepackage[raggedright]{titlesec}

\titlespacing\section{0pt}{12pt plus 4pt minus 2pt}{0pt plus 2pt minus 2pt}
\titlespacing\subsection{0pt}{12pt plus 4pt minus 2pt}{0pt plus 2pt minus 2pt}
\titlespacing\subsubsection{0pt}{12pt plus 4pt minus 2pt}{0pt plus 2pt minus 2pt}

\usepackage[font=small]{caption}
\usepackage{graphics,epsfig,amsmath,float,wrapfig}
\addto{\captionsenglish}{}

\newenvironment{changemargin}[2]{%
\begin{list}{}{%
\setlength{\leftmargin}{#1}%
\setlength{\rightmargin}{#2}%
}%

\item[]}{\end{list}}
\hypersetup{
    pdfauthor = {Adesh Singh}, 
    pdftitle = {}, 
    colorlinks,
    linkcolor={blue},
    citecolor={blue},
    linkcolor={blue},
    citecolor={blue}
}
\PassOptionsToPackage{unicode}{hyperref}
\begin{document}
\begin{changemargin}{0cm}{0cm}
\thispagestyle{empty}
\baselineskip25pt
\begin{center}
{\Large {\bf Quasiclassical electron transport in topological Weyl semimetals}}\\
\end{center}

\vfill
\baselineskip15pt
\begin{center}
{ \large \em \textbf{A Thesis}}
\vfill
\begin{center}
    \textit{\large Submitted for the award of the degree}
\end{center}
\large\textit{of}

\vskip .90
\baselineskip
{\large{\bf\em Doctor of Philosophy}}
\end{center}
\baselineskip15pt
\vfill
\begin{center} {\bf {\em by}} \\
{\large{\bf Azaz Ahmad}\\ (Reg. No. D19049)} \\
\end{center}

\vfill
\begin{center}
\begin{figure}[h!]
\centering
\includegraphics[scale=0.6]{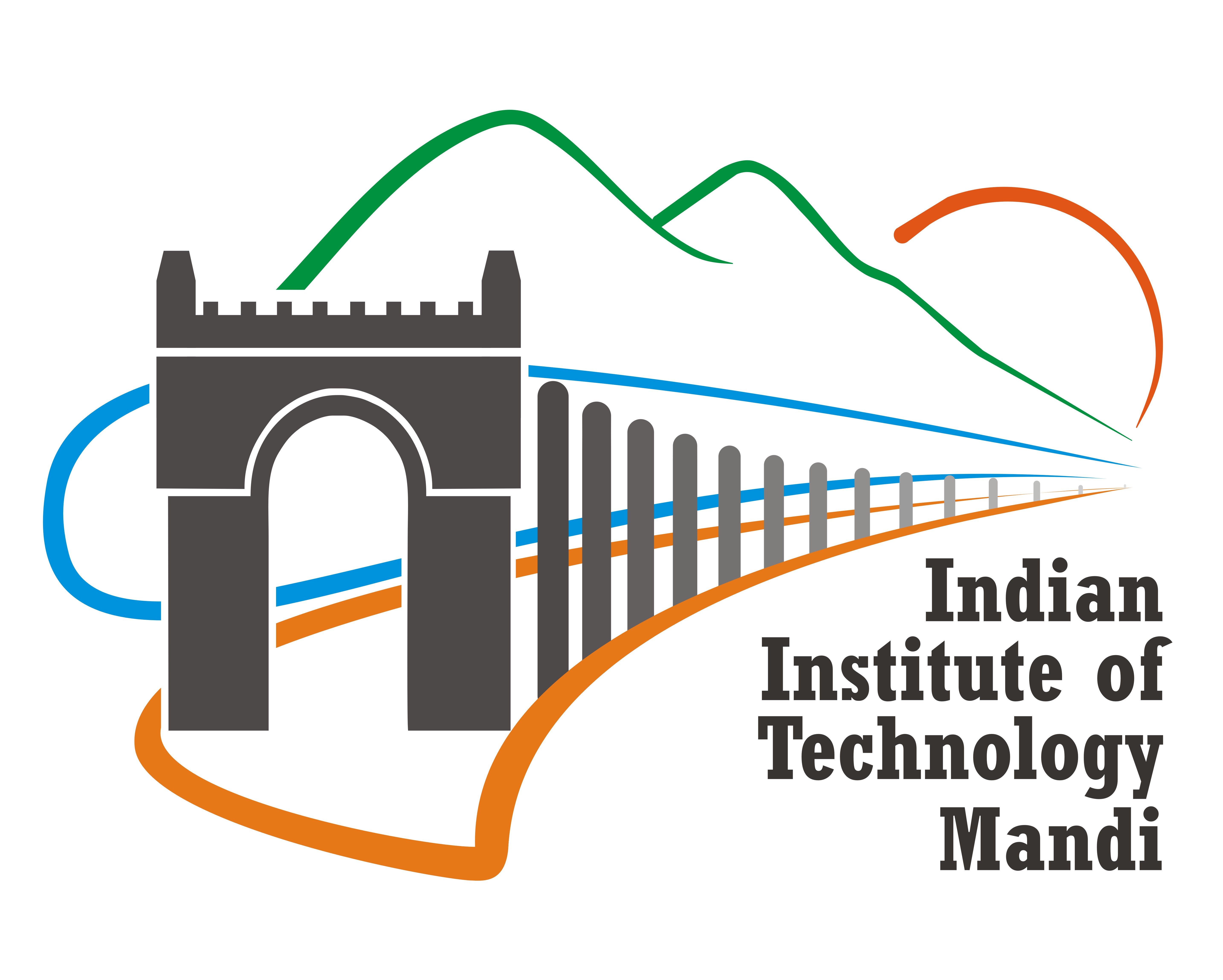}
\end{figure}
 {\bf {\large\em to the }} \\
{\bf {\large School of Physical Sciences}} \\
{\bf {\large Indian Institute of Technology Mandi}} \\
{\bf \large Kamand, Himachal Pradesh (175075), India} \\
{\bf 10th March, 2025} 
\end{center}
\end{changemargin}

\pagenumbering{roman}
\baselineskip=18pt
\vspace*{-2.5cm}
\begin{minipage}{0.2\textwidth}		
\includegraphics[width=1\textwidth]{IIT_Mandi_logo.jpg}
\end{minipage}
\begin{minipage}{0.75\textwidth}
\centering
		{\bf{INDIAN INSTITUTE OF TECHNOLOGY, MANDI}}	\\ [0.1em]
		{\bf{Kamand, Himachal Pradesh - 175075, INDIA}}						\\ [0.1em]
		www.iitmandi.ac.in								
	\end{minipage}\\\
	\rule{\textwidth}{1pt}\\\
	
\vspace*{1cm}
\centerline{\large \bf \underline{\smash{Declaration by the Research Scholar}}}
\vspace*{1cm}
\noindent 
I hereby declare that the entire work embodied in this Thesis is the result of investigations carried out by \textbf{Azaz Ahmad} in the \textbf{School of Physical Sciences}, Indian Institute of Technology Mandi, under the supervision of  \textbf{Dr. Gargee Sharma},  and  that  it  has  not  been  submitted  elsewhere  for  any  degree  or diploma. In keeping with the general practice, due acknowledgements have been made wherever the work described is based on finding of other investigators.\\[1cm]

Place: IIT Mandi  \hspace{7cm} Signature: \\\\
\vspace{1cm}
\hspace{-0.15cm}Date: \hspace{9cm} Name: \textbf{Azaz Ahmad} 
\vspace*{-2.5cm}
\begin{minipage}{0.2\textwidth}
		
		\includegraphics[width=0.9\textwidth]{IIT_Mandi_logo.jpg}
	\end{minipage}
	\begin{minipage}{0.75\textwidth}
		\centering
		{\bf{INDIAN INSTITUTE OF TECHNOLOGY, MANDI}}			\\ [0.1em]
		{\bf{Kamand, Himachal Pradesh - 175075, INDIA}}						\\ [0.1em]
		www.iitmandi.ac.in								
	\end{minipage}\\\
	\rule{\textwidth}{1pt}\\\
	
\vspace*{3cm}
\centerline{\large \bf \underline{\smash{Declaration by the Thesis Advisor}}}
\vspace*{1cm}
\noindent 
I hereby certify that the entire work in this thesis has been carried out by \textbf{Azaz Ahmad}, under my supervision in the \textbf{School of Physical Sciences, Indian Institute of Technology Mandi,}  and that no part of it has been submitted elsewhere for any Degree or Diploma.\\[3em]

Signature:\\[1em]
Name of the Guide: \textbf{Dr. Gargee Sharma}\\[1em]
\hspace*{0.05cm}Date:

\include{cert1}
\begin{center}
{\bf ACKNOWLEDGEMENTS} 
\end{center}
I am deeply grateful to my thesis advisor, \textbf{Dr. Gargee Sharma}, for her exceptional guidance and mentorship throughout my PhD journey. It is difficult to fully capture my appreciation in words, but with the few I have, I would say that I am committed to upholding and sharing the values and lessons she has imparted to me. Rational thinking and scientific temperament are fundamental to the progress of society, and she has nurtured these qualities in me. I am resolved to carry them forward to future generations.\\
I am grateful to my doctoral committee members, \textbf{Dr. Hari Varma}, \textbf{Dr. Sudhir Kumar Pandey}, \textbf{Dr. Prasanth P. Jose}, and \textbf{Dr. Bhaskar Mondal}, for their valuable assessments of my research progress throughout my PhD journey. I also extend my heartfelt appreciation to \textbf{Dr. Muslim Malik}, \textbf{Dr. Asif Khan}, and \textbf{Dr. Syed Abbas}, whose guidance and encouragement, much like that of elder brothers, inspired me to pursue new knowledge and remain dedicated to hard work. \\
I express my deepest gratitude to my elder brother, \textbf{Śrī Mo. Shaqib}, and father, \textbf{Śrī Manawwar Ali}, whose unwavering support has been the cornerstone of my life. They have guided me in every aspect of life, from navigating childhood struggles to supporting my academic aspirations. Their sacrifices, encouragement, and belief in my potential have provided me with the strength to pursue my dreams, even when the path seemed uncertain. I extend my heartfelt appreciation to my sisters, who have been my pillars of support at every stage of life. My elder sister, in particular, went above and beyond—she not only encouraged me academically but also shouldered my school bag during my primary years, ensuring that my focus remained on learning rather than on carrying heavy loads (a privilege I now look back on with a mix of gratitude and amusement). I am immensely grateful to my primary school teachers, \textbf{Śrī Jamal Ahmad} and \textbf{Śrī Satyaveer Singh}, who played a vital role in shaping my scientific temperament. Growing up in a village in Gonda district, Uttar Pradesh, where science education was not widespread, their passion for teaching ignited my curiosity and instilled in me a lifelong love for learning. Their encouragement laid the foundation for my journey from a small village to a research-intensive academic career. My grandfather, \textbf{Late Alhaj Śrī Nizamuddin}, himself a teacher, also had a profound influence on my early education. He introduced me to the fundamental concepts of science and mathematics, ensuring that my curiosity was nurtured even before I formally entered the academic world. His lessons, often simple yet profound, sparked my interest in logical reasoning and problem-solving, skills that have been invaluable throughout my research career.

Born into a farmer family, I was deeply connected to the land and the hard work that sustains it. My early education began in a charity-based primary school (Madrasa), where resources were limited, but my determination to learn was boundless. Despite the challenges of my environment, I pushed forward, eventually securing a place at Dr. Ram Manohar Lohia Avadh University in Ayodhya (Faizabad) for my Master’s degree in physics. My academic journey took a significant turn when I moved to New Delhi, where I achieved milestones such as clearing the CSIR-UGC-NET-June 2019 (All India Rank 38), GATE-2020, and Ph.D. entrance exams at premier institutions like Jawaharlal Nehru University (JNU) and IIT Mandi. In my preparation for the qualifying examinations, I was lucky to receive great assistance from my friends, Dr. Rashid M. Ansari, Javed Ahmad, and Mohd. Hashim Raza.  Their benevolence and direction were vital at each phase of the process, and I am profoundly appreciative of their steadfast assistance and support. My selection as a researcher at IIT Mandi opened doors to advanced studies in topological condensed matter physics. Throughout this journey, I have been fortunate to receive guidance and support from numerous mentors, colleagues, and friends who have shaped my academic and personal growth. Their encouragement and collaboration have been instrumental in refining my research, broadening my perspectives, and helping me aspire to contribute meaningfully to the scientific community.

I am grateful to my lab colleagues, Dr. Adesh Singh, Shubhanshu Karoliya, Arpan Gupta, and Gautham Varma K., for their insightful discussions and unwavering support. I also wish to thank my friends at IIT Mandi, including Kaushik P., Dr. Dheeraj Ranaut, Nasaru Khan, Umashankar Paridhi, Prakash Pandey, Vivek Pandey, Shivani Bhardwaj, Vinod Solet, Arishi, Himanshu Chaudhari, Shivam Mishra, Shamshad Ansari, Shahin Ansari, Vivek Kumar, Loki, and Jyotika. We shared many unforgettable experiences at IIT Mandi, especially during Saturday's cricket matches. Additionally, I extend my heartfelt gratitude to the mess staff, healthcare providers, bus drivers, security personnel, cleaning staff, and SPS school office team for contributing to a pleasant and memorable experience at IIT Mandi.

Finally, I express my sincere gratitude to all those—teachers, mentors, family members, and well-wishers—who have been part of my journey. From a small village in Uttar Pradesh to the corridors of IIT Mandi and beyond, this path has been one of resilience, learning, and relentless pursuit of knowledge. This acknowledgment is but a small token of my deep appreciation for everyone who has played a role in shaping the person I am today.

\newpage
\thispagestyle{empty} 
\vspace*{\fill} 
\begin{center}
    \Large 
    \textit{I dedicate this thesis to my cherished mother,} \\[1em]
    \textbf{\LARGE Śrīmatī Shamsunnisha}, \\[1em] 
    \textit{the most modest and kind-hearted individual I have ever encountered. She has exceeded her capabilities to bring joy into my life, making sacrifices that words cannot adequately express. Her unwavering affection and limitless benevolence have enriched me—not in financial terms, but in the most invaluable asset of all: happiness. Each of my accomplishments reflects her steadfast support, and this effort serves as a modest homage to her boundless love and dedication. {I am blessed to have a mother like her; her presence reminds me of the Quranic verse:}}
    
    \vspace{1.5em} 
    \includegraphics[width=0.6\textwidth]{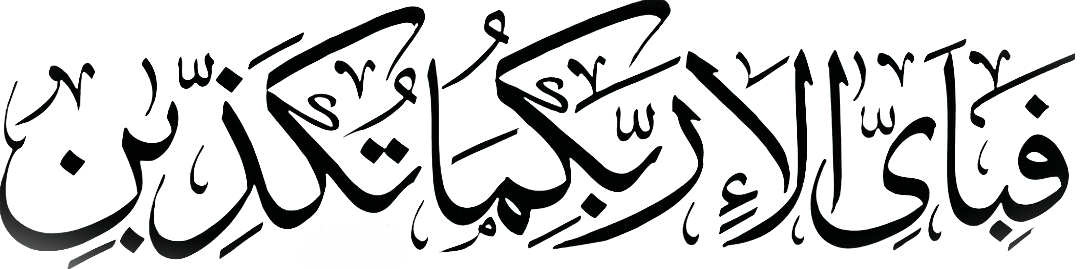} 
    \vspace{0.5em} 
    \textit{\large } \\[0.08em]
    \textbf{\large {``So, which of the bounties of your Lord will you deny?"}}
\end{center}
\vspace*{\fill} 

\begin{center}
{\large {\bf  ABSTRACT }}
\end{center}
Weyl fermions are simple yet powerful objects that connect ideas from geometry and topology to physics. Although scientists are still unsure if Weyl fermions exist as fundamental particles, there is growing evidence that they appear as particle-like excitations in certain advanced materials called Weyl semimetals (WSMs). These materials have unique electronic properties that make them exciting to study, with potential uses in future technologies. This thesis explores the latest discoveries about how electrons move through WSMs. \\
Chiral anomaly (CA) in Weyl semimetals continues to be a central theme in modern condensed matter physics, typically manifested through longitudinal magnetoconductance (LMC) and the planar Hall effect (PHE). Recent studies reveal that finite intervalley scattering can induce a sign reversal of LMC, but we identify an additional mechanism: a smooth lattice cutoff to the linear dispersion, inherent in real Weyl materials, introduces nonlinearity that drives negative LMC even with vanishing intervalley scattering. Employing a lattice model of tilted Weyl fermions within the Boltzmann approximation, we analyze LMC and PHE, mapping phase diagrams in relevant parameter spaces to elucidate the diagnostic features of CA. We further investigate the impact of elastic deformations (strain), which couple to electronic degrees of freedom as an axial magnetic field, influencing impurity-dominated diffusive transport. Our analysis shows that strain-induced chiral gauge fields lead to a 'strong sign-reversal' in LMC, unlike the external magnetic field which requires substantial intervalley scattering for similar behavior. The coexistence of external and chiral gauge fields introduces both strong and weak sign-reversals, enriching the LMC phase diagrams as functions of tilt, strain, and intervalley scattering. Additionally, we highlight distinct strain-induced features in the planar Hall conductance, offering experimentally testable predictions. Extending our study to nonlinear transport phenomena, we develop a comprehensive theory for the chiral anomaly-induced nonlinear Hall effect (CNLHE) in three-dimensional chiral quasiparticles, incorporating momentum-dependent chirality-preserving and chirality-breaking scattering processes. For Weyl semimetals, we observe nonmonotonic behavior of nonlinear Hall conductivity with Weyl cone tilt and a 'strong sign-reversal' with increasing internode scattering. In contrast, for spin-orbit coupled non-centrosymmetric metals, the CNLHE is predominantly governed by the orbital magnetic moment, showing consistently negative conductivity and quadratic magnetic field dependence. This contrast in nonlinear transport between Weyl semimetals and spin-orbit coupled systems unveils novel mechanisms in chiral quasiparticle dynamics. Finally, we expand the scope of CA to pseudospin-1 fermions, where multifold fermions exhibit distinct longitudinal magnetotransport properties. Our quasiclassical analysis reveals a transition from positive quadratic to negative LMC with increasing internode scattering, occurring at lower critical thresholds compared to Weyl fermions. Additionally, zero-field conductivity in pseudospin-1 systems shows enhanced sensitivity to internode scattering, underscoring the unique transport signatures of higher-pseudospin fermions. These insights provide a unified framework for diagnosing chiral anomaly across a spectrum of chiral quasiparticles and set the stage for future experimental explorations in Weyl semimetals, spin-orbit coupled metals, and multifold fermion systems.  

\clearpage
\fancyfoot{}      

\baselineskip=18pt
\doublespacing
\tableofcontents
\singlespacing
\clearpage
\doublespacing
\listoffigures
\singlespacing
\clearpage
\doublespacing
\listoftables 
\singlespacing
\clearpage
\pagenumbering{arabic}
\pagestyle{fancyplain}
 \renewcommand{\chaptermark}[1]{\markboth{\thechapter\ #1}{\thechapter\ #1}} 
\lhead[\fancyplain{}{}]{\fancyplain{}{}}
 \rhead[\fancyplain{}{}]{\fancyplain{}{\em\leftmark }}
 \renewcommand{\headrulewidth}{0.1pt}
 \cfoot{ \thepage \hspace{1pt} of \pageref{LastPage}} 
\baselineskip=18pt
\chapter{\label{intro}Introduction}{\small Portions of this chapter have been published in ``\textsc{Geometry, anomaly, topology, and transport in Weyl fermions"; Azaz Ahmad, Gautham Varma K., and Gargee Sharma, \textit{Journal of Physics: Condensed Matter} \textbf{37}, 043001 (2024).}}\\

The study of Weyl and Dirac fermions dates back nearly a century. In 1928, Dirac~\cite{dirac1928quantum} sought a quantum description of a relativistic electron and formulated the Dirac equation, which characterizes an electron with mass $m$ and momentum $\textbf{p}$ as a four-component spinor with the dispersion relation $\epsilon_\mathbf{p} = \sqrt{p^2c^2+m^2c^4}$. In the special case of $m=0$, Dirac's solution can be recast as two separate two-component fermions of opposite chiralities, known as Weyl fermions~\cite{weyl1929gravitation}. Over the last decade, Weyl fermions (WFs) have unexpectedly reemerged in condensed matter physics as quasiparticle excitations in certain (semi)metallic systems, known as Weyl semimetals (WSMs)~\cite{hosur2013recent,armitage2018weyl}. It has been found that the WSM phase can manifest at the transition point between a topological and a trivial insulator, acting as an intermediate state during the topological phase transition~\cite{murakami2007phase,murakami2007tuning,xu2011chern}. WSMs represent a robust and topological phase, characterized by massless Weyl fermionic excitations in the bulk that are safeguarded by translational symmetry. Additionally, they feature unique Fermi arc surface states, which correspond to the projections of the bulk gapless points in the Brillouin zone~\cite{wan2011topological,burkov2011topological,burkov2011weyl}. As a result, Weyl fermions reside at the confluence of topology, geometry, high-energy physics, and condensed matter, making their investigation highly valuable from various perspectives. In the past ten years, numerous theoretical predictions and experimental confirmations of the WSM phase have been reported in materials such as TaAs, NbAs, TaP, NbP, MoTe$_2$, and WTe$_2$~\cite{xu2015discovery,xu2016observation,shekhar2015extremely,soluyanov2015type,lv2015observation,yang2015weyl,zhang2016linear,weng2015weyl,arnold2016chiral,klotz2016quantum,liang2015ultrahigh,huang2015weyl,xu2016spin,hasan2021weyl}.

The study of electron transport in WSMs has attracted considerable attention. This interest stems from the intriguing interplay between the geometric and topological characteristics of WSMs and high-energy physics phenomena, leading to the emergence of exotic, anomalous, and topological behaviors absent in conventional metals. One of the most notable effects in WFs is the chiral anomaly (CA). This phenomenon originates from high-energy physics~\cite{adler1969axial,bell1969pcac}, where the conservation of left- and right-handed Weyl fermions is violated when electric and magnetic fields are not orthogonal. The chiral anomaly has resurfaced in the context of WSMs, drawing significant interest within the condensed matter community~\cite{armitage2018weyl,volovik2003universe,nielsen1981no,nielsen1983adler,wan2011topological,xu2011chern,zyuzin2012weyl,son2013chiral,goswami2013axionic,goswami2015axial,zhong2015optical,kim2014boltzmann,lundgren2014thermoelectric,cortijo2016linear,sharma2016nernst,zyuzin2017magnetotransport,das2019berry,kundu2020magnetotransport,knoll2020negative,sharma2020sign,bednik2020magnetotransport,he2014quantum,liang2015ultrahigh,zhang2016signatures,li2016chiral,xiong2015evidence,hirschberger2016chiral,ahmad2024nonlinear,mandal2022chiral,wu2018probing,Zhou_2013,onofre2024electric,mandal2024thermoelectric,trescher2015quantum,ferreiros2019mixed,udagawa2016field}. In WSMs, where Weyl fermions appear as quasiparticle excitations, CA is anticipated to manifest under the influence of external electromagnetic fields. Prominent transport signatures of this anomaly include positive longitudinal magnetoconductivity (LMC)~\cite{son2013chiral} and the planar Hall effect (PHE)~\cite{nandy2017chiral}. Extensive research has been dedicated to understanding these anomaly-driven conductivities in WSMs~\cite{zyuzin2012weyl,son2013chiral,goswami2013axionic,goswami2015axial,zhong2015optical,kim2014boltzmann,lundgren2014thermoelectric,cortijo2016linear,sharma2016nernst,zyuzin2017magnetotransport,das2019berry,kundu2020magnetotransport,knoll2020negative,sharma2020sign,bednik2020magnetotransport,das2019linear,sharma2019transverse,sharma2023decoupling,ahmad2021longitudinal,ahmad2023longitudinal,varma2024magnetotransport,sharma2017chiral,sharma2017nernst,nandy2017chiral,he2014quantum,zhang2016signatures,li2016chiral,xiong2015evidence,hirschberger2016chiral}. Moreover, non-electronic probes, such as optical processes, can also serve as indicators of CA~\cite{goswami2015optical,levy2020optical,parent2020magneto,song2016detecting,rinkel2017signatures,yuan2020discovery,cheng2019probing}. Interestingly, strain in Weyl semimetals induces axial vector fields, influencing both electronic and thermal transport properties, thus contributing to the unconventional behaviors observed in WSMs~\cite{grushin2016inhomogeneous,ahmad2023longitudinal}.

In this thesis, {we} discuss the research on the electronic transport properties of WSMs. In Subsec.~\ref{sec:geom}, {we} begin by providing an intuitive understanding of geometry, curvature, and topology within the framework of quantum transport. Subsequently, in Subsecs.~\ref{sec:qg and wf} and \ref{sec:wf and cond matt}, {we} establish connections between the seemingly distinct fields of quantum geometry and WFs, highlighting their significance in contemporary condensed matter physics. Finally, Subsec.~\ref{sec:ca and WF} explores the role of {CA} in Weyl fermions.

\section{Geometry and quantum physics}
\label{sec:geom}
\begin{figure}
    \centering
    \includegraphics[width=0.95\columnwidth]{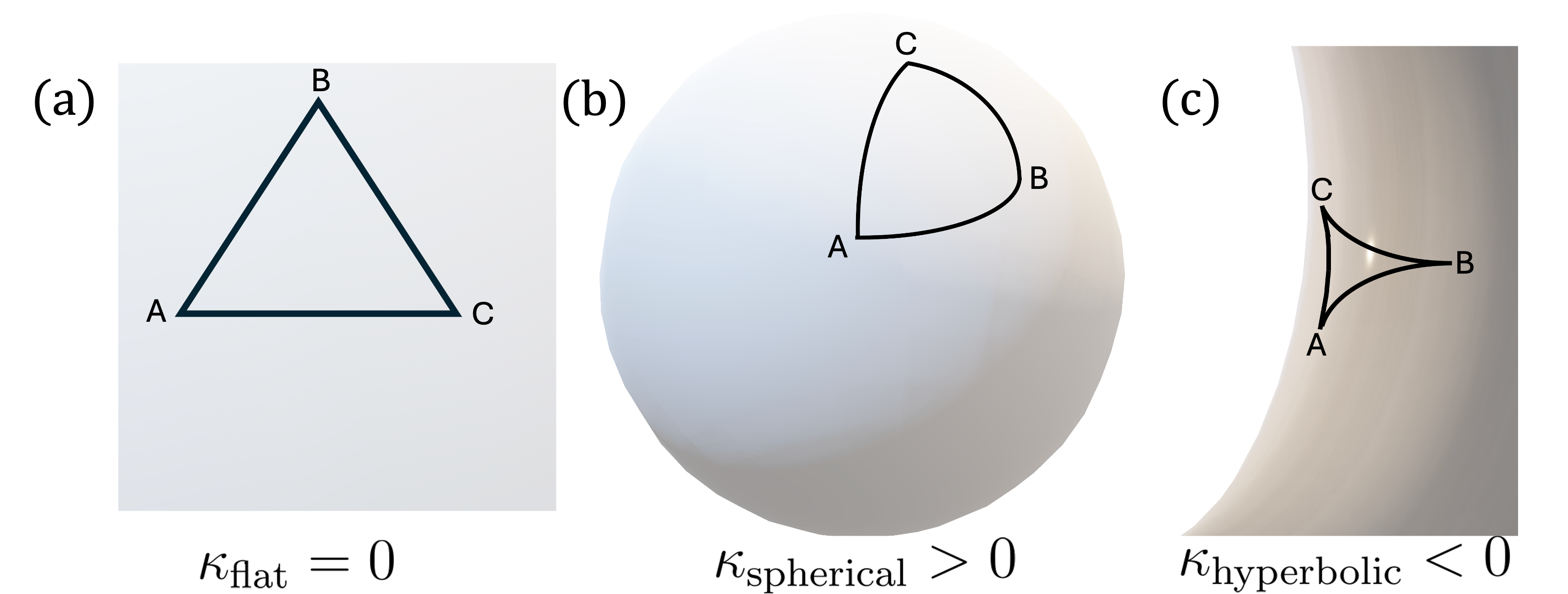}
    \caption{Geometry of different surfaces can be characterized by their curvature $\kappa$. The properties of objects existing on surfaces of different curvatures are different. For example, the sum of angles of a triangle drawn equals (a) $\pi$ for a flat surface, (b) greater than $\pi$ for a spherical, and (c) less than $\pi$ for a hyperbolic surface. Figure adapted from Ref.~\cite{ahmad2024geometry}.}
    \label{fig_triangles}
\end{figure}

\begin{figure}
    \centering
    \includegraphics[width=0.9\columnwidth]{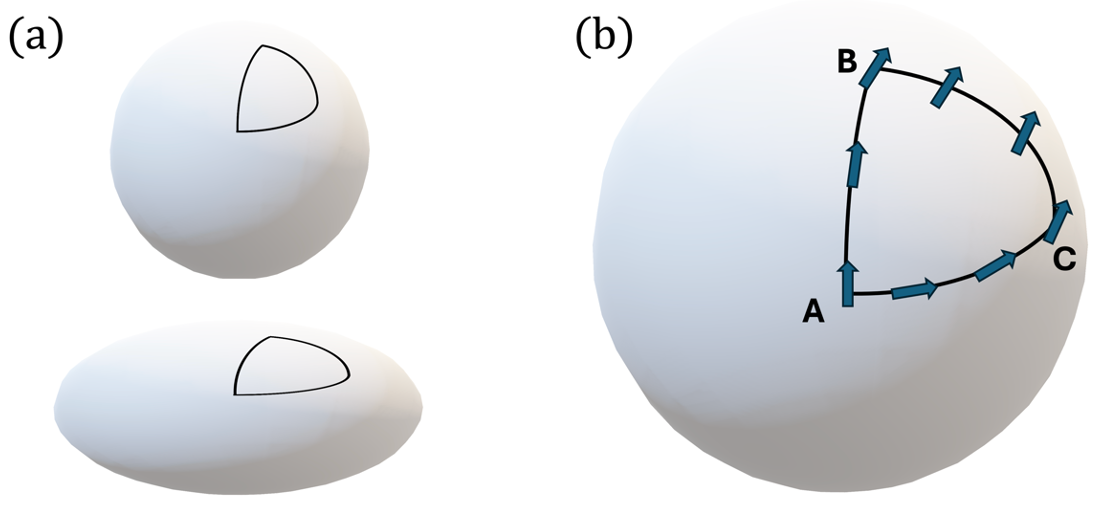}
    \caption{(a) Homeomorphism stretches a sphere to a spheroid but preserves the total curvature. (b) Parallel transporting a vector on the surface of a sphere along the path $A\rightarrow B\rightarrow C\rightarrow A$ results in an angular offset (holonomy). Figure adapted from Ref.~\cite{ahmad2024geometry}.}
    \label{fig_parallel}
\end{figure}

The geometry of space is fundamental in determining the properties of objects existing in it. The following example makes this more apparent. Consider a triangle in a flat space. Euclidean geometry predicts that the sum of angles of a triangle equals $\pi$. If one instead attempts to draw a triangle on a sphere, straight lines in flat space become geodesics, which connect two points by the shortest distance. In this case, the sum of the angles of the triangle is always greater than $\pi$, which is a characteristic of spherical geometry. If we repeat the exercise on a surface with hyperbolic geometry, the sum of the angles of a triangle is now less than $\pi$ (see Fig.~\ref{fig_triangles}). One can then define curvature ($\kappa$) as the angular excess $\epsilon(\triangle)$ (sum of the angles of a triangle minus $\pi$) per unit area $\mathcal{A}(\triangle)$~\cite{needham2021visual}: 
\begin{align}
    \kappa = \frac{\Sigma_{\mathrm{angles}} (\triangle)-\pi}{\mathrm{area}} \equiv \frac{\epsilon(\triangle)}{\mathcal{A}(\triangle)}.
\end{align}
The geometry of the three spaces discussed above can then be characterized by their respective curvatures (also see Fig.~\ref{fig_triangles}): 
\begin{align}
    \kappa_\mathrm{flat}&=0\nonumber\\
    \kappa_\mathrm{spherical}&>0\nonumber\\
    \kappa_\mathrm{hyperbolic}&<0\nonumber.
\end{align}
On more general surfaces the curvature may not be constant, and one can instead define a local (or Gaussian) curvature $\kappa_p$ at every point $p$ on the surface:
\begin{align}
    \kappa_p = \lim_{\triangle_p\rightarrow p}\frac{\epsilon(\triangle_p)}{\mathcal{A}(\triangle_p)}.
\end{align}
For example, the inner surface of a torus has $\kappa<0$ but the outer surface has $\kappa>0$. 
We can then imagine smoothly stretching a surface such that the lengths and angles of a triangle drawn on the surface are not preserved but no additional cuts or holes are introduced or removed from the surface (see Fig.~\ref{fig_parallel} (a)). Mathematically, such a transformation is termed \textit{homeomorphism}. A famous theorem by Gauss (the Gauss-Bonnet theorem) states that the total curvature of the surface is conserved under {homeomorphism}~\cite{needham2021visual, nakahara2018geometry}:
\begin{align}
\Sigma_p\kappa(p)\xrightarrow[]{\mathrm{homeomorphism}}\mathrm{constant}.
\end{align}
The Gauss-Bonnet theorem provides a deep and remarkable connection between geometry and topology and has influenced several branches of mathematics and physics. We will see this shortly, but for now we imagine parallel transporting a vector on the spherical surface and move it along the geodesic triangle (see Fig.~\ref{fig_parallel} (b)). It is easy to verify that the vector rotates when it returns to its starting point. The net rotation (mathematically termed as $\textit{holonomy}$) is equal to the angular excess of the geodesic triangle and is also equal to the total curvature of the triangle.
\begin{align}
\mathrm{holonomy}=\mathrm{angular}\hspace{1mm}{\mathrm{excess}}=\mathrm{total} \hspace{1mm}{\mathrm{curvature}}.
\end{align}
Although we discuss this for a triangular path in our example, a non-trivial holonomy may arise for any closed path. In other words, a vector $v=(a, b)$ acquires a phase (implemented by a rotation $\mathcal{R}(\theta)$) after parallel transport in a closed loop on a surface, i.e., 
\begin{align}
    v=\begin{pmatrix}
        a\\
        b
    \end{pmatrix}\xrightarrow[|a|^2+|b|^2=|a'|^2+|b'|^2]{\mathrm{parallel}\hspace{0.5mm} \mathrm{transport}}\begin{pmatrix}
        a'\\
        b'
    \end{pmatrix}=\mathcal{R}(\theta) v.
\end{align}

In quantum mechanics, the state of a system can be represented by a vector in Hilbert space. It is therefore natural to consider the effect of parallel transporting a quantum state $|n\rangle$ adiabatically in a closed loop on a curved surface. For example, this could be implemented by suitably varying a parameter $R_i$ of the underlying Hamiltonian $H(\textbf{R}(t))$ as the state evolves in time. The quantum state $|n\rangle$ rotates or in other words, acquires a phase [that can be referred as $\exp(i\gamma)$] at the end of the adiabatic evolution, where $\gamma$ is given by~\cite{berry1984quantal} 
\begin{align}
    \gamma = i\oint_c \langle n | \nabla n\rangle\cdot \mathbf{R},
\end{align}
where the gradient is taken with respect to the parameter $\mathbf{R}$. This is also known as the Berry phase~\cite{berry1984quantal}, which can also be expressed as an integral of the Gaussian curvature ($\boldsymbol{\Omega}(\mathbf{R})$) of the surface enclosed inside the closed loop:
\begin{align}
    \gamma = \int\boldsymbol{\Omega}(\mathbf{R})\cdot d\mathbf{S},
\end{align}
where $\boldsymbol{\Omega}(\mathbf{R})=i\nabla\times \langle n | \nabla n\rangle$ is also known as the Berry curvature. It turns out that the Berry phase is gauge invariant and is fully consistent with time-dependent quantum evolution. Since $\gamma$ is invariant under homeomorphism, it is also a topological invariant.

{ Before closing this section, here we would wish to clarify the geometric concepts relevant to quantum transport in Weyl semimetals, it is useful to distinguish between \textit{intrinsic} and \textit{extrinsic} curvature. Intrinsic curvature is a property of the manifold itself, independent of any embedding in higher-dimensional space, and can be measured entirely within the surface—such as through parallel transport or the angular sum of geodesic triangles. This is the type of curvature relevant for phenomena such as the Berry phase and Berry curvature, which play central roles in determining the geometrical properties of Bloch bands. In contrast, extrinsic curvature describes how a surface bends within an ambient space, which is less relevant in the context of momentum-space topology.
In our context, the “curved space” refers to a parameter manifold formed by the crystal momentum vector $\mathbf{k}$ in the Brillouin zone. The electronic band structure is described by a Hamiltonian $H(\mathbf{k})$, and the eigen-states $|n(\mathbf{k})\rangle$ form a vector bundle over this manifold. The curvature of this parameter space, quantified by the Berry curvature $\Omega_n(\mathbf{k}) = i \langle \nabla_{\mathbf{k}} n(\mathbf{k}) | \times | \nabla_{\mathbf{k}} n(\mathbf{k}) \rangle$, is intrinsic and encodes the geometric response of the system. For a general two-band model, such as $H(\mathbf{k}) = \mathbf{f}(\mathbf{k}) \cdot \boldsymbol{\sigma}$, where $\mathbf{f}(\mathbf{k})$ is a smooth vector-valued function of momentum, the degeneracy points (Weyl nodes) act as monopoles of Berry curvature, and the surrounding parameter space exhibits a non-trivial topology. As the momentum $\mathbf{k}$ traces a closed loop in this space, the quantum state acquires a Berry phase that reflects the intrinsic curvature of the manifold. This framework provides the geometric foundation for understanding chiral anomaly-related transport signatures in Weyl semimetals. We will discuss it in more details in upcoming sections.}
\section{Quantum geometry and Weyl fermions}
\label{sec:qg and wf}
To elucidate the non-trivial role of the Berry phase, we consider a simple two-band Hamiltonian of the form $H_\mathbf{k} = \mathbf{k}\cdot\boldsymbol{\sigma}$, where $\mathbf{k}$ is a vector in three-dimensions, and $\boldsymbol{\sigma}$ represents the vector of Pauli matrices.
\begin{align}
    H_\mathbf{k} = \mathbf{k}\cdot\boldsymbol{\sigma} = \begin{pmatrix}
       k_z & k_x - i k_y \\
        k_x + i k_y & -k_z
    \end{pmatrix}.
    \label{Eq_H_weyl}
\end{align}
\begin{figure}
    \centering
    \includegraphics[width=0.9\columnwidth]{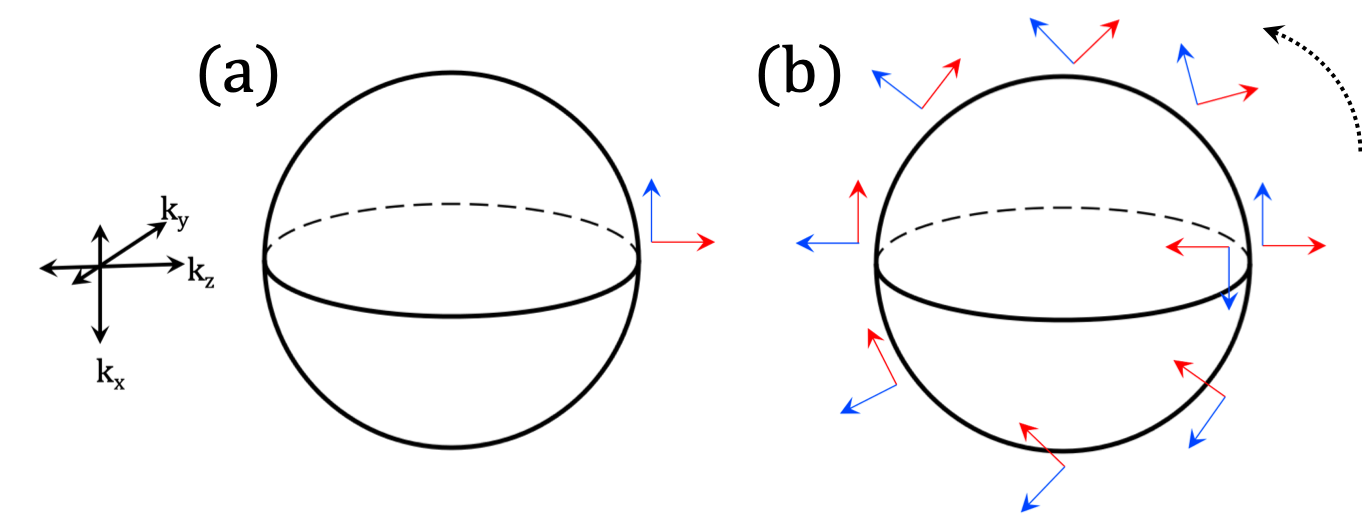}
    \caption{(a) Eigenstates of the Hamiltonian $H_\mathbf{k}$ (eigenframe) in Eq.~\ref{Eq_H_weyl} represented by the orthogonal red and blue arrows. (b) Evolving the eigenframe in the direction of the dotted arrow by varying the parameter $\theta$ from zero to $2\pi$ rotates it by $\pi$. Figure adapted from Ref.~\cite{ahmad2024geometry}.}
    \label{fig_sphere_weyl}
\end{figure}
In general, $\boldsymbol{\sigma}$ represents any suitable degree of freedom: orbital, spin, or sublattice, and $\mathbf{k}$  represents any suitable physical parameter. 
The eigenenergies of the above Hamiltonian are $E_\mathbf{k}=\pm k$ and the eigenstates are $|n\rangle^+ = (\cos(\theta/2) e^{-i\phi}, \sin(\theta/2))^\mathrm{T}$, $|n\rangle^- = (\sin(\theta/2) e^{-i\phi}, -\cos(\theta/2))^\mathrm{T}$, where $\theta$ is the polar angle $(\theta=\cos^{-1}k_z/k)$, and $\phi$ is the azimuthal angle $(\tan\phi=k_y/k_x)$. We evolve the eigenstates in a fixed gauge varying the polar angle $\theta = \cos^{-1}k_z/k$ (see Fig.~\ref{fig_sphere_weyl}). As $\theta$ varies from zero to $2\pi$, the eigenstates acquire a nontrivial $\pi$ phase. This is illustrated by the $\pi$ eigenframe rotation in Fig.~\ref{fig_sphere_weyl} (b).

The Berry curvature for the above Hamiltonian is easily evaluated to be
\begin{align}
    \boldsymbol{\Omega}_\mathbf{k}^{\pm} = \mp \frac{\hat{k}}{2 k^2}.
\end{align}
We note that this is the vector field with a singular point at the origin. This scenario is similar to the electric field generated by a point change, and thus we can say that the Berry curvature is generated by a monopole at the degeneracy point $\mathbf{k}=0$~\cite{dirac1931quantised,wu1975concept,xiao2010berry}. The degeneracy point thus acts as a source or a sink of the Berry curvature. Integrating the Berry curvature around the degeneracy point equals the number of monopoles (in this case one) in units of $2\pi$. As long as the degeneracy points are protected, homeomorphism on the Hamiltonian by employing small perturbations conserves the number of monopoles or in other words conserves the total curvature by Gauss-Bonnet theorem. This idea lies at the heart of topological protection of quantum states. 
\begin{figure*}
\floatbox[{\capbeside\thisfloatsetup{capbesideposition={right,top},capbesidewidth=0.35\columnwidth}}]{figure}[\FBwidth]
{\caption{Schematic diagram to show the Berry curvature field lines in momentum space for $k_z=0$. The red and green regions correspond to two points having opposite chirality. Here, $a$ is the lattice constant of the system. Figure adapted from Ref.~\cite{ahmad2024geometry}.}
\label{fig:BC_dipole}}
{\includegraphics[width=.99\linewidth, height = 5.5cm ]{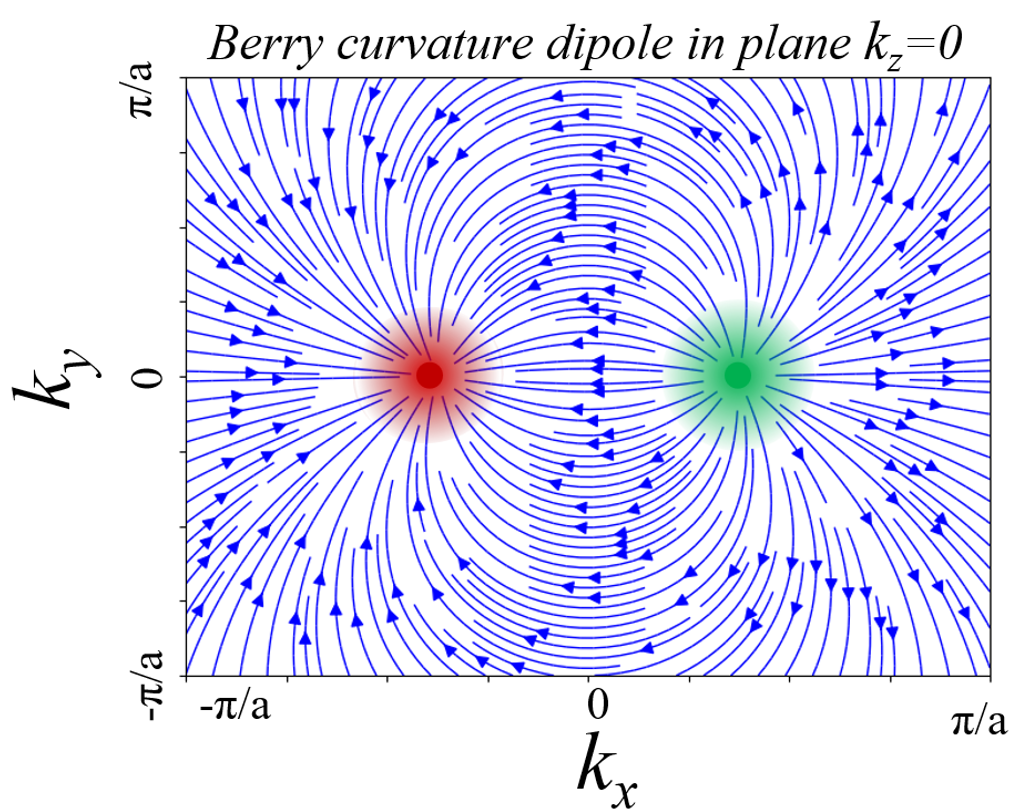}}
\end{figure*}

As an illustrative comparison, we consider the following Hamiltonian 
\begin{align}
    H_\mathbf{k}=\begin{pmatrix}
        k_z^2-k_x^2-k_y^2 & 2k_z \sqrt{k_x^2+k_y^2}\\
        2k_z \sqrt{k_x^2+k_y^2} & k_x^2+k_y^2-k_z^2
    \end{pmatrix},
\end{align}
which also has the same energy spectrum $E_\mathbf{k}=\pm k$, but the eigenframe in this case rotates by $2\pi$ as $\theta$ rotates by $2\pi$, entailing a trivial quantum geometry compared to Eq.~\ref{Eq_H_weyl}. 

Surprisingly, the Hamiltonian in Eq.~\ref{Eq_H_weyl} is identical to the Hamiltonian of Weyl fermions originating in high-energy physics~\cite{weyl1929gravitation}. Weyl fermions here refer to the massless solutions of the Dirac equation~\cite{dirac1928quantum}. When the mass term in the Dirac equation is set to zero, the Dirac Hamiltonian can be expressed in a block-diagonal form, also known as the Weyl Hamiltonian: 
\begin{align}
        H_\mathbf{k} = \begin{pmatrix}
            \mathbf{k}\cdot\boldsymbol{\sigma} & 0 \\
            0 & -\mathbf{k}\cdot\boldsymbol{\sigma}
        \end{pmatrix}.
\end{align}
The diagonal entries in the above Hamiltonian represent Weyl fermions of opposite chiralities (plus/minus or equivalently left/right)~\cite{peskin2018introduction}. 
In elementary particle physics neutrinos were initially thought to be Weyl fermions but it was later discovered that neutrinos have a finite mass. So far Weyl fermions have not been observed to exist as elementary particles. 
\section{Topological Weyl fermions in condensed matter physics}
\label{sec:wf and cond matt}
The occurrence of degenerate points in Bloch bands of crystals was first examined by Herring~\cite{herring1937accidental} who argued that band degeneracies could occur accidentally in solids. This raised the possibility of observing Weyl fermions as quasiparticle excitations around the degeneracy point where the Berry curvature in the periodic momentum space (Brillouin zone) can become singular. 
However, it was much later that Nielsen and Ninomiya studied chiral Weyl fermions on a lattice and pointed out that they must occur in pairs~\cite{nielsen1981no, nielsen1983adler}. The proof relies on the observation that every source of the Berry curvature vector field must have a sink so that the sum of the indices of the vector field on the Brillouin zone equals zero~\cite{smit2003introduction, friedan1982proof}. 

\begin{figure*}
    \centering
    \includegraphics[width=0.98\columnwidth]{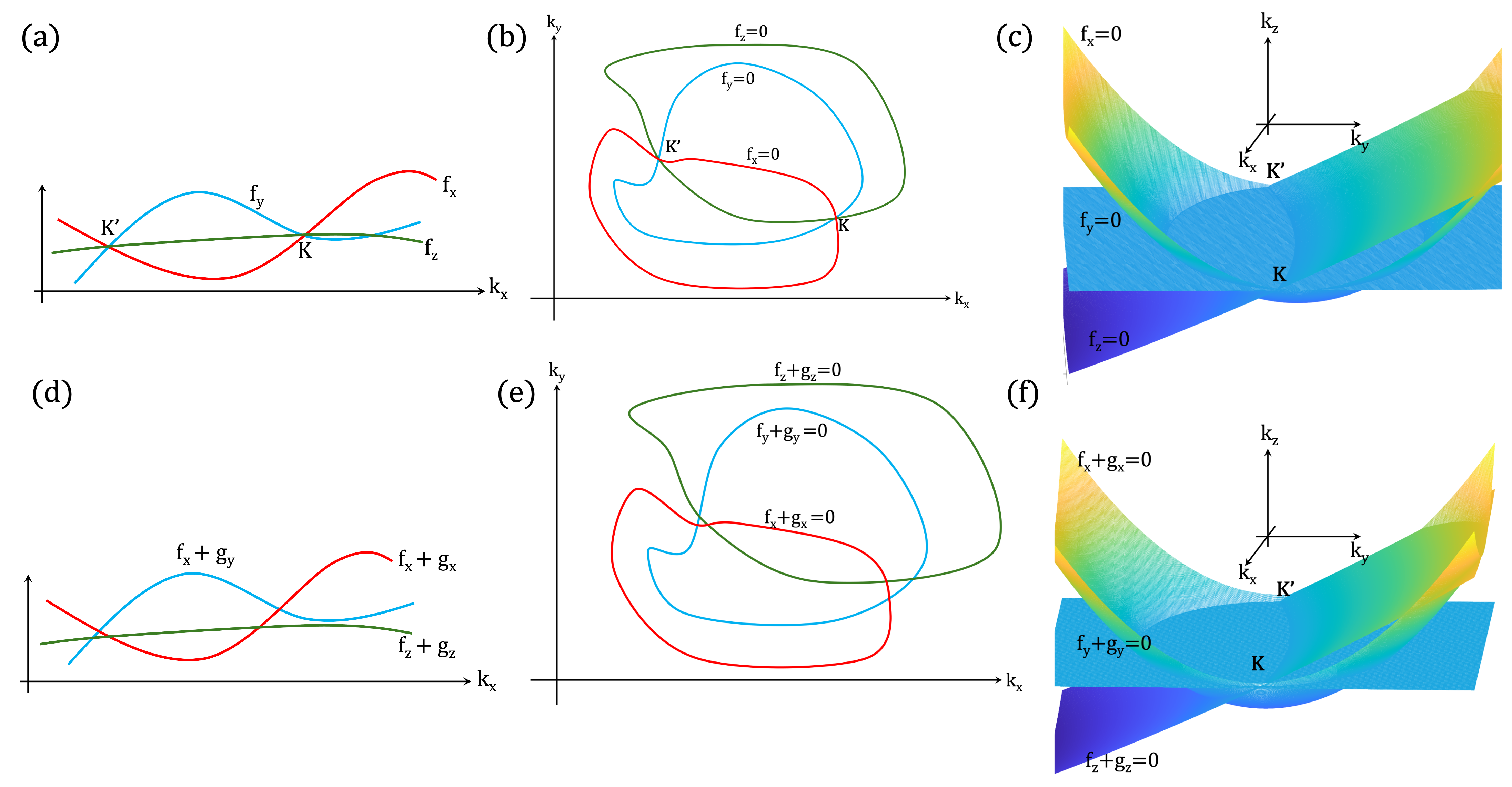}
    \caption{(a) Degenerate points in $d=1$ correspond to three functions intersecting at two points $K$ and $K'$. (b) When $d=2$, this condition is modified to three intersecting curves on a plane. (c) When $d=3$, the condition modifies to three intersecting surfaces meeting at two points $\mathbf{K}$ and $\mathbf{K}$'. (d) and (e) Small perturbations ($\mathbf{g}_\mathbf{k}$) move the curves and the fine-tuned Weyl points are destroyed. (f) Small perturbations move the surfaces and the Weyl points shift in momentum-space but remain protected. Figure adapted from Ref.~\cite{ahmad2024geometry}.}
    \label{fig_bandtouch}
\end{figure*}

The occurrence and topological protection of Weyl fermions as quasiparticles in three spatial dimensions can be understood by considering a general two-band Hamiltonian of the form:~\cite{armitage2018weyl} 
\begin{align}
    H_\mathbf{k} = \boldsymbol{\sigma}\cdot\mathbf{f}_\mathbf{k} = \begin{pmatrix}
        f_z(\mathbf{k}) & f_x(\mathbf{k}) - i f_y(\mathbf{k})\\
        f_x(\mathbf{k}) + i f_y(\mathbf{k}) & -f_z(\mathbf{k})
    \end{pmatrix}.
    \label{Eq_H_f_weyl}
\end{align}
In the above Hamiltonian $\mathbf{f}_\mathbf{k}$ is a general vector-valued function of crystal-momentum $\mathbf{k}$. The Hamiltonian can describe the low-energy band structure comprising two bands lying close to the Fermi energy in any spatial dimension $d$. The energy spectrum is given by $\epsilon_\mathbf{k} = \pm \sqrt{f_x(\mathbf{k})^2 + f_y(\mathbf{k})^2 + f_z(\mathbf{k})^2}$. Note that we measure energy relative to the midgap or the band degeneracy point. An energy shift can be easily accomplished by adding a term $\epsilon_0\mathbb{I}_{2\times 2}$ in the Hamiltonian. The condition for band-degeneracy is given by $f_x(\mathbf{k})^2 + f_y(\mathbf{k})^2 + f_z(\mathbf{k})^2=0$. Let us assume that when $d=1$,  band-degeneracy occurs at a points ${K}$ and $K'$, i.e., $f_x(K)=f_y(K)=f_z(K)=0$, and $f_x(K')=f_y(K')=f_z(K')=0$. This requires three scalar functions of a single variable to intersect at points $K$ and $K'$. This condition may be satisfied with some amount of fine-tuning (Fig.~\ref{fig_bandtouch} (a)). We now add a perturbative disorder to the Hamiltonian of the form $V_\mathbf{k} = \boldsymbol{\sigma}\cdot \mathbf{g}_\mathbf{k}$. The degeneracy condition at point $K$ is then modified to $f_x(K)+g_x(K)=f_y(K)+g_y(K)=f_z(K)+g_z(K)=0$, and a similar condition exists for the point $K'$ (Fig.~\ref{fig_bandtouch} (d)). In general, this is very hard to satisfy because of the random nature of the disorder. The band degeneracy points are therefore not protected and the degeneracy can be lifted by infinitesimal disorder. 
In two spatial dimensions ($d=2$), the band-degeneracy condition at points $\mathbf{K}=({K_x,K_y})$ and $\mathbf{K}'=({K'_x,K'_y})$ is $f_x(\mathbf{K})=f_y(\mathbf{K})=f_z(\mathbf{K})=0$, and $f_x(\mathbf{K}')=f_y(\mathbf{K}')=f_z(\mathbf{K}')=0$. This requires three curves to intersect precisely at two points, which again may be realized by fine-tuning (Fig.~\ref{fig_bandtouch} (b)). Small perturbations can move the curves (Fig.~\ref{fig_bandtouch} (e)) on the plane and thus the degenerate point is again not protected. Moving to $d=3$, the band-degeneracy condition at point $\mathbf{K}=({K_x,K_y,K_z})$ and $\mathbf{K}=({K_x',K_y',K_z'})$ is $f_x(\mathbf{K})=f_y(\mathbf{K})=f_z(\mathbf{K})=0$, and a similar condition for $\mathbf{K}'$. This condition amounts to three intersecting surfaces meeting at two points (Fig.~\ref{fig_bandtouch} (c)). Perturbations will change the surfaces and modify the degeneracy conditions (Fig.~\ref{fig_bandtouch} (f)). However, the degenerate points are just shifted in momentum space, unless the two points move closer and annihilate. Band touching is therefore robust in three spatial dimensions.  
Having found protected degenerate points in a three-dimensional solid, we can now expand the Hamiltonian in Eq.~\ref{Eq_H_f_weyl} around the point $\mathbf{K}$ as: 
\begin{align}
    H_\mathbf{k} = \begin{pmatrix}
        \nabla f_z(\mathbf{K}) &\nabla f_x(\mathbf{K})-i\nabla f_y(\mathbf{K})\\ 
        \nabla f_x(\mathbf{K})+i\nabla f_y(\mathbf{K})&-\nabla f_z(\mathbf{K})
    \end{pmatrix}\cdot\mathbf{k},
\end{align}
where $\mathbf{k}$ is now measured relative to the $\mathbf{K}$ point. The above Hamiltonian can be expressed as:
\begin{align}
    H_\mathbf{k} = \sum_{i=x,y,z}\sum_{j=x,y,z}v_{ij}k_i \sigma_j,
\end{align}
where $v_{ij}=\partial_i f_{j}$. This is the general form of the Weyl fermion Hamiltonian near the Weyl node. The nodal point is protected by the \textit{chirality} quantum number $\chi$, given by $\chi = \mathrm{sign} (\mathrm{det} [v_{ij}])$, which is also equal to the flux of the Berry curvature of a Bloch band: 
\begin{align}
    \chi = \frac{1}{2\pi}\oint{\boldsymbol{\Omega}_\mathbf{k}\cdot d\mathbf{S}}.
\end{align}
Here the integral is over the Fermi surface enclosing the nodal point and $d\mathbf{S}$ is the area element.
In an appropriate frame of reference we may write $v_{ij}=v_{i}\delta_{ij}$, and obtain a more familiar form of the Weyl Hamiltonian $H_{\mathbf{k}}=\Tilde{{\mathbf{k}}}\cdot\boldsymbol{\sigma}$, where $\Tilde{\mathbf{k}}={(v_xk_x, v_yk_y, v_zk_z)}$ similar to Eq.~\ref{Eq_H_weyl}. A similar expansion can be done around the $\mathbf{K}'$ point. Although $H_{\mathbf{k}}=\Tilde{{\mathbf{k}}}\cdot\boldsymbol{\sigma}$ is anisotropic, much of the fundamental Weyl physics can be understood by considering the isotropic version $H_\mathbf{k}=\mathbf{k}\cdot\boldsymbol{\sigma}$. The Berry curvature of a pair of Weyl nodes is schematically depicted in Fig.~\ref{fig:BC_dipole}.

{We note that the linear band structure for type-I WSMs doesn't have any cutoff for $\mathbf{k}$, and in some instances, it may lead to divergent results or theoretical artifacts that maybe unphysical. These divergences can be gotten rid of by employing a lattice model. A more general form of Hamiltonian, including non-linear dispersion away from the Weyl node, is possible to write by lattice regularization: $H^{\mathrm{Latt}}_{\mathrm{WSM}} = \chi E_0 \sin(a\mathbf{k}\cdot\mathbf{\sigma})$, where $E_0$ is an energy parameter providing energy bandwidth = $2E_0$ and $a$ lattice constant \cite{ahmad2021longitudinal, sharma2016nernst, trescher2015quantum}. This lattice regularization offers a physical ultraviolet cutoff to the low-energy spectrum. The high-energy cutoff that prevents the spectrum from diverging can be understood as follows: in a lattice system, the Brillouin zone (BZ) restricts the available momentum space suppressing the allowed $k$-values to the first BZ, and at the BZ boundary the energy spectrum of the lattice model of a WSM flattens. The lattice model of Weyl fermions may have important consequences. For instance, the lattice model of Weyl fermions results in a nonzero Nernst effect \cite{sharma2016nernst,sharma2017nernst} (which is also observed empirically \cite{liang2017anomalous}), despite the prediction that in the linear approximation, the effect disappears~\cite{lundgren2014thermoelectric}. Nevertheless, the linear approximation accounts for most of the effects in WSMs with closed Fermi surfaces.}

\section{Chiral anomaly of Weyl fermions}
\label{sec:ca and WF}
{CA} refers to the non-conservation of the left and right-handed chiral Weyl fermions in the presence of external gauge fields. It is also known as the Adler-Bell-Jackiw (ABJ) anomaly and originates in high energy physics \cite{adler1969axial}. This non-conservation of chiral charges results in a nonvanishing chiral current that may lead to chirality-dependent transport. In the context of WSMs, CA can be verified experimentally through the measurement of (but not limited to) magnetoconductance~\cite{son2013chiral}, Hall conductance~\cite{burkov2011weyl,burkov2014anomalous}, thermoelectric~\cite{lundgren2014thermoelectric} and Nernst effects~\cite{sharma2016nernst}, optical processes~\cite{goswami2015optical}, and non-local transport~\cite{parameswaran2014probing}. 

Nielson and Ninomiya were among the first to investigate {CA} in crystals~\cite{nielsen1981no,nielsen1983adler}. They derived the chiral anomaly or the ABJ anomaly from a physical point of view as the production of Weyl particles, and show that there is an absence of the net production of particles for local chiral invariant theories regularized on a lattice. They showed that fermion systems in lattice gauge theories are similar to electron systems in crystals, so as a result there should exist a mechanism in crystals that is similar to the ABJ anomaly. When two energy bands have point-like degeneracies, the Weyl particles can move from one degeneracy to the other in the presence of parallel electric and magnetic fields leading to a large longitudinal positive magnetoconductance~\cite{imran2018berry}, which is a manifestation of the anomaly in crystals. The exact sign of LMC is more nuanced and we discuss this in detail in a later section. 

\begin{figure*}
    \centering
    \includegraphics[width=.98\columnwidth]
    {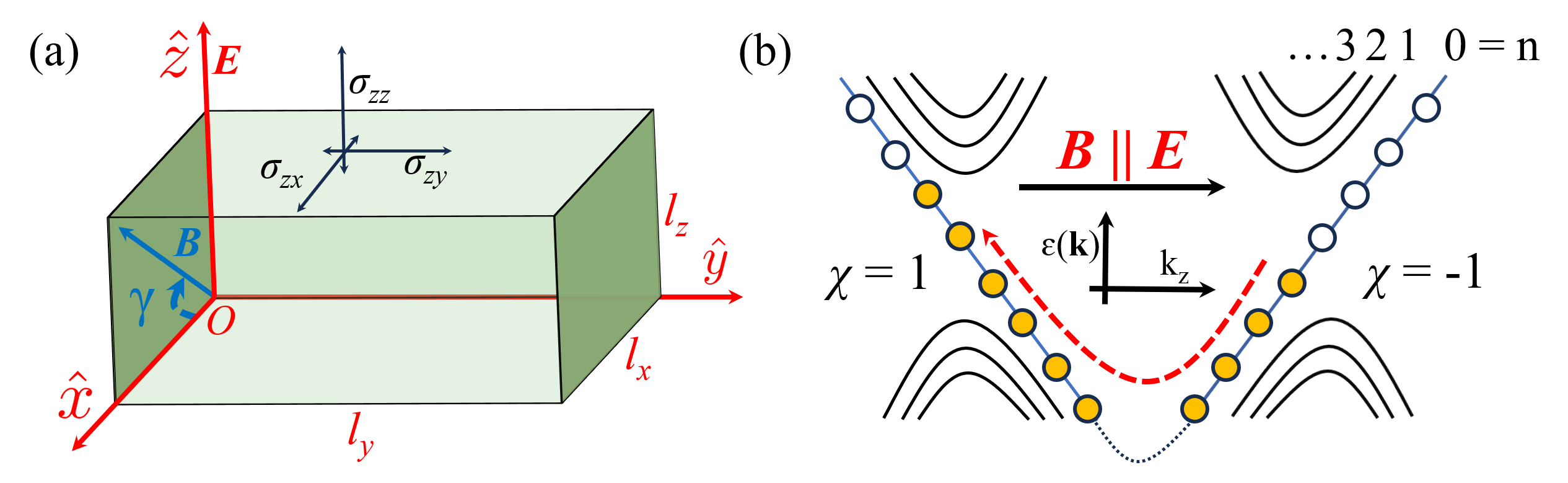}
    \caption{(a) Schematic illustration of the experimental setup to measure different conductivity responses of 3D WSMs under external electric and magnetic fields. $V=l_xl_y l_z$ is sample volume, $\gamma$ is to tune the direction of the magnetic field which is very helpful to study the chiral magnetic effect as chiral current is proportional to $\mathbf{E}\cdot\mathbf{B}$. (b) Landau level picture of the dispersion of WSMs having pair of Weyl nodes with chirality $\chi=\pm1$. Occupancy of the chiral Landau level ($n=0$) has been marked by the yellow circles and it is the only one to participate in the chiral pumping process under parallel electric and magnetic fields. Figure adapted from Ref.~\cite{ahmad2024geometry}. }
\label{fig:WSM_slab_nd_Landau_picture}
\end{figure*}

We present a brief overview of CA with help of a simple Landau-levels picture~\cite{hosur2013recent}.
In Fig.~\ref{fig:WSM_slab_nd_Landau_picture} (a), we sketch a WSM of finite size with volume $V=l_x l_y l_z$, subject to external electric ($\mathbf{E}$) and magnetic field ($\mathbf{B}$). We quantize the levels in presence of an external magnetic field. {The  corresponding} energy dispersion is plotted in Fig.~\ref{fig:WSM_slab_nd_Landau_picture} (b), where the energy levels disperse along the direction of magnetic field $\mathbf{B}$. Except $n=0$ level, all the Landau levels are degenerate with degeneracy $g={2\pi e B l_x l_y}/{\hbar}$. The zeroth Landau level is chiral in nature and its direction of dispersion depends upon the chirality $\chi$ of Weyl node. 
The general form of the dispersion may be explicitly written as ~\cite{hosur2013recent,sharma2023decoupling}:
\begin{align}
\epsilon(\mathbf{k})= \nonumber\\
&
\begin{cases}
    v_\mathrm{F} ~\mathrm{sign}(n) \sqrt{2\hbar|n| e B + (\hbar k_z)^2},& n = \pm1, \pm2...\\
    -\chi \hbar v_\mathrm{F} k_z,     & n = 0.
\end{cases}   
\end{align}
We will assume that $\mu \ll v_\mathrm{F} \sqrt{\hbar e B}$. For this case one focuses only on the zeroth Landau level physics; this is also dubbed as the \textit{quantum limit}. 
As depicted in Fig.~\ref{fig:WSM_slab_nd_Landau_picture} (a), for $\mathbf{E} = E \hat{z}$ and $\gamma=\pi/2$, i.e., $\mathbf{B} = B \hat{z}$, all the states are forced to mobilize along $\hat{z}-$direction: ${d \mathbf{k}}/{d t} = ({-e}/{\hbar} )\mathbf{E}$. Therefore, electrons in the zeroth Landau level at valley $\chi= 1$ move towards the left and at valley $\chi= -1$ towards the right. This mobilization of states appears as the disappearance of electrons from the band having a positive slope and reappearance to the band having a negative slope through a hidden channel marked by a black dotted curve in Fig.~\ref{fig:WSM_slab_nd_Landau_picture}(b). 
This process leads to non-conservation of charge at a particular chiral Landau level. In conclusion, chiral Landau levels have 1d {CA} which results in an imbalance of the population of charges at two valleys having opposite chiralities. The rate of change of charge particle at valley having chirality $\chi$ is given by:
\begin{align}
    \frac{\partial \mathcal{Q}^{\chi}_{\hat{z}}}{\partial t} = e \chi l_z \frac{|\dot{\mathbf{k}}|}{2\pi} = -2\pi e^2 \chi l_z \frac{|\mathbf{E}|}{\hbar}.
\end{align} 
Including degeneracy of the Landau levels ($g={2\pi e B l_x l_y}/{\hbar}$), this is generalised as follows:
\begin{align}
    \frac{\partial \mathcal{Q}^{\chi}_{3D}}{\partial t} = g \frac{\partial \mathcal{Q}^{\chi}_{\hat{z}}}{\partial t} = -\frac{e^3}{4\pi^2 \hbar^2} l_x l_y l_z \mathbf{E}\cdot\mathbf{B}.
\end{align}
We will see in the next section how the $\mathbf{E}\cdot\mathbf{B}$ term appears in the semiclassical equations of motion and allows us to study CA and CA-assisted transport through the Bloch-Boltzmann formalism. 
{\section{Weyl Semimetals with tilt: Type-I and Type-II}
The low energy Hamiltonian of a Weyl semimetal about the Weyl nodes can be written as:
\begin{align}
H_{\mathrm{WSM}}(\mathbf{k})=\sum_{\chi=\pm1} \chi\hbar v_{F}\mathbf{k}\cdot\boldsymbol{\sigma} + \hbar v_{F}(t_z^{\chi} k_z + t_x^{\chi} k_x )I_{2\times2},
\label{Hamiltonian}
\end{align}
where, $\chi$ is the chirality of the Weyl node, $\hbar$ is the reduced Plank constant, $v_{F}$ is the Fermi velocity, $\mathbf{k}$ is the wave vector measured from the Weyl node, $\boldsymbol{\sigma}$ is vector of Pauli matrices and $t_{x,z}$ are the tilt parameters. One may diagonalize this Hamiltonian to get energy dispersion, 
\begin{align}
\epsilon^{\chi}_k=\pm{\hbar v_{F}}k + \hbar v_{F}(t^{\chi}_z k_z+t^{\chi}_x k_x).
\label{Dispersion_ch1}
\end{align}
The type of the WSMs, i.e., whether it is type-I or type-II depends on the value of the tilt parameter in the Eq.~\ref{Dispersion_ch1}. For type-I WSM, the magnitude of the tilt parameter $t_i$ must be in the range $0\leq |{t_i}|/v_F < 1$. While for type II, the magnitude of the tilt parameter is greater than unity. For small tilt magnitudes ($<1$) the valence and conduction bands touch each other at only the Weyl points, but if the tilt is large ($>1$), the two bands overlap with the Fermi energy and make an electron-hole pocket at the Fermi level. In the latter case, the physical properties are calculated by summing the contribution from both bands. Several materials have been experimentally proved to have type-II Weyl point \cite{soluyanov2015type, chen2016magnetotransport,li2017evidence,yao2019observation, ghosh2024direction,udagawa2016field}. Type-II Weyl semimetals (WSMs) are anticipated to have distinct topological response functions compared to type-I WSMs. In particular, it has been proposed that if an external magnetic field is supplied perpendicular to tilt direction $\hat{t_i}$ in a type-II WSM, the zeroth chiral Landau level is not present. As a result, CA and the corresponding LMC are only expected to exist when the magnetic field is directed within a cone around the tilt axis $\hat{t_i}$ \cite{soluyanov2015type}. Therefore, CA and LMC exhibit a strong anisotropy in the direction of the applied magnetic field. However, using quasiclassical Boltzmann formalism, Sharma \textit{et al.} investigated the effects of chiral anomaly on longitudinal magnetotransport in a type-II WSM and demonstrated the existence of chiral-anomaly-induced positive LMC in all directions \cite{sharma2017chiral}, with no significant qualitative difference between the two. A $B-$linear component in LMC was shown to be present in the direction of tilt. Similar conclusions were reported in Ref.~\cite{das2019linear} as well.}
\section{Electron transport in Weyl fermions}
\label{sec2}
\subsection{Maxwell-Boltzmann transport
theory of Weyl fermions}
\label{sec:maxwellboltz}
\begin{figure*}
\floatbox[{\capbeside\thisfloatsetup{capbesideposition={right,top},capbesidewidth=0.47\columnwidth}}]{figure}[\FBwidth]
{\caption{Diagrammatic depiction of the classical route of a Bloch electron wavepacket in the phase space, which is governed by the classical equations of motion i.e., Eq.~\ref{Couplled_equation_ch1}. The position is indicated by the blue arrow and can be expressed as the coordinate $(\mathbf{r},\mathbf{k})$. The direction of motion is shown by the black arrow on the dotted route. Impurity sites that are naturally present in the system are denoted by A, B, and C. This semiclassical method enables us to follow an electron between two consecutive collisions. Figure adapted from Ref.~\cite{ahmad2024geometry}.}
\label{fig:Classical_path}}
{\includegraphics[width=.99\linewidth, height = 5.5cm ]{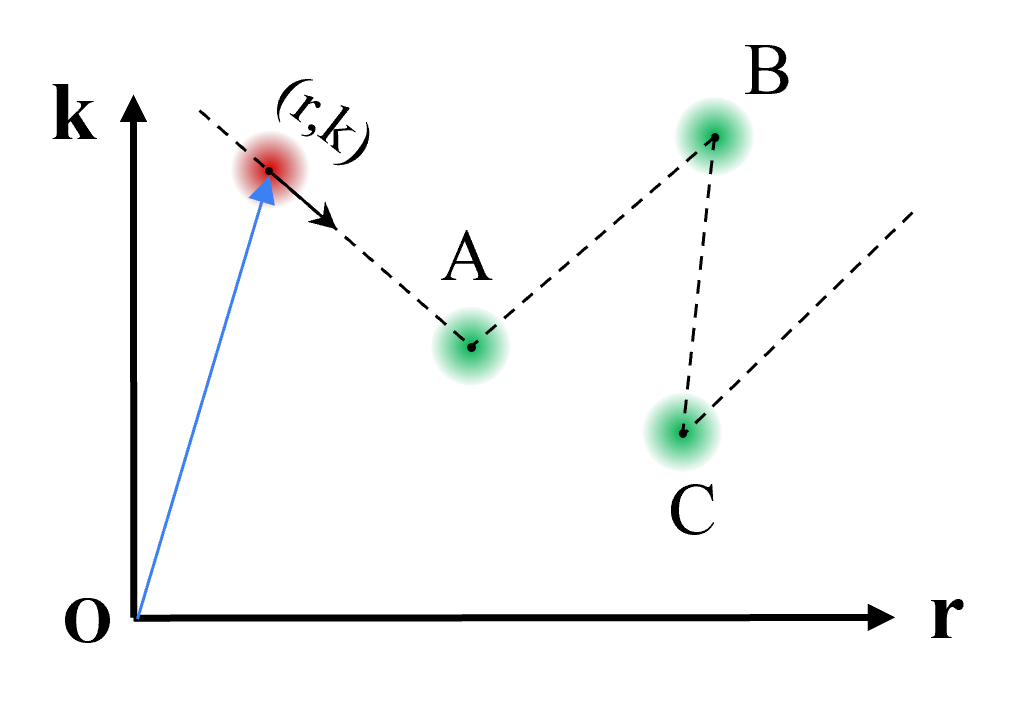}}
\end{figure*}

Electrons in solids are influenced by the periodic lattice potential of ions situated at well-defined basis points. The electronic conductivity of solids therefore must account for this periodic potential. A solution to this problem was given by Bloch~\cite{ashcroft1976nd} and thus noninteracting electrons in periodic potential are dubbed as {Bloch electrons}. Bloch electrons conserve the crystal momentum and have plane-wave solutions modulated by a periodic function: $\psi^n_{\mathbf{k}}(\mathbf{r})=e^{i\mathbf{k}\cdot\mathbf{r}} u^n_\mathbf{k}(\mathbf{r})$, where $n$ is the band index. Extending Sommerfield's theory to nonequilibrium cases in the presence of external perturbations, one then explores the conduction of solids. Here, the dynamics of Bloch electron wavepackets are considered classical. These classical equations describe the behavior of the {wave packet} of electron levels as shown in Fig.~\ref{fig:Classical_path}, which is forced to obey the uncertainty principle. 
The equations of motion to track the evolution of the position ($\mathbf{r}$) and wave vector ($\mathbf{k}$) of an electron in an external electromagnetic field ($\mathbf{E}$ and $\mathbf{B}$) are:
\begin{align}
    \mathbf{\dot{r}} \equiv \mathbf{v}=  \frac{1}{\hbar} \frac{\partial \epsilon(\mathbf{k})}{\partial\mathbf{k}},\nonumber\\
    \hbar\mathbf{\dot{k}} = e (\mathbf{E} + \mathbf{\dot{r}} \times \mathbf{B}).
    \label{eq:EOM_classical particle}
\end{align}
As we noted earlier, Weyl fermions in solids, due to the nontrivial topology of the bands, possess a Berry curvature which modifies the above equation to the one presented in the following equation~\cite{son2012berry}:
\begin{align}
\dot{\mathbf{r}}^\chi &= \mathcal{D}^\chi \left( \frac{e}{\hbar}(\mathbf{E}\times \boldsymbol{\Omega}^\chi) + \frac{e}{\hbar}(\mathbf{v}^\chi\cdot \boldsymbol{\Omega}^\chi) \mathbf{B} + \mathbf{v}_\mathbf{k}^\chi\right) \nonumber\\
\dot{\mathbf{p}}^\chi &= -e \mathcal{D}^\chi \left( \mathbf{E} + \mathbf{v}_\mathbf{k}^\chi \times \mathbf{B} + \frac{e}{\hbar} (\mathbf{E}\cdot\mathbf{B}) \boldsymbol{\Omega}^\chi \right),
\label{Couplled_equation_ch1}
\end{align}
where $\boldsymbol{\Omega}^\chi = -\chi \mathbf{k} /2k^3$ is the Berry curvature of Weyl fermions, and $\mathcal{D}^\chi = (1+e\mathbf{B}\cdot\boldsymbol{\Omega}^\chi/\hbar)^{-1}$.  The self-rotation of the Bloch wave packet, which has a finite spread in the phase space, also gives rise to an orbital magnetic moment (OMM) $\mathbf{m}^\chi_\mathbf{k}$~\cite{xiao2010berry}. In the presence of a magnetic field, the OMM shifts the energy dispersion as $\epsilon^{\chi}_{\mathbf{k}}\rightarrow \epsilon^{\chi}_{\mathbf{k}} - \mathbf{m}^\chi_\mathbf{k}\cdot \mathbf{B}$. Note that we have added the chirality index $\chi$ to distinguish Weyl fermions of different flavors. 

Returning to the quasiclassical formalism, it shows that a Bloch electron wavepacket has nonvanishing band velocity ($\mathbf{v}_n$) proportional to $({\partial \epsilon(\mathbf{k})}/{\partial \mathbf{k}} )$. So a perfect solid has infinite conductivity. The inclusion of the wave nature of the electron justifies this as constructive interference of scattered waves from an array of periodic potentials allowing it to propagate through solids without attenuation~\cite{ashcroft1976nd}. Since no crystal structure is perfect, imperfection leads to the degradation of current giving rise to finite conductivity. 
Using the semiclassical Boltzmann formalism, one can investigate charge transport in the presence of perturbative electric and magnetic fields and evaluate the conductivity. 
For spatially uniform fields, the formalism assumes the presence of a distribution function $f_\mathbf{k}$ of electrons which reduces to the Fermi-Dirac distribution in equilibrium. The nonequilibrium distribution function satisfies the following equation:~\cite{bruus2004many} 
\begin{align}
\dfrac{\partial f_{\mathbf{k}}}{\partial t}+ {\Dot{\mathbf{r}}_{\mathbf{k}}}\cdot \mathbf{\nabla_r}{f_{\mathbf{k}}}+\Dot{\mathbf{k}}\cdot \mathbf{\nabla_k}{f_{\mathbf{k}}}=I_{\mathrm{coll}}[f_{\mathbf{k}}].
\label{MB_equation_ch1}
\end{align}
We assume $f_\mathbf{k} = f_{0\mathbf{k}} + g_{\mathbf{k}}$, where $f_{0\mathbf{k}}$ is the standard Fermi-Dirac distribution and $g_{\mathbf{k}}$ is the deviation due to external fields. Perturbation theory allows us to go up to the order of choice to discuss the higher orders of different electromagnetic responses of the materials. When restricted to the first order in $\mathbf{E}$, then $g_\mathbf{k}\propto \mathbf{E}$.
The right-hand side in Eq.~\ref{MB_equation_ch1} (the collision integral term) accounts for the relaxation processes in the system. The simplest approach adapted in the literature is the `relaxation time approximation' with momentum-independent constant scattering time that assumes
\begin{align}
    I_{\mathrm{coll}}[f_{\mathbf{k}}] = -\frac{g_\mathbf{k}}{\tau},
\end{align}
where $\tau$ is the scattering time. Once the distribution function is obtained, conductivity can be straightforwardly obtained via the following relation:
\begin{align}
    \mathbf{j} = -e\sum_\mathbf{k} \dot{\mathbf{r}} g_\mathbf{k}.
\end{align}

In the case of Weyl fermions, we need to consider the scattering of an electron from an occupied state with momentum $\mathbf{k}$ and chirality $\chi$ to an unoccupied state with momentum $\mathbf{k}'$ and chirality $\chi'$. This leads to two processes: (i) if $\chi=\chi'$, this is termed as intranode (or intravalley) scattering, referring to the preservation of the chirality index of the particle, (ii) when $\chi\neq\chi'$, this is termed as internode (or intervalley) scattering referring to breaking of the chirality index. Therefore one typically needs to consider two scattering times: (i) $\tau_{\mathrm{inter}}$, which is the internode scattering time, and (ii) $\tau_\mathrm{intra}$, which is the intranode scattering time. Furthermore, the distribution functions at both nodes must have a chirality index as well. The collision integral can therefore be expressed as~\cite{son2013chiral,kim2014boltzmann} 
\begin{align}
    I_{\mathrm{coll}}[f^\chi_{\mathbf{k}}] = -\frac{f^\chi_{\mathbf{k}}-f^{\chi'\neq\chi}_{\mathbf{k}}}{\tau_{\mathrm{inter}}}-\frac{f^\chi_{\mathbf{k}}-f^{\chi'=\chi}_{\mathbf{k}}}{\tau_{\mathrm{intra}}}.
\end{align}
The internode scattering mechanism successfully captures the essence of {CA}-induced transport phenomena as first pointed out by Son and Spivak~\cite{son2013chiral}. It predicts \textit{negative} longitudinal magnetoresistance, which was initially considered a definitive signature of the manifestation of {CA} in solids. Recent works~\cite{knoll2020negative,sharma2020sign,ahmad2021longitudinal,sharma2023decoupling,ahmad2023longitudinal, mandal2024thermoelectric,medel2023planar} have evolved this understanding, which we will discuss later.

The linear response formalism for computing conductivity~\cite{mahan20089}, presupposes a timescale denoted by $\tau_{\phi}$, which signifies the interactions between the system and the external electric field. This timescale reflects the inelastic exchange of energy at a rate of $\tau_{\phi}$.
The rate is ideally assumed to be zero, meaning that $\tau_{\phi}$ is presumed to be the longest among all relevant timescales.
For weakly disordered Weyl semimetals, particularly when Landau quantization of energy levels becomes significant under intense magnetic fields, {CA} becomes apparent through a positive contribution to the {LMC}, expressed as $\mathbf{j}\propto B(\mathbf{E} \cdot \mathbf{B})$. The current flow is constrained by the internode scattering time ($\tau_\mathrm{inter}$), representing the timescale at which electrons scatter across the nodes and alter their chirality. Consequently, while the chiral charge isn't conserved, the global charge is conserved. For the linear response formalism to be valid, $\tau_\mathrm{inter}$ must significantly exceed $\tau_\phi$. However, when intranode scattering is the dominant scattering mechanism, the conservation of chiral charge along with global charge becomes important, and the calculations need to be re-examined.

When magnetic fields are weak and quantization of levels is unimportant, Son and Spivak~\cite{son2013chiral} predicted internode scattering induced positive longitudinal magnetoconductivity (LMC) in Weyl semimetals (WSMs) via the semiclassical Boltzmann approach. Later on, several studies proposed that positive LMC can arise solely from intranode scattering, as evidenced by a coupling term $\mathbf{E}\cdot\mathbf{B}$ (Eq.~\ref{Couplled_equation_ch1}) incorporated into the semiclassical equations of motion~\cite{kim2014boltzmann,lundgren2014thermoelectric,cortijo2016linear,sharma2016nernst,zyuzin2017magnetotransport,das2019berry,kundu2020magnetotransport,knoll2020negative}. This suggests that positive LMC can occur in WSMs even in the limit where $\tau_\mathrm{inter}/\tau_\mathrm{intra}\rightarrow\infty$. Notably, none of the studies differentiate between two parameter regimes: $\tau_\mathrm{intra}\ll\tau_\phi\ll\tau_\mathrm{inter}$ and $\tau_\mathrm{intra}\ll\tau_\mathrm{inter}\ll\tau_\phi$. This distinction carries significant consequences, as the chiral charge is conserved in the former case, while the latter indicates global charge conservation, occurring on a timescale larger than the intravalley timescale $\tau_\mathrm{intra}$ but smaller than $\tau_\phi$. The assertion that intranode scattering alone can yield positive LMC  assumes that $\tau_\mathrm{intra}\ll\tau_\phi\ll\tau_\mathrm{inter}$, which is incorrect. This issue was recently resolved in Ref.~\cite{sharma2023decoupling} where Sharma \textit{et al.} calculated transport properties including LMC correctly for various possible values of the parameters $\tau_\mathrm{intra}$, 
 $\tau_\mathrm{inter}$, and $\tau_\phi$. 

Furthermore, in the same work~\cite{sharma2023decoupling}, Sharma \textit{et al.}, showed that a constant relaxation time approximation in Weyl fermions is inherently inconsistent with charge conservation. This is understood by the following illustrative example. Consider one Weyl node with chirality $\chi$ and the following low-energy Hamiltonian: $H_\mathbf{k}=\chi \hbar v_F \mathbf{k}\cdot\boldsymbol{\sigma}$. The steady-state Boltzmann equation takes the following form in the relaxation-time approximation~\cite{sharma2023decoupling}:
\begin{align}
    e \mathcal{D}^\chi_\mathbf{k} \left(-\frac{\partial f_{0\mathbf{k}}}{\partial \epsilon_\mathbf{k}}\right) \left(\mathbf{v}^\chi_\mathbf{k}+ \frac{e}{\hbar}\mathbf{B} (\boldsymbol{\Omega}^\chi_\mathbf{k}\cdot\mathbf{v}^\chi_\mathbf{k})\right)\cdot\mathbf{E} = -\frac{g^\chi_\mathbf{k}}{\tau_\mathbf{k}}
    \label{Eq:Boltz1}
\end{align}
Charge conservation suggests that $
\sum\limits_\mathbf{k}g^\chi_\mathbf{k} = 0$. 
When both $\mathbf{E}$ and $\mathbf{B}$ are parallel to the $z-$axis, the charge conservation equation reduces to:
\begin{align}
    \int {\tau^\chi(\theta) \left({v}^\chi_{z}+ \frac{e}{\hbar}{B} (\boldsymbol{\Omega}^\chi_\mathbf{k}\cdot\mathbf{v}^\chi_\mathbf{k}) \right) \frac{k^3(\theta) \sin\theta}{|\mathbf{v}^\chi_\mathbf{k}\cdot\mathbf{k}|} d\theta} = 0. 
    \label{Eq:particle_consv_2}
\end{align}
Here, all quantities in the integrand are evaluated on the Fermi surface at zero temperature. A common simplification often utilized is to assume that the scattering time is independent of momentum $\mathbf{k}$, denoted as $\tau^\chi(\theta) = \tau^\chi$~\cite{kim2014boltzmann,lundgren2014thermoelectric,cortijo2016linear,sharma2016nernst,zyuzin2017magnetotransport,das2019berry,kundu2020magnetotransport}. However, it can be readily observed from Eq.~\ref{Eq:particle_consv_2} that when $\tau^\chi(\theta)$ is independent of $\theta$, the left-hand side of the equation does not reduce to zero. Therefore, a momentum-independent scattering time is inherently incompatible with particle number conservation. It is important to go beyond the constant relaxation-time approximation, which is reviewed next. 

\subsection{Beyond constant relaxation-time approximation}
\label{sec:beyond rta}
To correctly study magnetotransport, going beyond the constant relaxation-time approximation is inevitable~\cite{knoll2020negative,sharma2020sign,ahmad2021longitudinal,ahmad2023longitudinal,sharma2023decoupling}. We therefore write the collision integral as: 
\begin{align}
 I_{coll}[f^{\chi}_{\mathbf{k}}]=\sum_{\chi' \mathbf{k}'}{{W}^{\chi \chi'}_{\mathbf{k k'}}}{(f^{\chi'}_{\mathbf{k'}}-f^{\chi}_{\mathbf{k}})},
 \label{Collision_integral_ch1}
\end{align}
where the scattering rate ${{W}^{\chi \chi'}_{\mathbf{k k'}}}$ is evaluated in the first Born approximation (Fermi's golden rule) as:
\begin{align}
{\mathbf{W}^{\chi \chi'}_{\mathbf{k k'}}} = \frac{2\pi n}{\mathcal{V}}|\bra{u^{\chi'}(\mathbf{k'})}U^{\chi \chi'}_{\mathbf{k k'}}\ket{u^{\chi}(\mathbf{k})}|^2\delta(\epsilon^{\chi'}(\mathbf{k'})-\epsilon_F).
\label{Fermi_gilden_rule_ch1}
\end{align}
In the above expression $n$ is the impurity concentration, $\mathcal{V}$ is the system volume, $\ket{u^{\chi}(\mathbf{k})}$ is the Weyl spinor wave function obtained from diagonalization of the Hamiltonian in Eq.~\ref{Eq_H_weyl}, $U^{\chi \chi'}_{\mathbf{k k'}}$ is the scattering potential, and $\epsilon_F$ is the Fermi energy. We choose $U^{\chi \chi'}_{\mathbf{k k'}}$ to model non-magnetic point-like impurities. So, in general $U^{\chi \chi'}_{\mathbf{k k'}} = U^{\chi \chi'} \sigma_{0}$, where $\sigma_0$ is an identity matrix, and the parameter $U^{\chi \chi'}$ can distinguish the intervalley and intravalley scattering. We do not discuss magnetic impurities here but a recent work of ours discusses that as well~\cite{varma2024magnetotransport}.

Using Eq.~\ref{Couplled_equation_ch1},~\ref{Collision_integral_ch1} and keeping terms up to linear order in the applied fields, the Boltzmann transport equation can be written as:
\begin{align}
&\left[\left(\frac{\partial f_0^\chi}{\partial \epsilon^\chi_\mathbf{k}}\right) \mathbf{E}\cdot \left(\mathbf{v}^\chi_\mathbf{k} + \frac{e\mathbf{B}}{\hbar} (\boldsymbol{\Omega}^\chi\cdot \mathbf{v}^\chi_\mathbf{k}) \right)\right] = -\frac{1}{e \mathcal{D}^\chi}\sum\limits_{\chi'}\sum\limits_{\mathbf{k}'} W^{\chi\chi'}_{\mathbf{k}\mathbf{k}'} (g^\chi_{\mathbf{k}'} - g^\chi_\mathbf{k}).
 \label{Eq_boltz2_ch1}
\end{align}
This is a vector equation that can be simplified by fixing the electric field along a particular direction. We fix the electric field along increasing $x$-direction and the magnetic field can be rotated in $xz$-plane. Therefore, $\mathbf{E} = E(0,0,1)$ and  $\mathbf{B} = B (\cos{\gamma},0,\sin{\gamma})$, i.e., for $\gamma=\pi/2$ both the fields are parallel. By tuning the $\gamma$ we can control the component of the magnetic field along the electric field, i.e. $\mathbf{E}\cdot\mathbf{B}$ term in the Eq.~\ref{Couplled_equation_ch1} which is responsible for {CA}. This geometry is useful for studying the different electromagnetic responses, especially the {PHE} and {LMC} (we will return to this later). In this geometry, only the $z$-component of $\boldsymbol{\Lambda}^{\chi}_{\mathbf{k}}$  is relevant, so Eq.~\ref{Eq_boltz2_ch1} reduces to
\begin{align}
\mathcal{D}^{\chi}\left[{v^{\chi,z}_{\mathbf{k}}}+\frac{eB\sin{\gamma}}{\hbar}(\mathbf{v^{\chi}_k}\cdot\boldsymbol{\Omega}^{\chi}_k)\right] =\sum_{\chi' \mathbf{k}'}{{W}^{\chi \chi'}_{\mathbf{k k'}}}{(\Lambda^{\chi'}_{\mathbf{k'}}-\Lambda^{\chi}_{\mathbf{k}})}.
\label{boltzman_in_terms_lambda}  
\end{align}
We define a valley scattering rate:
\begin{align}
\frac{1}{\tau^{\chi}_{\mathbf{k}}(\theta,\phi)}=\sum_{\chi'}\mathcal{V}\int\frac{d^3\mathbf{k'}}{(2\pi)^3}(\mathcal{D}^{\chi'}_{\mathbf{k}'})^{-1}{W}^{\chi \chi'}_{\mathbf{k k'}}.
\label{Tau_invers_ch1}
\end{align}
On the right-hand side of the above equation, there is a sum over the chirality index $\chi$ which can run over multiple flavors. For example, time reversal symmetry broken WSM has a minimum of two Weyl cones, so the summation is over two nodes. But, inversion symmetry broken WSM has a minimum of four Weyl cones so it runs for all four nodes.  The overlap of the Bloch wave-functions $\bra{u^{\chi'}(\mathbf{k'})}U^{\chi \chi'}_{\mathbf{k k'}}\ket{u^{\chi}(\mathbf{k})}|^2$ in Eq.~\ref{Fermi_gilden_rule_ch1} is given by the following expression,
$\mathcal{G}^{\chi\chi'}(\theta,\phi) = [1+\chi\chi'(\cos{\theta}\cos{\theta'} + \sin{\theta}\sin{\theta'}\cos{\phi}\cos{\phi'} + \sin{\theta}\sin{\theta'}\sin{\phi}\sin{\phi'}]$. The overlap of the wave function includes the transition probabilities between fermions of the same chiralities as well as different chiralities. Taking Berry phase into account and the corresponding change in the density of states, $\sum_{k}\longrightarrow \mathcal{V}\int\frac{d^3\mathbf{k}}{(2\pi)^3}\mathcal{D}^\chi(k)$, Eq.~\ref{boltzman_in_terms_lambda} becomes:
\begin{align}
h^{\chi}_{\mu}(\theta,\phi) + &\frac{\Lambda^{\chi}_{\mu}(\theta,\phi)}{\tau^{\chi}_{\mu}(\theta,\phi)}=\sum_{\chi'}\mathcal{V}\int\frac{d^3\mathbf{k}'}{(2\pi)^3} \mathcal{D}^{\chi'}(k'){W}^{\chi \chi'}_{\mathbf{k k'}}\Lambda^{\chi'}_{\mu}(\theta',\phi')
\label{MB_in_term_Wkk'_ch1}
\end{align}
Here, the explicit form of the $h^{\chi}_{\mu}$ is $h^{\chi}_{\mu}(\theta,\phi)=\mathcal{D}^{\chi}[v^{\chi}_{z,\mathbf{k}}+eB\sin{\gamma}(\boldsymbol{\Omega}^{\chi}_{k}\cdot \mathbf{v}^{\chi}_{\mathbf{k}})]$ and is independent of the nature of the impurity sites. The momentum integral in Eq.~\ref{MB_in_term_Wkk'_ch1} has to be evaluated at the Fermi surface, and for that one has to change the momentum integration into the energy integration. 
In the zero temperature limit, for a constant Fermi energy surface, the Eq.~\ref{Tau_invers_ch1} and RHS of Eq.~\ref{MB_in_term_Wkk'_ch1} is reduced to the integration over $\theta'$ and $\phi'$:
\begin{align}
\frac{1}{\tau^{\chi}_{\mu,i}(\theta,\phi)} =  \mathcal{V}\sum_{\chi'} \Pi^{\chi\chi'}\iint\frac{(k')^3\sin{\theta'}}{|\mathbf{v}^{\chi'}_{k'}\cdot{\mathbf{k'}^{\chi'}}|}d\theta'd\phi' \mathcal{G}^{\chi\chi'}(\mathcal{D}^{\chi'}_{\mathbf{k'}})^{-1}
\label{Tau_inv_int_thet_phi_ch1}
\end{align}
The parameter $\tau$ in this expression is independent of radial distance but is dependent on the angles $\theta$,$\phi$, and the nature of the impurity sites.  By using this expression for $\tau$, and Eq.~\ref{Fermi_gilden_rule_ch1} the right-hand side of the Eq.~\ref{MB_in_term_Wkk'_ch1} takes the form:
\begin{multline}
[d^{\chi}+a^{\chi}\cos{\phi}+b^{\chi}\sin{\theta}\cos{\phi}+c^{\chi} \sin{\theta}\sin{\phi}]\\
=\sum_{\chi'}\mathcal{V}\Pi^{\chi\chi'}\iint f^{\chi'}(\theta',\phi')d\theta'd\phi'\\\times[d^{\chi'}-h^{\chi'}_{k'}+a^{\chi'}\cos{\theta'}+b^{\chi'}\sin{\theta'}\cos{\phi'}+c^{\chi'} \sin{\theta'}\sin{\phi'}],\\
\label{Boltzman_final_ch1}
\end{multline}
where $\Pi^{\chi \chi'} = N|U^{\chi\chi'}|^2 / 4\pi^2 \hbar^2$, $f^{\chi} (\theta,\phi)=\frac{(k)^3}{|\mathbf{v}^\chi_{\mathbf{k}}\cdot \mathbf{k}^{\chi}|} \sin\theta (\mathcal{D}^\eta_{\mathbf{k}})^{-1} \tau^\chi_\mu(\theta,\phi)$ and the exact form of the ansatz is $\Lambda^{\chi}_{\mathbf{k}}=[d^{\chi}-h^{\chi}_{k'} + a^{\chi}\cos{\phi} +b^{\chi}\sin{\theta}\cos{\phi}+c^{\chi}\sin{\theta}\sin{\phi}]\tau^{\chi}_{\mu}(\theta,\phi)$. 
When the aforementioned equation is explicitly put out, it appears as seven simultaneous equations that must be solved for eight variables.  
The particle number conservation provides the final restriction:
\begin{align}
\sum\limits_{\chi}\sum\limits_{\mathbf{k}} g^\chi_\mathbf{k} = 0.
\label{Eq_sumgk_ch1}
\end{align} 
For the eight unknowns ($d^{\pm 1}, a^{\pm 1}, b^{\pm 1}, c^{\pm 1}$), the equations \ref{Boltzman_final_ch1} and \ref{Eq_sumgk_ch1} are simultaneously solved with Eq.~\ref{Tau_inv_int_thet_phi_ch1}. Note that if we have more than two flavors of Weyl fermions, the number of equations and unknowns will increase. For example, in the case of inversion asymmetric WSMs, we need to numerically solve sixteen equations for sixteen coefficients~\cite{ahmad2021longitudinal,sharma2023decoupling}.
\section{Magnetoconductance in metals within the Drude model}
\label{Sec:Magnetoconductance in metals within the Drude model}
The impact of CA on the transport properties of WSMs was in the center of the discussion in the last decades. Initial calculations suggest that the positive LMC of WSMs can be explained using CA, now it is well steblished that the LMC can have either sign. This ambiguity of sign has to be addressed correctly. Before presenting the main part of the thesis, it is worth going through the magnetoconductance of the normal metals. The Drude model describes the electrical conductivity of metals using classical equations of motion \ref{eq:EOM_classical particle}. When a magnetic field $\mathbf{B}$ is applied perpendicular to the current, the charge carriers experience the Lorentz force ($=-e ~\mathbf{v} \times \mathbf{B}$), which modifies their motion and affects conductivity. Following derivation clarify this. \\
The equation of motion for an electron with charge $e$ and mass $m$ in an electric field $\mathbf{E}$ and magnetic field $\mathbf{B}$ is:
\begin{equation}
    m \frac{d\mathbf{v}}{dt} = -e (\mathbf{E} + \mathbf{v} \times \mathbf{B}) - \frac{m}{\tau} \mathbf{v}.
\end{equation}
Here, $\tau$ is the relaxation time, which accounts for scattering processes and $\mathbf{v}$ is the velocity of Bloch wave packet. In the steady-state regime, we set $d\mathbf{v}/dt = 0$, yielding:
\begin{equation}
-e (\mathbf{E} + \mathbf{v} \times \mathbf{B}) - \frac{m}{\tau} \mathbf{v} = 0.
\end{equation}
The above equation is written in matrix form:
\begin{equation}
\mathbf{v} = -\frac{e \tau}{m} (\mathbf{E} + \mathbf{v} \times \mathbf{B}).
\end{equation}
To further advancement, we have to fixed the direction of electric and magnetic field. For a magnetic field in the $z$-direction, $\mathbf{B} = (0,0,B)$ and Electric field along any general direction or $\mathbf{E} = (E_x,E_y,E_z)$, the velocity components satisfy:
\begin{equation}
    \begin{bmatrix}
        1 & \omega_c \tau & 0 \\
        -\omega_c \tau & 1 & 0 \\
        0 & 0 & 1
    \end{bmatrix}
    \begin{bmatrix}
        v_x \\
        v_y \\
        v_z
    \end{bmatrix} = -\frac{e \tau}{m}
    \begin{bmatrix}
        E_x \\
        E_y \\
        E_z
    \end{bmatrix},
\end{equation}
This is a system of equations to be solved for velocity components. Solving for $\mathbf{v}$:
\begin{equation}
    \begin{bmatrix}
        v_x \\
        v_y \\
        v_z
    \end{bmatrix} = -\frac{e \tau}{m}
    \begin{bmatrix}
        \frac{1}{1+(\omega_c\tau)^2} & \frac{\omega_c\tau}{1+(\omega_c\tau)^2} & 0 \\
        -\frac{\omega_c\tau}{1+(\omega_c\tau)^2} & \frac{1}{1+(\omega_c\tau)^2} & 0 \\
        0 & 0 & 1
    \end{bmatrix}
    \begin{bmatrix}
        E_x \\
        E_y \\
        E_z
    \end{bmatrix},
\end{equation}
where the cyclotron frequency is $\omega_c = eB/m$. The current density is given by $\mathbf{J} = -ne\mathbf{v}$, leading to the conductivity tensor:
\begin{equation}
    \begin{bmatrix}
        J_x \\
        J_y \\
        J_z
    \end{bmatrix} = \sigma_0
    \begin{bmatrix}
        \sigma_{xx} & \sigma_{xy} & \sigma_{xz} \\
        \sigma_{yx} & \sigma_{yy} & \sigma_{yz} \\
        \sigma_{zx} & \sigma_{zy} & \sigma_{zz}
    \end{bmatrix}
    \begin{bmatrix}
        E_x \\
        E_y \\
        E_z
    \end{bmatrix},
\end{equation}
where $\sigma_0 = ne^2 \tau / m$ is the Drude conductivity in zero field, and the conductivity tensor elements are defined as:
\begin{align}
    &\sigma_{xx} = \frac{\sigma_0}{1+(\omega_c\tau)^2}, \quad &\sigma_{xy} = \frac{\sigma_0 \omega_c\tau}{1+(\omega_c\tau)^2}, \quad &\sigma_{xz} = 0, \nonumber \\ 
    &\sigma_{yx} = -\frac{\sigma_0 \omega_c\tau}{1+(\omega_c\tau)^2}, \quad &\sigma_{yy} = \frac{\sigma_0}{1+(\omega_c\tau)^2}, \quad &\sigma_{yz} = 0, \nonumber \\ 
    &\sigma_{zx} = 0,  \quad &\sigma_{zy} = 0, \quad &\sigma_{zz} = \sigma_0.
\end{align}
The Drude model predicts a decrease in conductivity with increasing magnetic field due to the term $1 + (\omega_c \tau)^2$ in the denominator. However, the conductivity in the $z$-direction remains unaffected. Therefore, we can conclude that $\delta \sigma_{ij} (B) = \sigma_{ij} (B) - \sigma_{ij} (B=0)\leq 0$ for normal metals (Where, $i,j\in\{ x,y,z\}$) and it is not possible to have positive magnetoconductance for metals.

\setcounter{equation}{0}
\setcounter{table}{0}
\setcounter{figure}{0}

\section{Outline of the thesis}
\label{Outline}
The influence of the chiral anomaly (CA) on the sign of longitudinal magnetoconductance (LMC) and planar Hall conductance (PHC) in Weyl semimetals has long been a topic of debate among researchers. Early studies suggested that CA could lead to a positive LMC. However, it is now established that LMC can take either sign, depending on various factors such as the presence or absence of orbital magnetic moments arising from the self-rotation of the Bloch wave packet, the strength of intervalley scattering, the existence of chiral magnetic fields and tilted Weyl cones, and the nonlinearity of dispersion away from the Weyl node etc. This thesis investigates the sign of LMC and PHC under these different conditions. Additionally, it explores the nonlinear anomalous conductivity response and provides a systematic approach to extending the study of CA to multifold fermions. A chapter-wise summary of the thesis is presented below.

\textbf{Chapter 2}: This chapter explores the electronic transport properties of Weyl fermions using the semiclassical Boltzmann formalism, extending beyond the conventional constant relaxation time approximation. It incorporates the nonlinearity of the energy bands near the Weyl node. The chosen lattice model introduces a physical ultraviolet cutoff to the low-energy spectrum. We conduct a semi-analytical study of longitudinal magnetoconductance and the planar Hall effect in a lattice model of Weyl fermions with a `smooth' lattice cutoff. Here, a ‘smooth’ lattice cutoff refers to a dispersion that gradually shifts from linear to flat near the edges of the Brillouin zone, as opposed to a hard cutoff that simply excludes high-energy contributions. Additionally, we have included the influence of geometric properties such as Berry curvature and the orbital magnetic moment.

\textbf{Chapter 3}: This chapter explores the impact of strain-induced chiral gauge fields on the magnetotransport properties of Weyl semimetals, focusing on longitudinal magnetoconductance (LMC) and planar Hall conductance. Strain acts as an axial magnetic field, influencing impurity-dominated diffusive transport. We analyze the combined effects of strain, external magnetic fields, Weyl cone tilt, and intervalley scattering, incorporating momentum-dependent scattering and charge conservation. Our findings reveal that the chiral gauge field induces a ‘strong sign-reversal’ of LMC, characterized by a flipped magnetoconductance parabola, while an external magnetic field alone leads to strong sign-reversal only under strong intervalley scattering. When both fields are present, we observe both strong and weak sign-reversals, where weak sign-reversal depends on the relative orientations of the fields rather than the LMC parabola. This interplay produces distinct features in the LMC phase diagram as a function of tilt, strain, and intervalley scattering. Additionally, we examine the influence of strain-induced chiral gauge fields on planar Hall conductance, identifying unique characteristics that could be experimentally observed.

\textbf{Chapter 4}: This chapter develops a comprehensive theory of the chiral anomaly-induced nonlinear Hall effect (CNLHE) in three-dimensional chiral quasiparticles, incorporating momentum-dependent chirality-preserving and chirality-breaking scattering while ensuring global charge conservation. We analyze Weyl semimetals (WSMs) and spin-orbit coupled non-centrosymmetric metals (SOC-NCMs), uncovering that in WSMs, nonlinear Hall conductivity exhibits nonmonotonic behavior with tilt and undergoes strong sign-reversal with increasing internode scattering. In SOC-NCMs, the orbital magnetic moment alone drives a large CNLHE, showing distinct quadratic magnetic field dependence. Additionally, we reveal that spin Zeeman coupling mimics an effective tilt, enhancing the nonlinear Hall current. These predictions provide new experimental avenues for studying chiral transport.

\textbf{Chapter 5}: This chapter extends the study of chiral anomaly (CA) to pseudospin-1 fermions, exploring its impact on longitudinal magnetotransport. Using a rigorous quasiclassical approach, we go beyond conventional models by incorporating momentum-dependent relaxation times, the orbital magnetic moment, and global charge conservation. We find that magnetoconductance is positive and quadratic in weak internode scattering but becomes negative beyond a critical threshold, which is lower than in Weyl fermions. Additionally, internode scattering more strongly affects zero-field conductivity in pseudospin-1 systems. These results provide a framework for interpreting experiments on multifold fermions and identifying candidate materials with specific space group symmetries.

\textbf{Chapter 6}: This chapter provides a summary of the key findings of this research and explores potential directions for future studies.
\baselineskip 24pt

\chapter{\label{chap2}Longitudinal magnetoconductance and the planar Hall effect in a lattice model of tilted Weyl fermions}
{\small The contents of this chapter have appeared in ``\textsc{Longitudinal magnetoconductance and the planar Hall effect in a lattice model of tilted Weyl fermions}"; Azaz Ahmad and Gargee Sharma; \textit{Phys. Rev. B} \textbf{103}, 115146 (2021).}
\section{Abstract}
The experimental verification of chiral anomaly in Weyl semimetals is an active area of investigation in modern condensed matter physics, which typically relies on the  combined signatures of longitudinal magnetoconductance (LMC) along with the planar Hall effect (PHE). It has recently been shown that for weak non-quantizing magnetic fields, a sufficiently strong finite intervalley scattering drives the system to switch the sign of LMC from positive to negative. Here we unravel another  independent source that produces the same effect. Specifically, a smooth lattice cutoff to the linear dispersion, which is ubiquitous in real Weyl materials, introduces nonlinearity in the problem and also drives the system to exhibit negative LMC for non-collinear electric and magnetic fields even in the limit of vanishing intervalley scattering. We examine longitudinal magnetoconductivity and the planar Hall effect semi-analytically for a lattice model of tilted Weyl fermions within the Boltzmann approximation. We independently study the effects of a finite lattice cutoff and tilt parameters and construct phase diagrams in relevant parameter spaces that are relevant for diagnosing chiral anomaly in real Weyl materials. 
\section{Introduction}
As dictated by the well-known no-crossing theorem~\cite{von1993verhalten}, the Bloch bands in a solid typically do no cross each other at any point in the Brillouin zone. Some exceptions to this general rule are Dirac and Weyl materials, where non-trivial topology of the Bloch bands can stabilize the band degenerate point~\cite{volovik2003universe,chiu2016classification,armitage2018weyl,yang2018symmetry,murakami2007phase,murakami2007tuning,burkov2011topological,burkov2011weyl,wan2011topological,xu2011chern,yang2011quantum}.  In a Weyl semimetal (WSM), a band crossing point, also known as a Weyl node, can act as a source or sink of Abelian Berry curvature~\cite{xiao2010berry}. Since the net Berry flux through the Brillouin zone must vanish, the Weyl nodes must occur in multiples of two. The topological nature of the Bloch bands in a WSM gives rise to very interesting physics typically that is absent in conventional condensed matter systems. Some examples include the manifestation of anomalous Hall~\cite{yang2011quantum,burkov2014anomalous} and Nernst~\cite{sharma2016nernst,sharma2017nernst,liang2017anomalous} effects, open Fermi arcs~\cite{wan2011topological}, and the most prominent one being the manifestation of chiral or Adler-Bell-Jackiw anomaly~\cite{adler1969axial,nielsen1981no,nielsen1983adler,bell1969pcac,aji2012adler,zyuzin2012weyl,zyuzin2012weyl,son2012berry,goswami2013axionic,goswami2015optical, fukushima2008chiral}.

Weyl fermions have an associated chirality quantum number that is identical with the integral of the flux of the Berry curvature around a Weyl node. The number of Weyl fermions of a specific chirality remain conserved in the absence of an external gauge or gravitational field coupling. However, in the presence of background gauge fields, such as electric and magnetic fields, the separate number conservation laws for Weyl fermions is violated~\cite{adler1969axial,nielsen1981no,nielsen1983adler}. This is the result of chiral anomaly in Weyl fermions and has its origins rooted in high-energy physics. {The verification of chiral anomaly in  Weyl semimetals is an important area of investigation in condensed matter physics.}

Chiral anomaly in WSMs maybe verified by experimental probes such as that measure magnetoconductance~\cite{son2013chiral,kim2014boltzmann,zyuzin2017magnetotransport,he2014quantum,liang2015ultrahigh,zhang2016signatures,li2016chiral,xiong2015evidence,hirschberger2016chiral}, Hall effect~\cite{nandy2017chiral,kumar2018planar,yang2019w,li2018giant,chen2018planar,li2018giant2,yang2019frustration,pavlosiuk2019negative,singha2018planar}, thermopower~\cite{lundgren2014thermoelectric,sharma2016nernst,sharma2019transverse,das2019berry}, optical processes~\cite{goswami2015optical, levy2020optical, parent2020magneto, levy2020optical}, non-local transport~\cite{parameswaran2014probing}, optical phonons~\cite{song2016detecting,rinkel2017signatures,yuan2020discovery,cheng2019probing}. It was initially concluded that chiral anomaly in WSMs directly correlates with the observation of positive longitudinal magnetoconductance. For example, from elementary field-theory calculations~\cite{fukushima2008chiral}, the chiral chemical potential ($\mu_5$, which is difference between the chemical potential between Weyl nodes of two chiralities) created by the external parallel $\mathbf{E}$ and $\mathbf{B}$ fields in the presence of intervalley scattering is $\mu_5 = 3v_F^3 e^2 \tau_i {E}{B}/4\hbar^2 \mu^2$, where $v_F$, $\tau_i$, and $\mu$ denote the Fermi velocity, scattering time, and the chemical potential, respectively. The corresponding longitudinal current is given by ${j} = e^2 \mu_5 {B}/ 2\pi^2$, which immediately gives us positive longitudinal magnetoconductance. However, a detailed analysis shows that positive longitudinal
magnetoconductance is neither a necessary, nor a sufficient condition to prove the existence of chiral anomaly in WSMs. It has now been well established that both positive or negative magnetoconductance can arise from chiral anomaly in WSMs~\cite{goswami2015axial,lu2015high,chen2016positive,zhang2016linear,shao2019magneto,li2016weyl,ji2018effect,spivak2016magnetotransport,das2019linear,imran2018berry,dantas2018magnetotransport,johansson2019chiral,grushin2016inhomogeneous,cortijo2016linear,sharma2017chiral,knoll2020negative,xiao2020linear,sharma2020sign}. In the presence of strong magnetic field, when Landau quantization is relevant, the sign of magnetoconductance depends on the nature of scattering impurities~\cite{goswami2015axial,lu2015high,chen2016positive,zhang2016linear,shao2019magneto,li2016weyl,ji2018effect}. For weak magnetic fields, it was recently shown that sufficiently strong intervalley scattering can switch the sign of LMC~\cite{knoll2020negative,xiao2020linear}.

In this work we unravel another independent source that produces negative LMC for weak non-collinear electric and magnetic fields even for vanishing intervalley scattering strength. 
Around a Weyl node, the energy dispersion locally behaves as $\epsilon^\chi_\mathbf{k} = \hbar v_F k$, where $v_F$ is the Fermi velocity, while $k$ is the modulus of the wavevector measured from the nodal point. In practice, the linear energy dispersion around a Weyl node is only valid for a small energy window. In a realistic lattice model of Weyl fermions, the bands are no longer linear far apart from the nodal point, and the lattice regularization provides a physical ultraviolet cutoff to the low-energy spectrum. The lattice model of Weyl fermions introduces a source of non-linearity in the problem and has important implications in several physical properties. For example, the lattice model of Weyl fermions produces a non-zero Nernst effect~\cite{sharma2016nernst,sharma2017nernst} (as also observed experimentally~\cite{liang2017anomalous}), which is otherwise predicted to vanish in the linear approximation~\cite{lundgren2014thermoelectric}. Here, we semi-analytically examine longitudinal magnetoconductance and the planar Hall effect for a lattice model of Weyl fermions { that has a smooth lattice cutoff. By a `smooth' lattice cutoff we mean that the dispersion gradually transitions from being linear to becoming flat at the corners of the Brillouin zone. This is in contrast to imposing a hard cutoff to the linear spectrum by discarding the high energy contributions. {The lattice model we adopt here is also advantageous  over other continuum non-linear models} because (i) there is no need to impose a hard cutoff at higher energies, as the bands flatten out naturally at higher energies, (ii) includes non-linearities up to all orders, and (iii) the expressions for Berry curvature and orbital magnetic moment in the current model offer better analytical tractability than some other non-linear models.
It is also worthwhile to point out that the lattice model we adopt has exact analytical expressions for the Berry curvature, orbital magnetic moment, and band-velocities at all energies. This is in contrast to earlier works on a lattice model of Weyl semimetals mostly resort to numerical evaluation of various intrinsic quantities such as the Berry curvature and the orbital magnetic moment, as well as transport quantities such as longitudinal conductance or the Hall conductance~\cite{sharma2016nernst,sharma2017nernst,sharma2017chiral,nandy2017chiral,goswami2013axionic}. Therefore, in this work the associated transport quantities are also evaluated semi-analytically within the Boltzmann formalism. Further, it is not straightforward to incorporate internode scattering in lattice models of a WSM because the energy dispersion valid throughout the first Brillouin zone does not `see' any distinction between nodes. In contrast, here we  consider lattice model of an individual Weyl nodes and thus it is straightforward to incorporate intervalley scattering akin to the case of two Weyl nodes with linearized dispersion.}

We find that nonlinear lattice effects can produce negative LMC for non-collinear electric and magnetic fields even in the absence of intervalley scattering. Crucially, we note that it is important to account for orbital magnetic moment effects to obtain negative LMC.
We also find that in the presence of finite intervalley scattering, lattice effects drive the system to exhibit negative longitudinal magnetoconductance quickly at a lesser threshold of intervalley scattering as compared to the linearized approximation. 

\begin{figure}
	\includegraphics[width=1.0\columnwidth]{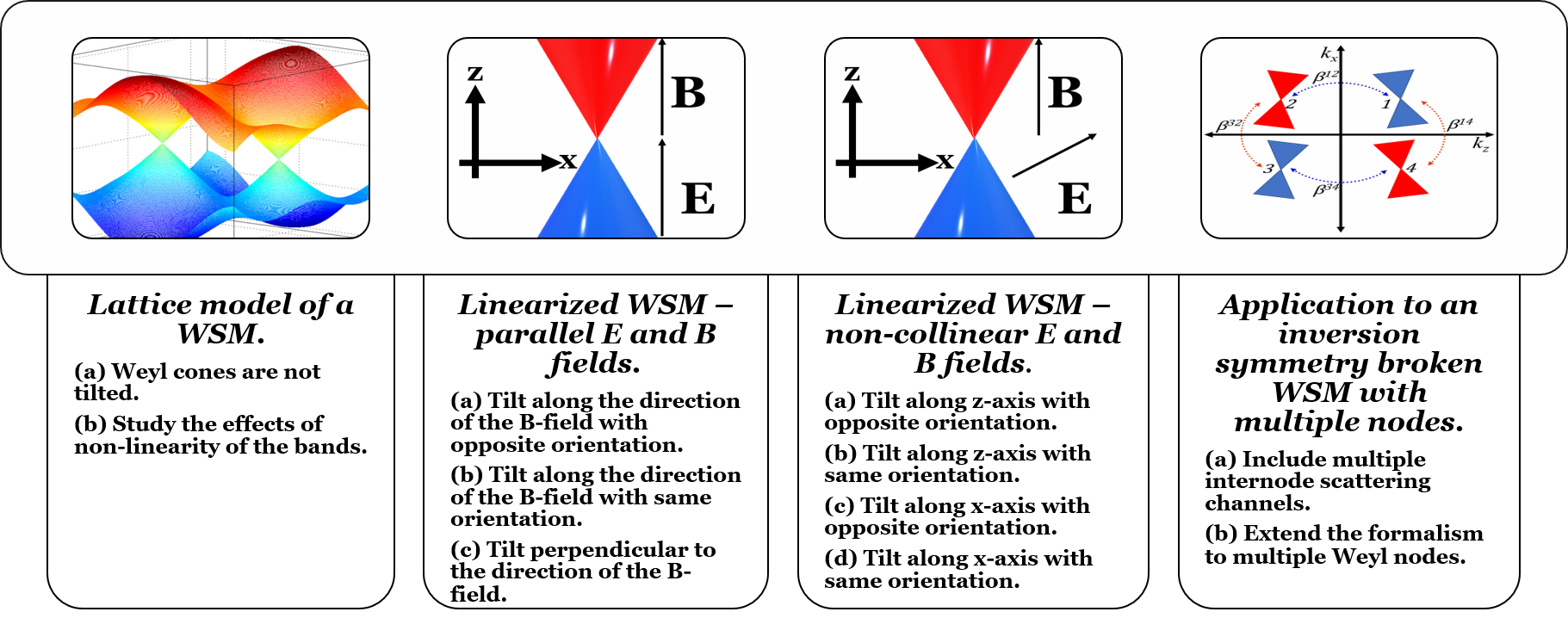}
	\caption{Schematic of the presentation of results in Section\ref{sec:ch2_results}.}
	\label{Fig_1}
\end{figure}

Further, in realistic materials the Weyl cones not only have a smooth lattice cutoff but are also in generally tilted along a particular direction~\cite{soluyanov2015type, sharma2017chiral, rostamzadeh2019large}. We also examine longitudinal magnetoconductance $\sigma_{zz}$ and the planar Hall conductance $\sigma_{xz}$ in the presence of a tilt parameter both parallel and perpendicular to the $z-$direction. When the electric and magnetic fields are aligned parallel to each other, and when the Weyl cones are tilted along the  direction of the magnetic field, LMC is  quadratic if the cones are oriented in the same direction, and the sign of LMC depends on the strength of intervalley scattering ($\alpha_i$). When the cones are tilted opposite to each other, LMC is found to be linear-in-$B$ with sign depending on the magnitude of the tilt as well as $\alpha_i$. When the cones are tilted perpendicular to the direction of the magnetic field, LMC is found to be quadratic, with the sign again depending on the value of intervalley scattering strength $\alpha_i$. However, more interesting features emerge when LMC is examined for non-collinear electric and magnetic fields, as demonstrated by several phase plots in the $\alpha_i-t_k$ space ($t_k$ being the tilt parameter). We also find that the planar Hall conductance also shows linear-in-$B$ behaviour for tilted Weyl cones oriented opposite to each other, and this linear-in-$B$ behavior is enhanced in the presence of intervalley scattering $\alpha_i$. { Lastly, we also discuss the applicability of our results to a scenario much relevant to actual Weyl materials, i.e., the case of a inversion symmetry broken Weyl semimetal by extending the Boltzmann formalism to tackle multiple nodes simultaneously. Interestingly, we find that despite the presence of internode scattering between nodes of opposite tilt orientation, the linear-in-$B$ LMC coefficient vanishes for our model. We find that the interplay of various internode scattering channels along with the magnitude of tilt parameter governs the sign of LMC.

This paper is organized as follows: In Section-\ref{Sec:Maxwell-Boltzmann transport theory}, we discuss the Boltzmann formalism for magnetotransport for a system of lattice Weyl nodes, that may also be tilted along a particular axis.  Section-\ref{sec:ch2_results} consists of our main results that is divides into four subsections as highlighted in Fig.~\ref{Fig_1}. Finally we conclude in Section \ref{sec:ch2_Discussions and Conclusions}. The technical details are relegated to the Appendices (PS: App.~\ref{appendix_ch2}).
}
\section{Boltzmann formalism for magnetotransport}
\label{ch2_Boltzmann formalism for magnetotransport}
We begin with the most general form of a tilted type-I Weyl node of a particular chirality $\chi$, including non-linear effects away from the Weyl node due to lattice regularization. The Hamiltonian expanded around each Weyl point can be expressed as 
\begin{align}
H_\mathbf{k} = \chi E_0 p(a \mathbf{k}\cdot \boldsymbol{\sigma}) + T^\chi_x q(a k_x) + T^\chi_z r(a k_z).
\label{ch2_Eq_H1weyl}
\end{align}
In the above expression, $E_0$ is an energy parameter, $T^\chi_x$ and $T^\chi_z$ are tilt parameters along the $x$ and $z$ directions, respectively, $\mathbf{k}$ is the momentum measured relative to the Weyl point, $\boldsymbol{\sigma}$ is the vector of the Pauli matrices. The functions, $p$, $q$, and $r$ are can assume any form as long as $p(0)=q(0)=r(0)=0$, but we choose $p(x) = q(x) = r(x) = \sin(x)$ as prototype of a lattice Weyl node. The corresponding energy dispersion is given by 
\begin{align}
\epsilon^\chi_k=\pm E_0 \sin(ka) +T^{\chi}_z \sin(a k_z) +T^{\chi}_x \sin(a k_x).
\label{ch2_Eq_E1kweyl}
\end{align}
Note that for a Weyl node without any tilt the energy bandwidth equals $2E_0$. 

We  study charge transport for weak electric and magnetic fields via the quasiclassical Boltzmann theory and thus the Landau quantization regime will not be relevant for our discussion. A phenomenological Boltzmann equation for the non-equilibrium distribution function $f^\chi_\mathbf{k}$ can be written as~\cite{bruus2004many} 
\begin{align}
\left(\frac{\partial}{\partial t} + \dot{\mathbf{r}}^\chi\cdot \nabla_\mathbf{r}+\dot{\mathbf{k}}^\chi\cdot \nabla_\mathbf{k}\right)f^\chi_\mathbf{k} = \mathcal{I}_{{col}}[f^\chi_\mathbf{k}],
\label{ch2_Eq_boltz1}
\end{align}
where the collision term on the right-hand side incorporates the effect of impurity scattering.
In the presence of electric ($\mathbf{E}$) and magnetic ($\mathbf{B}$) fields, the dynamics of the Bloch electrons is modified as~\cite{son2012berry} 
\begin{align}
\dot{\mathbf{r}}^\chi &= \mathcal{D}^\chi \left( \frac{e}{\hbar}(\mathbf{E}\times \boldsymbol{\Omega}^\chi + \frac{e}{\hbar}(\mathbf{v}^\chi\cdot \boldsymbol{\Omega}^\chi) \mathbf{B} + \mathbf{v}_\mathbf{k}^\chi)\right) \nonumber\\
\dot{\mathbf{p}}^\chi &= -e \mathcal{D}^\chi \left( \mathbf{E} + \mathbf{v}_\mathbf{k}^\chi \times \mathbf{B} + \frac{e}{\hbar} (\mathbf{E}\cdot\mathbf{B}) \boldsymbol{\Omega}^\chi \right),
\label{ch2_coupled_eqn}
\end{align}
where $\mathbf{v}_\mathbf{k}^\chi$ is the band velocity, $\boldsymbol{\Omega}^\chi = -\chi \mathbf{k} /2k^3$ is the Berry curvature, and $\mathcal{D}^\chi = (1+e\mathbf{B}\cdot\boldsymbol{\Omega}^\chi/\hbar)^{-1}$ is the factor by which the phase space volume is modified due to Berry phase effects. The self-rotation of Bloch wavepacket also gives rise to an orbital magnetic moment (OMM)~\cite{xiao2010berry} that is given by $\mathbf{m}^\chi_\mathbf{k} = -e \chi E_0 \sin(ak) \mathbf{k} /2\hbar k^3$ for the above lattice model (see Appendix-\ref{appendix_ch2} for details). In the presence of magnetic field, the OMM shifts the energy dispersion as $\epsilon^{\chi}_{\mathbf{k}}\rightarrow \epsilon^{\chi}_{\mathbf{k}} - \mathbf{m}^\chi_\mathbf{k}\cdot \mathbf{B}$. Note that the Berry curvature and the orbital magnetic moment are independent of the tilting of the Weyl cones.

The collision integral must take into account scattering between the two Weyl cones (internode, $\chi\Longleftrightarrow\chi'$), as well as scattering withing a Weyl cone (intranode, $\chi\Longleftrightarrow\chi$), and thus $\mathcal{I}_{{col}}[f^\chi_\mathbf{k}]$ can be expressed as 
\begin{align}
\mathcal{I}_{{col}}[f^\chi_\mathbf{k}] = \sum\limits_{\chi'}\sum\limits_{\mathbf{k}'} W^{\chi\chi'}_{\mathbf{k},\mathbf{k}'} (f^{\chi'}_{\mathbf{k}'} - f^\chi_\mathbf{k}),
\label{ch2_I_coll1}
\end{align}
where the scattering rate $W^{\chi\chi'}_{\mathbf{k},\mathbf{k}'}$ in the first Born approximation is given by~\cite{bruus2004many} 
\begin{align}
W^{\chi\chi'}_{\mathbf{k},\mathbf{k}'} = \frac{2\pi}{\hbar} \frac{n}{\mathcal{V}} |\langle \psi^{\chi'}_{\mathbf{k}'}|U^{\chi\chi'}_{\mathbf{k}\mathbf{k}'}|\psi^\chi_\mathbf{k}\rangle|^2 \delta(\epsilon^{\chi'}_{\mathbf{k}'}-\epsilon_F)
\label{ch2_Eq_W_1}
\end{align}
In the above expression $n$ is the impurity concentration, $\mathcal{V}$ is the system volume, $|\psi^\chi_\mathbf{k}\rangle$ is the Weyl spinor wavefunction (obtained by diagonalizing Eq.~\ref{ch2_Eq_H1weyl}), $U^{\chi\chi'}_{\mathbf{k}\mathbf{k}'}$ is the scattering potential profile, and $\epsilon_F$ is the Fermi energy. The scattering potential profile $U^{\chi\chi'}_{\mathbf{k}\mathbf{k}'}$ is determined by the nature of impurities (whether charged or uncharged or magnetic). Here we restrict our attention only to non-magnetic point-like scatterers, but particularly distinguish between intervalley and intravalley scattering that can be controlled independently in our formalism. Thus, the scattering matrix is momentum-independent but has a chirality dependence, i.e.,  $U^{\chi\chi'}_{\mathbf{k}\mathbf{k}'} = U^{\chi\chi'}\mathbb{I}$.

The distribution function is assumed to take the form $f^\chi_\mathbf{k} = f_0^\chi + g^\chi_\mathbf{k}$, where $f_0^\chi$ is the equilibrium Fermi-Dirac distribution function and $g^\chi_\mathbf{k}$ indicates the deviation from equilibrium. 
In the steady state, the Boltzmann equation (Eq.~\ref{ch2_Eq_boltz1}) takes the following form 
\begin{align}
\left[\left(\frac{\partial f_0^\chi}{\partial \epsilon^\chi_\mathbf{k}}\right) \mathbf{E}\cdot \left(\mathbf{v}^\chi_\mathbf{k} + \frac{e\mathbf{B}}{\hbar} (\boldsymbol{\Omega}^\chi\cdot \mathbf{v}^\chi_\mathbf{k}) \right)\right]
 = -\frac{1}{e \mathcal{D}^\chi}\sum\limits_{\chi'}\sum\limits_{\mathbf{k}'} W^{\chi\chi'}_{\mathbf{k}\mathbf{k}'} (g^\chi_{\mathbf{k}'} - g^\chi_\mathbf{k})
 \label{ch2_Eq_boltz2}
\end{align}
The deviation $g^\chi_\mathbf{k}$ is assumed to be small such that its gradient can be neglected and is also assumed to be proportional to the applied electric field 
\begin{align}
g^\chi_\mathbf{k} = e \left(-\frac{\partial f_0^\chi}{\partial \epsilon^\chi_\mathbf{k}}\right) \mathbf{E}\cdot \boldsymbol{\Lambda}^\chi_\mathbf{k}
\label{ch2_g1}
\end{align}
We will fix the direction of the applied external electric field to be along $+\hat{z}$, i.e., $\mathbf{E} = E\hat{z}$. Therefore only ${\Lambda}^{\chi z}_\mathbf{k}\equiv {\Lambda}^{\chi}_\mathbf{k}$, is relevant. Further, we rotate the magnetic field along the $xz$-plane such that it makes an angle $\gamma$ with respect to the $\hat{x}-$axis, i.e., $\mathbf{B} = B(\cos\gamma,0,\sin\gamma)$. When $\gamma=\pi/2$, the electric and magnetic fields are parallel to each other. When $\gamma\neq \pi/2$, the electric and magnetic fields are non-collinear and this geometry will be useful in analyzing the planar Hall effect, as well as LMC in a non-collinear geometry that has non-trivial implications in a lattice model as well as for tilted Weyl fermions even in the linear approximation.

Keeping terms only up to linear order in the electric field, Eq.~\ref{ch2_Eq_boltz2} takes the following form 
\begin{align}
\mathcal{D}^\chi \left[v^{\chi z}_{\mathbf{k}} + \frac{e B}{\hbar} \sin \gamma (\boldsymbol{\Omega}^\chi\cdot \mathbf{v}^\chi_\mathbf{k})\right] = \sum\limits_{\eta}\sum\limits_{\mathbf{k}'} W^{\eta\chi}_{\mathbf{k}\mathbf{k}'} (\Lambda^{\eta}_{\mathbf{k}'} - \Lambda^\chi_\mathbf{k})
\label{ch2_Eq_boltz3}
\end{align} 
In order to solve the above equation, we first define the valley scattering rate as follows
\begin{align}
\frac{1}{\tau^\chi_\mathbf{k}} = \mathcal{V} \sum\limits_{\eta} \int{\frac{d^3 \mathbf{k}'}{(2\pi)^3} (\mathcal{D}^\eta_{\mathbf{k}'})^{-1} W^{\eta\chi}_{\mathbf{k}\mathbf{k}'}}
\label{ch2_Eq_tau11}
\end{align}
One would assume that when $\gamma=\pi/2$, due to the electric and magnetic field both being parallel to the $\hat{z}$ axis the azimuthal symmetry is retained in the problem. However, due to the tilting of the Weyl cones the azimuthal symmetry is destroyed even for parallel electric and magnetic fields, and therefore the above integration (and all other subsequent integrations) must be performed both over $\theta$ and $\phi$ when either (i) the Weyl cones are tilted and/or (ii) $\gamma\neq\pi/2$. Note that finite lattice effects by themselves do not break azimuthal symmetry.  The radial integration is simplified due to the delta-function in Eq.~\ref{ch2_Eq_W_1}.

Substituting the scattering rate from Eq.~\ref{ch2_Eq_W_1} in the above equation, we have 
\begin{align}
\frac{1}{\tau^\chi_\mathbf{k}} = \frac{\mathcal{V}N}{8\pi^2 \hbar} \sum\limits_{\eta} |U^{\chi\eta}|^2 \iiint{(k')^2 \sin \theta' \mathcal{G}^{\chi\eta}(\theta,\phi,\theta',\phi') \delta(\epsilon^{\eta}_{\mathbf{k}'}-\epsilon_F)(\mathcal{D}^\eta_{\mathbf{k}'})^{-1}dk'd\theta'd\phi'},
\label{ch2_Eq_tau1}
\end{align}
where $N$ now indicates the total number of impurities, and $ \mathcal{G}^{\chi\eta}(\theta,\phi,\theta',\phi') = (1+\chi\eta(\cos\theta \cos\theta' + \sin\theta\sin\theta' \cos(\phi-\phi')))$ is the Weyl chirality factor defined by the overlap of the wavefunctions. Since quasiclassical Boltzmann theory is valid away from the nodal point such that $\mu^2\gg \hbar v_F^2 e B$, therefore without any loss of generality we will assume that the chemical potential lies in the conduction band. 

Including orbital magnetic moment effects, the energy dispersion $\epsilon^\chi_\mathbf{k}$ is in general a function of several parameters including the chirality index, i.e., $\epsilon^\chi_\mathbf{k}= \epsilon^\chi_\mathbf{k}(E_0,k,a,\chi,B,\theta,\gamma)$. This equation has to be inverted in order to find a constant energy contour $k^\chi = k^\chi(E_0,\epsilon^\chi_\mathbf{k},\\a,B,\theta,\gamma)$. For the case of lattice Weyl fermions, a closed-form analytical solution is not feasible and we will resolve to a numerical solution for $k^\chi$. For tilted Weyl fermions in the linearized spectrum approximation, it is possible to invert the equation as will be shown shortly. 

The three-dimensional integral in Eq.~\ref{ch2_Eq_tau1} is then reduced to just integration in $\phi'$ and $\theta'$. The scattering time ${\tau^\chi_\mathbf{k}}$ depends on the chemical potential ($\mu$), and is a function of the angular variables $\theta$ and $\phi$. 
\begin{align}
\frac{1}{\tau^\chi_\mu(\theta,\phi)} = \mathcal{V} \sum\limits_{\eta} \iint{\frac{\beta^{\chi\eta}(k')^3}{|\mathbf{v}^\eta_{\mathbf{k}'}\cdot \mathbf{k}'^\eta|}\sin\theta'\mathcal{G}^{\chi\eta}(\mathcal{D}^\eta_{\mathbf{k}'})^{-1} d\theta'd\phi'},
\label{ch2_Eq_tau2}
\end{align}
where the prefactor $\beta^{\chi\eta} = N|U^{\chi\eta}|^2 / 4\pi^2 \hbar^2$. The Boltzmann equation (Eq~\ref{ch2_Eq_boltz3}) assumes the form,
\begin{align}
&h^\chi_\mu(\theta,\phi) + \frac{\Lambda^\chi_\mu(\theta,\phi)}{\tau^\chi_\mu(\theta,\phi)} =\mathcal{V}\sum_\eta \iint {\frac{\beta^{\chi\eta}(k')^3}{|\mathbf{v}^\eta_{\mathbf{k}'}\cdot \mathbf{k}'^\eta|} \sin\theta'\mathcal{G}^{\chi\eta}(\mathcal{D}^\eta_{\mathbf{k}'})^{-1}\Lambda^\eta_{\mu}(\theta',\phi') d\theta'd\phi'}.
\label{ch2_Eq_boltz4}
\end{align}
We make the following ansatz for $\Lambda^\chi_\mu(\theta,\phi)$
\begin{align}
\Lambda^\chi_\mu(\theta,\phi)= (\lambda^\chi - h^\chi_\mu(\theta,\phi) + a^\chi \cos\theta +b^\chi \sin\theta\cos\phi + c^\chi \sin\theta\sin\phi)\tau^\chi_\mu(\theta,\phi),
\label{ch2_Eq_Lambda_1}
\end{align}
where we solve for the eight unknowns ($\lambda^{\pm 1}, a^{\pm 1}, b^{\pm 1}, c^{\pm 1}$). The L.H.S in Eq.~\ref{ch2_Eq_boltz4} simplifies to $\lambda^\chi + a^\chi \cos\theta + b^\chi \sin\theta\cos\phi + c^\chi \sin\theta\sin\phi$. The R.H.S of Eq.~\ref{ch2_Eq_boltz4} simplifies to
\begin{align}
\mathcal{V}\sum_\eta \beta^{\chi\eta} \iint f^{\eta} (\theta',\phi') \mathcal{G}^{\chi\eta} (\lambda^\eta - h^\eta_\mu(\theta',\phi') &+ a^\eta \cos\theta' + b^\eta \sin\theta'\cos\phi' \nonumber\\  &+ c^\eta \sin\theta'\sin\phi')d\theta'd\phi',
\label{ch2_Eq_boltz5_rhs}
\end{align}
where the function
\begin{align}
f^{\eta} (\theta',\phi') = \frac{(k')^3}{|\mathbf{v}^\eta_{\mathbf{k}'}\cdot \mathbf{k}'^\eta|} \sin\theta' (\mathcal{D}^\eta_{\mathbf{k}'})^{-1} \tau^\chi_\mu(\theta',\phi')
\label{ch2_Eq_f_eta}
\end{align}
The above equations, when written down explicitly take the form of seven simultaneous equations to be solved for eight variables (see Appendix-\ref{appendix_ch2} for details). The last constraint comes from the particle number conservation 
\begin{align}
\sum\limits_{\chi}\sum\limits_{\mathbf{k}} g^\chi_\mathbf{k} = 0
\label{ch2_Eq_sumgk}
\end{align}
Thus Eq.~\ref{ch2_Eq_Lambda_1}, Eq.~\ref{ch2_Eq_boltz5_rhs}, Eq.~\ref{ch2_Eq_f_eta} and Eq.~\ref{ch2_Eq_sumgk} can be solved together with Eq~\ref{ch2_Eq_tau2}, simultaneously for the eight unknowns ($\lambda^{\pm 1}, a^{\pm 1}, b^{\pm 1}, c^{\pm 1}$). Due to the complicated nature of the problem, the associated two dimensional integrals w.r.t \{$\theta'$, $\phi'$\}, and the solution of the simultaneous equations are all performed numerically. 
Before we proceed further, we will divide our results into two broad classes. The first class considers the effects of introducing a natural lattice cutoff for Weyl fermions without considering tilting of the Weyl cones. In the second class, we consider effects due to  tilting the Weyl cones in the linearized spectrum approximation, that is without considering effects due to a finite lattice cutoff. Although our formalism can handle the generic case of tilted lattice Weyl fermion, the reason for this division is because effects due to lattice and due to tilting of the Weyl cones can in fact be considered independent of each other, and linearized approximation speeds up the numerical computation. The combined effect from the two gives the net result.
\begin{table}[h!]
\centering
\begin{tabular}{|p{5.5cm}|p{2.cm}|p{5.5cm}|}
\hline
\textbf{Symmetry Setting} & \textbf{Existence of Weyl Nodes} & \textbf{Chiral Anomaly and LMC Effect} \\ \hline
Both TRS and IS are broken & allowed & CA is present; positive LMC due to anomaly-induced charge pumping \\ \hline
TRS is preserved, IS is broken & allowed & CA is present; LMC is enhanced under parallel electric and magnetic fields \\ \hline
IS is preserved, TRS is broken & allowed & CA is present; LMC is enhanced under parallel electric and magnetic fields \\ \hline
Both TRS and IS are preserved & forbidden & CA is absent; no anomaly-induced enhancement of LMC \\ \hline
\end{tabular}
\caption{{Effect of discrete symmetries on the existence of Weyl nodes and manifestation of chiral anomaly in longitudinal magnetoconductivity. In the presence of both symmetries, the system is either in a topologically trivial state or a Dirac state.}}
\label{table:symmetries}
\end{table}

\subsection{Non-collinear E and B fields without tilting of the Weyl cones for lattice Weyl fermions}
Including orbital magnetic moment effects, the energy dispersion assumes the form of the following transcendental equation,
\begin{align}
2\hbar k^2\epsilon^\chi_k&= 2\hbar k^2 E_0 \sin(ka) + e\chi E_0 \sin(ak) B(\cos(\theta)\sin\gamma + \sin\theta\cos\phi\cos\gamma).
\label{ch2_Eq_weyl_latt2}
\end{align}
The above equation has no closed-form solution for the momentum $k^\chi$, and therefore the constant Fermi energy contour in $k-$space is evaluated numerically. The semi-classical band velocities evaluated in spherical polar coordinates are,
\begin{align}
&v_k^\chi = \frac{E_0 a \cos(ak)}{\hbar} - \frac{u_2^\chi \cos(ak) \beta_{\theta\phi}}{\hbar ak^2} + \frac{2 u_2^\chi \sin(ak) \beta_{\theta\phi}}{\hbar a^2k^3},\nonumber\\
&v_\theta^\chi = \frac{u_2^\chi \sin(ak) (-d\beta_{\theta\phi}/d\theta)}{\hbar a^2 k^3},\\ \hspace{10mm}
&v^\chi_\phi = \frac{u_2^\chi (-d\beta_{\theta\phi}/d\phi)}{\hbar a^2 k^3 \sin\theta}.
\label{ch2_velocity_componenets}
\end{align}
where $\beta_{\theta\phi} = (\sin\theta\cos\phi\cos\gamma+\cos\theta\sin\gamma)$ and $u_2^\chi = -e \chi E_0 B a^2/2\hbar$.
\subsection{Non-collinear E and B fields with tilting of the Weyl cones in the linear approximation}
Since tilting and lattice cutoff effects are physically independent of each other, we treat these effects separately. Linearizing the Hamiltonian in Eq.~\ref{ch2_Eq_H1weyl} around the nodal point, we obtain 
\begin{align}
    H_\mathbf{k} = \chi \hbar v_F \mathbf{k}\cdot\boldsymbol{\sigma} + t^\chi_x k_x + t^\chi_z k_z,
    \label{ch2_Hk_liner_with_tilt}
\end{align}
where we define $v_F = aE_0/\hbar$, $t^\chi_i = T^\chi_i a$.
The expression for the constant energy contour becomes 
\begin{align}
{ k^\chi=\frac{\epsilon^{\chi}_{\mathbf{k}}+{\sqrt{(\epsilon_\mathbf{k}^{\chi})^2-l^\chi\chi \xi e v_F B\beta_{\theta\phi}}}}{l^\chi}},
\label{ch2_k_thph}
\end{align}
where $l\chi=2\hbar v_F + 2 t^{\chi}_z \cos{
\theta}+2t^{\chi}_x \sin{\theta}\cos{\phi}$, while the semiclassical velocities take the following form 
\begin{align}
v^{\chi}_x&=v_F\frac{k_x}{k}+\frac{t^{\chi}_x}{\hbar}+\frac{v_2^\chi}{k^2}\left(\cos\gamma\left({1}-\frac{2k^2_x}{k^2}\right)-\frac{2\sin\gamma k_x k_z}{k^2}\right),\nonumber\\
v^{\chi}_y&=v_F\frac{k_y}{k}+\frac{v_2^\chi}{k^2}\left(\cos{\gamma}\left(\frac{-2k_x k_y}{k^2}\right)+\sin{\gamma}\left(\frac{-2k_y k_z}{k^2}\right)\right),\nonumber\\
v^{\chi}_z&=v_F\frac{k_z}{k}+\frac{t^{\chi}_z}{\hbar}+\frac{v_2^\chi}{k^2}\left(\frac{-2\cos{\gamma}k_x k_z}{k^2}+\sin{\gamma}\left(1-\frac{2k^2_z}{k^2}\right)\right),\nonumber\\
v_2^\chi&=\frac{\chi e v_F B}{2 \hbar}.
\label{ch2_vel_components_linear_Hk}
\end{align}
\section{Results}
\label{sec:ch2_results}
We now present our main results in a format as schematically presented in Fig.~\ref{Fig_1}. {Here it is worth listing discrete symmetries that protect the topological properties of WSMs (please see table~\ref{table:symmetries}). In the presence of both symmetries, the system is either in a Dirac state or in a topologically trivial state; no anomaly-driven magnetoconductivity is expected.}

\subsection{LMC for lattice Weyl semimetal in the absence of tilt}
\begin{figure}
    \centering
    \includegraphics[width=0.328\columnwidth]{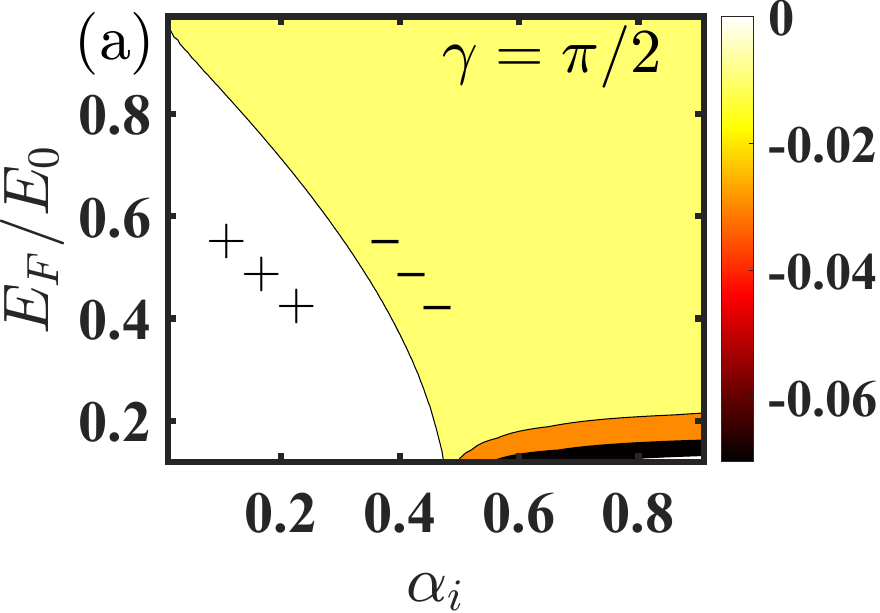}
    \includegraphics[width=0.328\columnwidth]{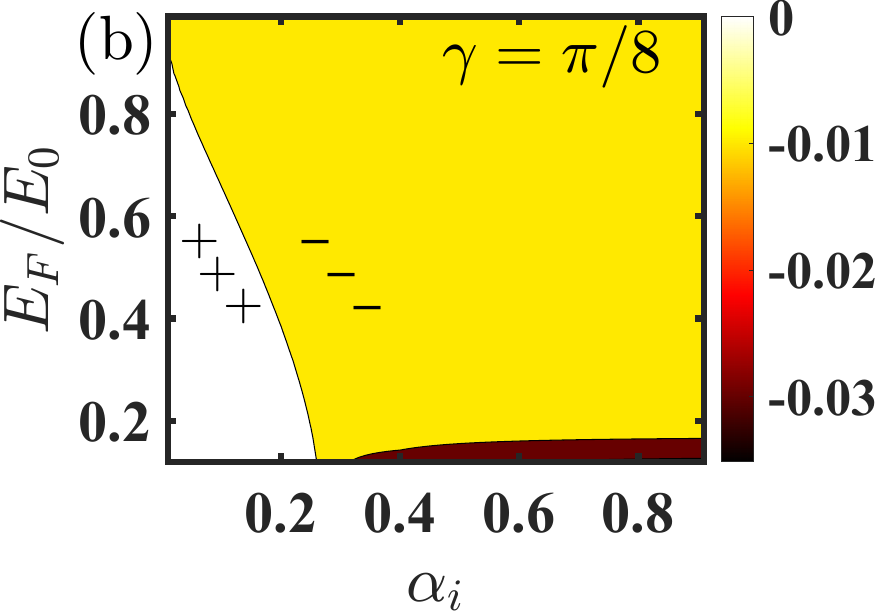}
    \includegraphics[width=0.328\columnwidth]{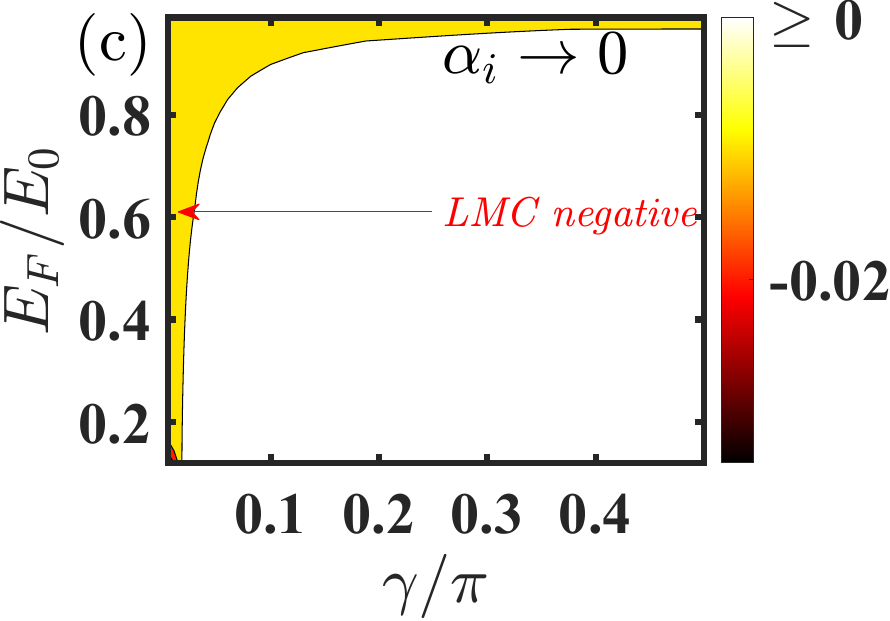}
    \caption{(a)-(b) Phase plot of the quadratic coefficient of the longitudinal magnetoconductance for a lattice model of untilted Weyl fermions as a function of Fermi energy and intervalley scattering strength $\alpha_i$ for various different angles of the magnetic field. We explicitly map the zero-LMC contour in the $E_F-\alpha_i$ space where the change in sign of LMC occurs. At higher Fermi energies the switching of LMC sign from positive to negative happens at a lower threshold of $\alpha_i=\alpha_i^c$ due to nonlinear lattice effects. Secondly, orienting the magnetic field direction away from the electric field also lowers the threshold value of $\alpha_i^c$. (c) Quadratic LMC coefficient in the limit of vanishing intervalley scattering strength $\alpha_i$ as a function of the Fermi energy and angle of the magnetic field.}
    \label{fig:lattice1}
\end{figure}

We first discuss the results for the lattice model of a Weyl semimetal without considering the effects of tilting of the Weyl cones. Since the effects of tilting of the Weyl cones are independent of lattice effects, tilting of the Weyl cones will be considered subsequently. 
The obtained LMC is found to be quadratic in magnetic field, and thus we expand the LMC as $\sigma_{zz}(B)=\sigma_{zz0} + \sigma_{zz2} B^2$. The linear-in-$B$ term $\sigma_{zz1}$, which is zero here will become crucial for our analysis when we introduce tilting of Weyl fermions, as discussed later on. The longitudinal magnetoconductance switches sign from positive to negative at a critical value of $\alpha_i^c(\gamma,E_F)$, i.e., the coefficient $\sigma_{zz2}$ becomes negative when $\alpha_i>\alpha_i^c(\gamma,E_F)$ as shown in Fig.~\ref{fig:lattice1}(a-c). At a fixed relative orientation of the magnetic field ($\gamma$), the threshold of $\alpha_i^c$ decreases as the Fermi energy is increased.
Within the linear approximation of a Weyl cone one obtains straight line contour separating positive and negative LMC areas with a constant $\alpha_i^c$ as a function of $E_F$ instead of a curved contour. Non-linear lattice effects lower the critical value of $\alpha_i^c$ highlighting the fact that lattice effects can assist driving the system to exhibit negative LMC . The explicit zero-LMC contour is plotted in Fig.~\ref{fig:lattice1}  separates positive and negative LMC region. 

A very interesting feature emerges when the magnetic field is oriented further away from the electric field, in which case the $\alpha_i$ becomes smaller. Now, from Fig.~\ref{fig:lattice1} (a)-(b), we note that even when $\alpha_i=0$, i.e., in the absence of any intervalley scattering, there is an upper energy cutoff  beyond which LMC becomes negative. This feature is specifically highlighted in Fig.~\ref{fig:lattice1} (c) where we plot $\sigma_{zz2}$ as a function of $E_F$ and angle $\gamma$, in the limit of vanishing intervalley scattering strength $\alpha_i$. This feature specifically points out the fact that lattice effects in Weyl fermions can \textit{independently} produce negative LMC even in the absence of a finite intervalley scattering, a previously unknown result. For parallel electric and magnetic fields, the LMC is primarily positive  even when lattice effects start to become important and becomes negative only at very high Fermi energies near the band edge. When the magnetic field is oriented away from the electric field, even small nonlinear lattice effects can produce negative LMC. {As one would expect, the non-linear effects matter more (less) when the Fermi energy is farther away (nearer) from the Weyl nodes, because conductivity is essentially a Fermi surface quantity. We recover this result of linearized dispersion~\cite{sharma2020sign} in the limit $E_F\ll E_0$.}

The planar Hall effect on the other hand does not display any sign change due to nonlinear lattice effects and displays the standard $\sin (2\gamma)$ trend as a function of the angle $\gamma$. Thus, we do not explicitly plot this behavior. PHE will be discussed in detail for tilted Weyl fermions subsequently. 


\begin{figure*}
\floatbox[{\capbeside\thisfloatsetup{capbesideposition={right,top},capbesidewidth=0.45\columnwidth}}]{figure}[\FBwidth]
{\caption{The quadratic coefficient of LMC is plotted as a function of $\alpha_i$ and $t_z^1$ when the Weyl cones are tilted in the direction of the magnetic field ($\hat{z}$) axis, and are oriented in the same direction  to each other ($t_z^1=t_z^{-1}$). The sign of the coefficient also corresponds to the sign of LMC. The contour separating positive and negative LMC regions is also clearly shown.}
\label{fig:szz_tiltz_same_ommon2}}
{\includegraphics[width=.95\linewidth, height = 4cm ]{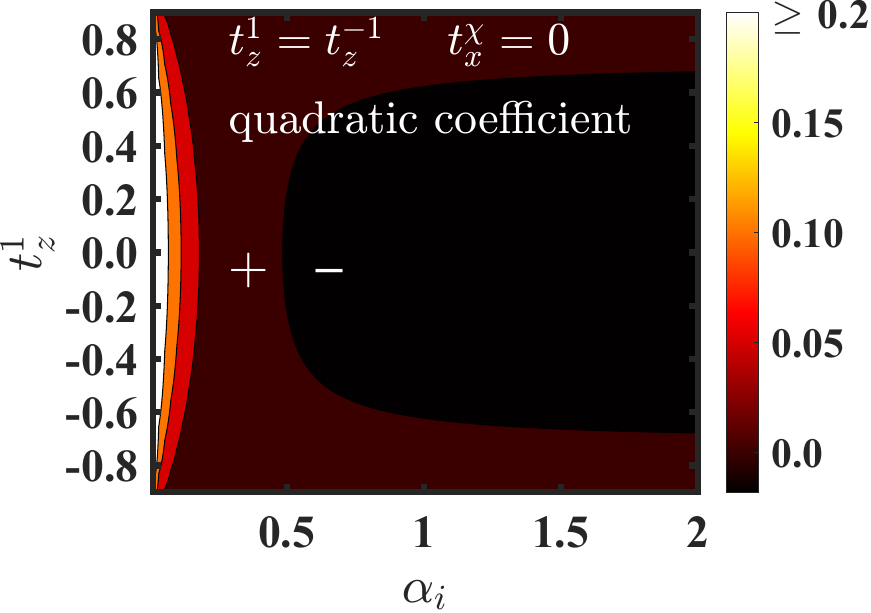}}
\end{figure*}

\begin{figure*}
\floatbox[{\capbeside\thisfloatsetup{capbesideposition={right,top},capbesidewidth=0.45\columnwidth}}]{figure}[\FBwidth]
{\caption{The sign of longitudinal magnetoconductance for non-collinear fields as a function of intervalley scattering strength and tilt parameter, when the cones are tilted along the same direction parallel to the $z$-axis.}
\label{Fig_lmc_sign_t1z_ai_tiltz_same_omm_on_gm}}
{\includegraphics[width=.85\linewidth, height = 4cm ]{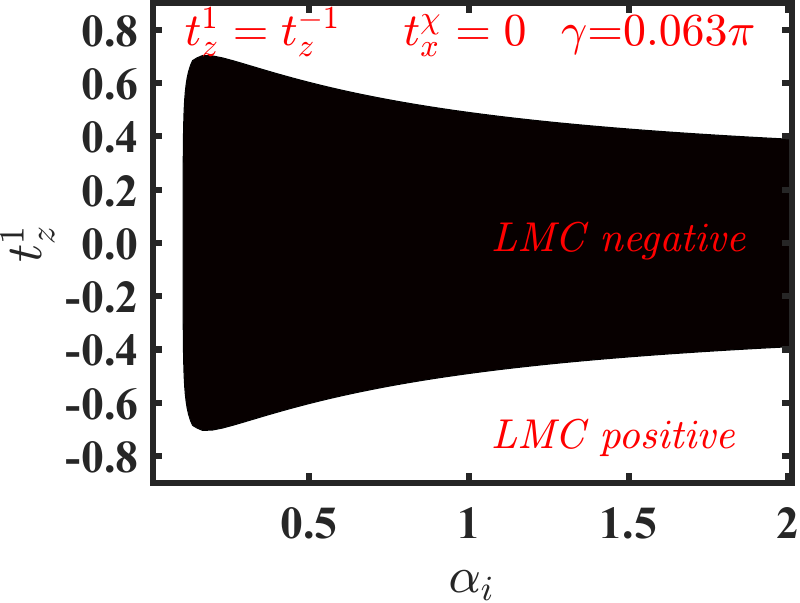}}
\end{figure*}

\subsection{LMC in tilted Weyl semimetal for parallel electric and magnetic fields }\label{Sec:TWBZ}
First we discuss the case of tilted Weyl fermions when the electric and magnetic fields are held parallel to each other, i.e., $\gamma=\pi/2$. In this case the PHE contribution is expected to vanish and hence only LMC is discussed. 

\subsubsection{Weyl cones titled along the magnetic field direction with opposite orientation}
\begin{figure}
    \centering   
    \includegraphics[width=0.485\columnwidth]{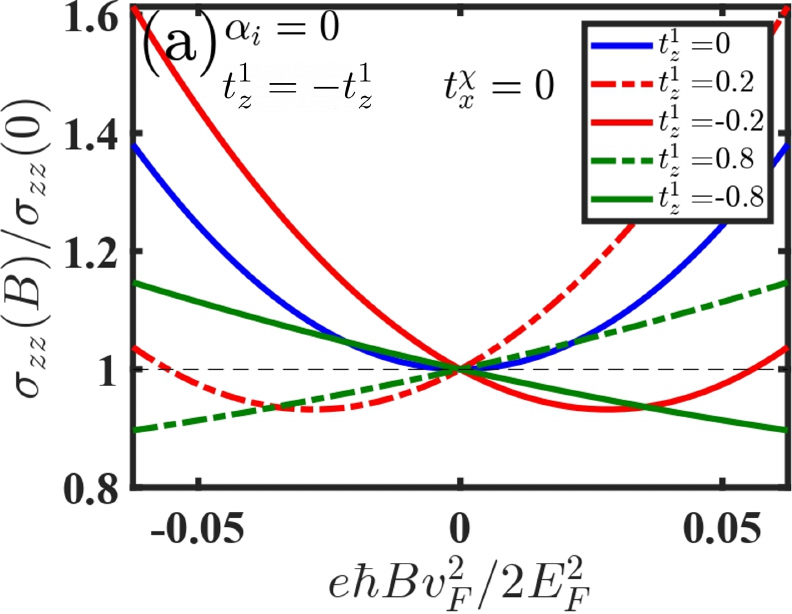}
    \includegraphics[width=0.485\columnwidth]{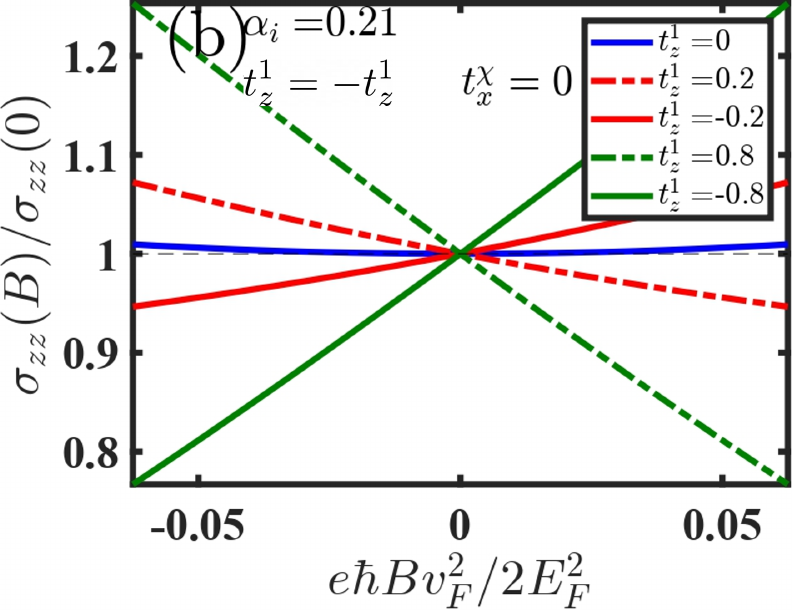}
    \caption{Longitudinal magnetoconductance $\sigma_{zz}(B)$ in the case when the Weyl cones are tilted in the direction of the magnetic field ($\hat{z}$) axis, but are oriented opposite to each other ($t_z^1=-t_z^{-1}$). (a) LMC as a function of magnetic field for various tilt parameters in the absence of intervalley scattering ($\alpha_i=0$). For a finite small tilt $t_z^1$ the LMC is asymmetric about zero magnetic field, but still appears to be quadratic. When the tilt is large, LMC is predominantly linear-in-$B$, (b) LMC in the presence of a finite intervalley scattering $\alpha_i$. }
     \label{fig:szz_tiltz_opp_ommon}
\end{figure}

Fig.~\ref{fig:szz_tiltz_opp_ommon} presents the results of LMC $\sigma_{zz}(B)$ as a function of magnetic field when the two Weyl cones are titled along the direction of the magnetic field but oriented opposite to each other, i.e. $t_z^1 = -t_z^{-1}$, and $t_x^\chi=0$. In the absence of any intervalley scattering and tilt, the LMC is always positive, quadratic in the magnetic field, and symmetric about $B=0$ as expected.  Now retaining the intervalley scattering to be zero, a finite tilt introduces a linear-in-$B$ term in the LMC and thus also introduces a corresponding asymmetry around $B=0$, i.e. now the value of the magnetoconductance depends on the direction of magnetic field, or more generally it is dependent on the orientation of the magnetic field with respect to the direction of the tilt. Note that the $B-$linear term survives because the tilts of the Weyl cones are opposite to each other. 
For higher tilt values ($\sim~\geq 0.4$) the linear-in-$B$ term dominates over the quadratic term and the LMC is observed to be linear in the relevant range of the magnetic field. 
\begin{figure}
    \includegraphics[width=0.33\columnwidth]{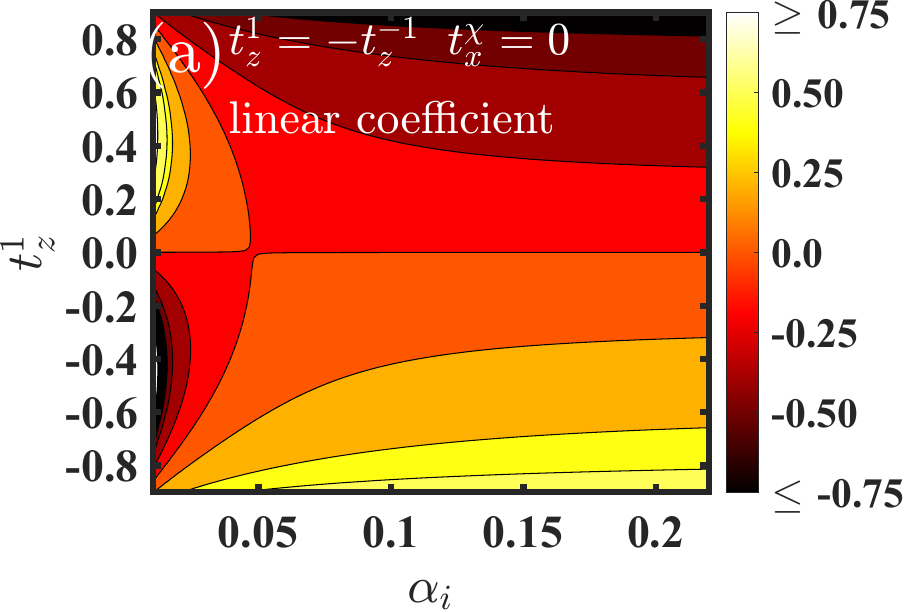}
    \includegraphics[width=0.33\columnwidth]{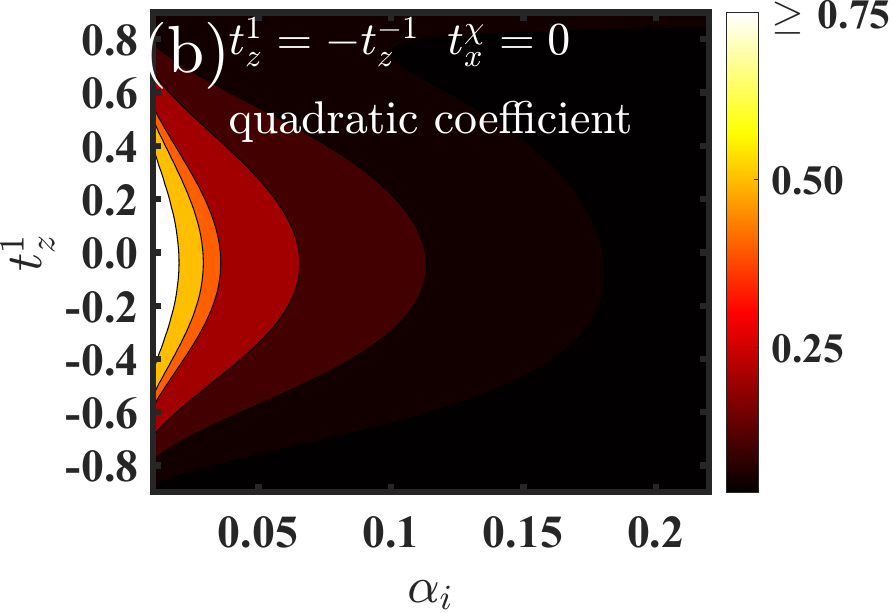}
    \includegraphics[width=0.315\columnwidth]{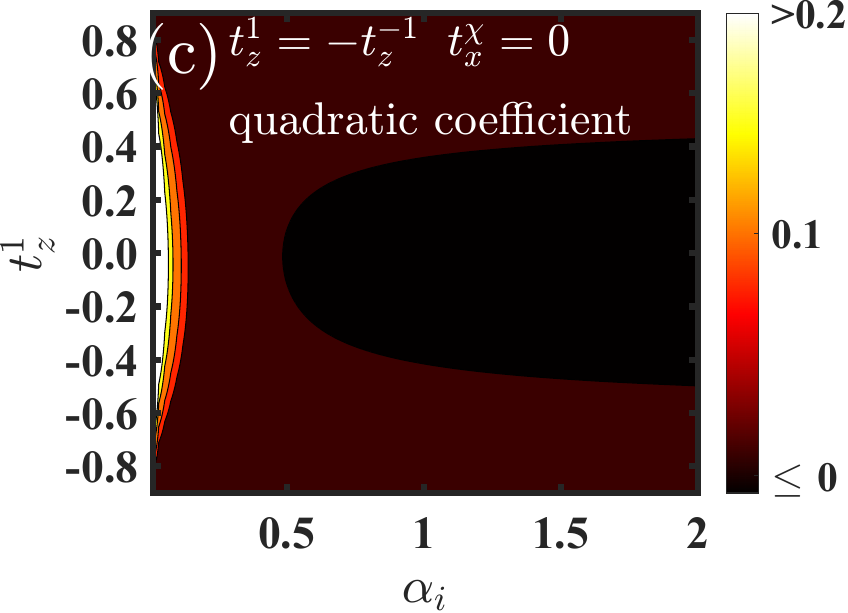}
    \caption{(a) and (b) Linear ($\sigma_{zz1}$) and quadratic ($\sigma_{zz2}$) coefficient of the LMC when the Weyl cones are titled in the direction of the magnetic field ($\hat{z}$) axis, but are oriented opposite to each other ($t_z^1=-t_z^{-1}$). Below $\alpha_i\sim 0.05$, the coefficients are similar in magnitude and LMC has an overall quadratic trend. for large enough $\alpha_i$ the linear coefficient dominates over the quadratic coefficient leading to an overall linear-in-$B$ LMC as well as a change in sign of LMC. (c) quadratic coefficient over a larger range exhibiting a change of sign.}
    \label{fig:szz_tiltz_opp_ommon2}
\end{figure}
In the presence of finite intervalley scattering, there is a qualitative change in the behavior of LMC, i.e., above a critical value $\alpha^c_i(t_z^1)$, LMC switches sign from positive to negative. 
In order to better understand this behavior, we expand the longitudinal magnetoconductance as $\sigma_{zz}(B) = \sigma_{zz0} + \sigma_{zz1} B + \sigma_{zz2}B^2$, where each coefficient $\sigma_{zzj}$ corresponds to the $j^{th}$ order in the magnetic field. The calculated LMC as a function of the magnetic field is then fit according to the above equation to obtain the coefficients $\sigma_{zzj}$. The linear and quadratic coefficients are plotted in Fig.~\ref{fig:szz_tiltz_opp_ommon2}.
For small intervalley scattering strength the linear and quadratic coefficients are similar in their magnitude, and therefore the behavior with respect to the magnetic field has an overall quadratic trend. When $\alpha_i$ crosses threshold value $\alpha^c_i$, the linear coefficient dominates and LMC switches sign as a function the magnetic field. Note that in the absence of any tilt, the linear coefficient is always zero and the LMC switches sign when $\alpha_i=0.5$~\cite{sharma2020sign}. However, for even small values of $t_z^1$, the linear coefficient dominates over the quadratic coefficient and the sign reversal in LMC occurs below $\alpha_i=0.5$. 
\begin{figure}
    \centering
    \includegraphics[width=0.485\columnwidth]{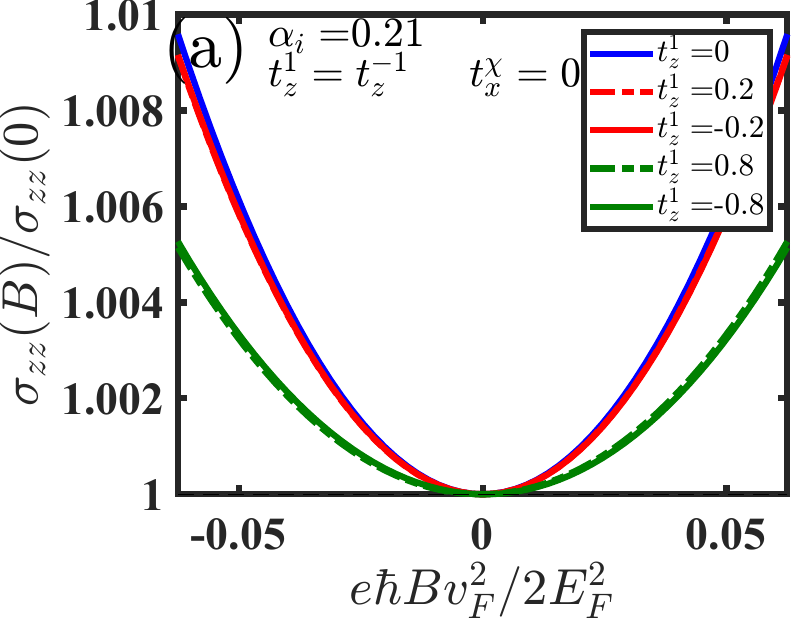}
    \includegraphics[width=0.485\columnwidth]{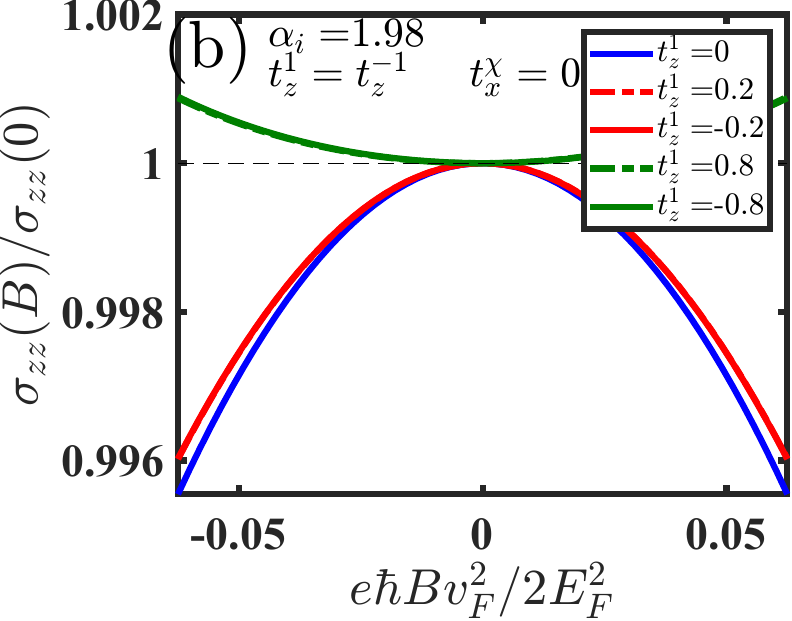}
    \caption{Longitudinal magnetoconductance $\sigma_{zz}(B)$ in the case when the Weyl cones are titled in the direction of the magnetic field ($\hat{z}$) axis, and are oriented in the same direction  to each other ($t_z^1=t_z^{-1}$).  LMC switches sign with the inclusion of $\alpha_i$ whenever $\alpha_i>\alpha_i^c(t_z)$}.
    \label{fig:szz_tiltz_same_ommon}
\end{figure}
\subsubsection{Weyl cones tilted along the magnetic field with same orientation}
Fig.~\ref{fig:szz_tiltz_same_ommon} presents the results of LMC as a function of magnetic field when the two Weyl cones are titled in the same direction with respect to each other in the direction of the magnetic field, i.e. $t_z^1 = t_z^{-1}$, and $t_x^\chi=0$. We first note that LMC is always quadratic in $B$, because the $B-$linear coefficients cancel out (as they appear with a chirality sign that is opposite for the two Weyl cones). 
When the intervalley scattering strength is small, LMC is always positive, however for large intervalley strength the behavior depends on the relative magnitude of the tilt parameter. If the magnitude of the tilt is small, the magnetoconductance switches sign, but this is opposed for large enough values of the tilt parameter. 

In Fig.~\ref{fig:szz_tiltz_same_ommon2} we present the phase plot of the quadratic coefficient $\sigma_{zz2}$. The sign of the quadratic coefficient corresponds to the sign of LMC in this case as the linear-in-$B$ term is absent. We also map out the contour in $\alpha_i-t_z^1$ space where the change in sign of LMC occurs. When $\alpha_i\gtrsim 0.5$ and $|t_z^1|\lesssim 0.6$, LMC is observed to be negative,  but remains positive and has a weak dependence on $\alpha_i$ when $|t_z^1|\gtrsim 0.6$. The LMC is determined by the interplay of $\alpha_i$ and $t_z^1$ and the tilt parameter opposes the change in sign of LMC due to intervalley scattering and its contribution dominates when $|t_z^1|\gtrsim 0.6$. 


\begin{figure*}
\floatbox[{\capbeside\thisfloatsetup{capbesideposition={right,top},capbesidewidth=0.45\columnwidth}}]{figure}[\FBwidth]
{\caption{LMC for non-collinear electric and magnetic fields when the Weyl cones are tilted along the $x$-axis and oppositely oriented to each other.  A finite tilt is noted to result in a small linear-in-$B$ contribution that enhances in the presence of intervalley scattering. The sign of LMC shows a striking change compared to Fig.~\ref{fig:szz_tiltz_same_ommon2}.}
\label{Fig_szz_vs_B_tiltx_opp_gamma}}
{\includegraphics[width=1\linewidth, height = 5.5cm ]{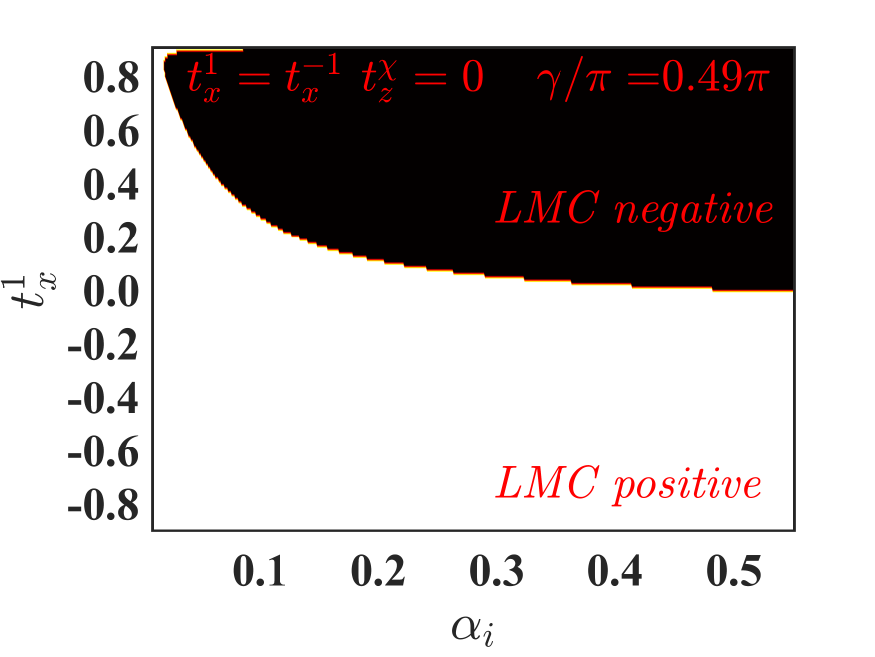}}
\end{figure*}

\begin{figure*}
\floatbox[{\capbeside\thisfloatsetup{capbesideposition={right,top},capbesidewidth=0.45\columnwidth}}]{figure}[\FBwidth]
{\caption{The sign of longitudinal magnetoconductance for non-collinear fields as a function of intervalley scattering strength and tilt parameter, when the cones are tilted along the same direction parallel to the $x$-axis.}
\label{Fig_lmc_sign_t1x_ai_tiltz_same_omm_on_gm}}
{\includegraphics[width=.95\linewidth, height = 5cm ]{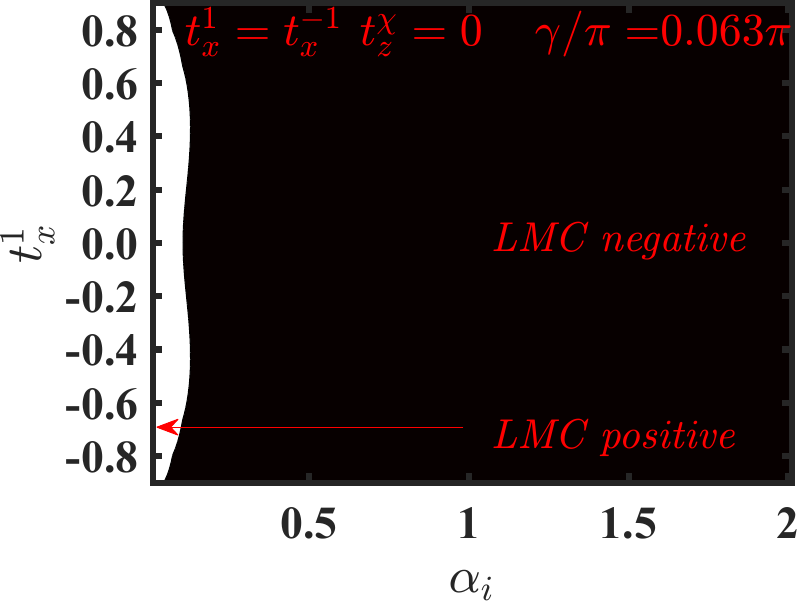}}
\end{figure*}

\begin{figure}
    \centering
    \includegraphics[width=0.458\columnwidth]{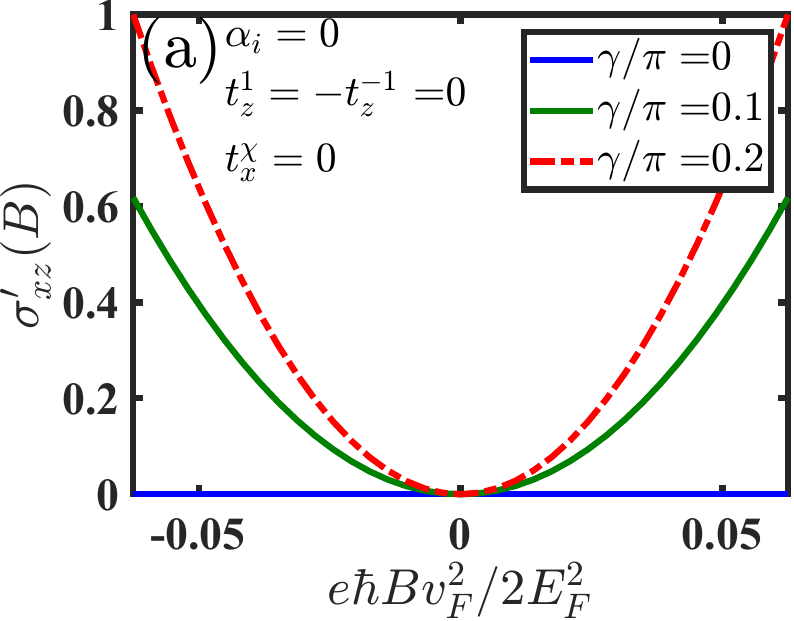}
     \includegraphics[width=0.458\columnwidth]{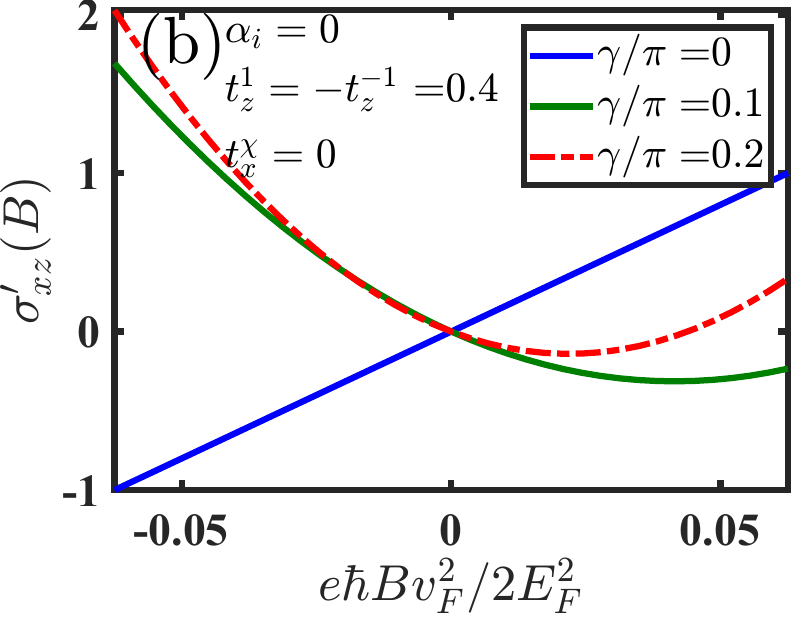}
    \caption{Normalized planar Hall conductivity $\sigma_{xz}'$ (prime indicating that the value is normalized with respect to the value at 0.5T) as a function of the magnetic field for different values of the tilt parameter $t_z^\chi$ (oppositely tilted Weyl cones) and at angles $\gamma$.  A finite tilt is observed to add a $B$-linear component that shifts the minima of $\sigma_{xz}'$ away from $B=0$. For a higher tilt value, the behavior is linear for all relevant range of magnetic field.}
    \label{fig:sxz_vs_B_tiltz_opp_omm_on}
\end{figure}
\begin{figure}
	\includegraphics[width=0.458\columnwidth]{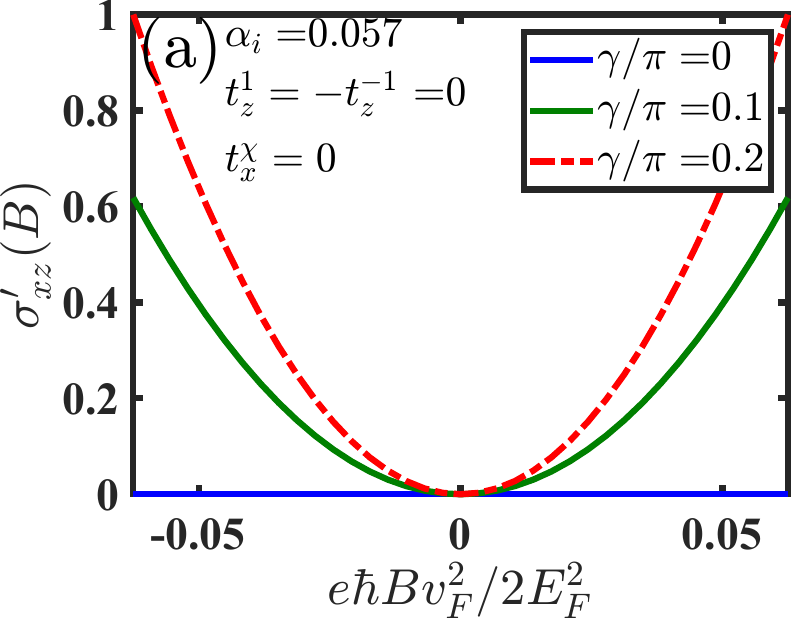}
	\includegraphics[width=0.458\columnwidth]{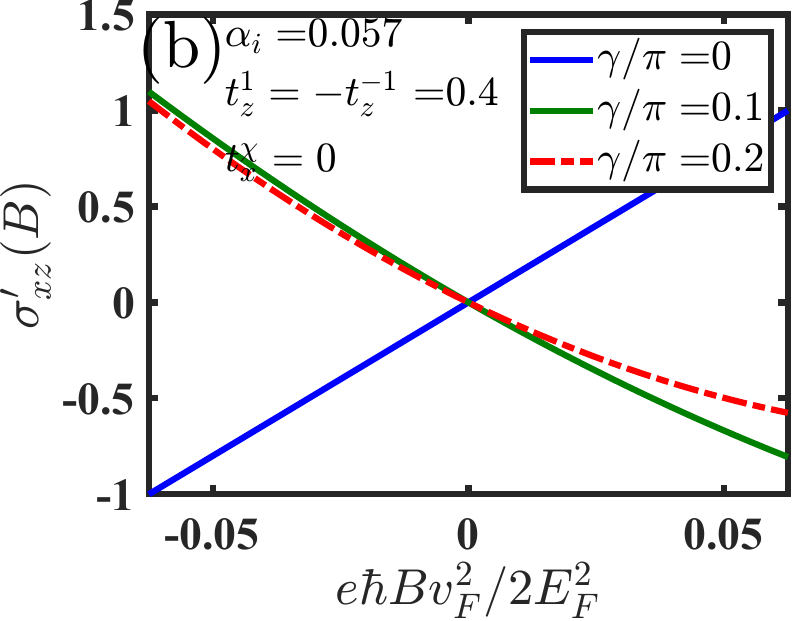}	
	 \caption{Normalized planar Hall conductivity $\sigma_{xz}'$ (prime indicating that the value is normalized with respect to the value at 0.5T) as a function of the magnetic field for different values of the tilt parameter $t_z^\chi$ (oppositely tilted Weyl cones) and at angles $\gamma$.  A finite   intervalley scattering strength $\alpha_i$ enhances the $B$-linear contribution, however only in the presence of a finite tilt.}
	  \label{fig:sxz_vs_B_tiltz_opp_omm_on2}
\end{figure}

\begin{figure}
    \centering
    \includegraphics[width=0.458\columnwidth]{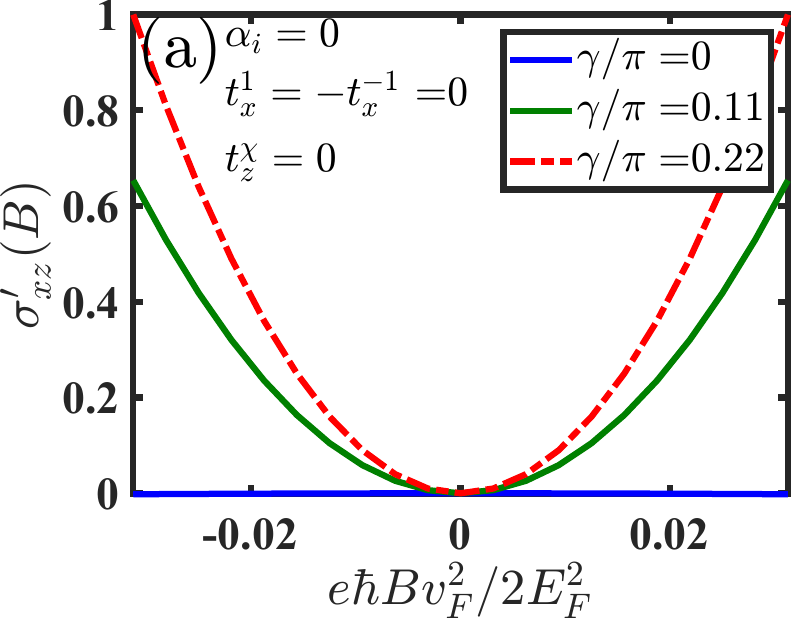}
    \includegraphics[width=0.458\columnwidth]{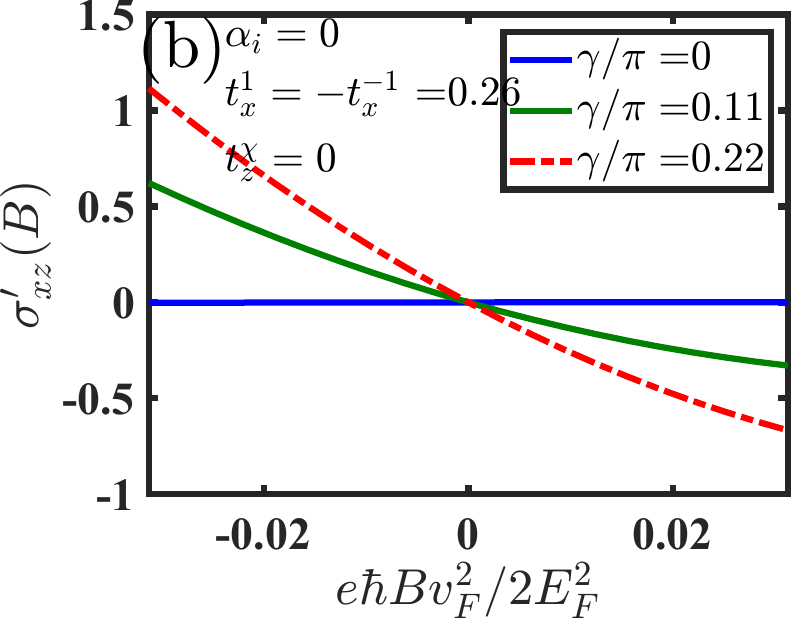}
    \caption{Normalized planar Hall conductivity $\sigma_{xz}'$ for oppositely tilted Weyl fermions along the $x$ direction (prime indicates normalization w.r.t. magnetic field at 0.5T). In the absence of intervalley scattering, a small tilt adds a linear-in-$B$ component.}
    \label{Fig_sxz_vs_B_tiltx_opp_gm}
\end{figure}
\begin{figure}
	\centering
	\includegraphics[width=0.458\columnwidth]{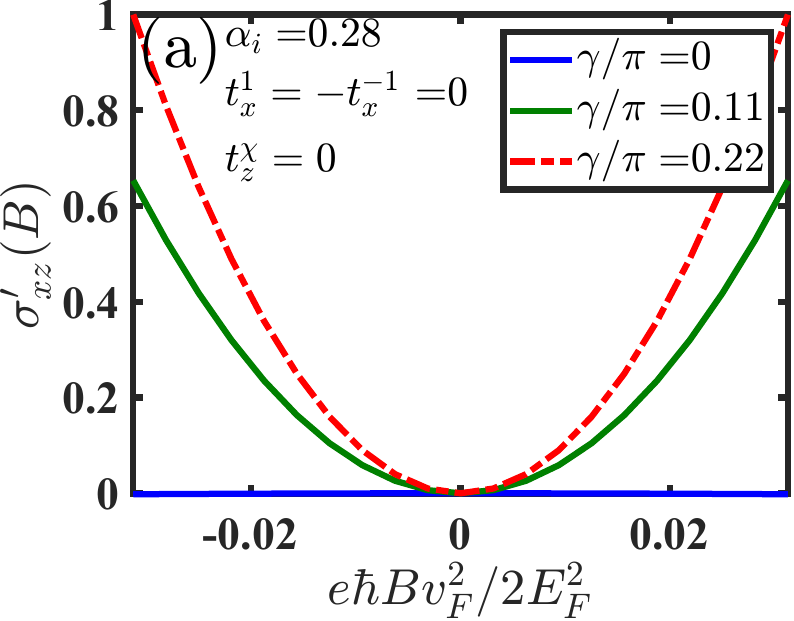}
	\includegraphics[width=0.458\columnwidth]{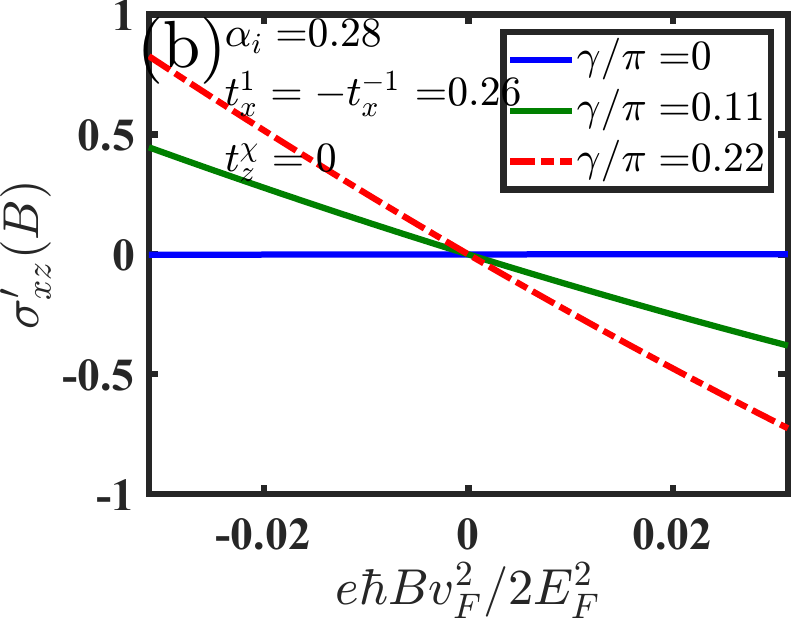}
	\caption{Normalized planar Hall conductivity $\sigma_{xz}'$ for oppositely tilted Weyl fermions along the $k_x$ direction (prime indicates normalization w.r.t. magnetic field at 0.5T). In the presence of intervalley scattering strength, the linear-in-$B$ component is enhanced, but only in the presence of a finite tilt.}
	\label{Fig_sxz_vs_B_tiltx_opp_gm2}
\end{figure}

\subsubsection{Weyl cones tilted perpendicular to the magnetic field}
When the Weyl coned are tilted orthogonal to the direction of the magnetic field, i.e. $t_x^1 = \pm t_x^{-1}\neq 0$ and $t_z^\chi=0$, we find the qualitative trend is very similar to the previously discussed case of $t_x^\chi=0$ and $t_z^1 = t_z^{-1}\neq 0$. Due to the qualitative similarities with Fig.~\ref{fig:szz_tiltz_same_ommon} and~\ref{fig:szz_tiltz_same_ommon2}, we do not explicitly plot the behavior. 

\subsection{LMC in tilted Weyl semimetal for non-collinear electric and magnetic fields}
Naively, one would expect that the longitudinal magnetoconductance will be qualitatively similar for non-collinear fields to collinear electric and magnetic fields. However, there are certain subtle non-trivial features that emerge. We will discuss each of these below. 

\subsubsection{Weyl cones tilted along the \textit{z}-axis with same orientation}
For same orientation of the Weyl cones, the linear-in-$B$ contribution to the LMC is absent. Thus the sign of the quadratic coefficient corresponds to the sign of LMC. One would therefore expect that the qualitative behavior in this case would again be similar to that observed in Fig.~\ref{fig:szz_tiltz_same_ommon}, however, we find that this is not the case. 
In Fig.~\ref{Fig_lmc_sign_t1z_ai_tiltz_same_omm_on_gm} we plot the sign of LMC as a function of the tilt $t_z^1$ and intervalley scattering $\alpha_i$ for a particular orientation of non-collinear electric and magnetic fields. 

As $\gamma\rightarrow\pi/2$ (parallel $\mathbf{E}$ and $\mathbf{B}$ fields), we recover the result presented in Fig.~\ref{fig:szz_tiltz_same_ommon}, i.e. the shape of zero-LMC contour resembles a ${U}$ (as in Fig.~\ref{Fig_lmc_sign_t1z_ai_tiltz_same_omm_on_gm}). Specifically, when $|t_z^1|\lesssim 0.6$ critical value of $\alpha_i$ where the sign change occurs is around 0.5. 
When $\gamma$ is directed away from $\pi/2$ the shape of the zero LMC contour looks like a curved trapezoid instead of $U$. The evolution from one to the other can be noted in Appendix-\ref{appendix_ch2}. 
This feature can be understood as a combination of two factors: for parallel field configuration, finite tilt and intervalley scattering drives the system to change the sign of LMC from positive to negative (as seen in Fig.~\ref{fig:szz_tiltz_same_ommon}), and secondly for non-collinear fields along with a finite $\alpha_i$ (even when $t_z^1=0$) drives the system to change LMC sign from positive to negative at a lower critical intervalley scattering strength~\cite{sharma2020sign}. The combination of these two assisting factors shapes the zero LMC contour in the current scenario.


\subsubsection{Weyl cones tilted along the \textit{x}-axis with same orientation}
From the previous discussions, it is suggested that the qualitative behavior of LMC for the three scenarios (a) \{$t_z^1=t_z^{-1}\neq 0, t_x^\chi=0$\}, (b) \{$t_x^1=t_x^{-1}\neq 0, t_z^\chi=0$\}, and (c) \{$t_x^1=-t_x^{-1}\neq 0, t_z^\chi=0$\} is similar to each other for collinear electric and magnetic fields. Therefore, rotating the magnetic field along the $xz$-plane (shifting $\gamma$ away from $\pi/2$) is  not expected to change any qualitative behavior. 
On the contrary, we find that this is not the case. 
Let us focus on cases (a) and (b). The qualitative similarity is for parallel electric and magnetic fields is given by the following properties: (i) LMC is quadratic in $B$, (ii) LMC switches sign from positive to negative above a critical intervalley scattering strength, (iii) LMC always remains positive if the tilt parameter is above a critical tilt value ($\gtrsim 0.6$), as also suggested by the shape of zero-LMC contour ($U$-shaped). 
When the fields are not parallel then in case (a) the zero-LMC contour assumes the form of an curved trapezoid (Fig.~\ref{Fig_lmc_sign_t1z_ai_tiltz_same_omm_on_gm}). The corresponding contour for the present case (b) assumes a different form as seen in Fig.~\ref{Fig_lmc_sign_t1x_ai_tiltz_same_omm_on_gm},. In this case, the region of negative LMC expands in the parameter space along with the reduction of the critical intervalley strength. The reduction of the critical intervalley strength can again be understood as a combination of the two factors like in the previous case (i) finite tilt and intervalley scattering drives the system to change the LMC sign from positive to negative, and secondly when the fields are non-collinear, finite intervalley scattering independently drives the system to change LMC sign from positive to negative much below. The different shape of the contour (negative LMC filling out the parameter space instead of a assuming a curved trapezoid) is essentially because the cones are now tilted along the $x$-direction and the magnetic field has an $x$-component to it, which is qualitatively different from the tilt occurring in the $z$-direction. The complete evolution can be noted in Appendix-\ref{appendix_ch2}.

\subsubsection{Weyl cones tilted along the \textit{x}-axis with opposite orientation}
In Sec.~\ref{Sec:TWBZ} we noted that when the Weyl cones are tilted perpendicular to the magnetic field axis, the qualitative behavior is independent of their mutual orientation. We find that when the field acquires even a small component along the direction of the tilt axis, the qualitative behavior of the LMC is strikingly different for different mutual orientations.
 Directing the magnetic field even slightly away from the $z$-axis changes the qualitative behavior when $t_x^1=-t_x^{-1}$, as a $B$-linear component is added in the LMC response. This is because the magnetic field now has a $x$-component and the tilts are oppositely oriented to each other (though tilted along the $x$-axis). Fig.~\ref{Fig_szz_vs_B_tiltx_opp_gamma} presents the plot of LMC $\sigma_{zz}$ as a function of the magnetic field when the angle of the magnetic field is slightly shifted away from the direction of the magnetic field. A finite tilt results in a small linear-in-$B$ contribution that is enhanced in the presence of intervalley scattering. The complete evolution is plotted in Appendix-\ref{appendix_ch2}.

\subsection{PHC in tilted Weyl semimetals for non-collinear electric and magnetic fields}
\subsubsection{Weyl cones tilted along the \textit{z}-axis with opposite orientation}
As mentioned earlier, planar Hall conductance will be non-zero when the electric and magnetic field are not parallel to each other. In Fig.~\ref{fig:sxz_vs_B_tiltz_opp_omm_on} and ~\ref{fig:sxz_vs_B_tiltz_opp_omm_on2} we plot the normalized planar Hall conductivity $\sigma_{xz}'$ as a function of the magnetic field for different values of the tilt parameter $t_z^\chi$ for oppositely tilted Weyl cones.
In the absence of intervalley scattering strength, a finite tilt is observed to add a $B$-linear component that shifts the minimum of the conductivity away from $B=0$. For higher values of tilt, the behavior is almost linear for all relevant range of magnetic field. On the other hand, a finite intervalley strength $\alpha_i$ enhances the $B$-linear contribution, however, only for tilted Weyl fermions. In Appendix D, we plot  the normalized planar Hall conductivity ($\sigma_{xz}'$) as a function of the angle $\gamma$ for several values of tilt parameter $t_z$ for oppositely tilted Weyl cones.

\subsubsection{Weyl cones tilted along the \textit{x}-axis with opposite orientation}

 Fig.~\ref{Fig_sxz_vs_B_tiltx_opp_gm} plots the normalized planar Hall conductance $\sigma_{xz}'$ as a function of the magnetic field. Even in the absence of intervalley scattering, a finite tilt of the Weyl cones along the $x$-direction causes the planar Hall conductivity to be linear in the magnetic field showing an asymmetry around $B=0$. The presence of intervalley scattering further enhances the $B$-linear contribution. A subtle but yet important difference between this and the previous case is that in the current scenario, the planar Hall conductivity remains zero when when  $\gamma=0$ i.e., when the magnetic field points along the $x$-direction because the direction is parallel with the direction of tilts in the Weyl cone. On the other hand, when $\gamma=0$, the planar Hall conductivity becomes finite and linear when the Weyl cones are tilted along the $z$-direction. This point {is} explicitly highlighted in Appendix D, where we plot $\sigma_{xz}'$ as a function of $\gamma$.
\subsection{LMC in an inversion symmetry broken WSM}
\begin{figure}
    \centering
    \includegraphics[width=0.65\columnwidth]{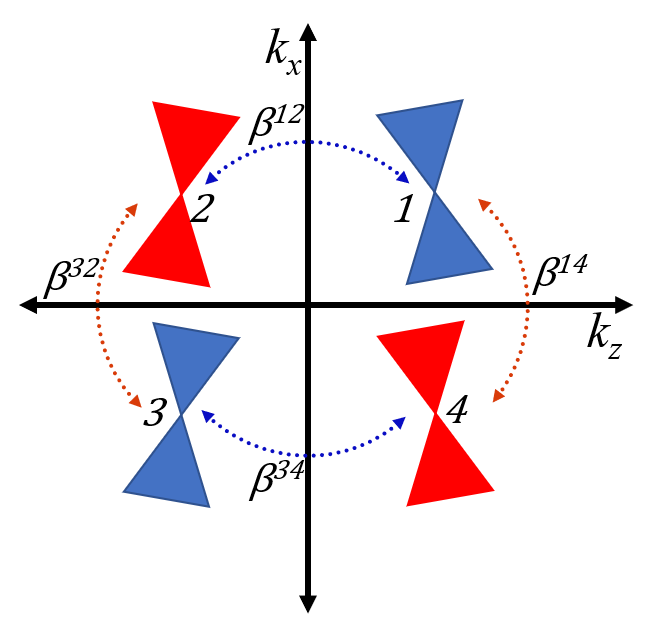}
    \caption{Model for an inversion asymmetric Weyl semimetal with tilted Weyl cones. The colors of the Weyl cones are indicative of chirality. The parameters $\beta^{mn}$ indicate internode scattering between different nodes. }
    \label{fig:fournodes}
\end{figure}
\begin{figure}
    \centering
    \includegraphics[width=0.328\columnwidth]{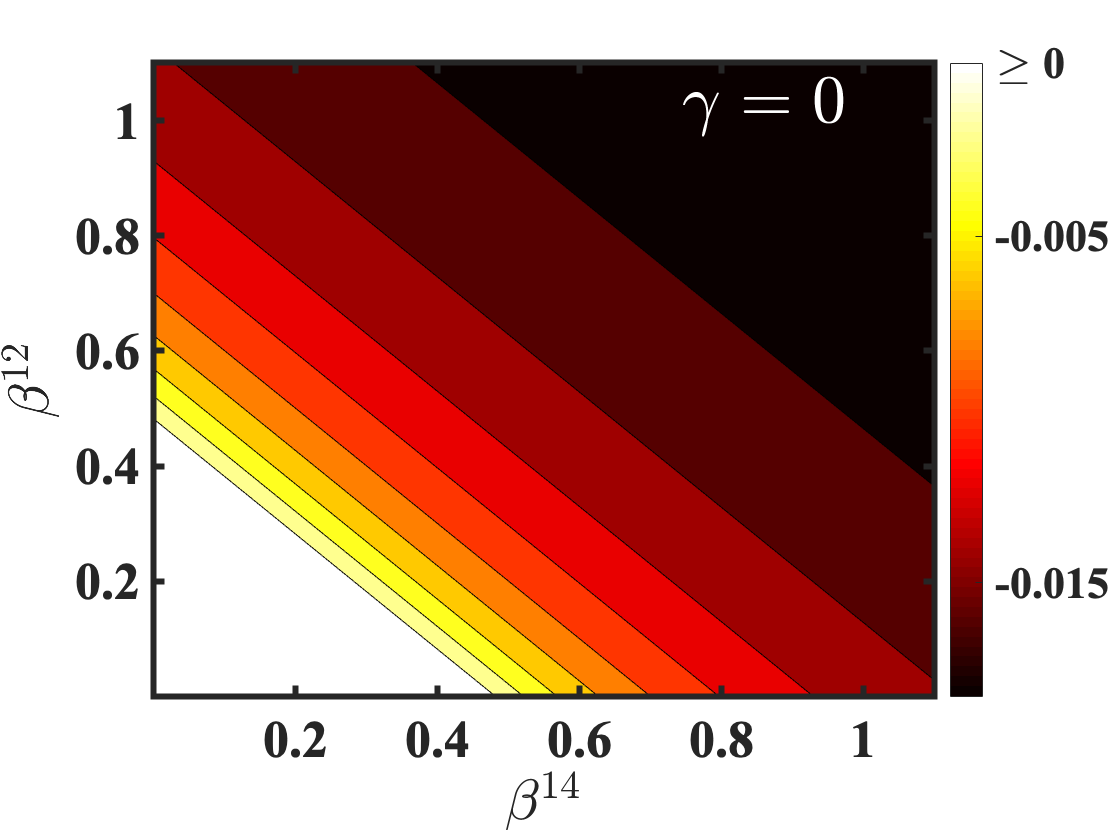}
    \includegraphics[width=0.328\columnwidth]{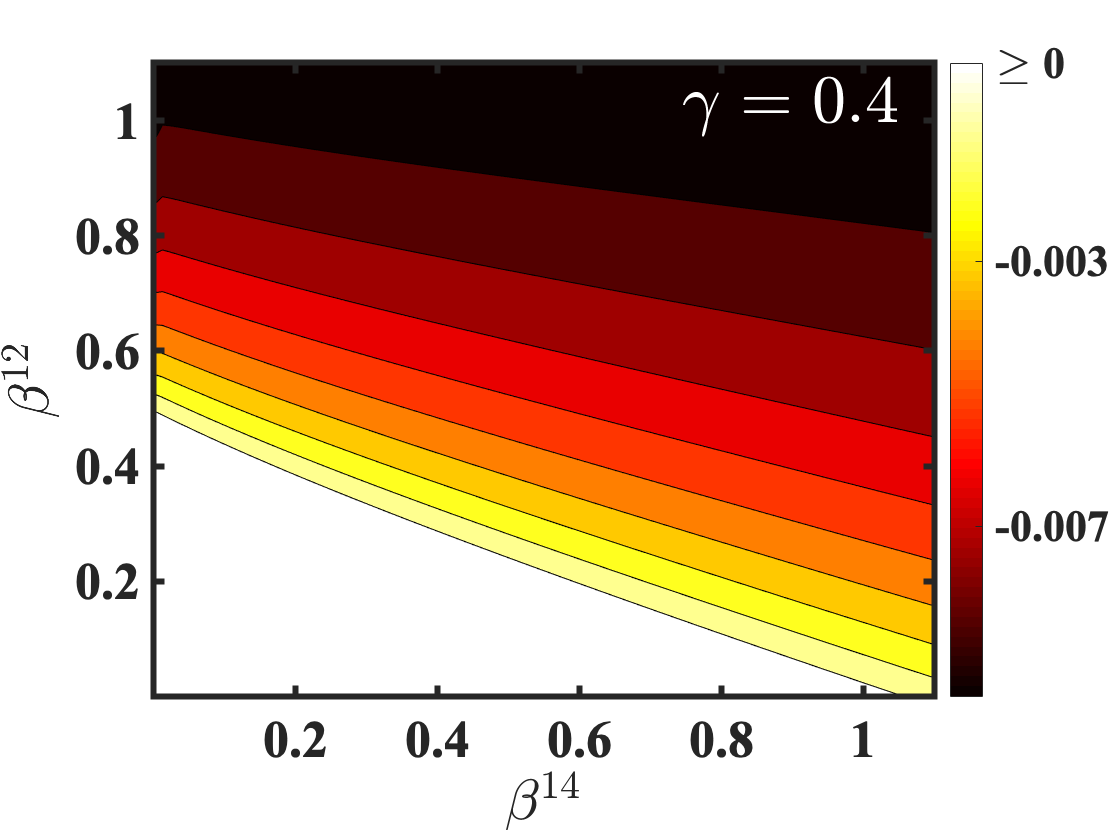}
    \includegraphics[width=0.328\columnwidth]{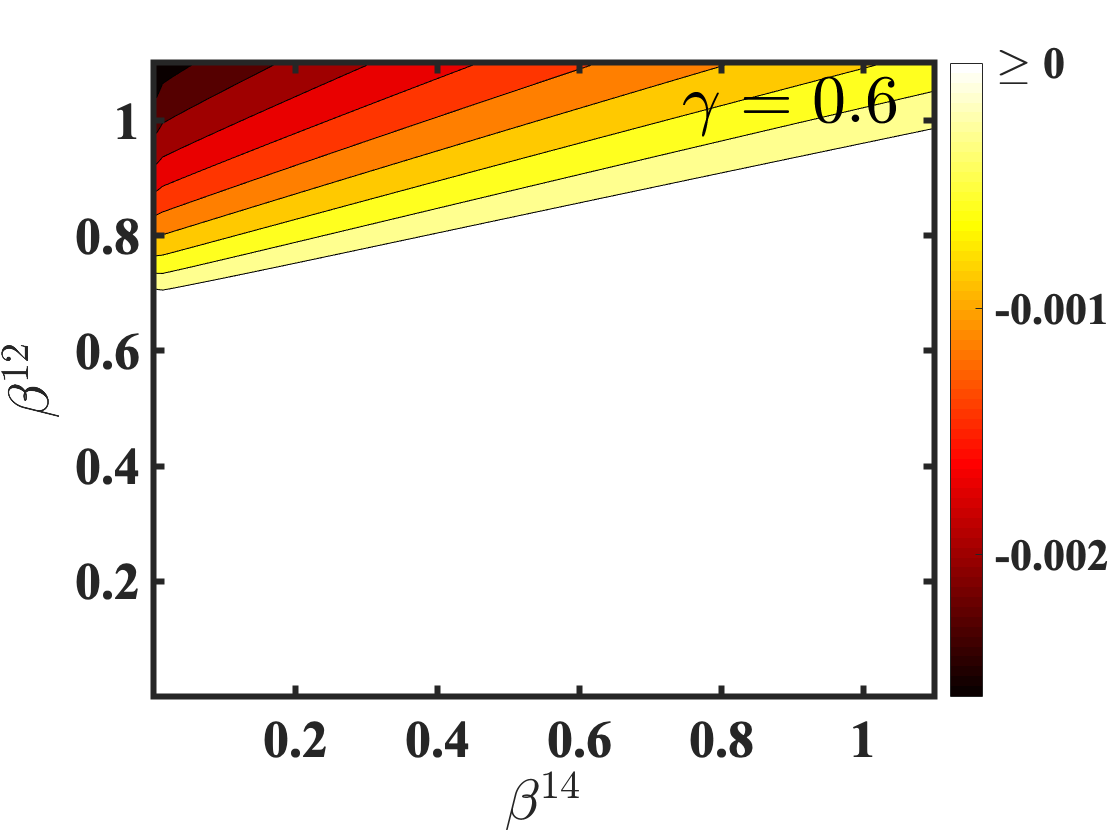}
    \caption{LMC for inversion symmetry broken WSM presented in Eq.~\ref{eq_H4nodes_ch2} for different tilt values. The interplay of internode scatterings $\beta^{12}$ and $\beta^{14}$ along with the tilt parameter governs the sign of LMC. White region corresponds to positive LMC.} 
    \label{fig:fn}
\end{figure}
Here we will discuss the applicability of our results to realistic Weyl materials. Specifically, our starting point is the following linearized model of a Weyl semimetal that preserves time-reversal symmetry but breaks inversion symmetry~\cite{mccormick2017minimal,bertrand2019complete}. For simplicity the tilting of the cones is considered only in one direction but the formalism can be generalized to tilting in multiple directions as well.
\begin{align}
    H_{ib} = \sum\limits_{n=1}^4 \left(\chi_{n}\hbar v_F \mathbf{k}\cdot\boldsymbol{\sigma} + \hbar v_F \gamma_n k_z \sigma_0\right).
    \label{eq_H4nodes_ch2}
\end{align}
The system consists of four Weyl nodes located at the points $\mathbf{K}=(\pm \pi/2,0,\pm \pi/2)$ in the Brillouin zone. In the above Hamiltonian $\chi_n$ is the chirality, and $\gamma_n$ is the tilt. Specifically, $(\chi_1,\gamma_1)=(-\chi_2,\gamma_2)=(\chi_3,-\gamma_3)=(-\chi_4,-\gamma_4)=(-1,\gamma)$, and the tilt parameter $\gamma$ is considered to be less than unity. This is also schematically represented in Fig.~\ref{fig:fournodes}. We consider four intranode scattering channels (node $n\Longleftrightarrow n$) and four internode scattering channels {(node $n\Longleftrightarrow (n+1)~\text{mod} 4$)}, as also highlighted in Fig.~\ref{fig:fournodes}. The internode scattering strength between node $m$ and node $n$ is $\beta^{mn}$. For simplicity, we  neglect scattering between nodes (4 $\Longleftrightarrow$ 2) and nodes (1 $\Longleftrightarrow$ 3) since they involve a large momentum transfer. Intranode scattering at various nodes is qualitatively same, while the internode scattering channels can be divided into two categories: (i) scattering between Weyl cones of opposite chirality and opposite tilt orientation (1 $\Longleftrightarrow$ 2) and (3 $\Longleftrightarrow$ 4), and (ii) scattering between Weyl cones of opposite chirality and same tilt orientation (1 $\Longleftrightarrow$ 4) and (2 $\Longleftrightarrow$ 3). Both of these cases have been discussed individually before, but here we consider the simultaneous effect of both categories. 

In order to solve for the longitudinal magnetoconductance for this system, the formalism presented in Section II for two Weyl nodes is generalized to a system of multiple nodes as well with arbitrary values of tilt and chirality (see Appendix E). The solution to the Boltzmann equation for the system presented in Eq.~\ref{eq_H4nodes_ch2} reduces to a system of sixteen linear equations that are solved numerically for the unknown coefficients $\{\lambda^n,a^n,b^n,c^n\}$. 

Interestingly, we find that despite the presence of internode scattering channels between cones of opposite tilt orientation, the linear-in-$B$ LMC vanishes. To understand this behaviour we note from Fig.~\ref{fig:szz_tiltz_opp_ommon2} that when tilt parameter changes sign ($t^1_z\rightarrow-t^1_z$), the linear-in-$B$ coefficient of LMC switches sign as well. From Fig.~\ref{fig:fournodes} we thus note that the linear-in-$B$ coefficient generated by internode scattering channel (1 $\Longleftrightarrow$ 2) will be cancelled by scattering channel (4 $\Longleftrightarrow$ 3). On the other hand the quadratic coefficient of LMC is an even function of the tilt parameter for all the scattering channels, making the behaviour quadratic in magnetic field. From Fig.~\ref{fig:szz_tiltz_same_ommon2} and Fig.~\ref{fig:szz_tiltz_opp_ommon2} we note that there is a quantitative difference in the quadratic coefficient for the two categories of internode scattering channel. 

In Fig.~\ref{fig:fn} we plot the quadratic coefficient of the LMC (also corresponding to the sign of LMC) as a function of internode scattering strengths for different values of the tilt parameter $\gamma$. In the absence of any tilting, LMC is symmetric in $\beta^{12}$ and $\beta^{14}$ exhibiting a change of sign from positive to negative along a straight line contour from (0.5,0) to (0,0.5) in the $\beta^{12}-\beta^{14}$ parameter space. For non-zero values of tilt parameter the curve first stretches along the $\beta^{14}$ axis and then along the $\beta^{12}$ axis as per the expectations from Fig.~\ref{fig:szz_tiltz_same_ommon2} and Fig..~\ref{fig:szz_tiltz_opp_ommon2}. The interplay of internode scatterings and the tilt parameter governs the sign of LMC. In the limit of large tilt parameter the LMC sign remains positive. 

\section{Discussions and Conclusions}
\label{sec:ch2_Discussions and Conclusions}
{ The linearity or nonlinearity of the bands is alone not a sufficient criteria to produce a finite longitudinal magnetoconductance (positive or negative) or planar Hall effect in Weyl semimetals. In WSMs, is in fact the topological nature of the bands that gives rise to finite LMC and PHE.} The topological nature of the bands is manifest in the Berry curvature and the orbital magnetic moment of the Bloch electrons. Even though the bands no longer disperse linearly away from the Weyl node, their topology is nevertheless preserved, as also demonstrated by exact expressions for Berry curvature and OMM in our prototype lattice model. 
We solved the Boltzmann equation semi-analytically for a lattice model of Weyl fermions and noted that the inclusion of orbital magnetic moment is crucial in obtaining negative LMC in the limit of vanishing intervalley scattering, just like it is crucial in obtaining negative LMC for strictly linearly dispersing Weyl fermions in the presence of intervalley scattering~\cite{knoll2020negative}. This points out to an important fact that nonlinear lattice effects can produce negative LMC for weak magnetic fields irrespective of the presence or absence of intervalley scattering. Therefore it is inconclusive to state that negative LMC for weak magnetic fields in a Weyl semimetals necessarily points out to the presence of intervalley scattering. 

Since nonlinear lattice effects are intrinsically present in real Weyl materials, likewise, the presence of a finite tilt is also inevitable. Finite lattice effects and effects due to tilting of the cones are largely independent of one another, and thus one can solve the Boltzmann equation for tilted Weyl fermions in the linearized approximation. The overall behaviour is given by a combination of both factors. 
We constructed several phase diagrams in relevant parameter space that are important for diagnosing chiral anomaly in Weyl materials. Specifically, we examine the longitudinal magnetoconductivity $\sigma_{zz}$ as well as the planar Hall conductivity $\sigma_{xz}$ for tilted Weyl fermions for the four relevant cases when the cones are tilted in the same or opposite direction along or perpendicular to the $z-$direction, i.e., (i) $t_x^1=t_x^{-1}$, and $t_z^\chi=0$, (ii) $t_x^1=-t_x^{-1}$, and $t_z^\chi=0$, (iii) $t_z^1=t_z^{-1}$, and $t_x^\chi=0$, (iv) $t_z^1=-t_z^{-1}$, and $t_x^\chi=0$. Crucially, the LMC is found to depend on the angle $\gamma$ that determines the orientation of the magnetic field w.r.t the electric field. When $\gamma=\pi/2$, the electric and magnetic fields are parallel, and the LMC has a linear-in-$B$ component only for case (iv) that results in its asymmetry around $B=0$. We found that LMC when evaluated in the limit $B\rightarrow 0^+$ switches sign as a function of intervalley scattering $\alpha_i$ and the tilt parameter. For cases (i), (ii), and (iii), LMC is symmetric around $B=0$ and quadratic in magnetic field, however, it changes sign from positive to negative depending on the magnitude of $\alpha_i$ and the tilt parameter. When $\gamma\neq \pi/2$, the phase plots for cases (i), (ii), and (iii) shows non-trivial behavior. In particular, the distinction between cases (i) and (iii) becomes evident due to qualitatively different phase plots in the $\alpha_i-t_x$ space separating negative and positive LMC regions, which however is quadratic in magnetic field. Specifically, the shape of the zero-LMC contour is distinct in the two cases.
Interestingly, for case (ii), a linear-in-$B$ component in LMC is added that vanishes in the limit of parallel electric and magnetic fields. This again results in qualitative different phase plots in the $\alpha_i-t_z$ space as a function of $\gamma$. To summarize, the shape of the zero-LMC contour in $\alpha_i-t$ space as a function of the angle $\gamma$ is qualitatively distinct in each of the four cases.

{ Next, we discussed the applicability of our results to a scenario much relevant to actual Weyl materials, i.e., the case of a inversion symmetry broken Weyl semimetal by extending the Boltzmann formalism to tackle multiple nodes simultaneously. Interestingly, we find that despite the presence of internode scattering between nodes of opposite tilt orientation, the linear-in-$B$ LMC coefficient vanishes for our model. We find that the interplay of various internode scattering channels along with the magnitude of tilt parameter governs the sign of LMC.}
Lastly, we also discuss the planar Hall conductivity $\sigma_{xz}$ for each of the above cases. A linear-in-$B$ component to $\sigma_{xz}$ is added in case (ii) and (iv), which is further enhanced by a finite $\alpha_i$. The distinction between cases (ii) and (iv) comes from the fact that in addition to $\sin (2\gamma)$, a $\cos \gamma$, and a $\sin\gamma$ trend to the planar Hall conductivity is as a function of the angle $\gamma$ for cases (iv) and (ii) respectively. The $\cos\gamma$ and $\sin\gamma$ trends are enhanced due to intervalley scattering.

\setcounter{equation}{0}
\setcounter{table}{0}
\setcounter{figure}{0}

\chapter{\label{chap3}Longitudinal  magnetoconductance and the planar Hall conductance in inhomogeneous Weyl semimetals}
{\small The contents of this chapter have appeared in ``\textsc{Longitudinal magnetoconductance and the planar Hall effect in a lattice model of tilted Weyl fermions}"; Azaz Ahmad, Karthik V. Raman, Sumanta Tewari, and Gargee Sharma; \textit{Phys. Rev. B} \textbf{107}, 144206 (2023).}
\section{Abstract}
Elastic deformations (strain) couple to the electronic degrees of freedom in Weyl semimetals as an axial magnetic field (chiral gauge field), which in turn affects their impurity dominated diffusive transport. Here we study the longitudinal magnetoconductance (LMC) in the presence of strain, Weyl cone tilt, and finite intervalley scattering, taking into account the momentum dependence of the scattering processes (both internode and intranode), as well as charge conservation. We show that strain induced chiral gauge field results in `strong sign-reversal' of the LMC, which is characterized by the reversal of orientation of the magnetoconductance parabola with respect to the magnetic field.  On the other hand, external magnetic field results in `strong sign-reversal', only for sufficiently strong intervalley scattering. When both external and chiral gauge fields are present, we observe both strong and weak sign-reversal, where in the case of weak sign-reversal, the rise and fall of magnetoconductivity depends on the direction of the magnetic field and/or the chiral gauge field, and is not correlated with the orientation of the LMC parabola. The combination of the two fields is shown to generate striking features in the LMC phase diagram as a function of various parameters such as tilt, strain, and intervalley scattering. We also study the effect of strain induced chiral gauge field on the planar Hall conductance and highlight its distinct features that can be probed experimentally.
\section{Introduction}
Fermions and the atomic lattice form the building blocks of condensed matter. While each of them are fundamentally different from the other, the interplay between the two leads to remarkable effects. In recent works, massless Dirac fermions, which have resurged in condensed matter, have been shown to couple to the elastic deformations of the lattice (strain) as an axial magnetic field (also known as chiral gauge field). Prominent examples where such fields can be realized include graphene~\cite{jackiw2007chiral,vozmediano2010gauge,guinea2010energy} and three-dimensional Weyl semimetals~\cite{cortijo2015elastic,pikulin2016chiral,grushin2016inhomogeneous}. For instance, in graphene, the generated field can be even as large as 300T, as observed via spectroscopic measurement of the Landau levels~\cite{levy2010strain}. A measurement of strain induced chiral magnetic field as well as its implications on electron transport in three-dimensional Weyl and Dirac semimetals materials is of high interest to the condensed matter community.

The reason why Weyl and Dirac semimetals also have been fascinating is due to some intriguing properties that are absent in conventional metals. Some examples include the anomalous Hall~\cite{yang2011quantum,burkov2014anomalous} and Nernst~\cite{sharma2016nernst,sharma2017nernst,liang2017anomalous} effects, open Fermi arcs~\cite{wan2011topological}, planar Hall and Nernst effects~\cite{nandy2017chiral,sharma2019transverse}, and the manifestation of chiral or Adler-Bell-Jackiw anomaly~\cite{adler1969axial,nielsen1981no,nielsen1983adler,bell1969pcac,aji2012adler,zyuzin2012weyl,zyuzin2012weyl,son2012berry,goswami2015optical, fukushima2008chiral,goswami2013axionic}. The origin of each of these effects can be traced down to the non-trivial topology of the Bloch bands. Specifically, the low-energy bandstructure of Weyl nodes comprise of pairs of non-degenerate massless Dirac cones that are topologically protected by the chirality quantum number (also known as the Chern number). Without any coupling to an external gauge field, the charge of a given chirality remain conserved. The conservation law is however broken when Weyl fermions are coupled to background gauge fields such as electric or magnetic fields~\cite{adler1969axial,nielsen1981no,nielsen1983adler}. This breakdown of conservation laws is known as `chiral anomaly', rooting its name from the particle physics literature. The verification of chiral anomaly in  Weyl semimetals is a very active area of investigation in condensed matter physics.

In a minimal model of Weyl semimetal, Weyl nodes must be separated in momentum space by a vector $\mathbf{b}$ to ensure topological protection. Alternatively, the vector $\mathbf{b}$ can also be interpreted as an axial gauge field since it couples with an opposite sign to Weyl nodes of opposite chirality~\cite{goswami2013axionic,volovik1999induced,liu2013chiral,grushin2012consequences,zyuzin2012topological}. Thus spatial variation of $\mathbf{b}$ generates an axial magnetic field $\mathbf{B}_5=\nabla\times \mathbf{b}$, which also couples oppositely to Weyl nodes of opposite chirality. An effective $\mathbf{B}_5$ field can emerge from an inhomogeneous strain profile in Weyl semimetals. In the presence of an effective chiral gauge field $\mathbf{B}_5$, the effective magnetic field experienced by Weyl fermions at a given node of chirality $\chi$ is $\mathbf{B}\longrightarrow\mathbf{B}+\chi\mathbf{B}_5$, where $\mathbf{B}$ is the external magnetic field. Therefore, the conservation laws are also modified accordingly  in the presence of the $\mathbf{B}_5$ field. Recent works have pointed out that even in the absence of an external magnetic field, the chiral gauge field influences the diffusive electron transport in Weyl semimetals by modifying its longitudinal magnetoconductance (LMC)~\cite{grushin2016inhomogeneous} as well as the planar Hall conductance (PHC)~\cite{ghosh2020chirality}. Although true in spirit, the drawback of these works is that they ignore the momentum dependence of scattering when the Weyl fermions scatter within a node (known as intranode scattering or intravalley scattering) conserving both the total charge and chiral charge, and also when they scatter to the other node (internode/intervalley scattering), in which case they conserve only the total charge. Moreover, intervalley scattering, which is the essence of `true chiral anomaly', has been neglected in Ref.~\cite{ghosh2020chirality}. In a recent work~\cite{sharma2022revisiting}, some of the co-authors of this work have pointed out that momentum dependence of scattering as well as charge conservation constraint can lead to drastic differences in the qualitative conclusions. It is therefore of immense importance to correctly treat the effect of strain induced gauge field on electron transport in Weyl semimetals, which is the focus of this work.

\begin{figure}
    \centering
    \includegraphics[width=\columnwidth]{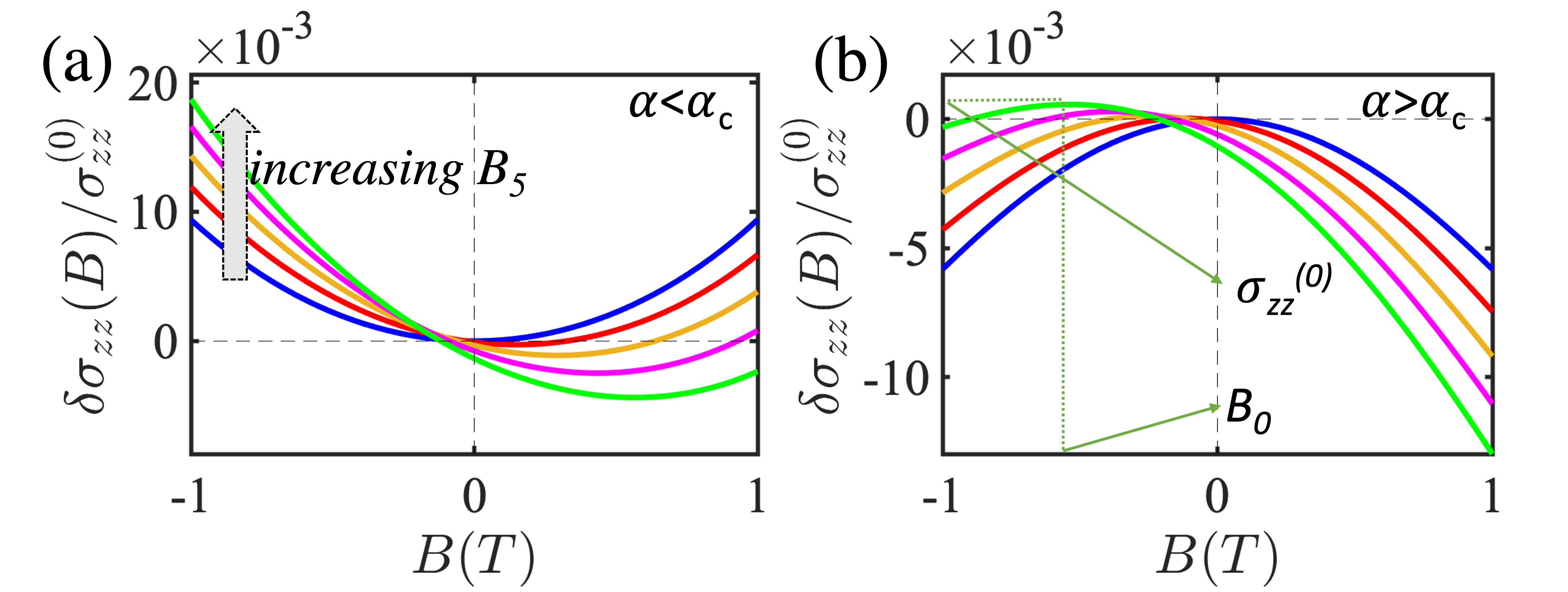}
    \caption{Change in LMC ($\delta\sigma_{zz}(B)$) with respect to the magnetic field for a minimal model of untilted TR broken WSM (Eq.~\ref{Eq:sigz1}). (a)  Weak intervalley scattering ($\alpha<\alpha_c$), and (b) strong (and weak) intervalley scattering ($\alpha>\alpha_c$). As we move from  blue to the green curve in both the plots (in the direction of the arrow), we increase $B_5$ from zero to 0.2T. The $B_5$-field is held parallel to the external magnetic field. The vertex $B_0$ and the corresponding $\sigma_{zz}^{(0)}$ is marked for the green curve in plot (b).}
    \label{fig:sigma001}
\end{figure}

In this work we critically examine the effect of strain induced chiral gauge field via the Boltzmann formalism (thus limiting ourselves to only weak perturbative fields) on two linear response quantities: the longitudinal magnetoconductance, and the planar Hall conductance. We study these effects in both time-reversal breaking WSM (with and without tilt) as well as inversion asymmetric Weyl semimetals. Earlier it was believed that positive longitudinal magnetoconductivity must manifest from chiral anomaly at least in the limit of weak external magnetic field, but this claim was corrected later on when sufficiently strong intervalley scattering was shown to switch the sign of longitudinal magnetoconductivity even in the weak-$\mathbf{B}$ limit~\cite{knoll2020negative}.  Typically, by positive (negative) longitudinal magnetoconductance we mean that $(\sigma(|\mathbf{B}|) - \sigma(\mathbf{B}=0)) >(<)\; 0$, i.e., the field dependent conductivity is greater (smaller) than the zero-field conductivity. 
Here we show that the presence of $\mathbf{B}_5$ field can also reverse the sign of LMC, but along a particular direction of the magnetic field (see Fig.~\ref{fig:sigma001}). This leads to an interesting scenario of the LMC being positive along one direction of the magnetic field, and negative when the direction of the magnetic field is reversed. To counter this  ambiguity in the sign of LMC, we introduce the idea of weak and strong sign-reversal, which depends on the orientation and the vertex of the parabola of magnetoconductivity with respect to the magnetic field (Eq.~\ref{eq:sigma_2}). We show that in the presence of only strain induced chiral gauge field (and absence of external magnetic field), the system shows signatures of strong sign-reversal for all values of intervalley scattering. In the presence of only the external magnetic field (and absence of chiral gauge field), the system shows strong sign-reversal only at sufficiently large values of scattering. In the presence of both chiral gauge and externally applied magnetic field, signatures of both weak and strong sign-reversal are observed, and furthermore very interesting features emerge in the phase diagram of LMC as a function of various system parameters such as the intervalley scattering, tilt, and strain. We point out that whenever external magnetic field is absent, we discuss weak and strong-sign reversal in context of the LMC parabola with respect to the chiral gauge field $\mathbf{B}_5$. When the external magnetic field is present (in either presence or absence of the chiral gauge field), weak and strong-sign reversal in discussed in context of the LMC parabola with respect to the external magnetic field $\mathbf{B}$.
We also extend the idea of weak and strong sign-reversal to the planar Hall conductance as well, and study the effect of strain induced gauge field on the same. Along with other features, we also unravel a very interesting behavior in the planar Hall conductance due to an interplay between the chiral gauge field and the external magnetic field. Specifically we observe a region in the parameter space where the planar Hall conductivity increases in magnitude upon increasing the scattering strength, which is counter-intuitive.  
In Section II, we introduce the concept of weak and strong sign-reversal using a minimal model of a TR broken WSM. We also study the interplay of strain, tilt, and intervalley scattering on LMC and PHC. In Section III, we present the results for inversion asymmetric Weyl semimetals. We conclude in Sec IV. All the calculations are relegated to the Appendix. 

\section{Time-reversal broken Weyl semimetals}
Consider a minimal model of a time-reversal symmetry broken Weyl semimetal, i.e., two linearly dispersing non-degenerate Weyl cones separated in momentum space. We also assume that there is no tilting of the Weyl cones in any direction.  The low-energy Hamiltonian is given by 
\begin{align}
    H = \sum\limits_\chi \sum\limits_\mathbf{k} {\chi\hbar v_F \mathbf{k}\cdot\boldsymbol{\sigma}}
    \label{Eq:HWeyl1}
\end{align}
Here $\chi=\pm 1$ is the chirality of the node, $\mathbf{k}$ is the momentum, $v_F$ is the velocity parameter, and $\boldsymbol{\sigma}$ is the vector of Pauli spin matrices. Both intranode and internode scattering processes are allowed, and the dimensionless intervalley scattering strength is denoted by $\alpha$ (see Appendix A for all the calculations). To study transport, we perturb the system with weak electric field that is fixed along the $\hat{z}-$axis. On application of a magnetic field parallel to the electric field, the longitudinal magnetoconductivity obtained in the semiclassical limit is expressed as
\begin{align}
\sigma_{zz}(B)= \sigma_{zz}^{(2)}B^2 + \sigma_{zz}^{(0)},
\label{Eq:sigz1}
\end{align}
where $\sigma_{zz}^{(0)}$ is the conductivity in absence of any magnetic field, while $\sigma_{zz}^{(2)}$ is the quadratic coefficient of magnetic field dependence. In contrast to earlier anticipation that the quadratic coefficient $\sigma_{zz}^{(2)}$ is always positive, it was recently realized that the coefficient can become negative if the intervalley scattering is sufficiently strong~\cite{knoll2020negative}. In other words, large intervalley scattering results in negative longitudinal magnetoconductivity or reverses its sign. Specifically this occurs above a critical intervalley scattering strength $\alpha_c$. The sign of the parameter $\sigma_{zz}^{(2)}$ also correlates with increasing or decreasing longitudinal magnetoconductivity.  We can call this as the usual `sign-reversal' of LMC, which refers to the fact that $\sigma_{zz}(|B|)-\sigma_{zz}(B=0)$ continuously changes sign from positive to negative. 
\begin{figure}
    \centering
    \includegraphics[width=\columnwidth]{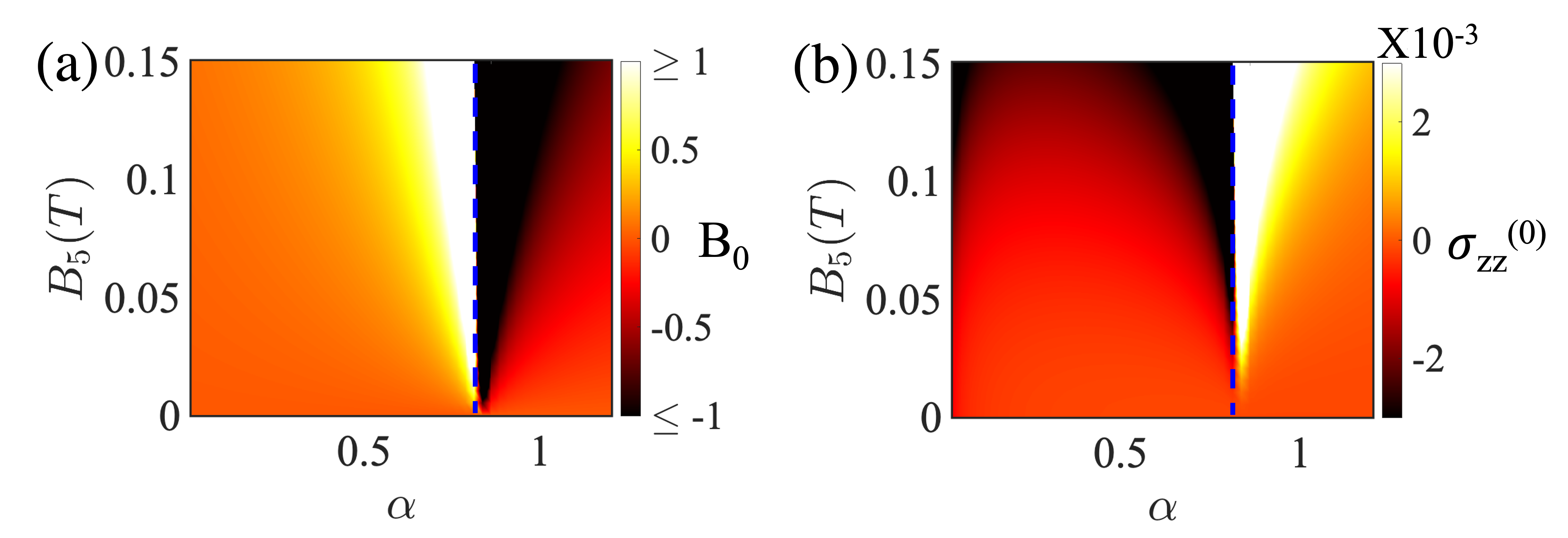}
    \caption{(a) The vertex of the parabola $B_0$, and (b) conductivity at $B_0$ for a minimal model of untilted TR broken WSM (Eq.~\ref{Eq:sigz1}). Around the blue dashed contour ($\alpha=\alpha_c$) we see `strong' sign-reversal. The parameters $B_0$ and $\sigma_{zz}^{(0)}$ show a striking change of sign as we move across the $\alpha_c$ contour. }
    \label{fig:sigma002}
\end{figure}
\begin{figure*}
    \centering
    \includegraphics[width=1\columnwidth]{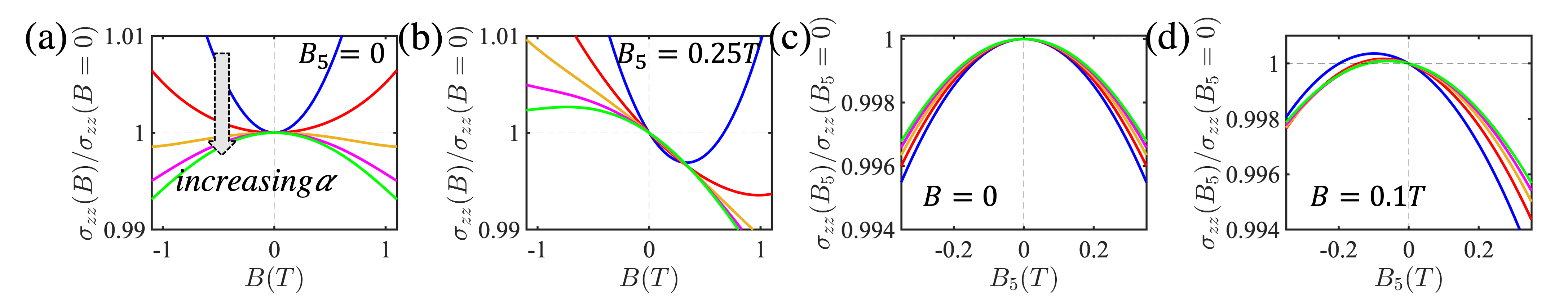}
    \caption{Longitudinal magnetoconductivity for a minimal model of TR broken untilted Weyl semimetal. (a) Increasing intervalley scattering strength results in strong sign-reversal. (b) In addition to this, infinitesimal strain now results in weak sign-reversal as well.  (c) When plotted as a function of the gauge field $B_5$, LMC is always strongly sign-reversed. (d) In the presence of an external magnetic field, we see signatures of weak-sign reversal as well. 
    In all the plots as we move from blue to the green curve we increase the intervalley scattering strength $\alpha$ from below $\alpha_c$ to above $\alpha_c$.}
    \label{fig:twonode_notilt_lmc_vs_B_vs_B5_vary_alpha}
\end{figure*}


\subsection{Longitudinal magnetoconductance \& strong and weak sign reversal}
Next, let us examine the behavior in the presence of an effective chiral gauge field ($B_5$) that may arise in inhomogeneous WSMs due to presence of strain. The chiral gauge field couples oppositely in opposite valleys, thus the net magnetic field becomes valley dependent, i.e., $B\rightarrow B+\chi B_5$. We first assume that $B_5$ is held parallel to the external magnetic field $B$. Fig.~\ref{fig:sigma001} plots the behavior of $\delta\sigma_{zz}(B)$, which is the change in LMC due to the magnetic field, i.e., $\delta\sigma_{zz}(B) = \sigma_{zz}(B) - \sigma_{zz}(B=0)$. We find that the increase or decrease of LMC depends on the direction of magnetic field, especially close to $B=0$. For example, when $\alpha<\alpha_c$, LMC decreases for positive values of magnetic field and increases for negative values of magnetic field. When $\alpha>\alpha_c$, the behavior is reversed. Furthermore, when $B$ is increased further away from zero  (in either direction), LMC increases (decreases) for both negative and positive values of B when $\alpha<\alpha_c$ ($\alpha>\alpha_c$).
Hence, it turns out that stating whether the longitudinal magnetoconductance is only positive or negative for a given scenario turns out to be rather ambiguous. 

To counter this, first we generalize the expression of magnetoconductivity to 
\begin{align}
\sigma_{zz}(B)= \sigma_{zz}^{(2)}(B-B_0)^2 + \sigma_{zz}^{(0)},
\label{eq:sigma_2}
\end{align}
The above definition allows us to shift the vertex of the parabola ($B_0$) away from origin, which is essential to fit the results presented in Fig.~\ref{fig:sigma001}. Now, in Fig.~\ref{fig:sigma001}(a) even though LMC is negative at low positive magnetic fields, it is in fact always positive when seen in reference to the vertex $B_0$, i.e., LMC is always positive when the change in the magnetic field and conductivity is seen with respect to the conductivity at $B_0$.
We call this as `weak' sign-reversal because the orientation of the parabola remains intact, and only the vertex is shifted from the origin, and also $\sigma_{zz}^{(2)}$ remains positive. Thus, when intervalley scattering is weak, strain in inhomogeneous WSMs drives the system to the `weak' sign-reversed state along a particular direction of the magnetic field. In summary, the characteristics defining weak sign-reversal are the following: (i) $B_0 \neq 0$, (ii) $\sigma_{zz}^{(0)} \neq \sigma_{zz}(B=0)$, (iii) $\mathrm{sign }\; \sigma_{zz}^{(2)}>0$. 

Now, when the strength of the intervalley scattering is greater than the critical value ($\alpha_c$), the orientation of the parabola is reversed, i.e., LMC does not again increase for $|B|>B_0$ unlike the earlier case, and $\sigma_{zz}^{(2)}$ becomes negative.
Due to this reason, we call this as `strong' sign-reversal. The only condition that we impose for strong sign-reversal is: (i)  $\mathrm{sign }\; \sigma_{zz}^{(2)}<0$, without any restriction to the values of $B_0$ and $\sigma_{zz}^{(0)}$. Therefore, the signatures of both strong and weak sign-reversal are:  (i) $B_0 \neq 0$, (ii) $\sigma_{zz}^{(0)} \neq \sigma_{zz}(B=0)$, (iii) $\mathrm{sign }\; \sigma_{zz}^{(2)}<0$. 
Since $B_0$ is shifted from the origin  due to infinitesimal strain even when $\alpha>\alpha_c$, we say that sufficiently strong intervalley scattering along with strain in inhomogeneous WSMs drives the system to show signatures of both weak and strong sign-reversal. This is demonstrated in Fig.~\ref{fig:sigma001} (b).
In general, the chiral gauge may be oriented away from the $z-$ axis and  rotated along the $xz$-plane. 
The variation of magnetoconductivity with respect to the angle $\gamma_5$ (the angle between $x$-axis and the $B_5$ field) is straightforward to understand. As $\gamma_5$ increases from zero to $\pi/2$, the contribution due to to the chiral gauge field increases in a sinusoidal fashion. We do not explicitly plot this behavior. 

In Fig.~\ref{fig:sigma002} we plot the parameters $B_0$ and $\sigma_{zz}^{(0)}$ as a function of the chiral gauge field and intervalley scattering strength. The transition from 'weak' to 'strong and weak' sign-reversed case (and vice-versa) is characterized by a sudden reversal in signs of the relative offset in conductivity $\sigma_{zz}^{(0)}$, as well as the vertex of the parabola $B_0$, i.e., $B_0\leq 0$ when $\sigma_{zz}^{(0)}\geq 0$, and vice-versa. In contrast, $\sigma_{zz}^{(2)}$ continuously interpolates across zero (not plotted). No discontinuity in $B_0$ or $\sigma^{(0)}$ is observed in the weak sign-reversed case, i .e., as the strain induced field is increased from zero for a constant intervalley scattering, the parameters $B_0$ and $\sigma_{zz}^{(0)}$ vary continuously. 

In Fig.~\ref{fig:twonode_notilt_lmc_vs_B_vs_B5_vary_alpha} we plot the the longitudinal magnetoconductivity as a function of magnetic field for different values of intervalley scattering. In the absence of chiral gauge field (Fig.~\ref{fig:twonode_notilt_lmc_vs_B_vs_B5_vary_alpha} (a)), as expected, we observe strong sign-reversal when $\alpha>\alpha_c$. In the presence of chiral gauge field field (Fig.~\ref{fig:twonode_notilt_lmc_vs_B_vs_B5_vary_alpha}(b)), we observe both strong and weak-sign reversal as also pointed out earlier. 
Typically, an increase in intervalley scattering strength decreases the magnetoconductivity, i.e., $|\sigma_{xz}(B,\alpha)|>|\sigma_{xz}(B,\alpha+\epsilon)|$, where $\epsilon$ is the infinitesimal increase in the scattering strength. We find this to be true even in the presence of strain induced chiral gauge field.
We particularly highlight this point here as this will be contrasted to the planar Hall conductivity that shows an anomalous increase in conductivity with increasing intervalley scattering strength. In Fig.~\ref{fig:twonode_notilt_lmc_vs_B_vs_B5_vary_alpha} (c) we plot the LMC in the presence of only chiral gauge magnetic field (i.e. $B=0$). Since, in this case the external magnetic field is zero, positive/negative LMC and weak/strong sign-reversal can only be defined with reference to the $B_5$ field. We find that the strain induced chiral gauge field by itself only results in strong sign-reversed phase irrespective of the intervalley scattering strength. We find this to be true even in the presence of external $B$-field (Fig.~\ref{fig:twonode_notilt_lmc_vs_B_vs_B5_vary_alpha} (d)).

\begin{figure}
    \centering
    \includegraphics[width=\columnwidth]{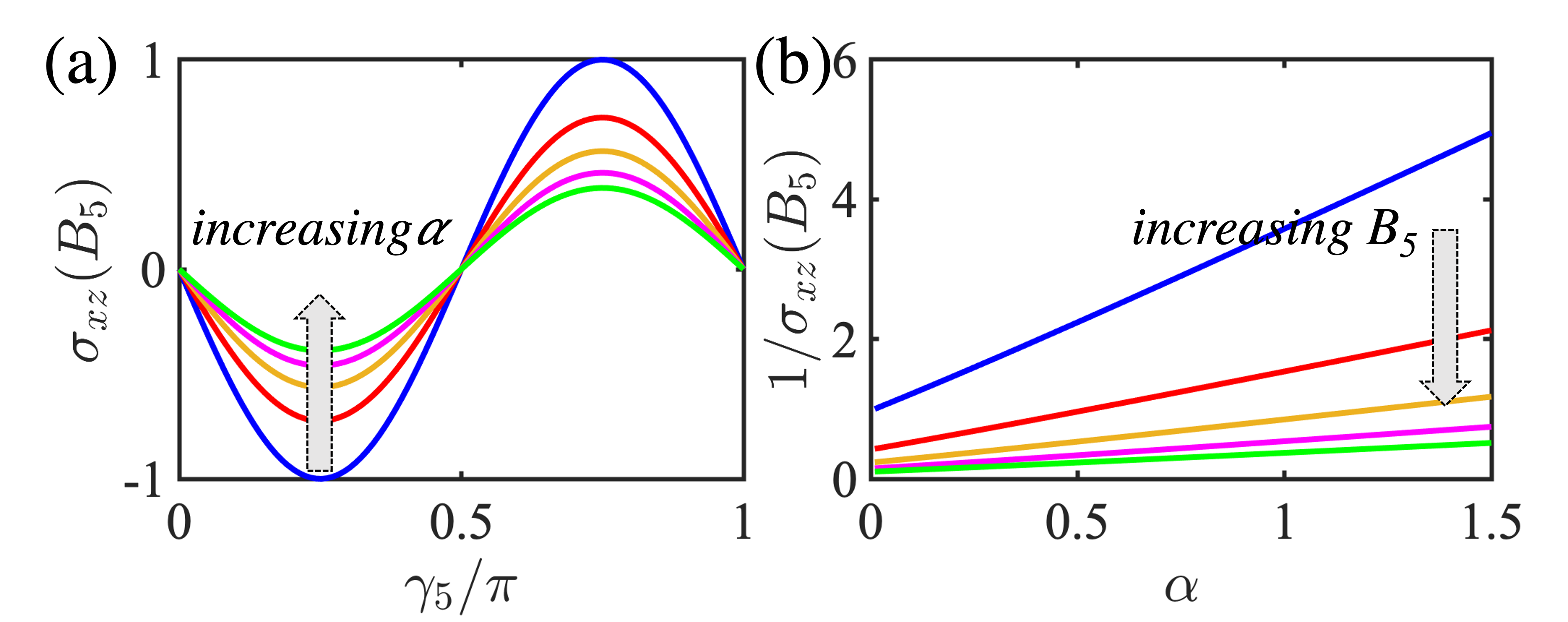}
    \caption{Planar Hall conductivity for a minimal model of untilted TR broken WSM in the absence of any magnetic field. (a) Variation with respect to the angle $\gamma_5$. Increasing $\alpha$ reduces the conductivity, as expected. (b) PHC behaves as the inverse of scattering strength. Since $\sigma_{xz}(B_5=0)=0$, we have normalized $\sigma_{xz}$ appropriately in both the plots. In creasing $B_5$ field increases the conductivity.}
    \label{fig:phc001}
\end{figure}
\begin{figure}
    \centering
    \includegraphics[width=\columnwidth]{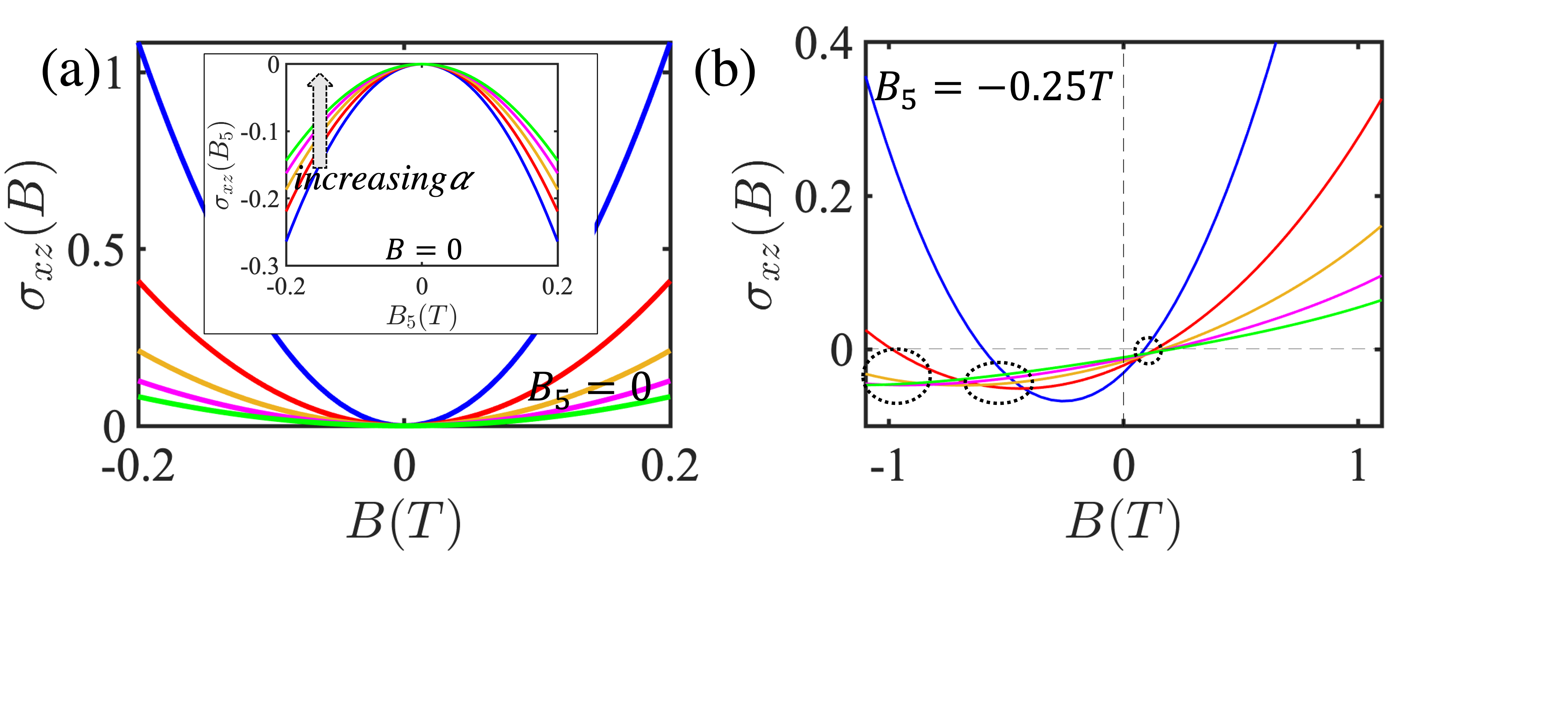}
    \caption{Planar Hall conductivity for a minimal model of untilted TR broken WSM. (a) PHC as a function of external magnetic field $B$ and no strain induced field ($B_5=0$) is compared with the inset where PHC has been plotted as a function of $B_5$ with no external field ($B=0$). The angle $\gamma$ was chosen to be equal to $\gamma_5$. Strain opposes the planar Hall effect albeit with different magnitude. (b) PHC in the presence of both magnetic field and strain. The chiral gauge field causes  weak sign-reversal. The dotted ellipses highlight regions that show an anomalous behavior with respect to intervalley scattering strength. The width of plots is reduced for better visibility. In all the curves, as we go from blue to green, we increase $\alpha$. All the plots are appropriately normalized.}
    \label{fig:phc002}
\end{figure}
\begin{figure}
    \centering
    \includegraphics[width=\columnwidth]{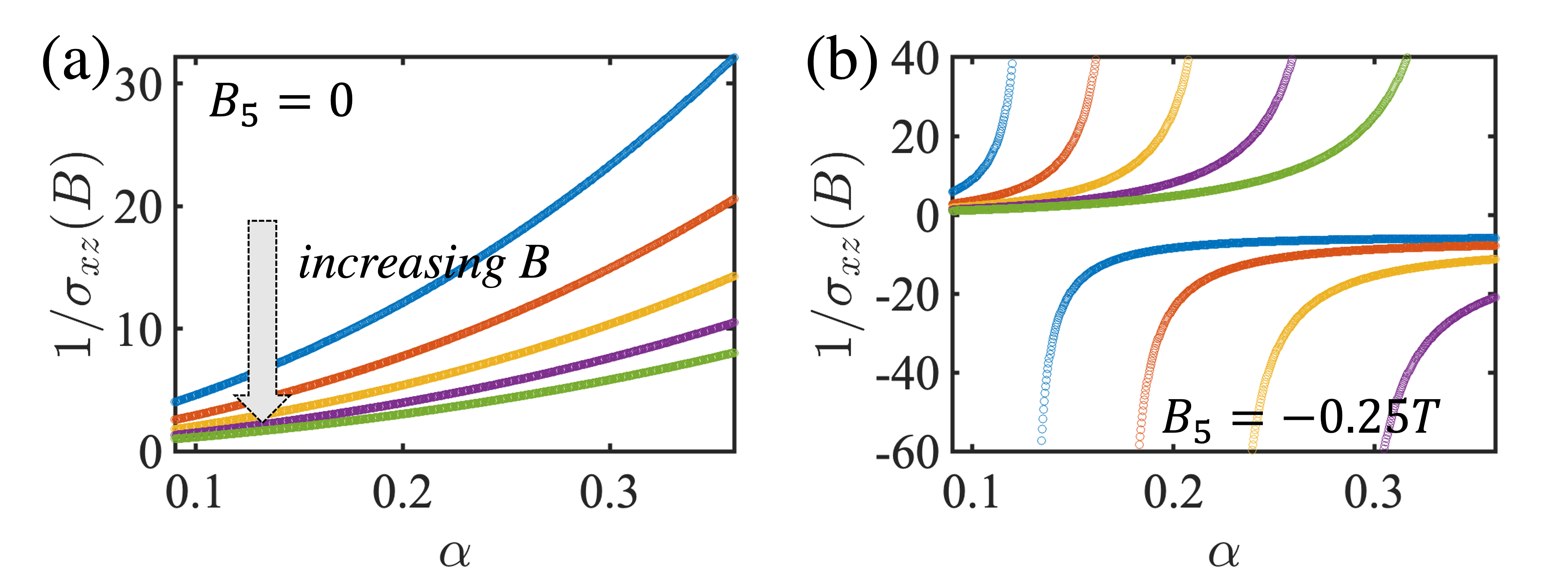}
    \caption{{Inverse of planar Hall conductance} for a minimal model of untilted WSM as a function of intervalley scattering strength. (a) in absence of $B_5$ field. (b) in presence of $B_5$ field. In all the curves, as we go from blue to green, we increase $B$. All the plots are appropriately normalized.}
    \label{fig:twonode_notilt_inv_sxz}
\end{figure}

\subsection{Planar Hall conductance}
Next, we study the effect of the chiral gauge field $B_5$ on the planar Hall conductance. The dependence on the magnetic field is typically quadratic and we may expand the planar Hall conductivity $\sigma_{xz}$ as
\begin{align}
\sigma_{xz}(B)= \sigma_{xz}^{(2)}(B-B_0)^2 + \sigma_{xz}^{(0)},
\label{eq:phc1}
\end{align} 
where $B_0$ is vertex of the parabola, and $\sigma_{xz}^{(2)}$ is the quadratic coefficient. 
The planar Hall conductivity depends on the angle of the applied magnetic field ($\sim \sin 2\gamma$), where $\gamma$ is the angle of the magnetic field with respect to the $x$-axis~\cite{nandy2017chiral}. To study the effect of strain, we first evaluate the planar Hall conductivity in the absence of any external magnetic field. In Fig.~\ref{fig:phc001} we plot the planar Hall conductivity $\sigma_{xz}(B_5)$ that is evaluated in the absence of external magnetic field. The angular behavior with respect to $\gamma_5$ is $\sim\sin 2\gamma_5$ as the case with the usual planar Hall conductivity. Here, we also explicitly examine the effect of intervalley scattering $\alpha$. Even though the conductivity is expected to decrease with increasing scattering, the functional form has still never been explicitly evaluated, especially when the scattering is momentum dependent. We numerically find that the planar Hall conductivity induced by the chiral gauge field  behaves as $\sim 1/\alpha$. 
\begin{figure}
    \centering
    \includegraphics[width=\columnwidth]{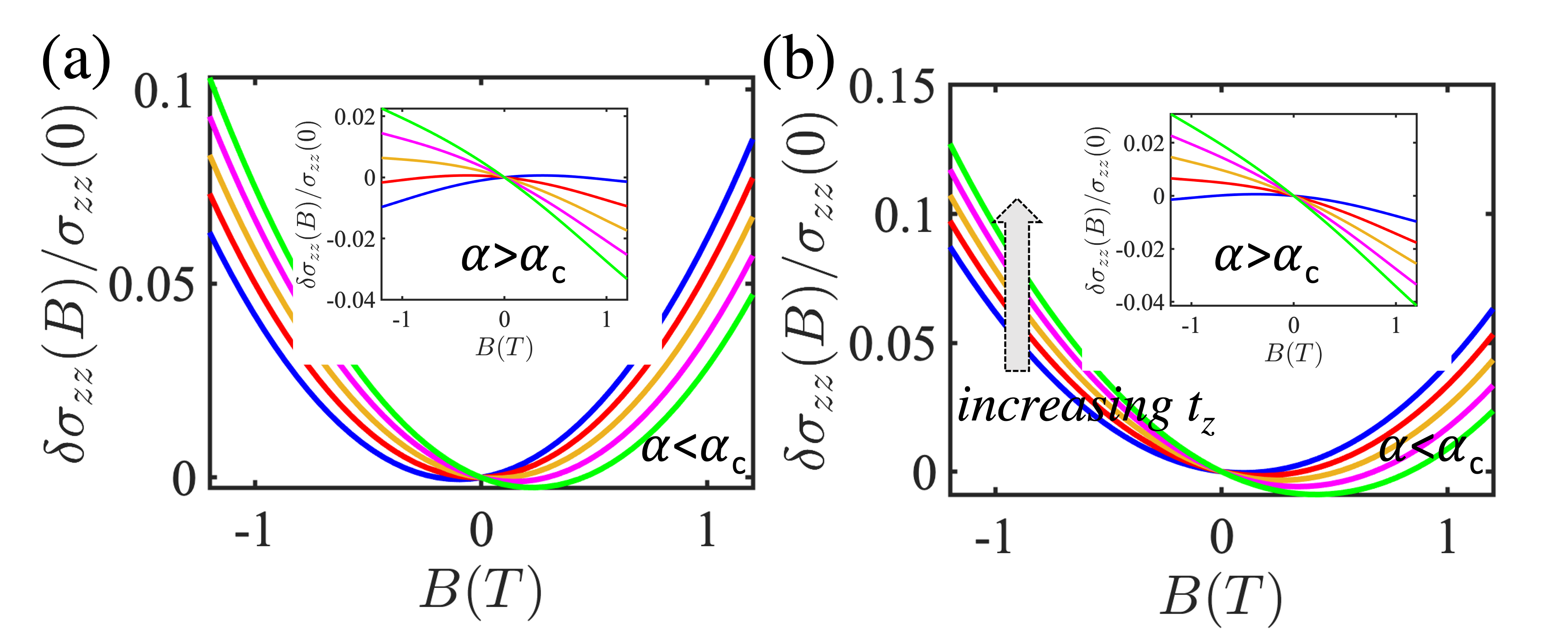}
    \caption{LMC for a tilted TR broken WSM (Eq.~\ref{Eq:HWeyl2}) with $t_z^{1} = -t_z^{(-1)}$. (a) When $B_5=0.1T$. (b) When $B_5=-0.1T$. The inset in both figures is for the case when $\alpha=1.2>\alpha_c$, while in the main figures $\alpha=0.2<\alpha_c$. As we move from blue to the green curve in both the plots, we increase $t_z/v_F$ from 0 to 0.06. The opposing effects and adding effects of strain and tilt are highlighted in (a) and (b) respectively.}
    \label{fig:tiltzLMC1}
\end{figure}
\begin{figure}
    \centering
    \includegraphics[width=\columnwidth]{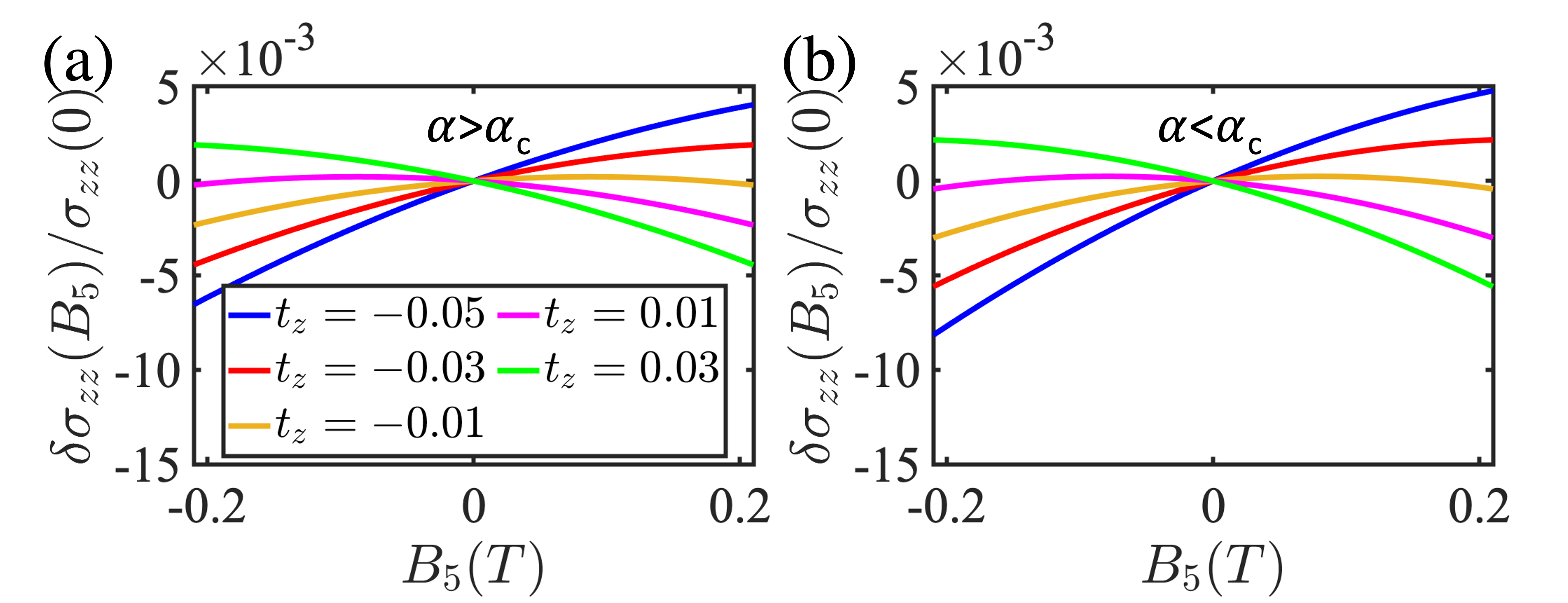}
    \caption{LMC for a tilted TR broken WSM when the tilts are oriented in the same direction. Both weak and strong sign-reversal is observed irrespective of the intervalley scattering strength. The legends are same in both the plots.}
    \label{fig:tiltzsameLMC2}
\end{figure}

We also compare and contrast the behavior of the planar Hall conductivity when (i) external magnetic field is applied and the strain induced field is absent, and (ii) when strain induced field is present but external magnetic field is absent. We find the contribution to the planar Hall conductivity to be different both in sign and magnitude, which is in contrast to earlier claims~\cite{ghosh2020chirality}. Specifically $\sigma_{xz} (B)$ increases with increasing $B$, while $\sigma_{xz} (B_5)$ decreases with increasing $B_5$.
This feature has been highlighted in Fig.~\ref{fig:phc002} (a). In other words, the chiral gauge field, alone, results in strong sign-reversal. We attribute this behavior to the inclusion of intervalley scattering, momentum dependence, as well as charge conservation that have been neglected in earlier works.

Finally, we also study the conductivity in the presence of both the external magnetic field and strain induced chiral magnetic field. In the presence of external magnetic field, the effect of strain is to shift and tilt the conductivity parabola, thereby resulting in weak sign-reversal of the conductivity as shown in Fig.~\ref{fig:phc002} (b). In contrast to the longitudinal magnetoconductivity, PHC never shows strong sign-reversal even on increasing the intervalley scattering above the critical value. However, interestingly, we find that in a certain window of the magnetic field, increasing intervalley scattering strength increases the magnitude of the planar Hall conductivity, which is counter-intuitive. We understand this behavior due to the opposing effects of strain induced PHC and magnetic field induced PHC. As discussed before, both of them individually have opposite and unequal contributions to the planar Hall conductivity. This is better visualized in Fig.~\ref{fig:twonode_notilt_inv_sxz}, where we plot the planar Hall conductivity as a function of the intervalley scattering strength $\alpha$. First, we notice that in the absence of $B_5$-field, the Hall conductivity shows some amount of non-linearity as a function of $1/\alpha$. This is contrasted to Fig.~\ref{fig:phc001}(b) (the case when $B=0$, $B_5\neq 0$) where linear behavior was observed for all ranges of $\alpha$. Second, in the presence of $B_5$ field, the behavior of $\sigma_{xz}$ with respect to $\alpha$ can be strikingly different. Due to the weak sign reversal, $\sigma_{xz}$ can switch sign, which explains the divergences in the plot in Fig.~\ref{fig:twonode_notilt_inv_sxz} (b). Furthermore, we find that when $\sigma_{xz}$ switches sign from positive to negative, the behavior with respect to $\alpha$ becomes anomalous, i.e., increasing $\alpha$, increases the magnitude of $\sigma_{xz}$. Such an anomalous behavior with respect to the intervalley scattering strength is not observed for longitudinal magnetoconductivity.

\subsection{Time-reversal broken WSM with tilt}
Having discussed the physics of strain induced gauge field in a minimal untilted model of Weyl fermions, we now discuss the case when there is a finite tilt in the Weyl cones. 
The Hamiltonian is given by 
\begin{align}
    H = \sum\limits_\chi \sum\limits_\mathbf{k} {\chi\hbar v_F \left( \mathbf{k}\cdot\boldsymbol{\sigma} + t^\chi_z k_z\right)}
    \label{Eq:HWeyl2}
\end{align}

\begin{figure}
    \centering
    \includegraphics[width=\columnwidth]{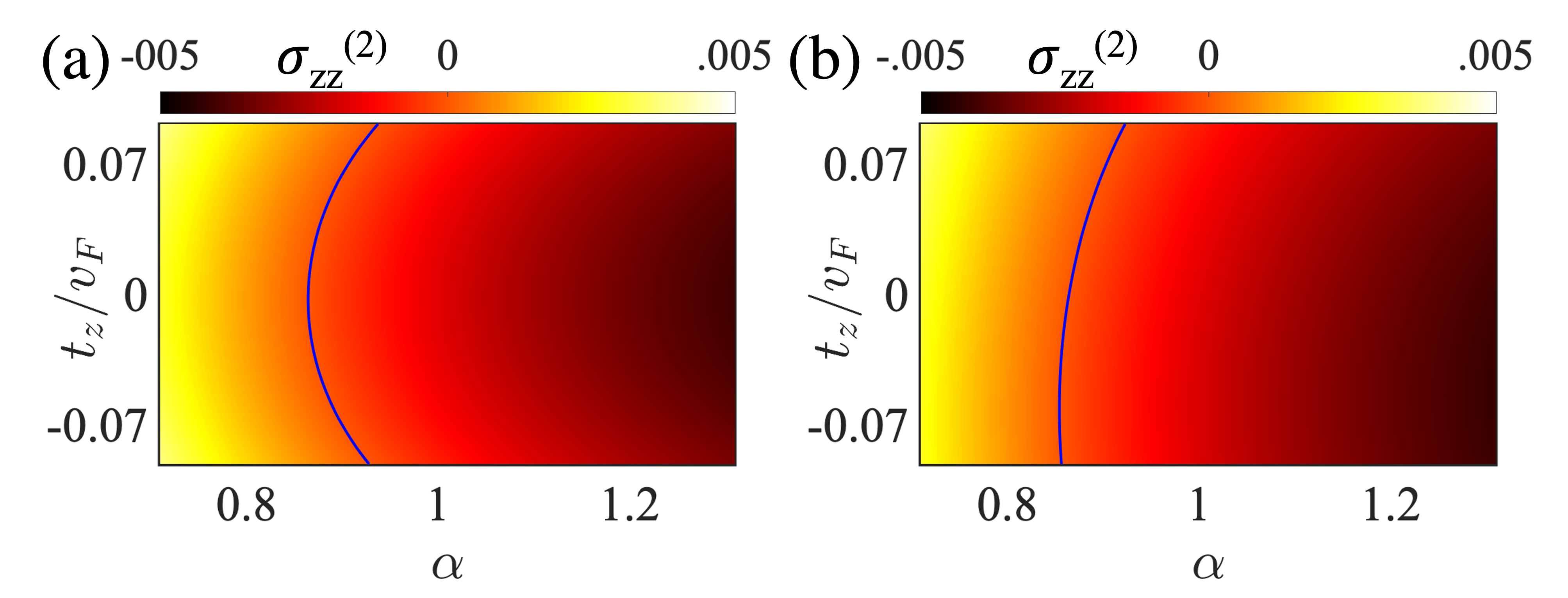}
    \caption{(a) The quadratic coefficient of the longitudinal magnetoconductivity $\sigma_{zz}^{(2)}$ for tilted TR broken WSM. (a) $t_z^{(1)} = -t_z^{(-1)}$. (b) $t_z^{(1)} = t_z^{(-1)}$. Strain induced chiral magnetic field was fixed to $B_5=0.1T$ in both the cases. The blue contour separates the regions when $\sigma_{zz}^{(2)}>0$ and when $\sigma_{zz}^{(2)}<0$ (strong sign-reversal).}
    \label{fig:tiltz_same_opp_lmccolorplot1}
\end{figure}

\begin{figure*}
    \centering
    \includegraphics[width=1\columnwidth]{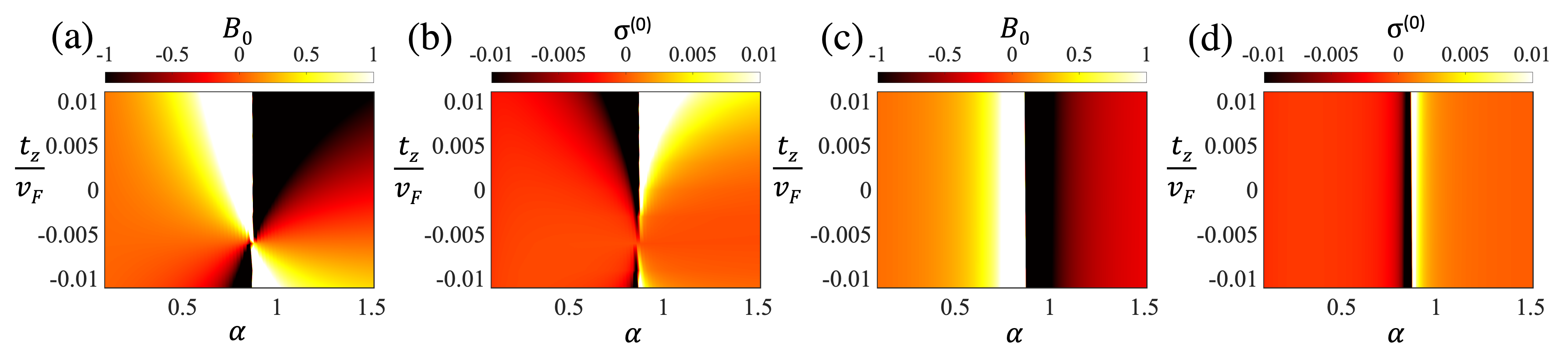}
    \caption{LMC parameters for tilted TR broken WSMs. The center of the parabola $B_0$ (a) and $\sigma^{(0)}$ as a function of the tilt parameter and intervalley scattering strength, in the presence of a fixed value of chiral gauge magnetic field $B_5=0.1T$. The tilts are oriented opposite to each other in plots (a) and (b). The plots (c) and (d) are for the case when the Weyl cone tilts are oriented in the same direction.}
    \label{fig:twonode_tiltz_colorplot_B_sigatB0}
\end{figure*}

Here $t_z$ is the tilting parameter along the $z$-axis. We only focus on the case when $t^\chi_{z}<v_F$, thus restricting ourselves to type-I Weyl semimetals. 
Depending on whether the two cones are tilted along the same or opposite direction, the behavior of both LMC and PHC can be different. In the absence of strain, if the cones are tilted in opposite directions, i.e., $t^\chi_z=-t^{\chi'}_z$, a linear in magnetic field term is added to the overall longitudinal magnetoconductivity, and the parabola is shifted and tilted along a particular direction. In other words, we can say that tilting results in weak sign-reversal, although this has never been explicitly pointed out in  earlier works~\cite{sharma2017chiral, das2019linear, ahmad2021longitudinal}. When the intervalley scattering strength is large, tilting the Weyl cones results in both weak and strong sign-reversal.
In the presence of both tilt and strain, we arrive at a very interesting scenario. Both of these parameters, i.e., $t_z$ and $B_5$, can tilt the LMC parabola either in the same direction or opposite direction, and this depends on the angle between the tilt direction and the strain induced gauge field. In Fig.~\ref{fig:tiltzLMC1} we plot the longitudinal magnetoconductivity for a tilted TR broken WSM presented in Eq.~\ref{Eq:HWeyl2} when the Weyl cones are tilted opposite to each other. Depending on the direction of the strain induced gauge field $B_5$, the effects of tilting and strain can either add up or even cancel out. In Fig.~\ref{fig:tiltzLMC1} (a), $B_5>0$, and the strain and tilting effects work in opposite directions, while in Fig.~\ref{fig:tiltzLMC1} (b), $B_5<0$, and the strain and tilting effects work in the same direction.

\begin{figure}
    \centering
    \includegraphics[width=\columnwidth]{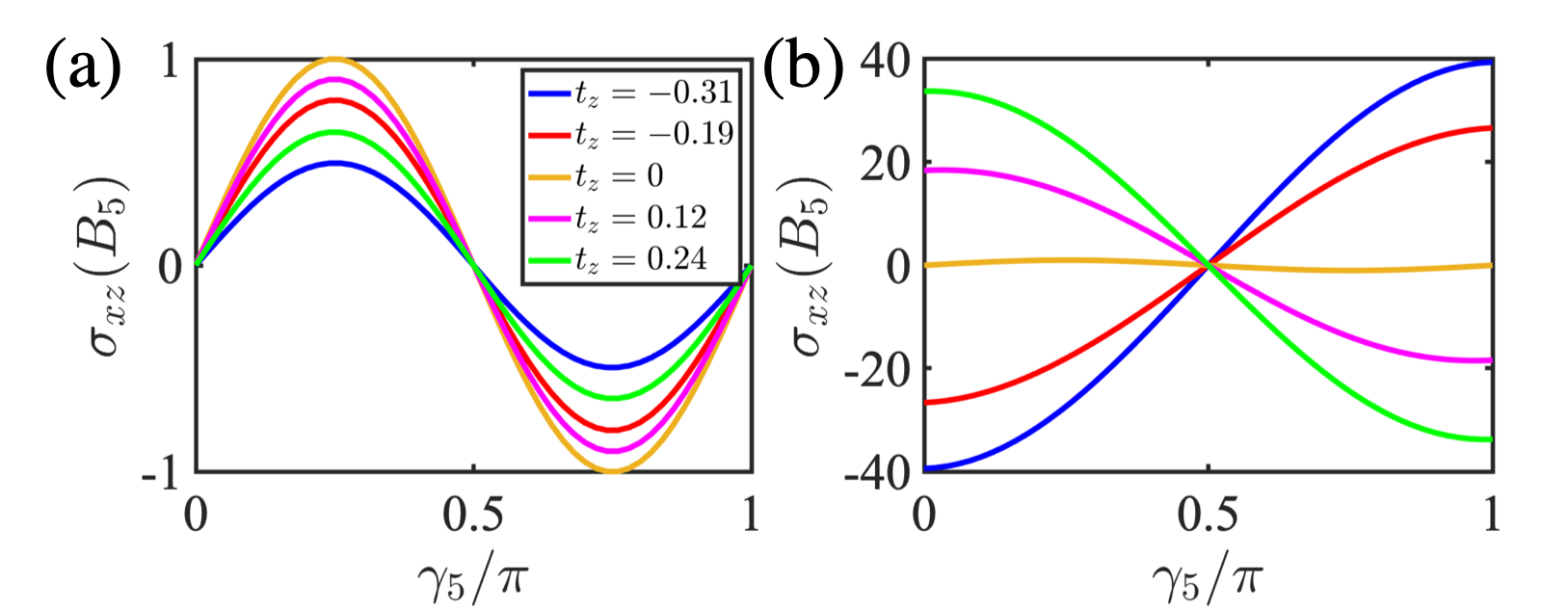}
    \caption{(a) The planar Hall conductance in TR broken tilted WSM as a function of the angle $\gamma_5$ when (a) the cones are tilted along opposite direction, and (b) cones are oriented along the same direction. The legends are the same in both the plots. Both plots are appropriately normalized such that the yellow curve is identical in both the figures as expected.}
    \label{fig:twonodes_tiltz_opp_same_phc_vs_gm5}
\end{figure}

\begin{figure}
    \centering
    \includegraphics[width=\columnwidth]{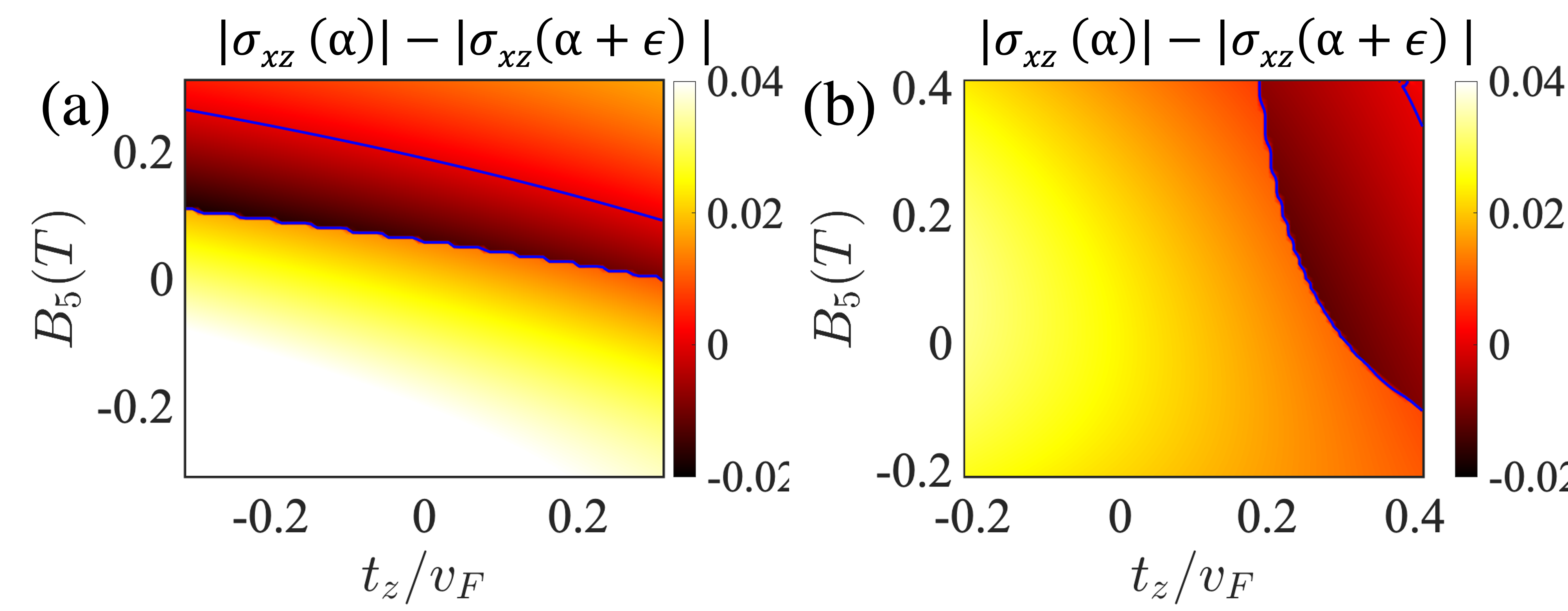}
    \caption{Change in the magnitude of the planar Hall conductivity ($|\sigma_{xz}(\alpha)|-|\sigma_{xz}(\alpha+\epsilon)|$) for a tilted TR broken WSM (Eq.~\ref{Eq:HWeyl2}) on infinitesimally increasing in the scattering strength (by $\epsilon$). (a) the Weyl cones are tilted in opposite direction. (b) the Weyl cones are tilted in the same direction. In the region enclosed within blue contours, we find anomalous behavior of conductivity with the scattering strength, i.e., the magnitude of the conductivity increases on the increase of scattering strength. We choose $\alpha=0.5$, and $\epsilon=0.01$.}
    \label{fig:twonodes_tiltz_opp_same_color_deltaphc}
\end{figure}

\begin{figure*}
    \centering
    \includegraphics[width=1\columnwidth]{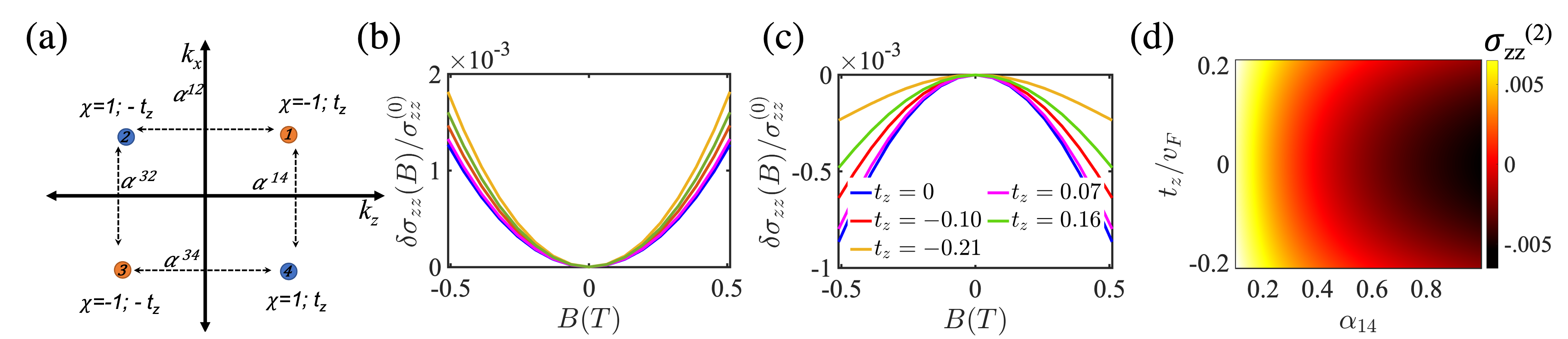}
    \caption{(a) Schematic of Weyl nodes in a prototype model of an inversion asymmetric Weyl semimetal. Here $\chi$ is the chirality, $t_z$ is the tilt, and $\alpha^{ij}$ are scattering rates from node $i$ to node $j$. (b) LMC as a function of magnetic field when the intervalley scattering rates are less than the critical value. (c) LMC as a function of magnetic field when the intervalley scattering rates are above the critical value. The legends in (b) and (c) are identical. (d) $\sigma_{zz}^{(2)}$ for a fixed value of $\alpha_{12}=0.19$. Plots (b), (c), and (d) are in the absence of strain, i.e., $B_5=0$.}
    \label{fig:fournodelmc1}
\end{figure*}
\begin{figure*}
    \centering
    \includegraphics[width=1\columnwidth]{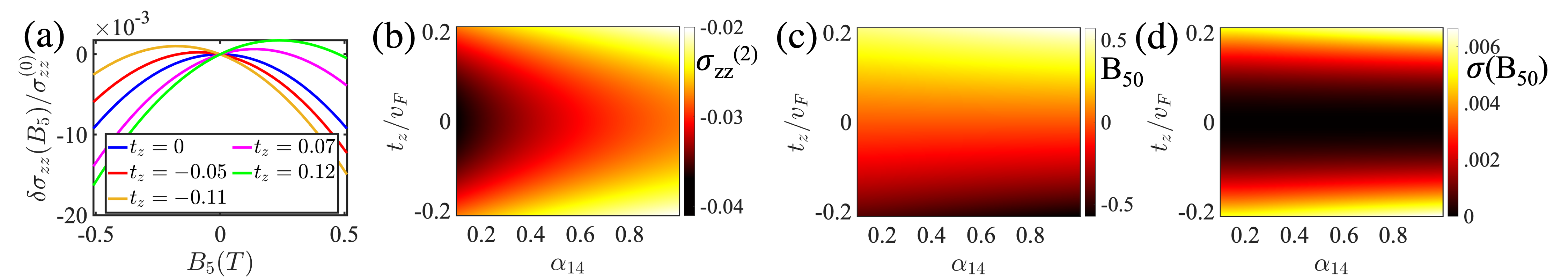}
    \caption{LMC for inversion asymmetric Weyl semimetal in the presence of strain induced chiral magnetic field ($B_5$) but absence of magnetic field. (a) A finite tilt can result in weak sign-reversal. The plot is for a fixed value of $\alpha_{12}=0.4$, but the qualitative behavior is independent of scattering strength. (b), (c), and (d) plot the parameters $\sigma_{zz}^{(2)}$, $B_{50}$, and $\sigma(B_{50})$ as a function of parameters $\alpha_{14}$ and $t_z$. We fixed $\alpha_{12}=0.19$.}
    \label{fig:fournodelmc2}
\end{figure*}
\begin{figure}
    \centering
    \includegraphics[width=\columnwidth]{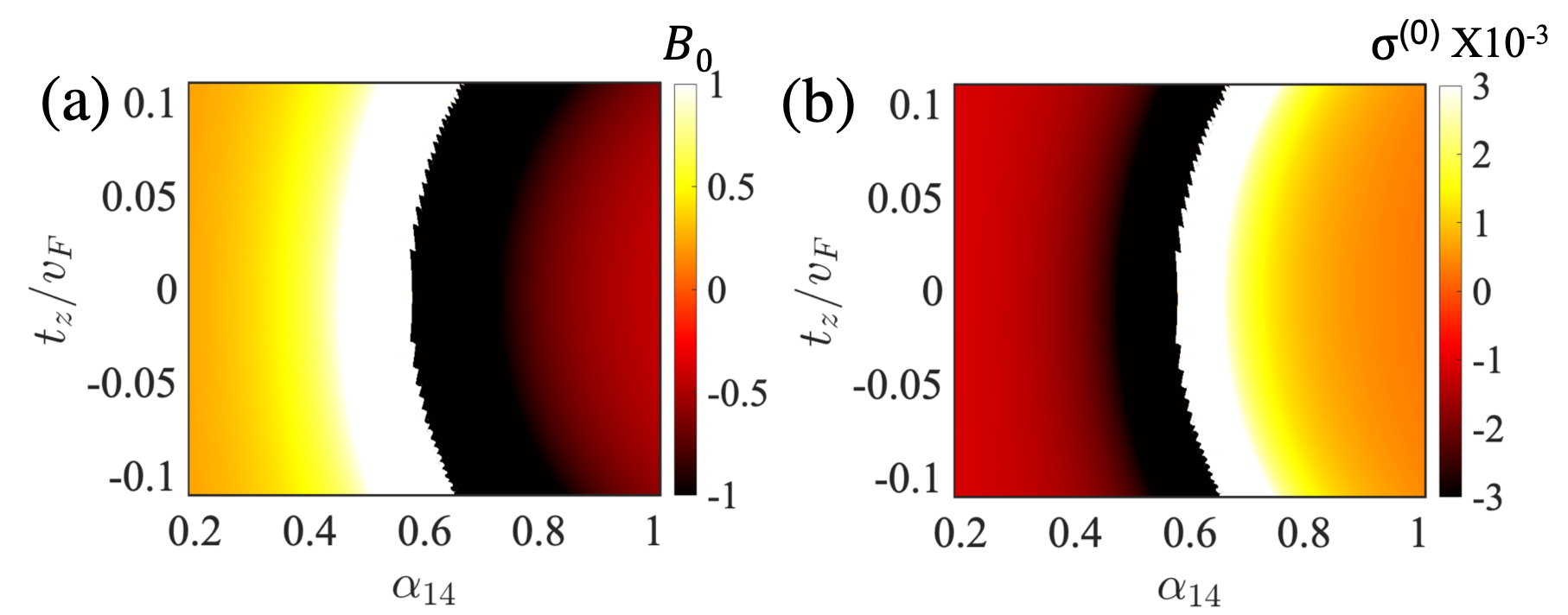}
    \caption{The parameters $B_0$ (a) and $\sigma^{(0)}$ (b) for inversion asymmetric Weyl semimetals (Eq.~\ref{eq_H4nodes}). We have fixed $\alpha_{12}=0.3$, $B_5=0.1T$. Weak sign reversal is not observed and strong sign-reversal occurs at $\alpha_{14}=\alpha_{14c}(t_z)$. }
    \label{fig:fournodes_lmc_colorplot_B0_sigatB0_1}
\end{figure}
\begin{figure*}
    \centering
    \includegraphics[width=1\columnwidth]{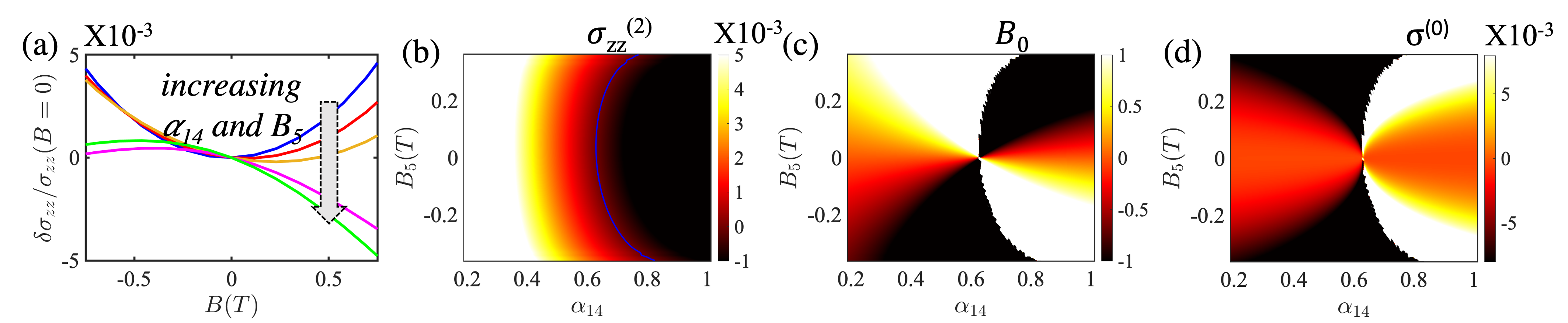}
    \caption{(a) LMC for inversion asymmetric Weyl semimetal. As we move from the blue to the green curve, we simultaneously increase $B_5$ as well as $\alpha_{14}$. Both weak and strong sign-reversal are exhibited. The plots (b), (c), and (d) plot the parameters $\delta\sigma_{zz}^{(2)}$, $B_0$, and $\sigma^{(0)}$ for fixed $\alpha_{12}$ and $t_z\neq 0$. The blue contour in plot (b) separates the phases where $\sigma_{zz}^{(2)}$ changes sign. Again, we see signatures of both weak and strong sign-reversal. The tilt parameter is fixed to $t_z/v_F=-0.1$.}
    \label{fig:fournodes_lmc_colorplot_varyB5_vary_a14}
\end{figure*}

\begin{figure}
    \centering
    \includegraphics[width=\columnwidth]{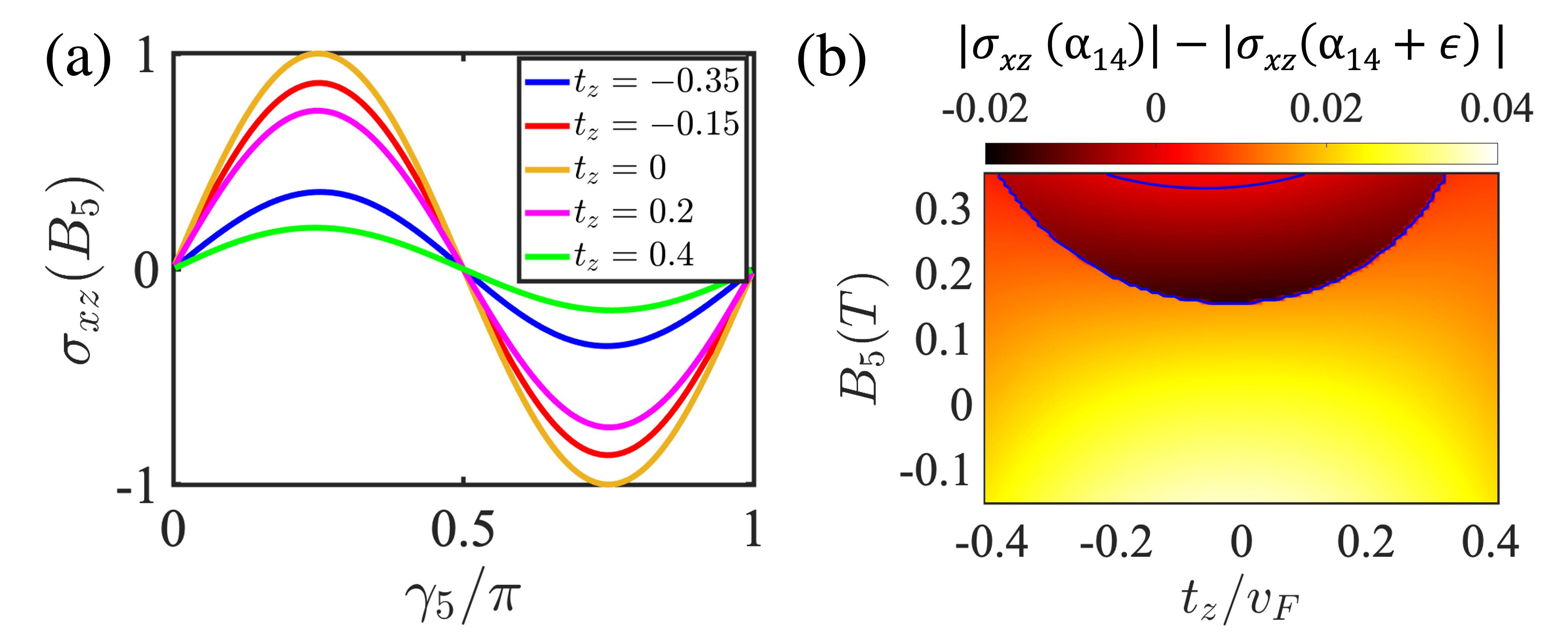}
    \caption{Planar Hall conductance for inversion asymmetric Weyl semimetal. (a) PHC as a function of $\gamma_5$, when $B=0$, and $B_5\neq 0$. (b) The change in the magnitude of the planar Hall conductivity on increasing $\alpha_{14}$ infinitesimally. In the region between the blue contours, we observe anomalous increase in conductivity. Here we fix, $B=1T$, $\alpha_{12}=0.4$, $\alpha_{14}=0.5$, and $\epsilon=0.01$. }
    \label{fig:fournodes_sxz_colorplot}
\end{figure}

In the absence of strain it is known that when the nodes are oriented along the same direction ($t^1_z = t^{-1}_z$), the linear component of the longitudinal magnetoconductivity does not survive as the contributions from both nodes cancel out~\cite{sharma2017chiral, das2019linear, ahmad2021longitudinal}. Hence, as expected, only strong sign-reversal is observed as a function of intervalley scattering strength. Now, in the presence of only strain induced field, such cancellation does not occur and one observes weak sign-reversal as a function of the tilt parameter. Furthermore, in the presence of $B_5$-field and absence of external magnetic field, we observe both strong and weak sign-reversal. To illustrate this, in Fig.~\ref{fig:tiltzsameLMC2} we plot LMC for a tilted TR broken WSM when the tilts are oriented in the same direction. When both magnetic field and strain induced chiral magnetic field are present, the combination of two can give rise to interesting features. In Fig.~\ref{fig:tiltz_same_opp_lmccolorplot1} we plot the quadratic coefficient $\sigma_{zz}^{(2)}$ as a function of both tilt and intervalley scattering strength in the presence of a $B_5$ field. We note that the presence of the tilt parameter curves the contour $\alpha_c$ separating the two strong sign-reversed regions, i.e., $\alpha_c = \alpha_c(t_z)$. The curvature is different when the Weyl cones are oriented opposite to each other or oriented along the same direction. 

Similarly, very striking features are observed for the parameters $B_0$ (the vertex of the parabola) as well as $\sigma_{zz}^{(0)}$. We demonstrate this in Fig.~\ref{fig:twonode_tiltz_colorplot_B_sigatB0}. We fix strain induced gauge field to be around $B_5=0.1T$. Let us first focus on the case when the Weyl cones are oriented opposite to each other.  When $\alpha<\alpha_c$, the sign of $B_0$ changes continuously from negative to positive as $t_z$ is varied from negative to positive. On the other hand, when $\alpha>\alpha_c$, the sign of $B_0$ changes from positive to negative as $t_z$ is varied from negative to positive. The effects of strain and tilt and strain can either add up or cancel out and the combination can tilt the parabola overall to the left or to the right resulting in weak sign-reversal. 
This is  demonstrated in the color plot  in Fig.~\ref{fig:twonode_tiltz_colorplot_B_sigatB0} (a). When $\alpha>\alpha_c$, the sign of $B_0$ changes discontinuously (feature of strong sign reversal).
Now, since weak sign-reversal does not change the sign of $\sigma^{(0)}$, we do not see a sign change in $\sigma^{(0)}$ as one varies the tilt for a given value of $\alpha$. The sign change in $\sigma^{(0)}$ only occurs as a result of strong sign-reversal (Fig.~\ref{fig:twonode_tiltz_colorplot_B_sigatB0} (b)). Now, when the cones are oriented along the same direction, the linear component arising from the tilt is canceled out and hence we do not observe any change in $B_0$ or $\sigma^{(0)}$ by varying the tilt. The only change occurs at $\alpha=\alpha_c$ due to strong sign-reversal. This is highlighted in Figs.~\ref{fig:twonode_tiltz_colorplot_B_sigatB0} (c) and (d). 

Next we discuss the strain induced planar Hall effect for tilted TR broken Weyl semimetals. 
When the cones are oriented along the opposite directions we observe a $\sim\sin 2\gamma_5$ behavior and the effect of the tilt is only quantitative, and so is the effect of varying intervalley scattering strength. On the other hand, when the cones are oriented along the same direction, the behavior changes to $\sim\sin\gamma_5$. Changing the tilt parameter can switch the sign of the planar Hall conductance as well, and result in qualitative changes in the behavior while changing the intervalley scattering strength only changes the overall magnitude. We demonstrate these features in Fig.~\ref{fig:twonodes_tiltz_opp_same_phc_vs_gm5}. 

Finally, we discuss the behavior of conductivity on changing the intervalley scattering strength $\alpha$. In Fig.~\ref{fig:twonodes_tiltz_opp_same_color_deltaphc}, we plot the change in the magnitude of the planar Hall conductivity ($|\sigma_{xz}(\alpha)|-|\sigma_{xz}(\alpha+\epsilon)|$) for an  infinitesimal increase in the scattering strength (by a small amount $\epsilon$). In both cases, i.e., when the Weyl cones tilted in opposite direction, and when the Weyl cones are tilted in the same direction, we find regions in the $B_5-t_z$ space where anomalous behavior of the Hall conductivity is observed, i.e., the magnitude of conductivity increases on increasing the intervalley scattering strength. We had already seen this behavior for untilted WSM as well (Fig.~\ref{fig:twonode_notilt_inv_sxz}), and here we calculate its dependence on the tilting of the Weyl cones. 
Before closing this section, we point out that in experiments where strain can be applied and manipulated on the inhomogeneous samples can test the above predictions.

\section{Inversion asymmetric Weyl semimetals}
Having discussed the effect of strain in time-reversal broken WSMs we now move on to the case of inversion asymmetric WSMs. To this end, we will restrict our attention to the following minimal model for an inversion asymmetric WSM that consists of four nodes as dictated by symmetry considerations:
\begin{align}
    H = \sum\limits_{n=1}^4 \left(\chi_{n}\hbar v_F \mathbf{k}\cdot\boldsymbol{\sigma} + \hbar v_F t_z^n k_z \sigma_0\right).
    \label{eq_H4nodes}
\end{align}
The system consists of four Weyl nodes located at the points $\mathbf{K}=(\pm k_0,0,\pm k_0)$ in the Brillouin zone. In Eq.~\ref{eq_H4nodes}, $\chi_n$ is the chirality, and we are also introducing the parameter $t_z^n$, that represents the tilting of the Weyl cone. The Weyl cones are assumed to be tilted only along the $z$ direction. Specifically, $(1,t_z)$=$(\chi_1,t_z^{(1)})=(-\chi_2,t_z^{(2)})=(\chi_3,-t_z^{(3)})=(-\chi_4,-t_z^{(4)})$, such that inversion symmetry is broken. The tilt parameter $t_z$ is considered to be less than unity. Fig.~\ref{fig:fournodelmc1}(a) plots the schematic diagram of this prototype inversion asymmetric Weyl semimetal. 
Specifically, we must consider four intranode scattering channels (node $n\Longleftrightarrow n$) and four internode scattering channels (node $n\Longleftrightarrow [n+1] \text{mod } 4$). The dimensionless scattering strength between node $m$ and node $n$ is denoted as $\alpha^{mn}$. For simplicity, we ignore the scattering between nodes (4 $\Longleftrightarrow$ 2) and nodes (1 $\Longleftrightarrow$ 3) since they involve a large momentum transfer compared to others. The four internode scatterings can be divided into two categories: (i) scattering between Weyl cones of opposite chirality and opposite tilt orientation (1 $\Longleftrightarrow$ 2) and (3 $\Longleftrightarrow$ 4), and (ii) scattering between Weyl cones of opposite chirality and same tilt orientation (1 $\Longleftrightarrow$ 4) and (2 $\Longleftrightarrow$ 3). Since both these categories result in different behaviors, it is of interest to see the interplay between the two. 
We first examine the behavior of longitudinal magnetoconductivity in the absence of any strain. Earlier, we examined that for a system of only two tilted cones (of opposite chirality), `weak' sign-reversal is possible only if the cones are oriented opposite to each other. However, in the current case, `weak' sign-reversal generated by internode scattering channel (1 $\Longleftrightarrow$ 2) is exactly cancelled by scattering channel (4 $\Longleftrightarrow$ 3). Second, the scattering (1 $\Longleftrightarrow$ 4) and (2 $\Longleftrightarrow$ 3) do not cause weak sign reversal as they involve Weyl cones with the same tilt. Therefore, in the absence of $B_5$ field, weak sign-reversal is not observed for the case of an inversion asymmetric WSM. 
In Fig.~\ref{fig:fournodelmc1} we plot longitudinal magnetoconductivity  for the inversion asymmetric Weyl semimetal (Eq.~\ref{eq_H4nodes}) in the absence of strain induced chiral gauge field $B_5$. As discussed, we do not observe any signature of weak sign-reversal, and there is only strong sign-reversal when $\alpha_{12}$ and/or $\alpha_{14}$ are large enough. Increasing tilt does not qualitatively change the behavior and increasing the magnitude of the  tilt in either direction is  only seen to increase the magnitude of magnetoconductivity. 

Next, we study the behavior in the absence of external magnetic field but in presence of strain induced gauge field $B_5$. First, similar to the case with TR broken Weyl semimetals, we find that strain induced chiral magnetic field $B_5$ always results in a negative LMC coefficient $\sigma_{zz}^{(2)}$. This results in in contradiction to earlier claims that find an increase in longitudinal magnetoconductivity with strain~\cite{grushin2012consequences,ghosh2020chirality}. The reason can be traced out to the non-inclusion of intervalley scattering, momentum dependent scattering, and charge conservation, all of which are included in the current work (see Appendix A).
Furthermore, we find that strain, by itself results in strong sign-reversal, while tilting results in weak sign reversal.
In Fig.~\ref{fig:fournodelmc2} (a) we plot LMC as a function of strain induced magnetic field $B_5$, which clearly demonstrates these features.
As before, we fit the magnetoconductivity via the following expression 
\begin{align}
\sigma_{zz}(B_5)= \sigma_{zz}^{(2)}(B-B_{50})^2 + \sigma_{zz}(B_{50}),
\label{eq:sigma_25}
\end{align}
where the slope of the conductivity $\sigma_{zz}^{(2)}$ is always found to be negative irrespective of the value of tilt, strain, intervalley scattering strengths across either nodes. The center of the parabola ($B_{50}$) directly correlates with the tilt parameter $t_z$. Depending on the sign of $t_z$, $B_{50}$ can be either positive or negative. The parameter $B_{50}$ is also found to have dependence on the scattering strength, but this dependence is relatively weak compared to the dependence on $t_z$.  In Figs.~\ref{fig:fournodelmc2} (b), (c), and (d), we plot the parameters $\sigma_{zz}^{(2)}$, $B_{50}$, and $\sigma_{zz}(B_{50})$ as a function of $\alpha_{14}$, and $t_z$, keeping $\alpha_{12}$ fixed, and $B=0$. No sharp discontinuities are observed in the parameters since the system is already in strong sign-reversed state.

In inversion asymmetric inhomogeneous Weyl semimetals, interesting effects can occur as a result of the interplay between the strain induced chiral gauge field, external magnetic field, and the tilt parameter. To study the same, we examine LMC as a function of external magnetic field for a fixed value of chiral gauge field, and use Eq.~\ref{eq:sigma_2} to evaluate the fit parameters $B_0$, $\sigma_{zz}^{(2)}$, and $\sigma_{zz}^{(0)}$. We do not find a signature weak sign-reversal, and only strong sign-reversal occurs when the intervalley scattering $\alpha_{14}>\alpha_{14c}$, where $\alpha_{14c}$ now is a function of tilt parameter.  Around $\alpha=\alpha_{14c}(t_z)$ we find a sharp change in the sign of the parameters $B_0$ and $\sigma_{zz}^{(0)}$ that corresponds to a continuous change of sign in $\sigma_{zz}^{(2)}$ as well. 
It is worthwhile pointing that by identifying the parameters $B_0$ and $\sigma_{zz}^{(0)}$ from the experimentally measured conductivity, their signs may help identify the dominant scattering mechanisms in the system, i.e., either internode or intranode scattering, and also provide us insight about the strain in the samples as well as the tilting if the Weyl cones.

Experimentally, one may also study LMC in inversion asymmetric Weyl semimetals by tuning the amount of strain in the system. Therefore it is of interest to study the effect of varying strain on LMC. 
In Fig.~\ref{fig:fournodes_lmc_colorplot_varyB5_vary_a14} (a) we plot $\delta\sigma_{zz} = \sigma_{zz}(B) - \sigma_{zz}(B=0)$ simultaneously varying the intervalley scattering strength $\alpha_{14}$ as well as the strain induced chiral gauge field $B_5$. We see signatures of both weak and strong sign-reversal. Increasing $\alpha$ beyond $\alpha_c$ results in strong sign-reversal, while change in the tilt parameter results in weak sign-reversal.
We fix the value of $\alpha_{12}$, and evaluate the fit parameters of $\sigma_{zz}(B)$ from Eq.~\ref{eq:sigma_2}. 
Fig.~\ref{fig:fournodes_lmc_colorplot_varyB5_vary_a14} (b) plots $\sigma_{zz}^{(2)}$ as a function of $B_5$ and $\alpha_{14}$. The contour $\alpha_{14c}$ where $\sigma_{zz}^{(2)}$ switches sign shows a dependence on $B_5$ as well. Therefore the contour $\alpha_{c}$ is in general a function of both $t_z$ and $B_5$. Fig.~\ref{fig:fournodes_lmc_colorplot_varyB5_vary_a14} (c) and (d) plot the parameters $B_0$ and $\sigma_{zz}^{(0)}$ obtained from Eq.~\ref{eq:sigma_2}, both of which display very interesting behavior as a result of varying $B_5$ and $\alpha_{14}$. In Fig.~\ref{fig:fournodes_lmc_colorplot_varyB5_vary_a14} (c), when $\alpha<\alpha_c(B_5)$, the sign of $B_0$ changes from negative to positive as $B_5$ changes sign from negative to positive. When $\alpha>\alpha_c(B_5)$, the change of sign is from positive to negative. At $\alpha=\alpha_c(B_5)$, there is strong sign-reversal resulting in sharp contrasting features on the both sides of $\alpha_c(B_5)$. On the other hand, in Fig.~\ref{fig:fournodes_lmc_colorplot_varyB5_vary_a14} (d), $\sigma_{zz}^{(0)}$ does not change sign as $B_5$ changes sign, but like $B_0$, it displays striking behavior around $\alpha_{c}(B_5)$ due to strong sign-reversal.  

Before closing this section, we also comment on the planar Hall effect in inversion asymmetric Weyl semimetals. Fig.~\ref{fig:fournodes_sxz_colorplot} (a) plots the planar Hall conductivity $\sigma_{xz}$ as a function of the angle $\gamma_5$ in the absence of an external magnetic field and presence of strain induced gauge field $B_5$. The PHC behaves as $\sim \sin(2\gamma_5)$ as in Fig.~\ref{fig:twonodes_tiltz_opp_same_phc_vs_gm5} (a). The contribution from the two time-reversed and opposite tilt Weyl node pairs adds up, while the contribution from two time-reversed and same tilt Weyl node pairs cancels out, and that is why we do not get a $\sim\sin(\gamma_5)$ trend as in Fig.~\ref{fig:twonodes_tiltz_opp_same_phc_vs_gm5} (b). In Fig.~\ref{fig:fournodes_sxz_colorplot} (b), we plot the change in the magnitude of the planar Hall conductivity upon infinitesimally increasing the intervalley strength $\alpha_{14}$. We again notice a region in the $B_5-t_z$ space where the variation of conductivity is anomalous, i.e. increasing intervalley scattering increases the magnitude of the conductivity. A similar plot is observed when we instead fix $\alpha_{14}$ and vary $\alpha_{12}$, therefore we do not explicitly plot this here. 

\section{Conclusions}
The sign of longitudinal magnetoconductivity in Weyl semimetals due to chiral anomaly has been a subject of intense research~\cite{spivak2016magnetotransport, das2019linear,imran2018berry,kim2014boltzmann, dantas2018magnetotransport,johansson2019chiral,das2019berry,cortijo2016linear,zyuzin2017magnetotransport,knoll2020negative,sharma2020sign,ahmad2021longitudinal,sharma2022revisiting, sharma2017chiral}. Almost unanimously, the sign of longitudinal magnetoconductivity has been agreed upon to be positive, at least in the limit of weak magnetic fields. However, various factors, such as tilting of the Weyl cones, strain and inhomogeneties in the material, qualitatively affect the LMC in Weyl semimetals. The interplay between various parameters, such as intervalley scattering, tilt, strain induced chiral gauge field, and the external magnetic field, leads to many striking features in both the longitudinal magnetoconductance and the planar Hall conductance of Weyl semimetals, which has been the focus of this work.  

In this work, we first show that the conventional method of assigning sign to magnetoconductivity, i.e., comparing the magnitude of conductivity for field $B$ with $B\pm\epsilon$ ($\epsilon$ being arbitrary),  leads to ambiguities when the system is subjected to strain. Specifically, the sign of magnetoconductivity could depend on the direction of the magnetic field. 
Thus there is a necessity to define weak sign-reversal and strong sign-reversal, both of which are qualitatively different, and result in qualitatively different responses. Weak sign-reversal, in general, leads to smooth changes in the fit parameters of the conductivity, while strong sign-reversal leads to very sharp changes. Weak sign-reversal is specifically is characterized by a change in the vertex and the axis of the parabola of conductivity with respect to the magnetic field, while strong sign-reversal is characterized by an opposite orientation, i.e., the direction in which the parabola opens is reversed.
Broadly speaking: (i) when strain induced chiral gauge field is absent and external magnetic field is present, strong intervalley scattering results in strong-sign reversal, (ii) when chiral gauge field is present and magnetic field is absent, the system, by default, shows strong sign-reversed state for both weak and strong intervalley scattering, (iii) when both chiral gauge and external magnetic field are present, there is both weak and strong sign-reversal. The latter is also experimentally the most relevant scenario, and we show that it leads to very striking phase plots that can be explored experimentally in current and upcoming experiments in Weyl semimetals. In practice, the parameters could be evaluated by fitting the conductivity from the experiments and that could give us insight into the strain, tilt, and dominant scattering mechanism in the system. We have also studied the effect of strain on the planar Hall conductance.
Another striking feature of anomalous variation of the planar Hall conductivity is also unraveled due to the rich interplay between the chiral gauge and external magnetic field, where the magnitude of conductivity can increase on increasing scattering strength.

{This study acknowledges the limitations of the results and proposes additional experiments to address these shortcomings.  The theoretical model posits idealized, spatially uniform strain-induced fields that produce homogeneous axial magnetic fields ($B_5$). However, real experimental conditions frequently present non-uniform strain profiles resulting from defects, sample geometry, or strain relaxation effects.  Experimental implementations of $B_5$ fields have been demonstrated across multiple platforms.  In Dirac and Weyl semimetals like $Cd_3As_2$, $TaAs$, and $WTe_2$, controlled strain has been applied via uniaxial or biaxial pressure through methods such as piezoelectric stacks, substrate bending techniques, and thermal expansion mismatch on patterned substrates \cite{pikulin2016chiral,cortijo2015elastic}.  Furthermore, focused ion beam (FIB) sculpting and strain gradient engineering in hetero-structures have facilitated inhomogeneous yet adjustable strain distributions that resemble axial gauge fields. Strain magnitudes on the order of $0.1\% - 1\%$ lattice deformation ($\sim 10^{-3} - 10^{-2}$) have been demonstrated to generate measurable pseudo-magnetic fields in graphene and topological semimetals, which correspond to effective axial magnetic fields in the range of $B_5 \sim 0.1 - 1$ Tesla.  Experimentalists should anticipate that, although perfect uniformity may not be attainable, the realistic strain magnitudes are adequate to capture qualitative characteristics, including the sign reversal induced by chiral anomalies in longitudinal magnetoconductance (LMC) and alterations in the planar Hall effect (PHE).  Strain gradients can result in spatial averaging of $B_5$, which may diminish the magnitude of observed effects without eliminating their occurrence.  Consequently, our theoretical predictions are both robust and amenable to experimental testing within realistic material constraints and fabrication conditions.}

\setcounter{equation}{0}
\setcounter{table}{0}
\setcounter{figure}{0}

\chapter{\label{chap4}Chiral anomaly-induced nonlinear Hall effect in three-dimensional chiral fermions}
{\small The contents of this chapter have appeared in ``\textsc{Chiral anomaly-induced nonlinear Hall effect in three-dimensional chiral fermions}"; Azaz Ahmad, Gautham Varma K, and Gargee Sharma; \textit{Phys. Rev. B} \textbf{111}, 035138 (2025).}\vspace{0.5in}
\section{Abstract}
Chiral fermionic quasiparticles emerge in certain quantum condensed matter systems such as Weyl semimetals, topological insulators, and spin-orbit coupled noncentrosymmetric metals.  Here, a comprehensive theory of the chiral anomaly-induced nonlinear Hall effect (CNLHE) is developed for three-dimensional chiral quasiparticles, advancing previous models by rigorously including momentum-dependent chirality-preserving and chirality-breaking scattering processes and global charge conservation. Focusing on two specific systems--Weyl semimetals (WSMs) and spin-orbit coupled non-centrosymmetric metals (SOC-NCMs), we uncover that the nonlinear Hall conductivity in WSMs shows nonmonotonic behavior with the Weyl cone tilt and experiences a `strong-sign-reversal' with increasing internode scattering, diverging from earlier predictions. For SOC-NCMs, where nonlinear Hall conductivity has been less explored, we reveal that unlike WSM, the orbital magnetic moment alone can drive a large CNLHE with distinctive features: the CNLH conductivity remains consistently negative regardless of interband scattering intensity and exhibits a quadratic dependence on the magnetic field, contrasting the linear dependence in WSMs. Furthermore, we discover that in SOC-NCMs the spin Zeeman coupling of the magnetic field acts like an effective tilt term which can further enhance the CNLH current. These findings offer fresh insights into the nonlinear transport dynamics of chiral quasiparticles and can be verified in upcoming experiments on such materials. 
\section{Introduction}
The concept of chiral particles originates from high-energy physics~\cite{peskin1995introduction}. 
While electrons, protons, and neutrons have chiral aspects in their interactions and internal structure, they are not fundamentally chiral particles because of their finite mass. On the other hand, the existence of massless chiral fermions is now well-established in condensed matter systems~\cite{hasan2021weyl}. They emerge in certain materials as quasiparticles exhibiting behavior analogous to the theorized chiral fermions in particle physics. Two prominent examples of these materials are topological insulators (TIs)~\cite{hasan2010colloquium,qi2011topological} and Weyl semimetals (WSMs)~\cite{hosur2013recent,armitage2018weyl}. TIs have gapped bulk states, while their boundary states are massless and chiral. 
In contrast, WSMs have gapless bulk chiral states (Weyl fermions) that are topologically protected by a non-vanishing Chern number, which is equivalent to the chirality quantum number. 
Nielsen \& Ninomiya, who first studied the regularization of Weyl fermions on a lattice, showed that they must occur in pairs of opposite chiralities~\cite{nielsen1981no,nielsen1983adler}, thus leading to the conservation of both chiral charge and global charge in absence of any gauge fields. Probing the chirality of the emergent Weyl fermions in WSMs has been of utmost theoretical and experimental interest since the past decade~\cite{Yan_2017,hasan2017discovery,burkov2018weyl,ong2021experimental,nagaosa2020transport,lv2021experimental,ahmad2024geometry,mandal2022chiral}. 

In the presence of external electromagnetic fields, the chiral charge is not conserved, which is the celebrated chiral anomaly (CA) or the Adler-Bell-Jackiw (ABJ) anomaly~\cite{adler1969axial} of Weyl fermions. The non-conservation of chiral charges leads to an anomaly-induced current that may be verified in WSMs by measuring its transport and optical properties~\cite{parameswaran2014probing,hosur2015tunable,goswami2015optical,goswami2013axionic,son2013chiral,burkov2011weyl,burkov2014anomalous,lundgren2014thermoelectric,sharma2016nernst,kim2014boltzmann,zyuzin2017magnetotransport,cortijo2016linear}. Interestingly, CA has been proposed to occur in systems that are not WSMs~\cite{gao2017intrinsic,dai2017negative,andreev2018longitudinal,wang2018intrinsic,nandy2018berry,fu2020quantum,pal2021berry,wang2021helical,sadhukhan2023effect,PhysRevB.105.L180303,das2023chiral,varma2024magnetotransport}. The quasiparticles, in this case, are not necessarily massless but have a notion of chirality due to their underlying spinor structure.
This has led to the generalization that CA may manifest in any system with nonzero Berry flux through the Fermi surface, irrespective of the energy dispersion, number of Weyl nodes, or the underlying symmetries of the Hamiltonian~\cite{PhysRevB.105.L180303}. A specific example is that of spin-orbit-coupled (SOC) non-centrosymmetric metals (NCMs) that host nonrelativistic fermions but have multiple Fermi surfaces with fluxes of opposite Berry curvature~\cite{PhysRevB.105.L180303}. A few recent studies have investigated CA-induced electronic and thermal transport in SOC-NCMs~\cite{verma2019thermoelectric,PhysRevB.105.L180303,das2023chiral,varma2024magnetotransport}.
While some band properties in SOC-NCMs may be similar to those of WSMs, their transport responses are strikingly different~\cite{varma2024magnetotransport}.

The transport of chiral quasiparticles in condensed matter systems is affected by two key scattering processes: (i) chirality-breaking, and (ii) chirality-preserving scattering. In the context of WSMs, these are also known as (i) internode scattering (chirality-breaking), and (ii) intranode scattering (chirality-preserving), respectively, as Weyl fermions of opposite chiralities live at different Weyl nodes (or valley points) in the momentum space. Since SOC-NCMs have just one relevant nodal point, but with multiple Fermi surfaces with opposing fluxes of the Berry curvature, the two types of scattering mechanisms refer to  (i) interband (chirality-breaking) and (ii) intraband (chirality-preserving) scattering, respectively.  
Notably, the chiral anomaly manifests by the first process--the chirality-breaking scattering, which is governed by corresponding scattering timescale $\tau_\mathrm{inter}$. The second process that preserves chirality, which is not directly related to the anomaly, is governed by a timescale $\tau_\mathrm{intra}$. Nevertheles, a series of earlier works~\cite{kim2014boltzmann,lundgren2014thermoelectric,cortijo2016linear,sharma2016nernst,zyuzin2017magnetotransport,das2019berry,kundu2020magnetotransport,mandal2022chiral} have primarily focused on the role of $\tau_\mathrm{intra}$ while investigating CA-induced transport, while neglecting $\tau_\mathrm{inter}$. This is sometimes justified by stating that the chirality preserving scattering often dominates, i.e., $1/\tau_\mathrm{inter}\ll 1/\tau_\mathrm{intra}$. 
But even in the approximation that $\tau_\mathrm{inter}\gg\tau_\mathrm{intra}$, the analysis in most of the previous studies is flawed for two main reasons: (i) they neglect global charge conservation, and (ii) they assume a momentum-independent scattering time, which was recently shown to be inaccurate for chiral quasiparticles~\cite{sharma2023decoupling}. Although several earlier works on WSMs also incorporate both scattering times~\cite{zeng2023quantum,mandal2022chiral,cheon2022chiral,deng2019quantum,zyuzin2017magnetotransport,yip2015kinetic}, they neglect momentum dependent scattering and mainly rely on a constant relaxation-time approximation. The nontrivial role of momentum-dependent scattering and going beyond the relaxation-time approximation was highlighted in Ref.~\cite{sharma2023decoupling}, where it was shown that for two fully decoupled Weyl nodes, momentum-independent scattering time is inconsistent with chiral and global charge conservation. This is not unexpected because just like in graphene, fermions in a Weyl semimetal exhibit momentum-independent scattering, which is highly anisotropic in nature. Consider $|u^\chi_\mathbf{k}\rangle$, the chiral Weyl spinor and $|u^{\chi'}_{\mathbf{k}'}\rangle$, the scattered Weyl spinor. The overlap has an angular dependence:
$|\langle u^\chi_\mathbf{k}|u^{\chi'}_\mathbf{{k}'}\rangle|^2 = 1+\chi\chi'(\cos{\theta}\cos{\theta'} + \sin{\theta}\sin{\theta'}\cos{\phi}\cos{\phi'} + \sin{\theta}\sin{\theta'}\sin{\phi}\sin{\phi'}$. It turns out that integration over momentum-space to conserve global charge fails if one assumes isotropic scattering~\cite{sharma2023decoupling}.
Furthermore, only momentum-dependent scattering yields longitudinal magnetoconductance that switches sign as a function of the internode scattering strength~\cite{knoll2020negative}, and yields positive magnetoresistance in the absence of any chiral charge transfer, as expected from chiral anomaly. Recent studies have refined the understanding and analysis of transport in chiral Weyl fermions by moving beyond previous assumptions, leading to some striking and significant predictions in linear magnetotransport~\cite{knoll2020negative,sharma2020sign,sharma2023decoupling,ahmad2021longitudinal,ahmad2023longitudinal}. 

Apart from inducing currents proportional to the applied field (linear response), chirality-violating processes can induce nonlinear effects as well, such as the nonlinear Hall effect~\cite{PhysRevB.103.045105, cheon2022chiral,PhysRevB.105.L180303}. In an inversion symmetry-broken Weyl semimetal with tilted Weyl cones, a nonlinear Hall effect can be induced by the chiral anomaly, known as the chiral anomaly-induced nonlinear anomalous Hall effect (CNLHE), which is the combined effect of the Berry curvature-induced anomalous velocity $\mathbf{v}_{\mathrm{anom}}=({e}/{\hbar})\mathbf{E}\times \mathbf{\Omega_{k}}$~\cite{xiao2010berry}  and the chiral anomaly. The effect is nonzero when the Fermi surface is asymmetric and the Hamiltonian exhibits broken inversion symmetry. In WSMs, the tilt of the Weyl cone creates an asymmetric Fermi surface around the projection of the Weyl node on the Fermi surface. It is important to note that the chiral anomaly-induced nonlinear Hall effect (CNLHE) is distinct from the CNLHE caused by the Berry curvature dipole (BCD) \cite{sodemann2015quantum}, as the latter can occur even without an external magnetic field.
Previous works on CNLHE~\cite{li2021nonlinear, nandy2021chiral,zeng2022chiral,park2022nondivergent} assume that the internode scattering rate is much lower than the intranode scattering rate, or in other words $\tau_\mathrm{inter}\gg\tau_\mathrm{intra}$, thereby neglecting the role of internode scattering. Furthermore, the analysis suffers from the aforementioned shortcomings: (i) neglecting global charge conservation, and (ii) assumption of a momentum-independent scattering time, both of which breakdown for chiral quasiparticles of multiple flavors.

In this work, we present a complete theory of the nonlinear anomalous Hall effect, correctly including the effects of chirality-breaking and chirality-preserving scattering, retaining their full momentum dependence, and incorporating global charge conservation. Our theory is generic and works for any system with chiral quasiparticles of multiple flavors, however, we focus on two particular systems of experimental interest: (i) Weyl semimetal, and (ii) spin-orbit coupled noncentrosymmetric metal. We find that in Weyl semimetals the nonlinear anomalous Hall conductivity is a nonmonotonic function of the Weyl cone tilt, which is in contrast to earlier studies~\cite{li2021nonlinear,nandy2021chiral,zeng2022chiral}. Furthermore, we also find that sufficiently strong internode scattering (not considered in earlier works) flips the sign of conductivity leading to `strong-sign-reversal'. Additionally, we also examine the effect of strain (also not considered in prior works), and find that strain-induced chiral gauge field also gives rise to nonlinear anomalous Hall effect but without any `strong-sign-reversal'. 
The nonlinear anomalous Hall conductivity has not been earlier analyzed in spin-orbit coupled non-centrosymmetric metals, which forms another important focus of this work. While, the chiralities of quasiparticles in WSMs and SOC-NCMs is exactly the same, their nonlinear current response is remarkably distinct from each other. 
Interestingly, we discover that unlike WSMs, the anomalous orbital magnetic moment in SOC-NCMs can drive a large CNLH current when the electric and magnetic fields are noncollinear. We further find that including the effect of spin Zeeman coupling of the magnetic field acts like an effective tilt term which tilts the Fermi surfaces of both the chiral flavors in the same direction, thereby further enhancing the CNLH current. We highlight significant differences between the nonlinear conductivity obtained for WSMs and SOC-NCMs. First, CNLHE can be driven in SOC-NCMs by anomalous orbital magnetic moment, unlike WSMs where the cones must be necessarily tilted. Second, CNLH conductivity in WSMs flips its sign with sufficiently strong internode scattering, unlike SOC-NCMs where CNLH conductivity remains always negative even for sufficiently high interband scattering (although both these processes break the quasiparticle chirality). Third, CNLH conductivity is linear in $B$ for WSMs but is quadratic in $B$ for SOC-NCMs. Lastly, the angular dependence of CNLH conductivity is strikingly different from that of WSMs. 

In Section II, we present the Boltzmann theory where an analytical ansatz to the electron distribution function is derived. Secion III and IV discuss the CNLH conductivity is WSMs and SOC-NMCs, respectively. We conclude in Section V. 
\section{Maxwell-Boltzmann transport theory}
\label{Sec:Maxwell-Boltzmann transport theory}
We use the semiclassical Maxwell-Boltzmann formalism to describe the dynamics of three-dimensional chiral fermions in the presence of external electric and magnetic fields. The non-equilibrium distribution function $f^{\chi}_{\mathbf{k}}$ describing fermions with chirality $\chi$, evolves as:
\begin{align}
\dfrac{\partial f^{\chi}_{\mathbf{k}}}{\partial t}+ {\Dot{\mathbf{r^{\chi}_{\mathbf{k}}}}}\cdot \mathbf{\nabla_r}{f^{\chi}_{\mathbf{k}}}+\Dot{\mathbf{k^{\chi}}}\cdot \mathbf{\nabla_k}{f^{\chi}_{\mathbf{k}}}=I_{coll}[f^{\chi}_{\mathbf{k}}],
\label{MB_equation_ch4}
\end{align}
with $f^{\chi}_\mathbf{k} = f_{0} + g^{\chi}_{\mathbf{k}}$ + $h^{\chi}_{\mathbf{k}}$, where $f_{0}$ is standard Fermi-Dirac distribution, and $g^{\chi}_{\mathbf{k}}$ and $h^{\chi}_{\mathbf{k}}$ are deviations up to the first and second order in electric field ($E$), respectively. Without loss of generality, we fix the electric field along the $z-$direction and express the deviations as:
\begin{align}
g^{\chi}_\mathbf{k}&= -e\left({\dfrac{\partial f_{0}}{\partial {\epsilon}}}\right){\Lambda^{\chi}_\mathbf{k}} E\nonumber\\
h^{\chi}_\mathbf{k}&= -e\left({\dfrac{\partial g^{\chi}_\mathbf{k}}{\partial {\epsilon}}}\right){\Gamma^{\chi}_\mathbf{k}} E\nonumber\\
&=e^2\left(\left(\frac{\partial^2 f_0}{\partial \epsilon^2}\right)\Lambda^\chi_\mathbf{k}+\left(\frac{\partial \Lambda^\chi_\mathbf{k}}{\partial \epsilon}\right)\left(\frac{\partial f_0}{\partial \epsilon}\right)\right) \Gamma^{\chi}_\mathbf{k} E^2
\label{Eq:g1_ch4},
\end{align}
where ${\Lambda}^{\chi}_\mathbf{k}$ and $\Gamma^\chi_\mathbf{k}$ are the unknown functions to be evaluated, and all their derivatives with respect to energy are taken at the Fermi surface in the limit $T\rightarrow 0$. 

The right-hand side in Eq.~\ref{MB_equation_ch4}, i.e., collision integral term incorporates both chirality-breaking and chirality-preserving scattering and is expressed as:
\begin{align}
 I_{coll}[f^{\chi}_{\mathbf{k}}]=\sum_{\chi' \mathbf{k}'}{\mathbf{W}^{\chi \chi'}_{\mathbf{k k'}}}{(f^{\chi'}_{\mathbf{k'}}-f^{\chi}_{\mathbf{k}})},
 \label{Collision_integral}
\end{align}
where, the scattering rate ${\mathbf{W}^{\chi \chi'}_{\mathbf{k k'}}}$ calculated using Fermi's golden rule:
\begin{align}
\mathbf{W}^{\chi \chi'}_{\mathbf{k k'}} = \frac{2\pi n}{\mathcal{V}}|\bra{u^{\chi'}(\mathbf{k'})}U^{\chi \chi'}_{\mathbf{k k'}}\ket{u^{\chi}(\mathbf{k})}|^2\times\delta(\epsilon^{\chi'}(\mathbf{k'})-\epsilon_F).\nonumber\\
\label{Fermi_gilden_rule_ch4}
\end{align}
In the above expression \lq n\rq~is impurity concentration, \lq $\mathcal{V}$\rq~is system volume, $\ket{u^{\chi}(\mathbf{k})}$ is chiral spinor, $U^{\chi \chi'}_{\mathbf{k k'}}$ is scattering potential profile, and $\epsilon_F$ is the Fermi energy. We choose $U^{\chi \chi'}_{\mathbf{k k'}}= I_{2\times2}U^{\chi \chi'}$ for elastic impurities, where, $U^{\chi \chi'}$ distinguishes chirality-breaking and chirality-preserving scatterings, which can be controlled in our formalism. We denote the relative magnitude of chirality-breaking to chirality-preserving scattering by the ratio $\alpha = U^{\chi\chi'\neq\chi}/U^{\chi\chi}$ in our formalism. In the context of WSMs, $\alpha$ denotes the ratio of internode to intranode scattering strength, while for SOC-NCMs it denotes the ratio of interband to intraband scattering strength.

\begin{figure*}
    \includegraphics[width=0.98\columnwidth]{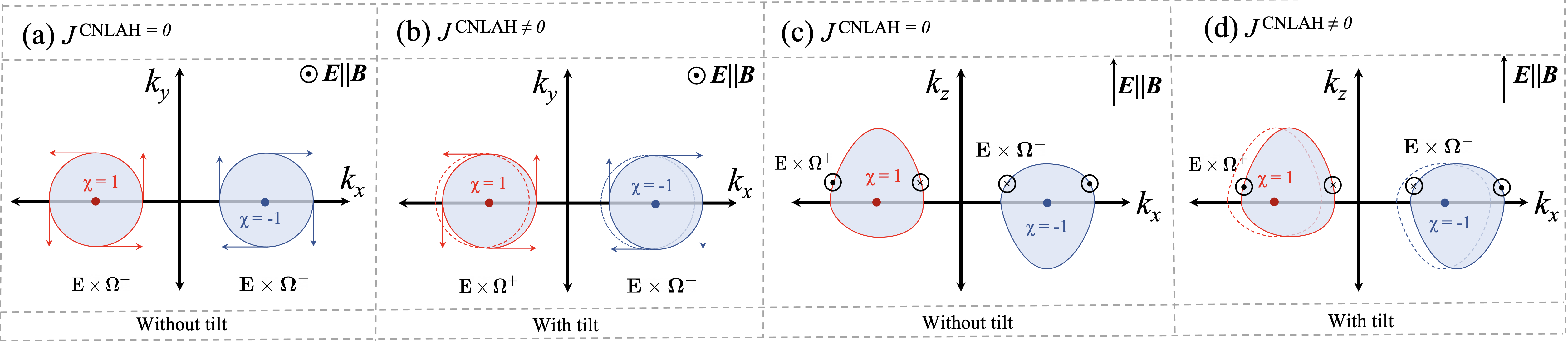}
    \caption{(a) Cross-sectional views of Fermi surface of a WSM at a constant  $k_z=0$ plane in the absence of tilt, and (b) in the presence of tilt. The anomalous velocity vectors $\sim (\mathbf{E}\times \boldsymbol{\Omega}^\chi_\mathbf{k})$ at each $\mathbf{k}$-point on the Fermi surface are indicated by blue and red arrows. (c) Cross-sectional view of Fermi surface of a WSM at a constant $k_y=0$ plane in the absence of tilt. The anomalous velocity vectors now point in and out of the $k_x-k_z$ plane. Due to the effect of orbital magnetic moment, the Fermi surface becomes egg-shaped, and the anomalous velocity vector has a nonuniform magnitude on it. The CNLH current on each valley is non-zero and opposite in sign and thus the total CNLH remains zero. In this particular case $J^{\chi,\mathrm{CNLH}}\propto -\chi B^2$ for $\alpha\leq\alpha_c$ ($\alpha_c$= critical value above which sign reversal occurs).(d) Cross-sectional view of Fermi surface of a WSM at a constant $k_y=0$ plane in the presence of tilt. The CNLH current is of unequal magnitudes on both nodes and therefore the net current does not vanish. In all the plots $\mathbf{E}\parallel\mathbf{B}\parallel \hat{z}$, and the effects of OMM are included.}
\label{fig:WSM_Schematic_Fermi_Surface}
\end{figure*}

In the presence of electric ($\mathbf{E}$) and magnetic ($\mathbf{B}$) fields, semiclassical dynamics of the quasiparticles are modified and governed by the following equation~\cite{son2012berry,knoll2020negative}:
\begin{align}
\dot{\mathbf{r}}^\chi &= \mathcal{D}^\chi_\mathbf{k} \left( \frac{e}{\hbar}(\mathbf{E}\times \boldsymbol{\Omega}^\chi) + \frac{e}{\hbar}(\mathbf{v}^\chi\cdot \boldsymbol{\Omega}^\chi) \mathbf{B} + \mathbf{v}_\mathbf{k}^\chi\right) \nonumber\\
\dot{\mathbf{p}}^\chi &= -e \mathcal{D}^\chi_\mathbf{k} \left( \mathbf{E} + \mathbf{v}_\mathbf{k}^\chi \times \mathbf{B} + \frac{e}{\hbar} (\mathbf{E}\cdot\mathbf{B}) \boldsymbol{\Omega}^\chi \right),
\label{Couplled_equation}
\end{align}
where, $\mathbf{v}_\mathbf{k}^\chi = ({\hbar}^{-1}){\partial\epsilon^{\chi}(\mathbf{k})}/{\partial\mathbf{k}}$ is band velocity, $\boldsymbol{\Omega}^\chi_\mathbf{k} = i \nabla_{\mathbf{k}} \times \langle u^{\chi}(\mathbf{k}) | \nabla_{\mathbf{k}} | u^{\chi}(\mathbf{k}) \rangle$ is the Berry curvature, and $\mathcal{D}^\chi_\mathbf{k} = (1+e\mathbf{B}\cdot\boldsymbol{\Omega}^\chi_\mathbf{k}/\hbar)^{-1}$ is the factor by which density of states is modified due to presence of the Berry curvature. For WSMs, the Berry curvature is evaluated to be $\boldsymbol{\Omega}^\chi_\mathbf{k}=-\chi \mathbf{k} /2k^3$. Unlike a classical point particle, the Bloch wave packet has a finite spatial spread. As a result, it exhibits self-rotation around its center of mass, leading to an orbital magnetic moment (OMM), which is given by  $\mathbf{m}^{\chi}_\mathbf{k}= -\frac{ie}{2\hbar} \text{Im} \langle \nabla_{\mathbf{k}} u^{\chi}|[ \epsilon_0(\mathbf{k}) - \hat{H}^{\chi}(\mathbf{k}) ]| \nabla_{\mathbf{k}} u^{\chi}\rangle$~\cite{xiao2010berry,hagedorn1980semiclassical,chang1996berry}. The orbital magnetic moment (OMM) is an intrinsic property of the band, unaffected by the wave packet's particular size or shape, and depends only on the Bloch functions, much like the Berry curvature. It also triggers current responses that are unconventional and otherwise not expected from Bloch bands with a trivial topology. Therefore it is termed an `anomalous orbital magnetic moment'. For WSMs, this is evaluated to be $\mathbf{m}^{\chi}_\mathbf{k}=-{\chi e v_{F} \mathbf{k}}/{2k^2}$. Due to the orbital magnetic moment, the energy dispersion is modified in the presence of the external magnetic field: $\epsilon^\chi_k \rightarrow \epsilon_k - \mathbf{m}^{\chi}_\mathbf{k} \cdot \mathbf{B}$. This changes the spherical Fermi surface to an egg-shaped Fermi surface as schematically displayed in Fig.~\ref{fig:WSM_Schematic_Fermi_Surface}.
We rotate the magnetic field along the $xz-$plane:  $\mathbf{B} = B (\cos{\gamma},0,\sin{\gamma})$, i.e., for $\gamma=\pi/2$ both the fields are parallel to each other. 

The focus of this work is to investigate the effect of chiral-anomaly term, and we therefore neglect the Lorentz force term. This also allows us to make analytical progress. We point out that this approximation becomes exact in the limit $\gamma\rightarrow\pi/2$. Even if $\gamma<\pi/2$, the Lorentz force magnitude is comparatively smaller~\cite{ahmad2021longitudinal}.
Keeping terms up to the second order in the electric field, the Boltzmann transport equation reduces to the following set of equations:
\begin{align}
&\mathcal{D}^{\chi}_\mathbf{k}\left[{v^{\chi,z}_{\mathbf{k}}}+\frac{eB\sin{\gamma}}{\hbar}(\mathbf{v^{\chi}_k}\cdot\boldsymbol{\Omega}^{\chi}_k)\right]
 = \sum_{\chi' \mathbf{k}'}{\mathbf{W}^{\chi \chi'}_{\mathbf{k k'}}}{(\Lambda^{\chi'}_{\mathbf{k'}}-\Lambda^{\chi}_{\mathbf{k}})}.\label{Eq_boltz_E_1}
 \\
&\mathcal{D}^{\chi}_\mathbf{k}\frac{\partial}{\partial \epsilon^\chi_\mathbf{k}} \left(\frac{\partial f_0}{\partial \epsilon^\chi_\mathbf{k}} ~\Lambda^{\chi}_{\mathbf{k}}\right)
\left[{v^{\chi,z}_{\mathbf{k}}}+\frac{eB\sin{\gamma}}{\hbar}(\mathbf{v^{\chi}_k}\cdot\boldsymbol{\Omega}^{\chi}_k)\right]=\nonumber\\
& \sum_{\chi' \mathbf{k}'}{\mathbf{W}^{\chi \chi'}_{\mathbf{k k'}}}\left(\Gamma^{\chi'}_{\mathbf{k}'}\frac{\partial}{\partial \epsilon^{\chi'}_\mathbf{k'}} \left(\frac{\partial f_0}{\partial \epsilon^{\chi'}_{\mathbf{k}'}}\Lambda^{\chi'}_{\mathbf{k}'}\right)-\Gamma^{\chi}_{\mathbf{k}}\frac{\partial}{\partial \epsilon^{\chi}_\mathbf{k}} \left(\frac{\partial f_0}{\partial \epsilon^{\chi}_{\mathbf{k}}}\Lambda^{\chi}_{\mathbf{k}}\right)\right)
\label{Eq_boltz_E_2}
\end{align}

Eq.~\ref{Eq_boltz_E_1} can be solved for $\Lambda^\chi$, which can then be used to solve for $\Gamma^\chi$ in Eq.~\ref{Eq_boltz_E_2}, and then the distribution function is evaluated using Eq.~\ref{Eq:g1_ch4}. Once the distribution function is evaluated, the current density can be evaluated as:
\begin{align}
    \mathbf{J}=-e\sum_{\chi,\mathbf{k}} f^{\chi}_{\mathbf{k}} \dot{\mathbf{r}}^{\chi}.
    \label{Eq:J_formula}
\end{align}
We primarily focus on the second-order anomalous Hall response induced by the chiral anomaly, which is given by 
\begin{align}
    \mathbf{J}^\mathrm{CNLH}=-\frac{e^2}{\hbar}\sum_{\chi,\mathbf{k}} \mathcal{D}^\chi_\mathbf{k} g^{\chi}_{\mathbf{k}}  (\mathbf{E}\times \boldsymbol{\Omega}^\chi_\mathbf{k})
    \label{Eq:CNLH_formula}
\end{align}
To evaluate all the different responses, $\mathbf{J}^{\mathrm{CNLH}}$ is written as \cite{yao2024geometrical}:
\begin{align}
    J^{\mathrm{CNLH}}_{\alpha} = \sum_{\chi=\pm 1}\sum_{\beta \gamma}\sigma^{\chi}_{\alpha\beta \gamma} E_{\beta}E_{\gamma},
    \label{Eq:CNLH_componetns_formula}
\end{align}
with, $\alpha, \beta, \gamma = \{x,y,z\}$. Comparison of Eq.~\ref{Eq:CNLH_formula} and Eq.~\ref{Eq:CNLH_componetns_formula} gives different components of nonlinear conductivity 3-rank tensor ($\sigma^{\chi}_{\alpha\beta\gamma}$). For, $\mathbf{E} =E \hat{z}$, the anomalous velocity ($\mathbf{v}^{\chi}_{\mathrm{anom}} \sim \mathbf{E} \times \boldsymbol{\Omega}^{\chi}_{\mathbf{k}}$) has components in $xy$-plane. Since we rotate the magnetic field in $xz$-plane, we measure the Hall response along the $y$-direction, i.e., we evaluate $\sigma_{yzz}$. A component of the nonlinear current is also generated along the $x-$direction ($\sigma_{xzz}$), which contributes to the planar nonlinear Hall effect, and is seen to vanish. 

Moving on, we define the chiral scattering rate as follows:
\begin{align}
\frac{1}{\tau^{\chi}(\theta,\phi)}=\sum_{\chi'}\mathcal{V}\int\frac{d^3\mathbf{k'}}{(2\pi)^3}(\mathcal{D}^{\chi'}_{\mathbf{k}'})^{-1}\mathbf{W}^{\chi \chi'}_{\mathbf{k k'}}.
\label{Tau_invers_ch4}
\end{align}
$\mathbf{W}^{\chi \chi'}_{\mathbf{k k'}}$ is defined in Eq.~\ref{Fermi_gilden_rule_ch4} and the corresponding overlap of the Bloch wave function is given by the following expression:
$\mathcal{G}^{\chi\chi'}(\theta,\phi) = [1+\chi\chi'(\cos{\theta}\cos{\theta'} + \sin{\theta}\sin{\theta'}\cos{\phi}\cos{\phi'} + \sin{\theta}\sin{\theta'}\sin{\phi}\sin{\phi'}]$. Note that this expression for $\mathcal{G}^{\chi\chi'}(\theta,\phi)$ holds for both our systems of interest: WSM and SOC-NCM. For chiral particles with a different spinor structure, $\mathcal{G}^{\chi\chi'}(\theta,\phi)$ should be appropriately modified. 
\begin{figure}
    \centering
    \includegraphics[width=\columnwidth]{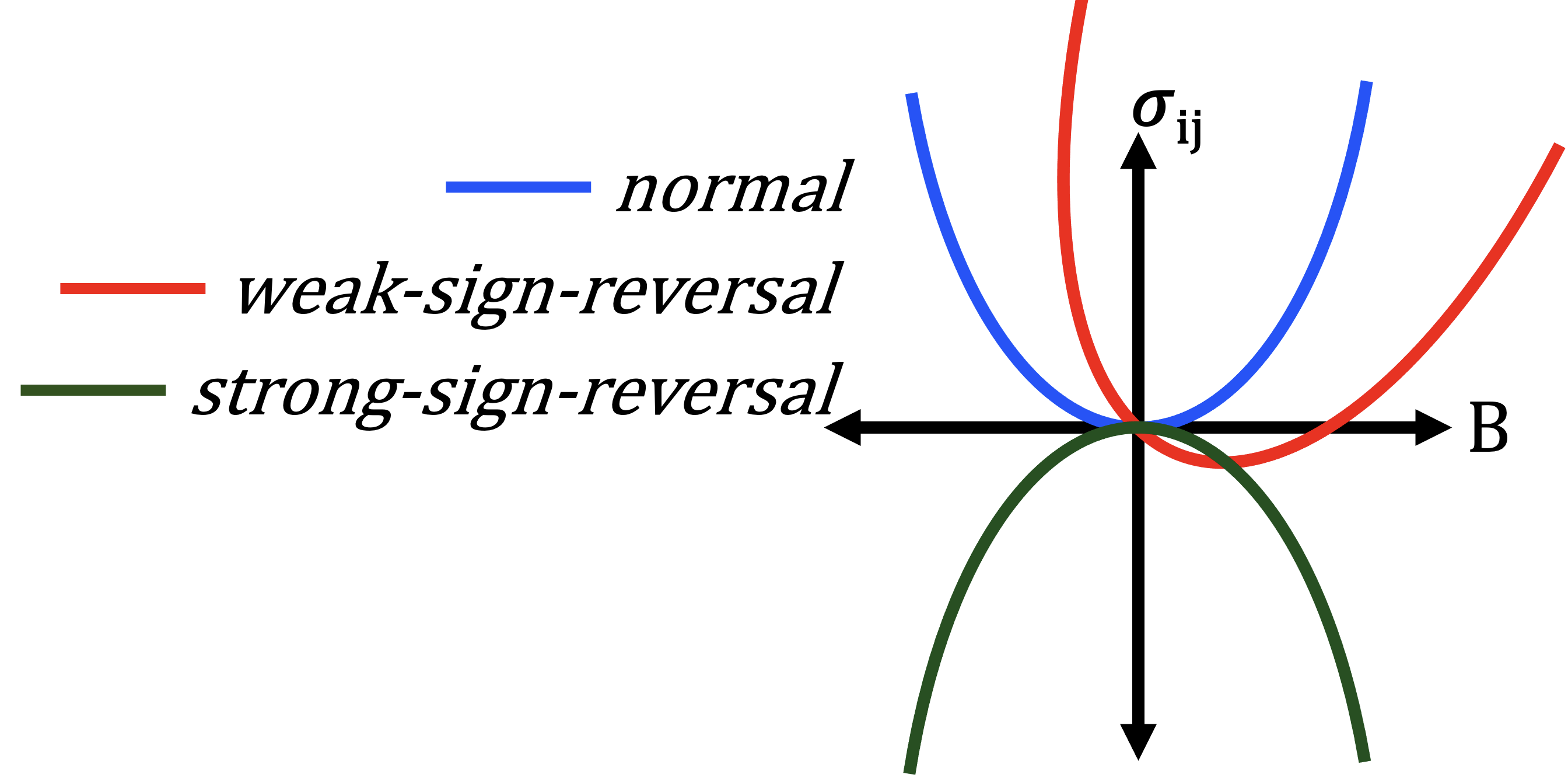}
    \caption{A schematic representation of weak-sign-reversal and strong-sign-reversal of conductivity $\sigma_{ij}$ compared to normal quadratic-in-$B$ conductivity in chiral Weyl systems~\cite{ahmad2023longitudinal,varma2024magnetotransport,ahmad2024geometry}. }
    \label{fig:signreverseschematic}
\end{figure}
Taking Berry phase into account and corresponding change in density of states, $\sum_{k}\longrightarrow \mathcal{V}\int\frac{d^3\mathbf{k}}{(2\pi)^3}\mathcal{D}^\chi_\mathbf{k}$, Eq.~\ref{Eq_boltz_E_1} becomes:
\begin{align}
l^{\chi}(\theta,\phi) + \frac{\Lambda^{\chi}(\theta,\phi)}{\tau^{\chi}(\theta,\phi)}=\sum_{\chi'}\mathcal{V}\int\frac{d^3\mathbf{k}'}{(2\pi)^3} \mathcal{D}^{\chi'}_{\mathbf{k}'}\mathbf{W}^{\chi \chi'}_{\mathbf{k k'}}\Lambda^{\chi'}(\theta',\phi').
\label{MB_in_term_Wkk'_ch4}
\end{align}
Here, $l^{\chi}(\theta,\phi)=\mathcal{D}^{\chi}_{\mathbf{k}}[v^{\chi}_{z,\mathbf{k}}+eB\sin{\gamma}(\boldsymbol{\Omega}^{\chi}_{k}\cdot \mathbf{v}^{\chi}_{\mathbf{k}})/
\hbar]$ evaluated at the Fermi surface. Eq.~\ref{Tau_invers_ch4} and right-hand side of Eq.~\ref{MB_in_term_Wkk'_ch4} is reduced to integration over $\theta'$ and $\phi'$,
\begin{align}
\frac{1}{\tau^{\chi}(\theta,\phi)} =  \mathcal{V}\sum_{\chi'} \Pi^{\chi\chi'}\iint\frac{(k')^3\sin{\theta'}}{|\mathbf{v}^{\chi'}_{k'}\cdot{\mathbf{k'}^{\chi'}}|}d\theta'd\phi' \mathcal{G}^{\chi\chi'}(\mathcal{D}^{\chi'}_{\mathbf{k'}})^{-1}.
\label{Tau_inv_int_thet_phi}
\end{align}
\begin{multline}
\mathcal{V}\sum_{\chi'} \Pi^{\chi\chi'}\iint f^{\chi'}(\theta',\phi') \mathcal{G}^{\chi \chi'} d\theta' d\phi'\times[d^{\chi'} - l^{\chi'}(\theta',\phi') \\+ a^{\chi'} \cos\theta' + b^{\chi'} \sin\theta' \cos{\phi'} + c^{\chi'}\sin{\theta'} \cos{\phi'}],
\end{multline}
where, $\Pi^{\chi \chi'} = N|U^{\chi\chi'}|^2 / 4\pi^2 \hbar^2$ and $f^{\chi} (\theta,\phi)=\frac{(k)^3}{|\mathbf{v}^\chi_{\mathbf{k}}\cdot \mathbf{k}^{\chi}|} \sin\theta (\mathcal{D}^{\chi'}_{\mathbf{k}})^{-1} \tau^\chi(\theta,\phi)$. Using the ansatz $\Lambda^{\chi}_{\mathbf{k}}=[d^{\chi}-l^{\chi}(\theta,\phi) + a^{\chi}\cos{\phi} +b^{\chi}\sin{\theta}\cos{\phi}+c^{\chi}\sin{\theta}\sin{\phi}]\tau^{\chi}(\theta,\phi)$, the above equation can be written in following form:
\begin{align}
&d^{\chi}+a^{\chi}\cos{\phi}+b^{\chi}\sin{\theta}\cos{\phi}+c^{\chi} \sin{\theta}\sin{\phi}
=\mathcal{V}\sum_{\chi'}\Pi^{\chi\chi'}\iint f^{\chi'}(\theta',\phi')\mathcal{G}^{\chi\chi'}d\theta'd\phi'\nonumber \\&\times[d^{\chi'}-l^{\chi'}(\theta',\phi')+a^{\chi'}\cos{\theta'}+b^{\chi'}\sin{\theta'}\cos{\phi'}+c^{\chi'} \sin{\theta'}\sin{\phi'}].
\label{Boltzman_final_ch4}
\end{align}
When this equation is explicitly written, it appears as seven simultaneous equations that must be solved for eight variables. The particle number conservation provides an additional restriction.
\begin{align}
\sum\limits_{\chi}\sum\limits_{\mathbf{k}} g^{\chi}_\mathbf{k} = 0
\label{Eq_sumgk_ch4}
\end{align} 
For the eight unknowns ($d^{\pm 1}, a^{\pm 1}, b^{\pm 1}, c^{\pm 1}$), Eq.~\ref{Boltzman_final_ch4} and Eq.~\ref{Eq_sumgk_ch4} are simultaneously solved with Eq.~\ref{Tau_inv_int_thet_phi}. The nonlinear Hall conductivity induced by chiral anomaly is then evaluated using Eq.~\ref{Eq:CNLH_formula} and Eq.~\ref{Eq:CNLH_componetns_formula}. The current equation is written in the integral form as:
\begin{align}
    \mathrm{J}^\mathrm{CNLH} = \sum_{\chi'=\pm 1} E_{z} E_{z}\iint d\theta'd\phi' \mathcal{L}^{\chi'} \Lambda^{\chi'} \Omega^{\chi'}_{x},
    \label{Current_B_depen_gm_depend}
\end{align}
with, $\mathcal{L}^{\chi}(B,\theta,\phi,\gamma) = \frac{(k^{\chi})^3 \sin\theta}{|\mathbf{v}^\chi_{\mathbf{k}}\cdot \mathbf{k}^{\chi}|}$, $\Lambda^{\chi}$ is ansatz defined above and $\Omega^{\chi}_{x}$ is the $x$-component of the BC. 


\begin{figure*}
\floatbox[{\capbeside\thisfloatsetup{capbesideposition={left,top},capbesidewidth=0.55\columnwidth}}]{figure}[\FBwidth]
{\caption{CNLH conductivity $\sigma_{yzz}$ as a function of magnetic field $B$ for untitled Weyl nodes with chiralities $\chi=\pm1$. In this particular case $J^{\chi,\mathrm{CNLH}}\propto -\chi B^2$ for $\alpha\leq\alpha_c$ (where $\alpha_c$ is the critical value above which sign reversal occurs). The magnitude of $J^{\chi,\mathrm{CNLH}}$ is the same at each node but of opposite sign. This leads to zero CNLH in absence of any tilt, corresponding to the configuration in Fig.~\ref{fig:WSM_Schematic_Fermi_Surface} (c). This highlights the necessity of the asymmetric Fermi surface to obtain a non-zero net CNLH in WSMs.}
\label{fig:Single_node_CNLAH_WSM}}
{\includegraphics[width=.90\linewidth, height = 6cm ]{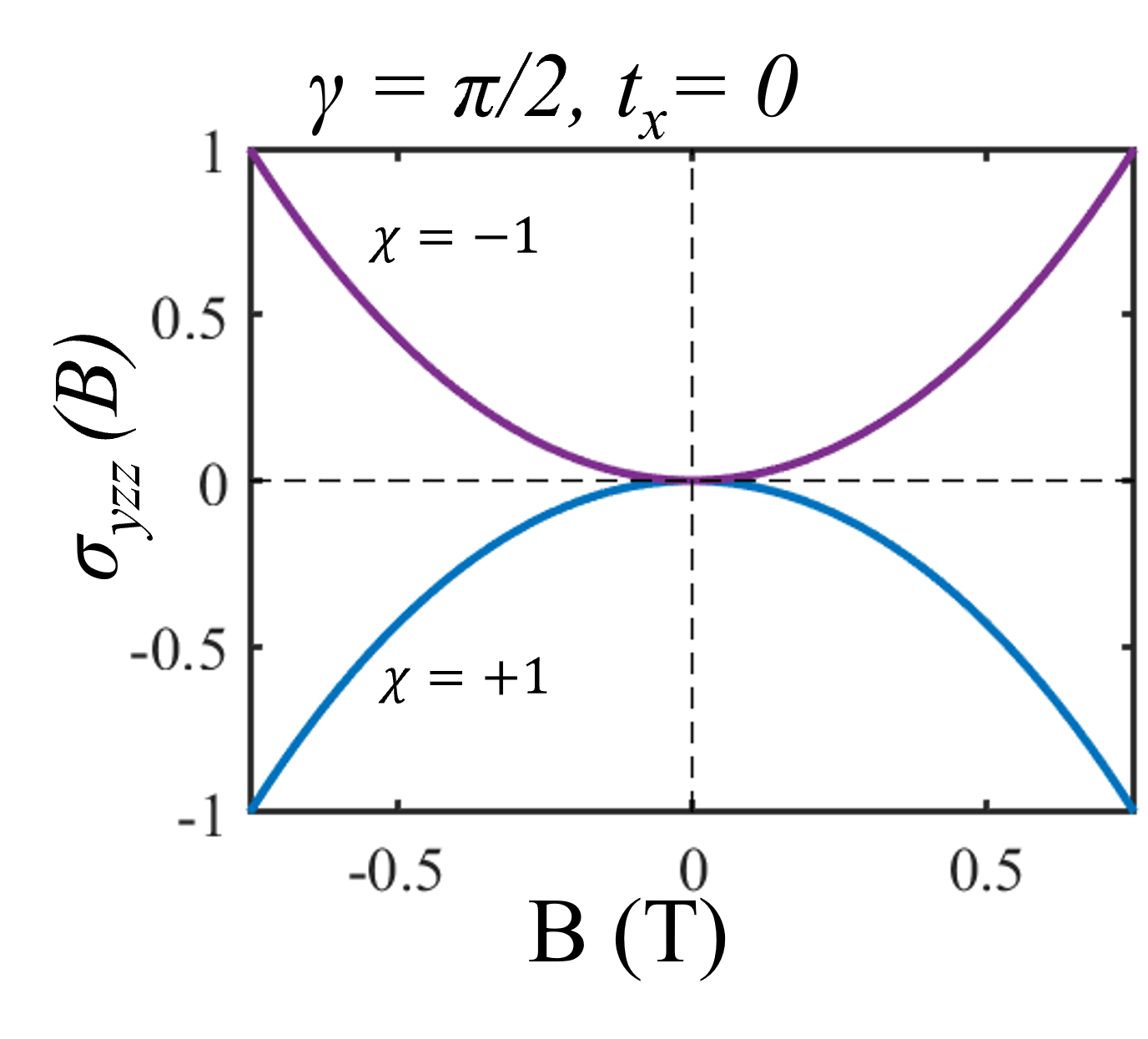}}
\end{figure*}

\begin{figure*}
    \centering   \includegraphics[width=1\columnwidth]{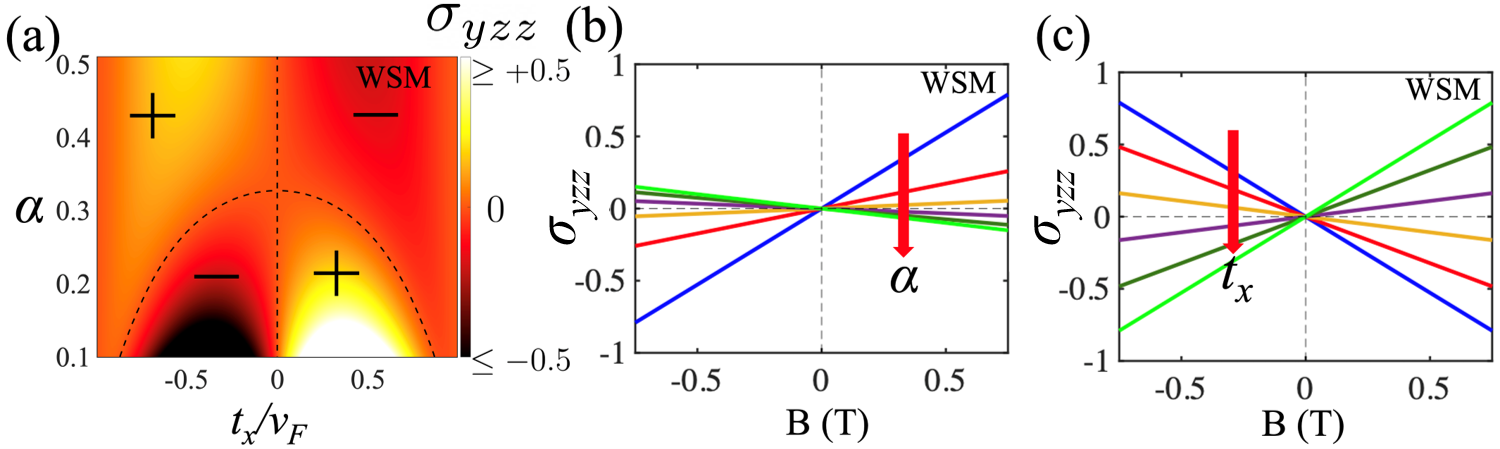}
    \caption{(a) CNLH conductivity $\sigma_{yzz}$ as a function of the relative intervalley scattering strength $\alpha$ and the Weyl cone tilt along $x$-direction ($t_x$), for a constant value of magnetic field. Regions of positive and negative magnetoconductivity are separated by black dashed contours and are marked by $+$ and $-$ signs, respectively. (b) $\sigma_{yzz}$ as a function of $B$ for different values of $\alpha$ for a constant $t_x$. The red arrow indicates the direction of the increment of $\alpha$, leading to `strong-sign-reversal'. (c) $\sigma_{yzz}$ as a function of $B$ for different values of tilt $t_x$ and a constant value of $\alpha$. Here, along the direction of the arrow, we vary tilt $t_x$ from $-0.25~v_F$ to $0.25~v_F$. In all the plots $\gamma=\pi/2$, i.e., the electric and magnetic fields are parallel to each other.}
\label{fig:CNLH_Normal_phase_tx_vary}
\end{figure*}

\begin{figure}
    \includegraphics[width=1\columnwidth]{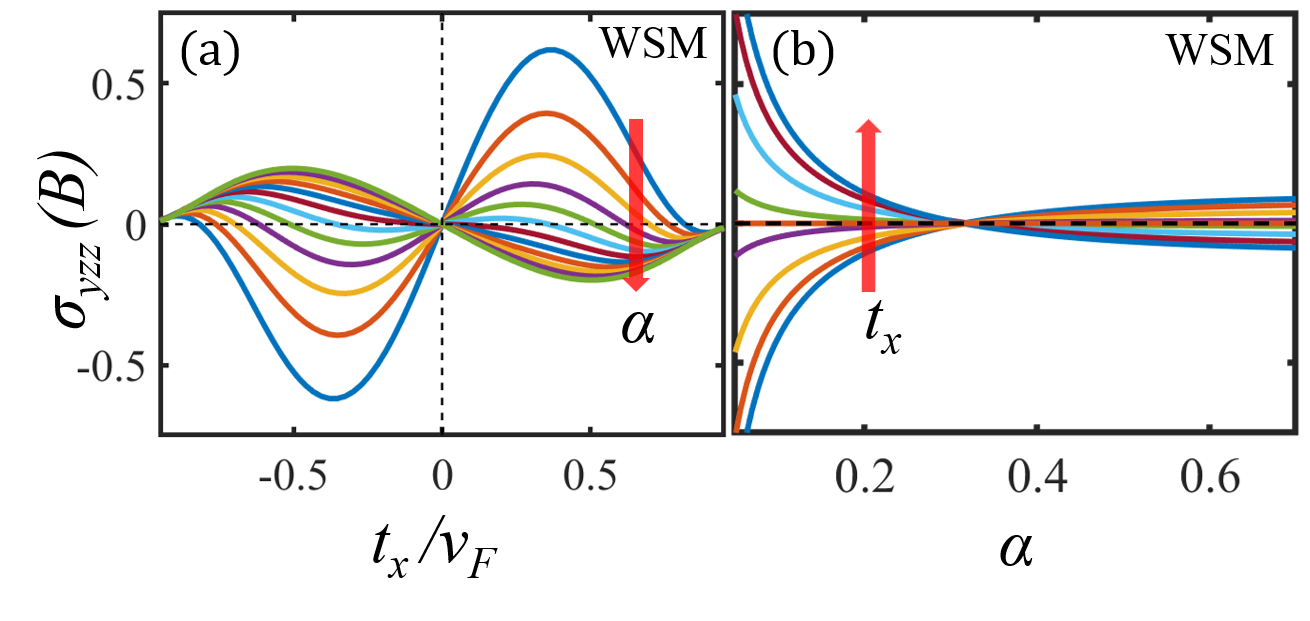}
    \caption{(a) CNLH conductivity $\sigma_{yzz}$ as a function of $t_x$ in WSMs at different values of $\alpha$. The red arrow shows the increment of $\alpha$ from $0.05$ to $1.00$. It is evident that $\sigma_{yzz}$ is linear and monotonic only for small values of $\alpha$ and $t_x$. Increasing $\alpha$ leads to sign-reversal of the conductivity. (b) $\sigma_{yzz}$ as a function of $\alpha$ for WSMs at different values of $t_{x}$. Here, we have chosen $B\simeq 0.50 T$ and $\gamma=\pi/2$, i.e., the electric and magnetic fields are parallel to each other. The red arrow indicates increment of $t_{x}$ from $-0.25~ v_F$ to {$+0.25 ~v_F$}. The sweetspot at $\alpha\approx 1/3$, where $\sigma_{yzz}\approx0$ is noteworthy. In all the plots $\sigma_{yzz}$ has been appropriately normalized.}
\label{fig:CNLH_tx_and_alp_normal_plt}
\end{figure}
\begin{figure*}
    \centering   \includegraphics[width=1\columnwidth]{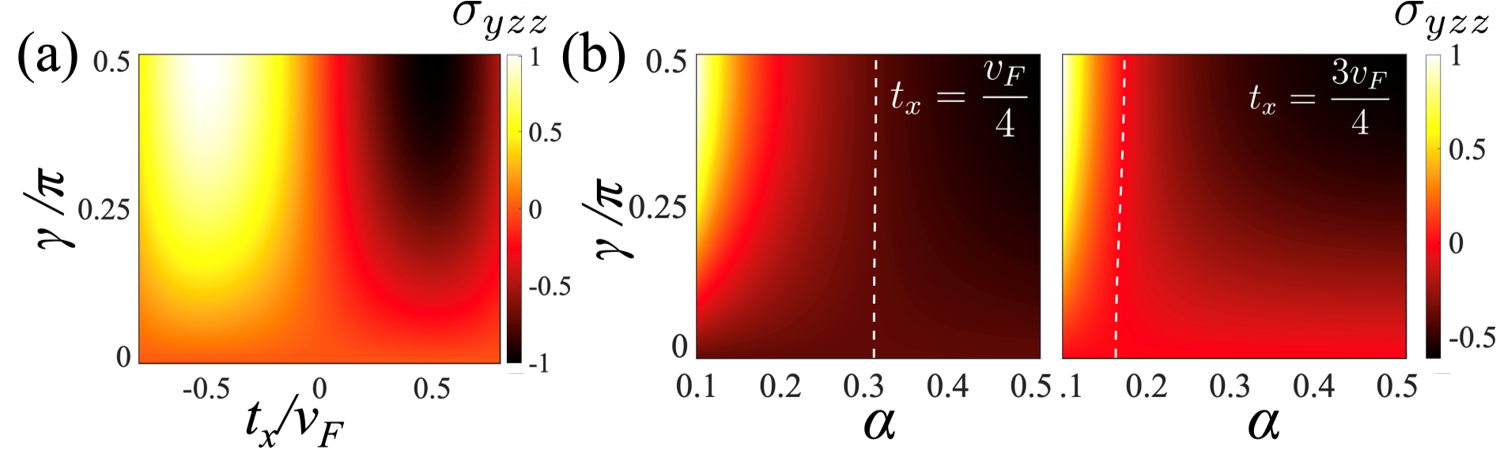}
    \caption{(a) CNLH conductivity as a function of $t_x$ and $\gamma$ for a constant value of $\alpha$. (b) CNLH conductivity as a function of $\gamma$ and $\alpha$ for two different values of the tilt parameter $t_x$. The white dashed contour separates the region of positive and negative conductivity. We note that the zero-conductivity contour shows weak dependence on the angle of the magnetic field and a stronger dependence on $t_x$.}
\label{fig:CNLH_Normal_phase_gm_vary}
\end{figure*}

\section{Chiral nonlinear anomalous Hall effect in WSMs}
\subsection{Low-energy Hamiltonian}
We begin with the following low-energy Hamiltonian of a Weyl semimetal:
\begin{align}
H_{\mathrm{WSM}}(\mathbf{k})=
\sum_{\chi=\pm1} \chi\hbar v_{F}\mathbf{k}\cdot\boldsymbol{\sigma} + \hbar v_{F}(t_z^{\chi} k_z + t_x^{\chi} k_x )I_{2\times2},
\label{Hamiltonian_ch4}
\end{align}
where $\chi$ is the chirality of the Weyl node, $\hbar$ is the reduced Plank constant, $v_{F}$ is the Fermi velocity, $\mathbf{k}$ is the wave vector measured from the Weyl node, $\boldsymbol{\sigma}$ is vector of Pauli matrices, and $t_{x,z}$ are the tilt parameters {($\mathbf{t}=t_x\hat{x}+t_z\hat{z}$)}. Unless otherwise stated, we choose $t^\chi_z = t^{-\chi}_z$, and $t^\chi_x = t^{-\chi}_x$. We justify this as follows: In an inversion-asymmetric Weyl semimetal, there are at least four Weyl points (two of each chirality) positioned in the Brillouin zone in a way that maintains time-reversal symmetry.
Now, the chiral nonlinear Hall effect (CNLH) in a Weyl semimetal (WSM) is driven by an asymmetric Fermi surface. To achieve this effect, one needs to break time-reversal symmetry. Including such a suitable term causes a relative energy shift between Weyl nodes of the same chirality~\cite{li2021nonlinear}. 
Under these conditions, with an asymmetric Fermi surface and broken inversion symmetry, we have the ideal environment to produce a non-vanishing chiral nonlocal Hall effect (CNLH)~\cite{li2021nonlinear}. This 
justifies the use of Eq.~\ref{Hamiltonian_ch4} as a starting point. We point out that although we have chosen $t^\chi_x = t^{-\chi}_x$, any alternative tilt configuration other than $t^\chi_x + t^{-\chi}_x=0$ will result in a nonvanishing CNLH conductivity. 
Diagonalization of Hamiltonian in Eq.~\ref{Hamiltonian_ch4} leads to the energy dispersion given by:
\begin{align}
\epsilon^{\chi}_k=\pm{\hbar v_{F}}|k| + \hbar v_{F}(t^{\chi}_z k_z+t^{\chi}_x k_x).
\label{Dispersion}
\end{align}
Due to coupling with the external magnetic field, the energy dispersion becomes
\begin{align}
    \epsilon^{\chi}_k=\pm{\hbar v_{F}}|k| + \hbar v_{F}(t^{\chi}_z k_z+t^{\chi}_x k_x) + \frac{\chi e v_{F} \mathbf{k}\cdot\mathbf{B}}{2k^2}.
    \label{Eq:Fermi_contour_WSM}
\end{align}
The constant energy Fermi contour, which is the locus of all points with constant energy $\epsilon$, can be then evaluated to be
\begin{align} k^{\chi} = \frac{\epsilon+\sqrt{\epsilon^{2} - n^{\chi} \chi e v_{F} B \beta_{\theta \phi}}}{n^{\chi}}.
\end{align}
Here, $n^{\chi} = 2 \hbar v_{F} + 2 t_{x} \hbar v_{F} \sin{\theta} \cos{\phi} + 2 t_{z} \hbar v_{F} \cos{\theta}$, and ${\beta_{\theta \phi} = \sin{\theta} \cos{\phi} \cos{\gamma} + \cos{\theta} \sin{\gamma}}$.
The topological nature of following Bloch states of Hamiltonian in Eq.~\ref{Hamiltonian_ch4}: $\ket{u^{+}}^{T}=[e^{-i\phi}\cos(\theta /2),\sin(\theta/2)], \ket{u^{-}}^{T}=[-e^{-i\phi}\sin(\theta /2),\cos(\theta/2)]$, gives rise nonzero flux of the Berry curvature $\boldsymbol{\Omega}^{\chi}_\mathbf{k}=-{\chi \mathbf{k}}/{2k^3}$. 
Due to change in the dispersion, the band velocity components are also altered, which we evaluate to be:
\begin{align}
&v^{\chi}_x=v_{F}\frac{k_x}{k}+v_Ft^{\chi}_x\nonumber
+\frac{u^{\chi}_{2}}{k^2} \left(\cos{\gamma}\left(1-\frac{2k^2_x}{k^2}\right)+\sin{\gamma}\left(\frac{-2k_x k_z}{k^2}\right)\right), \nonumber\\
&v^{\chi}_y=v_{F}\frac{k_y}{k}\nonumber +\frac{u^{\chi}_{2}}{k^2}\left(\cos{\gamma}\left(\frac{-2k_x k_y}{k^2}\right)+\sin{\gamma}\left(\frac{-2k_y k_z}{k^2}\right)\right),\nonumber\\
&v^{\chi}_z=v_{F}\frac{k_z}{k}+v_Ft^{\chi}_z +\frac{u^{\chi}_{2}}{k^2}\left(\cos{\gamma}\left(\frac{-2k_x k_z}{k^2}\right)+\sin{\gamma}\left(1-\frac{2k^2_z}{k^2}\right)\right),
\label{velocity_components}
\end{align}
with $u^\chi_2={\chi e v_{F} B}/{2 \hbar}$.
\subsection{Weak and strong sign-reversal}
The sign of longitudinal magnetoconductivity in Weyl materials has been intensely investigated in prior literature. While chiral anomaly in untilted Weyl semimetals is predicted to show positive LMC for weak internode scattering, it reverses sign for sufficiently strong internode scattering~\cite{knoll2020negative,sharma2020sign,sharma2023decoupling}. On the other hand, even a small amount of tilting in the Weyl cone can result in negative LMC along a particular direction of the magnetic field even for weak internode scattering. However, the reversal in sign in these two cases is fundamentally quite different, which leads to the classification of `strong-sign-reversal' and `weak-sign-reversal' as defined in Ref.~\cite{ahmad2023longitudinal,varma2024magnetotransport}. We briefly review it here. A general expression for the magnetoconductivity tensor can be written as~\cite{ahmad2023longitudinal}
\begin{align}
\sigma_{ij}(B)=  \sigma_{ij}^{(0)} + (B-B_0)^2 \sigma_{ij}^{(2)},
\label{Eq-sij-fit}
\end{align}
which incorporates (i) normal quadratic $B-$dependence, (ii) linear-in-$B$ dependence and sign change along a particular direction of the magnetic field, and (iii) quadratic-in-$B$ dependence with negative sign, in a single framework.
The features characterizing `weak-sign-reversal' include (i) $B_0 \neq 0$, (ii) $\sigma_{ij}^{(0)} \neq \sigma_{ij}(B=0)$, and (iii) $\mathrm{sign }\; \sigma_{ij}^{(2)}>0$. In this case, the vertex of the magnetoconductivity parabola is shifted from the origin, and the conductivity is of different signs for small positive and negative magnetic fields. However, the orientation of the parabola is still positive, i.e., $\mathrm{sign }\; (\sigma_{ij}^{(2)})>0$.
`Strong-sign-reversal' is characterized by $\mathrm{sign }\; (\sigma_{ij}^{(2)})<0$, which implies a complete reversal of the orientation of the parabola. Tilting of Weyl cones can result in `weak-sign-reversal' while intervalley scattering or strain is generally expected to result in `strong-sign-reversal'~\cite{ahmad2023longitudinal,varma2024magnetotransport}. Fig.~\ref{fig:signreverseschematic} schematically explains the distinction between the two cases. 
\begin{figure*}
    \centering \includegraphics[width= 0.98 \columnwidth]{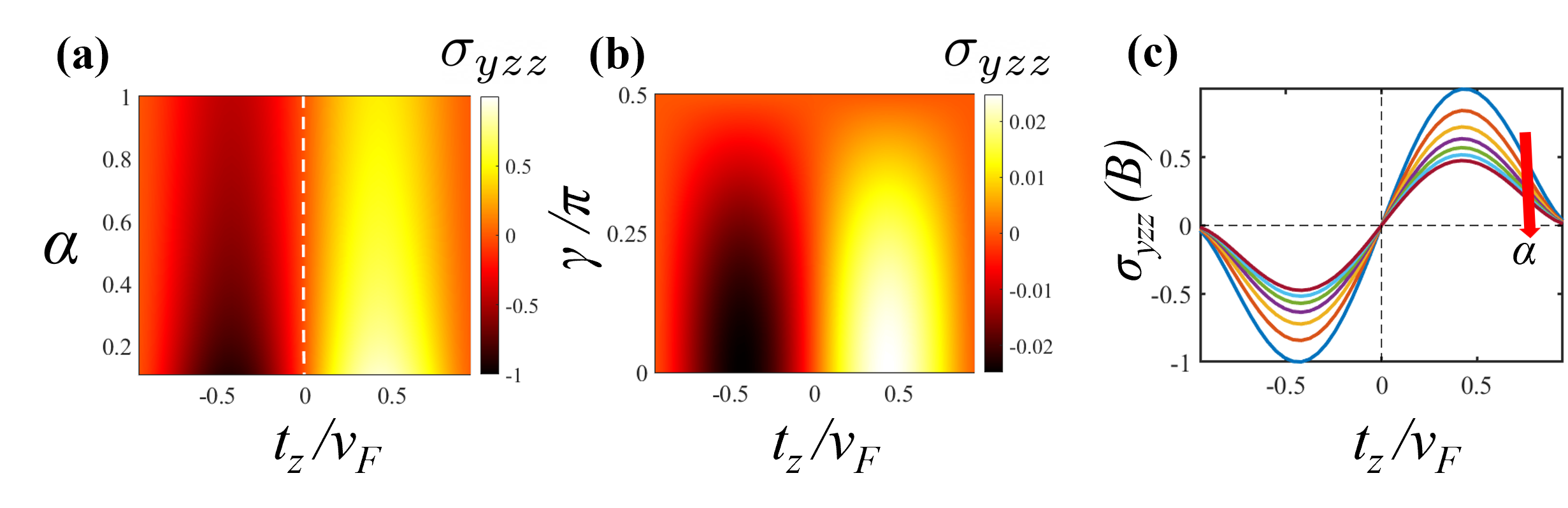}
    \caption{(a) CNLH conductivity $\sigma_{yzz}$ as a function of the relative intervalley scattering strength $\alpha$ and the Weyl cone tilt along $z$-direction ($t_z$), for a constant value of magnetic field ($B=0.50 ~T$). Unlike Fig.~\ref{fig:CNLH_Normal_phase_tx_vary}, no sign reversal of CNLH with respect to $\alpha$ is seen. (b) Phase plot of $\sigma_{yzz}$ as a function of $t_z$ and $\gamma$. (c) $\sigma_{yzz}$ as a function of $t_z$ for different values of tilt $\alpha$ and a constant value of magnetic field $B$. Here, along the direction of the arrow, we vary $\alpha$ from $0.10$ to $1.0$.}
\label{fig:WSM_NLH_tz_vary}
\end{figure*}

\begin{figure}
    \centering
    \includegraphics[width= 0.98 \columnwidth]{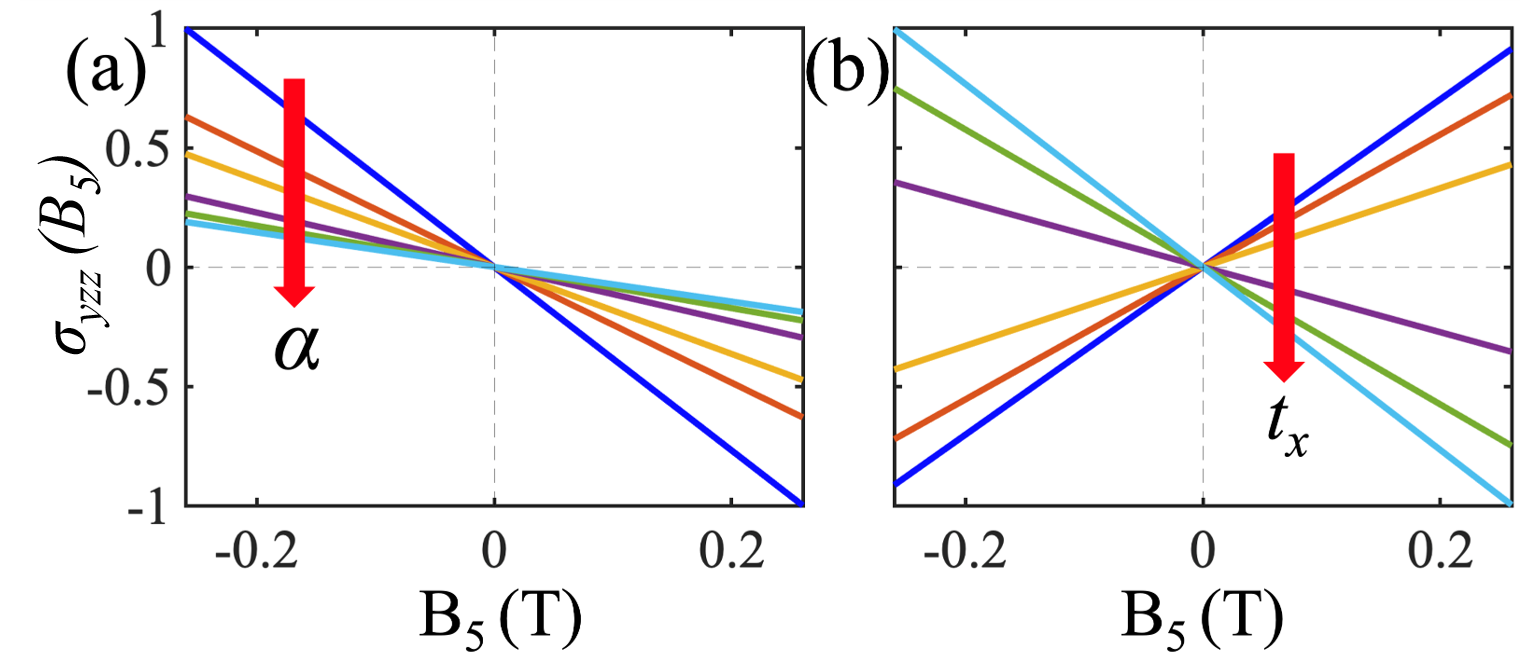}
    \caption{The anomalous nonlinear Hall conductivity as a function of the strain-induced magnetic field $\mathbf{B}_5$. (a) For different intervalley scattering strengths $\alpha$ but fixed $t_x$. (b) For different values of tilt $t_x$ but fixed $\alpha$. The parameters $\alpha$ and $t_x$ increase in the direction of the arrow in both the respective plots. The $\mathbf{B}_5$ field here was chosen parallel to the electric field.}
    \label{fig:wsm_cnlh_B5}
\end{figure}
\begin{figure*}
    \includegraphics[width= 0.98 \columnwidth]{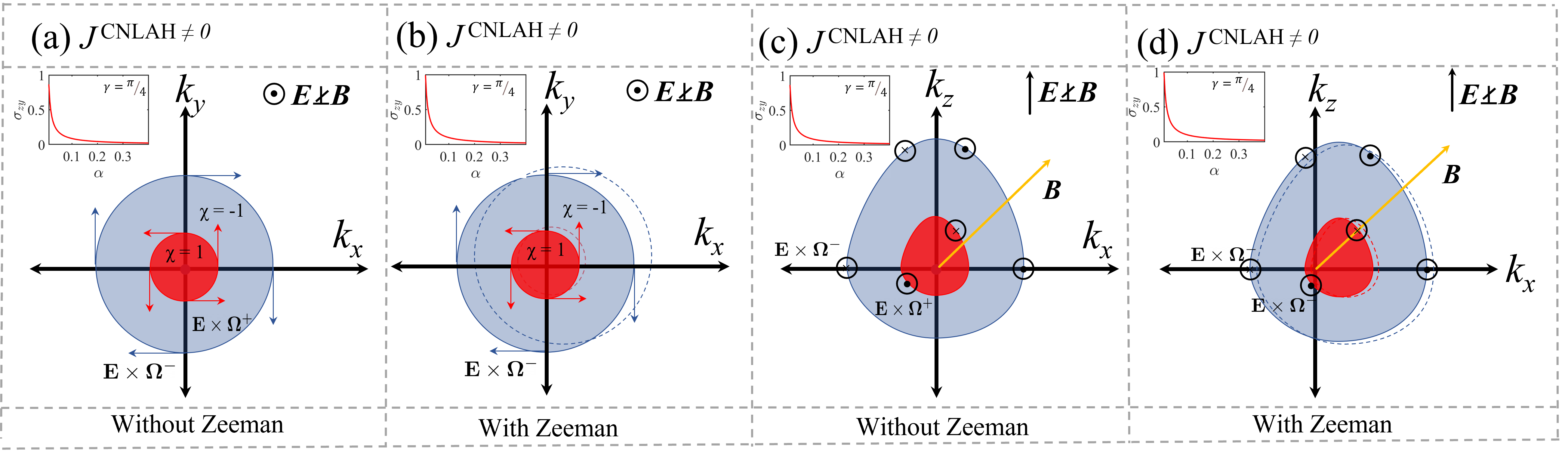}
    \caption{Schematic illustration of the origin of CNLH in SOC-NCMs. (a) Cross-sectional view of the Fermi surface in the $k_z=0$ plane without the effect of spin Zeeman coupling. (b) Cross-sectional view of the Fermi surface in the $k_z=0$ plane with the effect of spin Zeeman coupling. 
    The anomalous velocity at two Fermi surfaces having opposite chirality is marked by blue and red arrows. 
    (c) Cross-sectional view in the $k_y=0$ plane without the effect of spin Zeeman coupling (d) The cross-sectional view in the $k_y=0$ plane with the effect of spin Zeeman coupling.
    The anomalous velocity vectors now point in and out of the $k_z-k_x$ plane. The spin Zeeman field effectively acts like a $B-$dependent tilt term in the Hamiltonian. In all the plots $\mathbf{E}\parallel \hat{z}$, $\mathbf{B} = B(\cos\gamma,0,\sin\gamma)$, and the effect of the OMM has been considered. The insets show the relative magnitudes of the total CNLH current with respect to the interband scattering parameter $\alpha$. The current is seen to enhance including the effect of spin Zeeman coupling.}
\label{fig:SOC_Schematic_Fermi_Surface}
\end{figure*}

\subsection{Nonlinear anomalous Hall conductivity}
We are now in a position to discuss the chiral anomaly-induced nonlinear anomalous Hall conductivity in WSMs.
In the absence of any Weyl cone tilting and effects of orbital magnetic moment, the net anomalous velocity vector vanishes for each node resulting in zero nonlinear anomalous Hall conductivity. When the Weyl cones are untilted and when the effect of orbital magnetic moment is excluded, from symmetry considerations it is easy to conclude that the net CNLH current vanishes at each node. 
When effects of the orbital magnetic moment are included, the Fermi surface becomes egg-shaped (see Fig.~\ref{fig:WSM_Schematic_Fermi_Surface}) and the net anomalous velocity vector does not vanish, resulting in a nonzero CNLH current at each node. However, the net current vanishes when the contribution from both nodes is added up because the current at both nodes is of equal magnitudes and opposite signs. This is plotted in Fig.~\ref{fig:Single_node_CNLAH_WSM} for better clarification. Now, when the Weyl cones are further tilted, the nonlinear Hall current at both nodes is unequal in magnitude, resulting in a net non-zero CNLH current. Fig.~\ref{fig:WSM_Schematic_Fermi_Surface} schematically presents cross-sectional views of the Fermi surface of a Weyl semimetal, highlighting the mechanism resulting in a nonvanishing CNLH current.

In Fig.~\ref{fig:CNLH_Normal_phase_tx_vary}, we plot the CNLH conductivity $\sigma_{yzz}$ as function of $t_x$ and $\alpha$ for $\gamma=\pi/2$ (parallel electric and magnetic fields). We first note that the CNLH conductivity is an odd function of tilt $t_x$, in accordance with the findings of Ref.~\cite{li2021nonlinear}. However, we find that CNLH conductivity is non-monotonic as a function of tilt $t_x$, which is in striking contrast to Ref.~\cite{li2021nonlinear} that reports monotonic behavior with respect to $t_x$. 
We find that even small intervalley scattering results in non-monotonicity. The CNLH conductivity first increases as a function of $t_x$ and then decreases after reaching a maximum. Furthermore, for a fixed value of $t_{x}$, increasing the internode scattering strength $\alpha$, the magnitude of the CNLH conductivity decreases and eventually flips sign after a critical value $\alpha_c$, i.e. displays `strong-sign-reversal'. These twin effects cause a prominent `half-lung' like pattern as shown in Fig.~\ref{fig:CNLH_Normal_phase_tx_vary}(a). In Fig.~\ref{fig:CNLH_Normal_phase_tx_vary}(b), we plot CNLH conductivity as a function of $B$ at different values of the internode scattering strength $\alpha$ for fixed tilt. It is clear that increasing $\alpha$ results in sign-reversal of $\sigma_{yzz}$. Now, fixing $\alpha$ and increasing the amount of tilt, $\sigma_{yzz}$ also changes sign as shown in Fig.~\ref{fig:CNLH_Normal_phase_tx_vary}(c). The non-monotonicity of the CNLH effect and the existence of `strong-sign-reversal' are the prominent features we discover, which have been unreported so far. We attribute these to the effects of chirality-violating scattering and global charge conservation that have not been correctly accounted for in earlier studies.  

To gain further insight, we plot $\sigma_{yzz}$ as a function of $t_x$ in Fig.~\ref{fig:CNLH_tx_and_alp_normal_plt} (a) for different values of the internode scattering strength $\alpha$. The conductivity is highly non-monotonic--it first increases as a function of $t_x$ and decreases and eventually becomes close to zero when $t_x\approx 1$. Interestingly, we discover that as $\alpha$ is increased, (i) the conductivity increases, (ii) then quickly falls to zero for some value of $t_x<1$, (iii) then becomes negative, and (iv) finally approaches zero again when $t_x\approx 1$. When $\alpha$ is increased, $\sigma_{yzz}$ falls to zero and becomes negative at smaller and smaller values of $t_x$. When $\alpha$ is large enough, the conductivity $\sigma_{yzz}$ eventually reverses sign at $t_x\approx 0$.
In Fig.~\ref{fig:CNLH_tx_and_alp_normal_plt} (b), we plot $\sigma_{yzz}$ as a function of $\alpha$ for different values of the $t_x$. Remarkably, we find a sweet-spot at $\alpha\approx 1/3$, where $\sigma_{yzz}\approx 0$ for all values of $t_x\lesssim 0.25 v_F$. 

Having discussed the CNLH conductivity for collinear electric and magnetic fields and the effect of $\alpha$, we now discuss the case when $\mathbf{E}$ and $\mathbf{B}$ are noncollinear, since in many experimental setups, the effect of rotating the magnetic field is investigated. In Fig.~\ref{fig:CNLH_Normal_phase_gm_vary}(a) we plot $\sigma_{yzz}$ as a function of tilt $t_{x}$ and $\gamma$ for a finite  value of $\alpha$. The conductivity is an odd function of $t_x$, and is a non-monotonic function of the tilt for all values of $\gamma$. In Fig.~\ref{fig:CNLH_Normal_phase_gm_vary}(b), we plot $\sigma_{yzz}$ as a function of tilt $\alpha$ and $\gamma$ for two different fixed values of  value of $t_x$. For all angles of the magnetic field $\gamma$, the conductivity shows strong-sign reversal as a function of the intervalley scattering strength $\alpha$, however, the dependence of the zero-conductivity contour (separating the regions of positive and negative conductivity) is seen to be weak, unlike linear longitudinal magnetoconductivity that shows a stronger dependence on $\gamma$~\cite{ahmad2021longitudinal}. The dependence of the zero-conductivity contour is stronger on $t_x$, as we also note from Fig.~\ref{fig:CNLH_Normal_phase_tx_vary}. 

For completeness, we now discuss the case of tilt along the $z-$direction, i.e., the tilt-vector is $\mathbf{t=(0,0,t_z)}$. In Fig.~\ref{fig:WSM_NLH_tz_vary} (a), we plot CNLH in relevant parameter space for the case when Weyl cones are tilted along $z$-direction. This appears similar to Fig.~\ref{fig:CNLH_Normal_phase_tx_vary}(a) rotated by $\pi$. This is because the distribution function depends on the relative orientation of the tilt, magnetic field direction, and the electric field direction. When the Weyl nodes are tilted along the $x-$direction, the vector $\mathbf{t}\times\mathbf{B}$ is maximum for $\gamma=\pi/2$, i.e., when the electric and magnetic field are parallel to each other. When the nodes are tilted along the $z-$direction, this vector is maximum when $\gamma=0$. 
{Thus, the direction of the nonlinear current is given by given by $\mathbf{E}\times\boldsymbol{\Omega}_\mathbf{k}$, and the magnitude is proportional to $(\mathbf{t} \times \mathbf{B})\cdot (\mathbf{E}\times\boldsymbol{\Omega}_\mathbf{k})$. On expanding the cross product, we note that the magnitude is proportional to ($\mathbf{t}\cdot\mathbf{E}$)($\mathbf{B}\cdot\boldsymbol{\Omega}_\mathbf{k}$) - ($\mathbf{t}\cdot\boldsymbol{\Omega}_\mathbf{k}$)($\mathbf{B}\cdot\mathbf{E}
$). The second term is the usual chiral anomaly term, while the first term survives even when $\mathbf{E}\cdot\mathbf{B}=0$. It arises from the topological $(\mathbf{B}\cdot\boldsymbol{\Omega}_\mathbf{k})$ term, and contributes via (i) the density of states factor ($\mathcal{D}^\chi_\mathbf{k} = (1+e\mathbf{B}\cdot\boldsymbol{\Omega}^\chi_\mathbf{k}/\hbar)^{-1}$), and (ii) the -$\mathbf{m}.\mathbf{B}$ coupling due to anomalous OMM that induces an intrinsic Hall effect~\cite{das2021intrinsic}.}
This finishes the symmetry analysis of how CNLH behaves along different tilting directions.

\subsection{Effects of strain}
We next discuss the effect of strain on the nonlinear anomalous Hall conductivity. In a topological protected Weyl semimetal, Weyl nodes are separated in the momentum space by a finite vector $\mathbf{b}$. The vector $\mathbf{b}$ is also interpreted as an axial gauge field because of its opposite coupling to Weyl nodes of opposing chiralities~\cite{goswami2013axionic,volovik1999induced,liu2013chiral,grushin2012consequences,zyuzin2012topological}. A position-dependent $\mathbf{b}$ vector generates an axial magnetic field (denoted as $\mathbf{B}_5=\nabla\times \mathbf{b}$), which also couples oppositely to Weyl nodes of opposite chirality. Such a scenario can arise if Weyl semimetals are subjected to an inhomogeneous strain profile. The effective magnetic field experienced by a fermion at node $\chi$ is therefore $\mathbf{B}\longrightarrow\mathbf{B}+\chi\mathbf{B}_5$. Recent works have studied the role of strain in longitudinal and planar Hall conductivity~\cite{grushin2016inhomogeneous,ghosh2020chirality,ahmad2023longitudinal}, however, its role in the nonlinear anomalous Hall conductivity remains unexplored. 

In Fig.~\ref{fig:wsm_cnlh_B5} we plot the nonlinear Hall conductivity $\sigma_{yzz}$ as a function of the strain induced $\mathbf{B}_5$ field. As the intervalley scattering strength is increased the conductivity is suppressed. However, unlike Fig.~\ref{fig:CNLH_Normal_phase_tx_vary}, where we examined the effect of an external magnetic field, there is no strong-sign-reversal for large values of the intervalley scattering strength. 
This can be traced to the opposite coupling of the chiral gauge field at the two nodes of opposite chirality.
As mentioned above, the effective magnetic field on Weyl nodes becomes $\mathbf{B}\longrightarrow\mathbf{B}+\chi\mathbf{B}_5$, which in the absence of external magnetic field is reduced to $\chi\mathbf{B}_5$. When this effective magnetic field coupled to the OMM, the energy dispersion remains chirality independent, i.e., $-\mathbf{m^\chi(\mathbf{k}) \cdot\mathbf{B}}$ has the same sign and magnitude for both the nodes. This reflects in the absence of sign reversal for WSMs with respect to $B_5$. Similar conclusions for the longitudinal magnetoconductivity were found in Refs.~\cite{knoll2020negative,sharma2023decoupling}. Furthermore, we find that strain-induced nonlinear $\sigma_{yzz}$ also changes sign as a function of the tilt parameter as seen in Fig.~\ref{fig:wsm_cnlh_B5} (b). 
The measurement of nonlinear anomalous Hall conductivity in Weyl semimetals (i) in the presence of strain but the absence of magnetic field, and (ii) in the absence of strain but the presence of external magnetic field, can provide us crucial insights into the role and strength of internode scattering. For instance, if the measured nonlinear conductivity is negative in both scenarios, it is strongly suggestive of large internode scattering. Conversely, if the conductivity is positive in one case and negative in the other, it is indicative of weak internode scattering.

 \begin{figure*}
    \centering
    \includegraphics[width=  0.98\columnwidth]{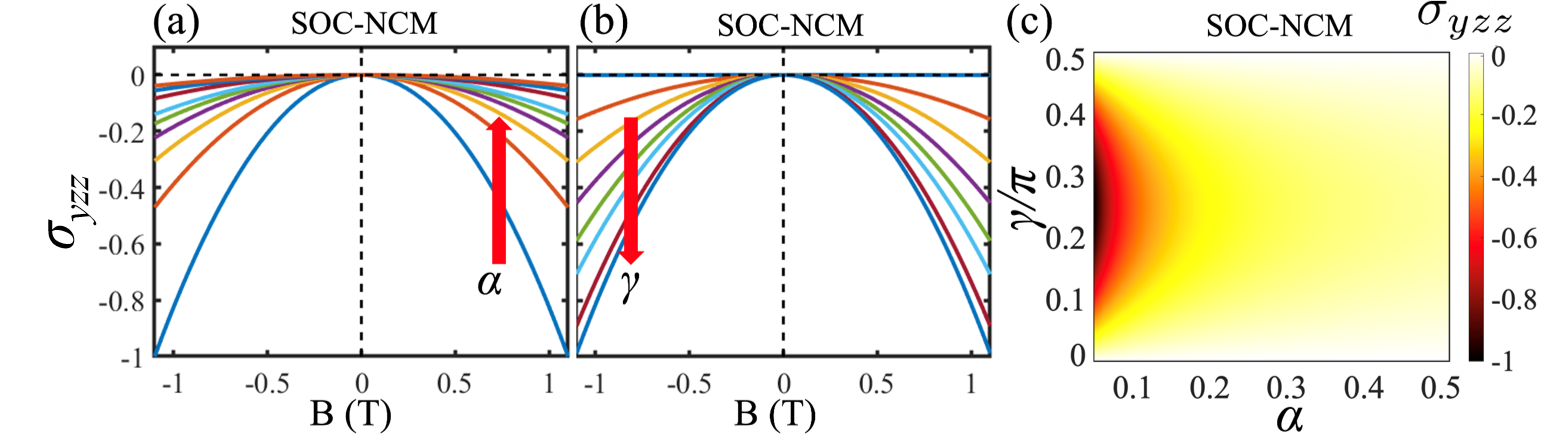}
    \caption{CNLH conductivity for spin-orbit coupled noncentrosymmetric metals for (a) different values of the interband scattering strength $\alpha$, and (b) for different values of $\gamma$ increasing from zero to $\pi/4$. In both plots, the corresponding parameters increase in the direction of the arrow. (c) The corresponding color plot indicates that the conductivity peaks at $\pi/4$ and decreases with increasing $\alpha$.}
    \label{fig:cnlahe_socnsm}
\end{figure*}

\section{CNLH in spin-orbit coupled noncentrosymmetric metals}
\subsection{Low-energy Hamiltonian}
We begin with the following low-energy Hamiltonian of a spin-orbit coupled noncentrosymmetric metal expanded near the high-symmetry point~\cite{mukherjee2012order,varma2024magnetotransport}
\begin{align}
    H(\mathbf{k}) = \frac{\hbar^2 k^2}{2m} + \hbar \vartheta \mathbf{k}\cdot\boldsymbol{\sigma} 
    \label{Eq_H_socncm}
\end{align}
where $m$ is the effective mass, $\vartheta$ incorporates the spin-orbit coupling parameter. We couple the spin degrees of freedom of the system to a Zeeman field given by~\cite{cano2017chiral,sharma2017nernst,gadge2022anomalous} 
\begin{align}
    H_z = -\mathbf{M}\cdot\boldsymbol{\sigma},
\end{align}
where $\mathbf{M}$ is related to the external magnetic field by $\mathbf{M} = -g\mu_B \mathbf{B}/2$, where $\mu_{B}$ is the Bohr magneton and $g$ is Landé g-factor ($g \sim 50$ \cite{xie2021kramers,yakunin2010spin}). Since $\boldsymbol{\sigma}$ does not in general refer to pure spin degrees of freedom in WSMs, we have not considered this term in WSMs. The resultant Hamiltonian, including the spin Zeeman coupling, becomes 
\begin{align}
    H(\mathbf{k}) = \frac{\hbar^2 k^2}{2m} + \hbar \vartheta \left(\mathbf{k}+\frac{\mathbf{M}}{\hbar \vartheta}\right)\cdot\boldsymbol{\sigma},
\end{align} 
A change of variables ($\mathbf{k}\rightarrow \mathbf{k}-\mathbf{M}/\hbar\vartheta$) yields 
\begin{align}
    H(\mathbf{k}) = \frac{\hbar^2k^2}{2m} + \hbar\vartheta\mathbf{k}\cdot\boldsymbol{\sigma} + \hbar\vartheta(k_x t_x + k_zt_z) + E_0,
\end{align}
where $E_0 = M^2/2m\vartheta^2$ is an irrelevant constant energy shift, $t_x = -M_x/m\vartheta^2$, $t_z = -M_z/m\vartheta^2$. Remarkably, in the new reference frame, the Hamiltonian resembles that of a tilted Weyl semimetal! The effect of the spin Zeeman coupling is therefore to tilt the Fermi surfaces just like the tilted Weyl cones of a Weyl semimetal. Note that the effective tilt is proportional to the amount of spin Zeeman coupling and inversely proportional to the effective mass $m$. Therefore, for a purely relativistic $\mathbf{k}\cdot\boldsymbol{\sigma}$ Hamiltonian, where the effective mass term $\sim k^2/m\rightarrow 0$, the effective tilt term vanishes. Hence this property of the noncentrosymmetric metal is distinct from an inversion asymmetric Weyl semimetal where the effective mass term $\sim k^2/m\rightarrow 0$ is absent. 
The energy spectrum is evaluated to be:
\begin{align}
    \epsilon_\mathbf{k}^\lambda = \frac{\hbar^2 k^2}{2 m} + \lambda \hbar \vartheta {k}
    +\hbar \vartheta (k_{x} t_{x} + k_{z} t_{z}) + E_{\mathrm{0}},
\label{eq:E_Eigen_value_SOC_3_in_q}
\end{align}
with, $\lambda=\pm1$ representing two spin-orbit split bands. We note that both the Fermi surfaces are tilted along the same direction as a result of the spin Zeeman field. To obtain the constant energy Fermi contour, we need to add the orbital magnetic moment coupling to the energy spectrum and invert Eq.~\ref{eq:E_Eigen_value_SOC_3_in_q}. This yields a cubic equation in $k$ that needs to be solved for $k=k(\theta,\phi)$. Since the analytical expression is lengthy and uninteresting, we do not provide it here. The change of variables is implemented straightforwardly in the Boltzmann equation. The Jacobian remains invariant, but the constant energy Fermi contour is appropriately modified while integrating over a constant energy surface in the Boltzmann equation.

Without loss of generality, we assume that the chemical potential lies above the nodal point $\mathbf{k}=0$, and hence the Fermi surface is composed of two disjointed surfaces as shown in Fig.~\ref{fig:SOC_Schematic_Fermi_Surface}. Both the surfaces enclose a nontrivial flux of Berry curvature, which is of the same magnitude but opposite sign. This is similar to the case of a Weyl semimetal where the Berry curvature is of the same magnitude but opposite signs at the two valleys. Interestingly, in SOC-NCM, the anomalous orbital magnetic moment 
($\mathbf{m}^\lambda_\mathbf{k}$) has the same sign and magnitude, which is different from a Weyl semimetal where the signs are reversed at the two nodes. With the application of an external magnetic field, the orbital magnetic moment couples to it as $-\mathbf{m}^\lambda_\mathbf{k}\cdot\mathbf{B}$, leading to the oval-shaped Fermi surfaces as shown in Fig.~\ref{fig:SOC_Schematic_Fermi_Surface}. In Weyl semimetal, the coupling is opposite in the two valleys, and thus, the shapes of Fermi the surfaces are reversed (see Fig.~\ref{fig:WSM_Schematic_Fermi_Surface}).

\subsection{Nonlinear anomalous Hall conductivity}
In Fig.~\ref{fig:SOC_Schematic_Fermi_Surface} (a) and (c), we plot a cross-sectional view of the Fermi surface of the SOC-NCM including the effect of the orbital magnetic moment but without considering spin Zeeman coupling (which is the effective tilt term). The net anomalous velocity vector ($\sim \mathbf{E}\times\boldsymbol{\Omega}_\mathbf{k}$) at both the nodes does not vanish since the Fermi surface is no longer symmetrical around the $k_x-k_y$ plane. Furthermore, the magnitudes of the CNLH current at the two nodes are not of equal magnitudes (unlike the case of a WSM). This results in a net nonzero CNLH current, unlike the WSM where the orbital magnetic moment alone does not result in a nonzero current. Fig.~\ref{fig:SOC_Schematic_Fermi_Surface} (b) and (d) depict the effect of including spin Zeeman coupling, which further introduces an asymmetry in the Fermi surface and enhances the total CNLH current. 
In Fig.~\ref{fig:cnlahe_socnsm}, we plot the nonlinear anomalous Hall conductivity for spin-orbit noncentrosymmetric metal described in Eq.~\ref{Eq_H_socncm}. We find that the nonlinear conductivity $\sigma_{yzz}$ is quadratic in the magnetic field, in contrast to a Weyl semimetal where $\sigma_{yzz}$ is seen to be linear in $B$. An additional $B-$dependence enters in SOC-NCMs because (i) the current here is driven by anomalous orbital magnetic moment unlike in WSM where it is driven by a finite tilt, and (ii) the generated effective tilt due to the spin Zeeman coupling is field-dependent, unlike in WSMs, where a constant tilt that is inherent to the band-structure is assumed. 

To further analyze the magnetic field dependence, we examine how different quantities that contribute to the current vary with the magnetic field. In the case of WSMs, the magnetic field couples to the dispersion relation through the orbital magnetic moment (OMM) term ($-\mathbf{m^{\chi}_{\mathbf{k}}} \cdot \mathbf{B}$). This coupling term causes the Fermi contour $k_F^\chi$ to depend on $\chi$, $\gamma$, and $B$, which makes the Berry curvature field dependent $\Omega^\chi_k\sim (k^\chi(B))^{-2}$. While the CNLH current in Eq.~\ref{Eq:CNLH_componetns_formula} appears in a simplified form, incorporating the magnetic field dependence into the relevant quantities leads to the general form of the conductivity tensor shown in Eq.~\ref{Eq-sij-fit}. The presence of constant tilt terms in the dispersion introduces a linear-in-$B$ term in the conductivity tensor. For untilted WSMs, the CNLH conductivity is quadratic in $B$ and opposite in sign (Fig.~\ref{fig:Single_node_CNLAH_WSM}), canceling contributions from both bands. However, when constant tilt terms are present, the CNLH conductivity includes a linear-in-$B$ term with the same sign in both valleys, producing a non-zero linear-in-$B$ CNLH conductivity. This is in agreement with the symmetry considerations that both inversion and time-reversal symmetry are broken in Eq.~\ref{Hamiltonian_ch4}. 
In SOC-NCMs, we find that $\sigma_{yzz} \propto B^2$. Furthermore, it is quadratic in $B$ in both bands, but their magnitudes are different. The band with the smaller Fermi surface (see Fig.~\ref{fig:SOC_Schematic_Fermi_Surface}) has the greater contribution of the two. In the absence of Zeeman coupling, time-reversal symmetry is preserved, generating a quadratic response. A Zeeman coupling acts like a tilt term, and in principle allows for a linear-in-$B$ term, but since the effective tilt is itself a function of the magnetic field, the overall response is still quadratic. {An additional physical insight can be obtained by focusing on the first order deviation of the equilibrium distribution function ($f_0$). In presence of an anomalous OMM and Berry curvature, there exists an anomalous velocity ($\mathbf{v}^{\chi}_{\mathrm{anom}} \sim \mathbf{E} \times \boldsymbol{\Omega}^{\chi}_{\mathbf{k}}$) and a first-order effect from the magnetic field in the equilibrium distribution function $f^{\chi}_{\mathbf{k}}$, represented by -$\mathbf{m^{\chi}_{\mathbf{k}}}.\mathbf{B}$ $\frac{df_0}{d\epsilon^{\chi}_{\mathbf{k}}}$, which induces an intrinsic Hall effect \cite{das2021intrinsic}. Therefore, chiral charge pumping and the intrinsic Hall effect work in tandem, producing a current that displays a $E^2 B^2$ behavior.} Furthermore, in SOC-NCMs, we find the conductivity $\sigma_{yzz}$ to be negative, which is suppressed with increasing intervalley scattering strength, but importantly does not flip its sign. This is contrasted to the case of Weyl semimetal where a strong-sign-reversal is observed. This crucial difference is attributed to the different nature of the orbital magnetic moment in both cases. 
\begin{figure}
    \centering   \includegraphics[width= 0.98\columnwidth]{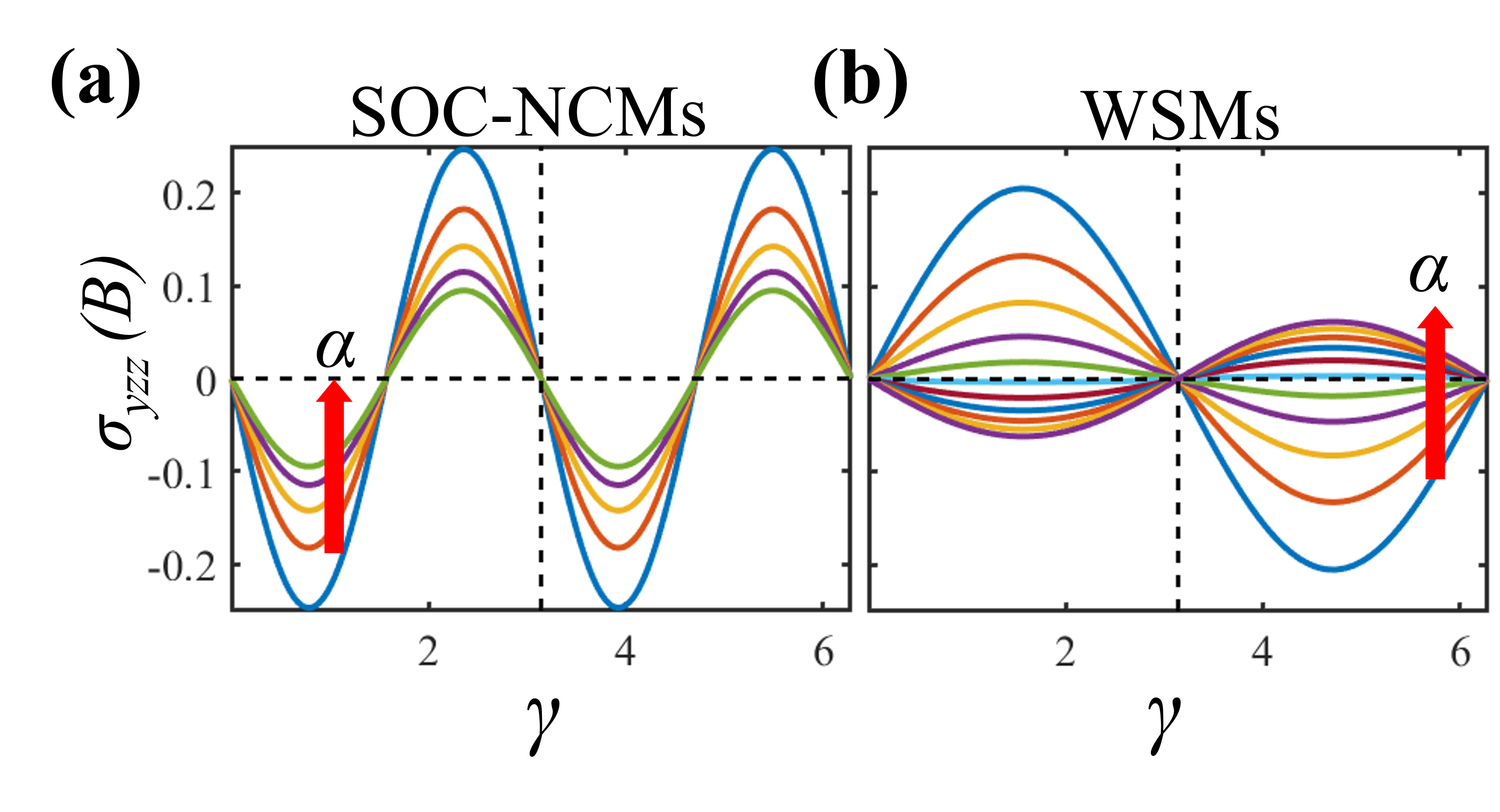}
    \caption{CNLH conductivity $\sigma_{yzz}$ as a function of magnetic field angle $\gamma$ for both the system, i.e., WSMs and SOC-NCMs. For WSMs, $\sigma_{yzz} \propto \sin(\gamma)$, and CNLH current has shown to be extremum at $\gamma=\pi/2$. For SOC-NCMs, $\sigma_{yzz} \propto \sin(2\gamma)$ and maxima occurs at $\gamma=\pi/4$. This figure highlights the fact that both systems don't show the same angular dependence due to different symmetries of the Hamiltonian.}
\label{fig:NLH_vs_gm}
\end{figure}
The behavior of $\sigma_{yzz}$ with the angle of the magnetic field $\gamma$ in the current case is also of special interest. The angular dependence of the CNLH conductivity for both systems is illustrated in Fig.~\ref{fig:NLH_vs_gm} at various intervalley (interband) scattering strengths $\alpha$. Unlike the case of Weyl semimetal, where $\sigma_{yzz}$ is maximum when $\gamma=\pi/2$ (when the electric and magnetic field are parallel to each other), in SOC-NCMs, the conductivity is maximum when $\gamma=\pi/4$, i.e., the electric and magnetic field are at $\pi/4$ angle with respect to each other. This important distinction is understood as follows. First, we observe that in WSMs, the current is driven by a finite tilt of the Weyl cones along the $k_x$ direction (as we need an asymmetric Fermi surface). Therefore the current is maximum when the magnetic field points along $\hat{z}-$direction, parallel to the electric field. In SOC-NCMs, the current is driven by anomalous magnetic moment. 
Now, when $\mathbf{E}\parallel\mathbf{B}\parallel \hat{z}$, integral of the anomalous velocity vector $({v}_\textrm{anom}\propto \Omega_x\propto\cos\phi)$ vanishes due to azimuthal symmetry, and therefore the net anomalous current vanishes as well. In WSMs, azimuthal symmetry is destroyed due to a finite $t_x$, even when $\mathbf{E}\parallel\mathbf{B}\parallel \hat{z}$. The different angular dependencies are similar to the case of a Weyl semimetal, which shows a $\sin \gamma$ dependence for the planar Hall effect in the presence of tilt and $\sin 2\gamma$ in the absence of tilt, which can also be explained from symmetry arguments~\cite{nandy2017chiral}.\\

\section{Conclusions}
In this work, we advance the theoretical understanding of the chiral anomaly-induced nonlinear anomalous Hall effect (CNLHE) in three-dimensional chiral fermionic systems, with a particular focus on Weyl semimetals (WSMs) and spin-orbit coupled noncentrosymmetric metals (SOC-NCMs). By rigorously incorporating momentum-dependent chirality-preserving and chirality-breaking scattering processes, as well as global charge conservation, we address critical gaps in the existing models, thereby providing a more robust and comprehensive framework for analyzing CNLHE.
In the context of Weyl semimetals, we uncover a complex, nonmonotonic relationship between the nonlinear anomalous Hall conductivity and the Weyl cone tilt. This behavior is notably sensitive to the strength of internode scattering, leading to a `strong-sign-reversal' of the conductivity. Moreover, we also investigate the effects of strain-induced chiral gauge fields on CNLHE, demonstrating that while such strain can indeed generate nonlinear Hall effects, it does so without inducing a sign reversal in the conductivity. Experiments performed with and without external strain in WSMs can shed light on the role of internode scattering by comparing the nonlinear anomalous Hall conductivity (NLAHC) in both scenarios. 
For spin-orbit coupled noncentrosymmetric metals, we reveal that the anomalous orbital magnetic moment is sufficient to drive a large nonlinear conductivity, which is distinguished by its negative sign, regardless of the strength of interband scattering, and its quadratic dependence on the magnetic field. This behavior starkly contrasts with the linear magnetic field dependence observed in WSMs and highlights the fundamental differences between these two classes of materials. We also identify the spin Zeeman coupling of the magnetic field as a crucial factor that acts as an effective $B-$dependent tilt term, further amplifying the CNLHE in SOC-NCMs. The theoretical insights presented in this work extend the current understanding of CNLHE in chiral quasiparticles and provide a critical foundation for current and upcoming experimental investigations. 
\subsection{{Quantum Oscillations as a Probe of Chiral Anomaly and Nonlinear Transport}}

{Our study primarily examines nonlinear Hall transport signatures, particularly the chiral anomaly-induced nonlinear Hall effect (CNLHE). However, it is important to note that quantum oscillations, including the Shubnikov-de Haas (SdH) oscillations, serve as an independent and sensitive probe of Landau level quantization and Berry curvature phenomena in Weyl semimetals.  SdH oscillations in magnetoresistance specifically result from the periodic modulation of the density of states as Landau levels intersect the Fermi energy in the presence of an applied magnetic field.}

{ The oscillations convey precise information regarding the Fermi surface geometry, Berry phase, and chiral Landau level structure, which are essential for understanding the chiral anomaly's manifestation.  They offer a complementary method to experimentally validate the theoretical predictions presented in this thesis concerning LMC, PHE, and particularly CNLHE.  Recent research \cite{zeng2023quantum} has shown that the interaction between nonlinear response and Landau quantization can lead to quantum oscillations in nonlinear transport coefficients, which are directly associated with chiral anomaly physics.}

 {Although the current model does not explicitly incorporate quantum oscillatory behavior, future research may expand this framework to include quantized magnetic fields and analyze the modulation of nonlinear Hall conductivity.  One could examine the influence of Berry curvature and orbital magnetic moment on the phase and amplitude of Shubnikov-de Haas oscillations in the nonlinear regime.  We expect that in materials characterized by low carrier density and clean limit conditions—such as $TaAs$, $NbAs$, or $WTe2$—quantum oscillations of nonlinear Hall signals may act as reliable indicators of chiral anomaly.  We propose SdH measurements as a valuable experimental tool for the future investigation of CNLHE, especially in systems where traditional signatures may be masked by disorder or symmetry limitations.}

\chapter{\label{chap5}Chiral anomaly and longitudinal magnetoconductance in pseudospin-1 fermions}
{\small The contents of this chapter have appeared in ``\textsc{Chiral anomaly and longitudinal magnetoconductance in pseudospin-1 fermions}"; Azaz Ahmad and Gargee Sharma; \ \textit{arXiv, cond-mat.mes-hall} \textbf {2412}, 10500 (2024). This paper is currently under review.}
\section{Abstract}
Chiral anomaly (CA), a hallmark of Weyl fermions, has emerged as a cornerstone of condensed matter physics following the discovery of Weyl semimetals. While the anomaly in pseudospin-1/2 (Weyl) systems is well-established, its extension to higher-pseudospin fermions remains a frontier with critical implications for transport phenomena in materials with multifold fermions.  We present a rigorous quasiclassical analysis of CA and longitudinal magnetotransport in pseudospin-1 fermions, advancing beyond conventional models that assume constant relaxation times and neglect the orbital magnetic moment and global charge conservation. Our study uncovers a magnetic-field dependence of the longitudinal magnetoconductance: it is positive and quadratic-in-$B$ for weak internode scattering and transitions to negative values beyond a critical internode scattering strength. Notably, the critical threshold is lower for pseudospin-1 fermions compared to their pseudospin-1/2 counterparts. We show analytically that the zero-field conductivity is affected more strongly by internode scattering for pseudospin-1 fermions than conventional Weyl fermions. 
These insights provide a foundational framework for interpreting recent experiments on multifold fermions and offer a roadmap for probing CA in candidate materials with space group symmetries 
199, 214, and 220. 
\section{Introduction}
Chiral anomaly (CA) of Weyl fermions was first discovered as a contributing factor in the decay of pions~\cite{bertlmann2000anomalies}. Over the past decade, chiral anomaly has seen a remarkable resurgence in condensed matter physics, driven by the definitive discovery of Weyl fermions in solids~\cite{hosur2013recent,armitage2018weyl,Yan_2017,hasan2017discovery,burkov2018weyl,ong2021experimental,nagaosa2020transport,lv2021experimental,mandal2022chiral,ahmad2024geometry}. Theoretical advances, however, date back to the 1980s when Nielsen \& Ninomiya, who first studied lattice Weyl fermions, proved that they must occur in pairs of opposite chiralities~\cite{nielsen1981no,nielsen1983adler}. Such pairing ensures the conservation of both the chiral and global charge. When external gauge fields are present, chiral charge is not conserved, a phenomenon that is now well known as the chiral anomaly or the Adler-Bell-Jackiw (ABJ) anomaly~\cite{adler1969axial} of Weyl fermions. The manifestation of this anomaly is investigated through transport, thermoelectric, and optical experiments in systems hosting Weyl fermions, known as Weyl semimetals (WSMs)~\cite{parameswaran2014probing,hosur2015tunable,goswami2015optical,goswami2013axionic,son2013chiral,burkov2011weyl,burkov2014anomalous,lundgren2014thermoelectric,sharma2016nernst,kim2014boltzmann,zyuzin2017magnetotransport,cortijo2016linear,das2019berry,kundu2020magnetotransport}.

While Weyl fermions have made an entry from high-energy physics to condensed matter, certain symmetries of condensed matter systems make it possible to realize free fermionic excitations which are not allowed by Poincar\'{e} symmetry in high-energy physics~\cite{bradlyn2016beyond,tang2017multiple,chang2018topological}. Specifically, the low-energy Hamiltonian of a Weyl fermion is $H_\mathbf{k}\sim\chi\mathbf{k}\cdot\boldsymbol{\sigma}$, where $\boldsymbol{\sigma}$ is the vector of Pauli matrices, and $\chi$ is the chirality of the fermion. Fermionic excitations of the type $H_\mathbf{k}\sim \chi\mathbf{k}\cdot\mathbf{S}$ are allowed in solids, where $\mathbf{S}$ is a higher (pseudo)spin generalization of the vector of Pauli matrices. These excitations are multifold degenerate chiral quasiparticles, carry a nontrivial Chern number $|\mathcal{C}|>1$, and are sources and sinks of the Berry curvature. In contrast, Weyl fermions have a twofold degeneracy with Chern number $|\mathcal{C}|=1$.

Although CA has been heavily discussed in the context of Weyl fermions or even Kramer-Weyl fermions~\cite{cheon2022chiral,das2023chiral,varma2024magnetotransport}, its generalization beyond fermions of pseudospin$-1/2$ has received little attention. On generic grounds, chiral particles respond differently to external magnetic fields, and
one does expect the anomaly to persist. When external fields are applied such that $\mathbf{E}\cdot\mathbf{B}\neq 0$, the chiral charge is not conserved, and a chiral current is generated. There have been theoretical advances based on this idea, aimed at understanding chiral anomaly and associated transport features in multifold fermions~\cite{ezawa2017chiral,lepori2018axial,nandy2019generalized}. Still, the quasiclassical analysis suffers from shortcomings such as imposing a constant relaxation time, neglecting orbital magnetic moment, internode scattering effects, and global charge conservation. Going beyond these standard assumptions is indispensable to correctly reproduce the physics of chiral anomaly and magnetotransport in chiral fermions in the experimentally accessible low-field limit as shown in some recent works~\cite{knoll2020negative,sharma2020sign,ahmad2021longitudinal,ahmad2023longitudinal}.

This Letter presents a complete quasiclassical analysis of chiral anomaly and longitudinal magnetotransport in pseudospin-1 fermions. Moving beyond the conventional constant relaxation-time approximation and taking into account the effects of orbital magnetic moment and global charge conservation, we discover that chiral anomaly in pseudospin-1  fermions generates a positive and quadratic-in-$B$ longitudinal magnetoconductance for low internode scattering, which becomes negative as the strength of internode scattering ($\alpha$) is increased beyond a critical value ($\alpha^{(1)}$). Interestingly, this critical intervalley scattering strength is lower in pseudospin-1 fermions than the regular Weyl-fermions ($\alpha^{(1)}<\alpha^{(1/2)}$). We also show analytically that the zero-field conductivity decreases sharply with internode scattering for pseudospin-1 fermions than conventional Weyl fermions.
This study becomes even more pertinent in light of recent experiments that have probed this anomaly in multifold fermionic systems~\cite{balduini2024intrinsic}, and upcoming experiments that could be performed on relevant materials belonging to space groups $199$, $214$, and $220$~\cite{bradlyn2016beyond}.  

\section{Semimetals with 3-fold degeneracy} The low-energy Hamiltonian of the pseudospin-1 fermion expanded about a nodal point is~\cite{bradlyn2016beyond}:
\begin{equation}
H(\mathbf{k})=\hbar v_{F}\mathbf{k}\cdot\mathbf{S}.
\label{Hamiltonian1_ch5}    
\end{equation}
where $v_{F}$ is a material-dependent velocity parameter, $\mathbf{k}$ is the wave vector measured from the nodal point, and $\mathbf{S}$ is the vector of spin-1 Pauli matrices. One may diagonalize this Hamiltonian to get energy eigenvalues:
\begin{align}
\epsilon_k = 0, ~\pm{\hbar v_{F}}k.
\label{Dispersion_s1_ch5}
\end{align}
The dispersion consists of three bands, including a flat band with zero energy.
The Bloch-states corresponding to the non-zero energies are calculated to be:
\begin{align}
|u^+\rangle = \left[\cos^2(\theta) e^{-2i\phi}, ~ \frac{1}{\sqrt{2}}\sin(\theta) e^{-i\phi},  ~\sin^2(\theta/2)\right]^\mathrm{T}, \nonumber\\
|u^-\rangle = \left[\sin^2(\theta/2) e^{-2i\phi}, ~ \frac{-1}{\sqrt{2}}\sin(\theta) e^{-i\phi},  ~\cos^2(\theta/2)\right]^\mathrm{T},
\label{Eq:wave_function}
\end{align}
with $\theta$ and $\phi$ being the polar and azimuthal angles respectively. The Chern numbers of the bands are $\nu=\mp 2$, while the flat band is trivial. 
Such fermions may emerge in a body-centered cubic lattice with space groups 199, 214, and 220 at the $\mathbf{P}$ point of the BZ~\cite{bradlyn2016beyond}. The $\mathbf{P}$ point does not map to its time-reversal partner ($\mathbf{P}\neq -\mathbf{P}$).
Due to the Nielsen-Ninomiya theorem, the Brillouin zone must compensate for sources and sinks of the Berry curvature~\cite{nielsen1981no,nielsen1983adler}. Therefore, the low-energy minimal Hamiltonian for pseudospin-1 semimetal may be written as 
\begin{align}
    H(\mathbf{k})=\sum_{\chi=\pm 1} \chi \hbar v_{F}\mathbf{k}\cdot\mathbf{S}.
    \label{Eq:Hamiltonian2_ch5}
\end{align}
The Berry curvature of the conduction bands of the Hamiltonian is evaluated to be:
\begin{align}
  \boldsymbol{\Omega}^\chi_\mathbf{k} = i \nabla_{\mathbf{k}} \times \langle u^{\chi}(\mathbf{k}) | \nabla_{\mathbf{k}} | u^{\chi}(\mathbf{k}) \rangle \equiv -\chi \mathbf{k} /k^3.
    \label{Eq:BC}
\end{align}
In contrast to a classical point particle, a Bloch wave packet in a crystal possesses a finite spatial extent. Consequently, it undergoes self-rotation around its center of mass, resulting in an orbital magnetic moment (OMM), expressed as ~\cite{xiao2010berry,hagedorn1980semiclassical,chang1996berry},
\begin{align}
\mathbf{m}^{\chi}_\mathbf{k}&= -\frac{ie}{2\hbar} \text{Im} \langle \nabla_{\mathbf{k}} u^{\chi}|[ \epsilon_0(\mathbf{k}) - \hat{H}^{\chi}(\mathbf{k}) ]| \nabla_{\mathbf{k}} u^{\chi}\rangle \nonumber\\
&=-{\chi e v_{F} \mathbf{k}}/{k^2}.
\label{Eq:OMM}
\end{align}
In the presence of an external magnetic field ($\mathbf{B}$), the orbital magnetic moment couples to it, and the dispersion relation is modified as $\epsilon^\chi_\mathbf{k} \rightarrow \epsilon^\chi_\mathbf{k} - \mathbf{m}^{\chi}_\mathbf{k} \cdot \mathbf{B}$. Consequently, the Fermi contour becomes anisotropic: 
\begin{align} k_F^{\chi}(\theta) = \frac{\epsilon_{F} + \sqrt{\epsilon^{2}_{F} - \chi \eta ~\epsilon^2_0 \cos(\theta)}}{2 v_{F}\hbar}.
\label{Eq:K_chi}
\end{align}
with $\epsilon_F$ being the Fermi energy, $\epsilon_0 = \sqrt{4 eB  v^2_{F}\hbar}$ and the variable $\eta\in\{0, 1\}$ is used to toggle the effect of the orbital magnetic moment, such that its effect can be studied separately. 

\section{Quasiclassical transport} The dynamics of the quasiparticles in the presence of electric ($\mathbf{E}$) and magnetic ($\mathbf{B}$) fields, are described by the following equation~\cite{son2012berry,knoll2020negative}:
\begin{align}
\dot{\mathbf{r}}^\chi &= \mathcal{D}^\chi_\mathbf{k} \left( \frac{e}{\hbar}(\mathbf{E}\times \boldsymbol{\Omega}^\chi) + \frac{e}{\hbar}(\mathbf{v}^\chi\cdot \boldsymbol{\Omega}^\chi) \mathbf{B} + \mathbf{v}_\mathbf{k}^\chi\right) \nonumber\\
\dot{\mathbf{p}}^\chi &= -e \mathcal{D}^\chi_\mathbf{k} \left( \mathbf{E} + \mathbf{v}_\mathbf{k}^\chi \times \mathbf{B} + \frac{e}{\hbar} (\mathbf{E}\cdot\mathbf{B}) \boldsymbol{\Omega}^\chi \right).
\label{Couplled_equation_ch5}
\end{align}
To describe the dynamics of three-dimensional pseudospin-1 fermions under external electric and magnetic fields, we employ the quasiclassical Boltzmann formalism, where the evolution of the non-equilibrium distribution function $f^{\chi}_{\mathbf{k}}$ is given by:
\begin{align}
\dfrac{\partial f^{\chi}_{\mathbf{k}}}{\partial t}+ {\Dot{\mathbf{r}}^{\chi}_{\mathbf{k}}}\cdot \mathbf{\nabla_r}{f^{\chi}_{\mathbf{k}}}+\Dot{\mathbf{k}}^{\chi}\cdot \mathbf{\nabla_k}{f^{\chi}_{\mathbf{k}}}=I_{\mathrm{coll}}[f^{\chi}_{\mathbf{k}}],
\label{MB_equation}
\end{align}
with $f^{\chi}_\mathbf{k} = f_{0} + g^{\chi}_{\mathbf{k}}$, where $f_{0}$ is the Fermi-Dirac distribution, and $g^{\chi}_{\mathbf{k}}$ is the deviation. We fix the electric and magnetic fields along the $z-$direction and the deviation up to the first order in perturbation is expressed as:
\begin{align}
g^{\chi}_\mathbf{k}&= -e\left({\dfrac{\partial f_{0}}{\partial {\epsilon}}}\right){\Lambda^{\chi}_\mathbf{k}} E,
\label{Eq:g1}
\end{align}
where ${\Lambda}^{\chi}_\mathbf{k}$ is unknown function to be evaluated. The collision integral considers two impurity-dominated distinct scattering processes: (i) scattering between $\chi$ and $\chi'\neq\chi$, and (ii) scattering between $\chi$ to $\chi'=\chi$. These are also known as internode (intervalley) scattering, and intranode (intravalley) scattering, respectively. The collision integral is expressed as~\cite{son2013chiral,kim2014boltzmann}:
\begin{align}
 I_{coll}[f^{\chi}_{\mathbf{k}}]=\sum_{\chi' \mathbf{k}'}{\mathbf{W}^{\chi \chi'}_{\mathbf{k k'}}}{(f^{\chi'}_{\mathbf{k'}}-f^{\chi}_{\mathbf{k}})},
 \label{Collision_integral_ch5}
\end{align}
where the scattering rate ${\mathbf{W}^{\chi \chi'}_{\mathbf{k k'}}}$ calculated using Fermi's golden rule:
\begin{align}
\mathbf{W}^{\chi \chi'}_{\mathbf{k k'}} = \frac{2\pi n}{\mathcal{V}}|\bra{u^{\chi'}(\mathbf{k'})}U^{\chi \chi'}_{\mathbf{k k'}}\ket{u^{\chi}(\mathbf{k})}|^2\times\delta(\epsilon^{\chi'}(\mathbf{k'})-\epsilon_F).\nonumber\\
\label{Fermi_gilden_rule}
\end{align}
Here $n$ represents the impurity concentration, $\mathcal{V}$ represents the system volume, and $U^{\chi \chi'}_{\mathbf{k k'}}$ describes the scattering potential profile. For elastic impurities, we set $U^{\chi \chi'}_{\mathbf{k k'}}= I_{3\times3}U^{\chi \chi'}$, where $U^{\chi \chi'}$ controls scattering between electrons of different ($\chi\neq\chi'$) and same chirality ($\chi=\chi'$). The relative magnitude of chirality-breaking to chirality-preserving scattering is expressed by the ratio $\alpha = U^{\chi\chi'\neq\chi}/U^{\chi\chi}$. The overlap $\mathcal{T}^{\chi \chi'}({\theta,\theta',\phi,\phi'})=|\bra{u^{\chi'}(\mathbf{k'})}U^{\chi \chi'}_{\mathbf{k k'}}\ket{u^{\chi}(\mathbf{k})}|^2$ is generally a function of both the polar and azimuthal angles, making the scattering strongly anisotropic. However, for the chosen arrangement of the fields, i.e., $\mathbf{E}=E\hat{z}$, and  $\mathbf{B} = B \hat{z} $, the terms involving $\phi$, and $\phi'$ are irrelevant due to azimuthal symmetry as they vanish when the integral with respect to $\phi'$ is performed. We drop such terms and write the the overlap function as 
\begin{align}
\mathcal{T}^{\chi \chi'}_{\theta,\theta}
=\begin{cases}
			[\cos^4(\theta/2) \cos^4(\theta'/2) + \sin^4(\theta/2) \sin^4(\theta'/2)\\+ \frac{1}{4} \sin^2(\theta) \sin^2(\theta')]
            \delta_{\chi,\chi'}+\\
            [\sin^4(\theta/2) \cos^4(\theta'/2) + \cos^4(\theta/2) \sin^4(\theta'/2)\\+ \frac{1}{4} \sin^2(\theta) \sin^2(\theta')]\delta_{\chi,-\chi'}
		 \end{cases}
   \label{Overlap_of_spinor}
\end{align}

Using Eq's.~\ref{Couplled_equation_ch5}, \ref{Eq:g1} and \ref{Collision_integral_ch5}, Eq.~\ref{MB_equation} is written in the following form:
\begin{align}
&\mathcal{D}^{\chi}_\mathbf{k}\left[{v^{\chi,z}_{\mathbf{k}}}+\frac{eB}{\hbar}(\mathbf{v^{\chi}_k}\cdot\boldsymbol{\Omega}^{\chi}_k)\right]
 = \sum_{\chi' \mathbf{k}'}{\mathbf{W}^{\chi \chi'}_{\mathbf{k k'}}}{(\Lambda^{\chi'}_{\mathbf{k'}}-\Lambda^{\chi}_{\mathbf{k}})}.
 \label{Eq_boltz_E1}
 \end{align}
 Before further simplifying the above equation, we define the chiral scattering rate as follows:
\begin{align}
\frac{1}{\tau^{\chi}(\theta)}=\sum_{\chi'}\mathcal{V}\int\frac{d^3\mathbf{k'}}{(2\pi)^3}(\mathcal{D}^{\chi'}_{\mathbf{k}'})^{-1}\mathbf{W}^{\chi \chi'}_{\mathbf{k k'}}.
\label{Tau_invers}
\end{align}
Eq.~\ref{Eq_boltz_E1} then transforms to:
\begin{align}
h^{\chi}(\theta) + \frac{\Lambda^{\chi}(\theta)}{\tau^{\chi}(\theta)}=\sum_{\chi'}\mathcal{V}\int\frac{d^3\mathbf{k}'}{(2\pi)^3} \mathcal{D}^{\chi'}_{\mathbf{k}'}\mathbf{W}^{\chi \chi'}_{\mathbf{k k'}}\Lambda^{\chi'}(\theta').
\label{MB_in_term_Wkk'}
\end{align}
Here, $h^{\chi}(\theta)=\mathcal{D}^{\chi}_{\mathbf{k}}[v^{\chi}_{z,\mathbf{k}}+eB(\boldsymbol{\Omega}^{\chi}_{k}\cdot \mathbf{v}^{\chi}_{\mathbf{k}})/
\hbar]$, evaluated on the Fermi surface. Employing azimuthal symmetry, Eq.~\ref{Tau_invers} and Eq.~\ref{MB_in_term_Wkk'} simplify to an integration over $\theta'$:
\begin{align}
\frac{1}{\tau^{\chi}(\theta)} =  \mathcal{V}\sum_{\chi'} \Pi^{\chi\chi'}\int\frac{(k')^3\sin{\theta' d\theta'}}{|\mathbf{v}^{\chi'}_{k'}\cdot{\mathbf{k'}^{\chi'}}|} \mathcal{T}^{\chi\chi'}_{\theta \theta'}(\mathcal{D}^{\chi'}_{\mathbf{k'}})^{-1}.
\label{Tau_inv_int_theta}
\end{align}
\begin{align}
&h^{\chi}(\theta) + \frac{\Lambda^{\chi}(\theta)}{\tau^{\chi}(\theta)}=\mathcal{V}\sum_{\chi'} \Pi^{\chi\chi'}\int d\theta' f^{\chi'}(\theta') \mathcal{T}^{\chi\chi'}_{\theta \theta'} \Lambda^{\chi'}(\theta')/\tau^{\chi'}(\theta'),
\label{Eq_h_tau_boltz}
\end{align}
 where, $\Pi^{\chi \chi'} = N|U^{\chi\chi'}|^2 / 4\pi^2 \hbar^2$, $f^{\chi} (\theta)=\frac{(k)^3}{|\mathbf{v}^\chi_{\mathbf{k}}\cdot \mathbf{k}^{\chi}|} \sin\theta (\mathcal{D}^\eta_{\mathbf{k}})^{-1} \tau^\chi(\theta)$ and $\mathcal{T}^{\chi\chi'}_{\theta \theta'}$ is defined in Eq.~\ref{Overlap_of_spinor}. With the ansatz $\Lambda^{\chi}(\theta)= [\lambda^{\chi}-h^{\chi}(\theta) + a^{\chi}\cos^4(\theta/2) +b^{\chi}\sin^4(\theta/2)+c^{\chi}\sin^2(\theta)]\tau^{\chi}(\theta)$, Eq.~\ref{Eq_h_tau_boltz} is expressed as:
 \begin{align}
\lambda^{\chi} + &a^{\chi}\cos^4(\theta/2) +b^{\chi}\sin^4(\theta/2)+c^{\chi}\sin^2(\theta)
=\mathcal{V}\sum_{\chi'}\Pi^{\chi\chi'}\int f^{\chi'}(\theta')d\theta' ~\mathcal{T}^{\chi\chi'}_{\theta \theta'}\nonumber \\&\times[\lambda^{\chi'}-h^{\chi'}(\theta') + a^{\chi'}\cos^4(\theta'/2) +b^{\chi'}\sin^4(\theta'/2)+c^{\chi'}\sin^2(\theta')].
\label{Boltzman_final}   
 \end{align}
When explicitly written, this equation consists of eight simultaneous equations that need to be solved for eight variables, and particle number conservation serves as an additional constraint.  
The current is calculated using the expression :
\begin{align}
    \mathbf{J}=-e\sum_{\chi,\mathbf{k}} f^{\chi}_{\mathbf{k}} \dot{\mathbf{r}}^{\chi}.
    \label{Eq:J_formula_ch5}
\end{align}
\begin{figure}
    \centering
    \includegraphics[width=0.98\columnwidth]{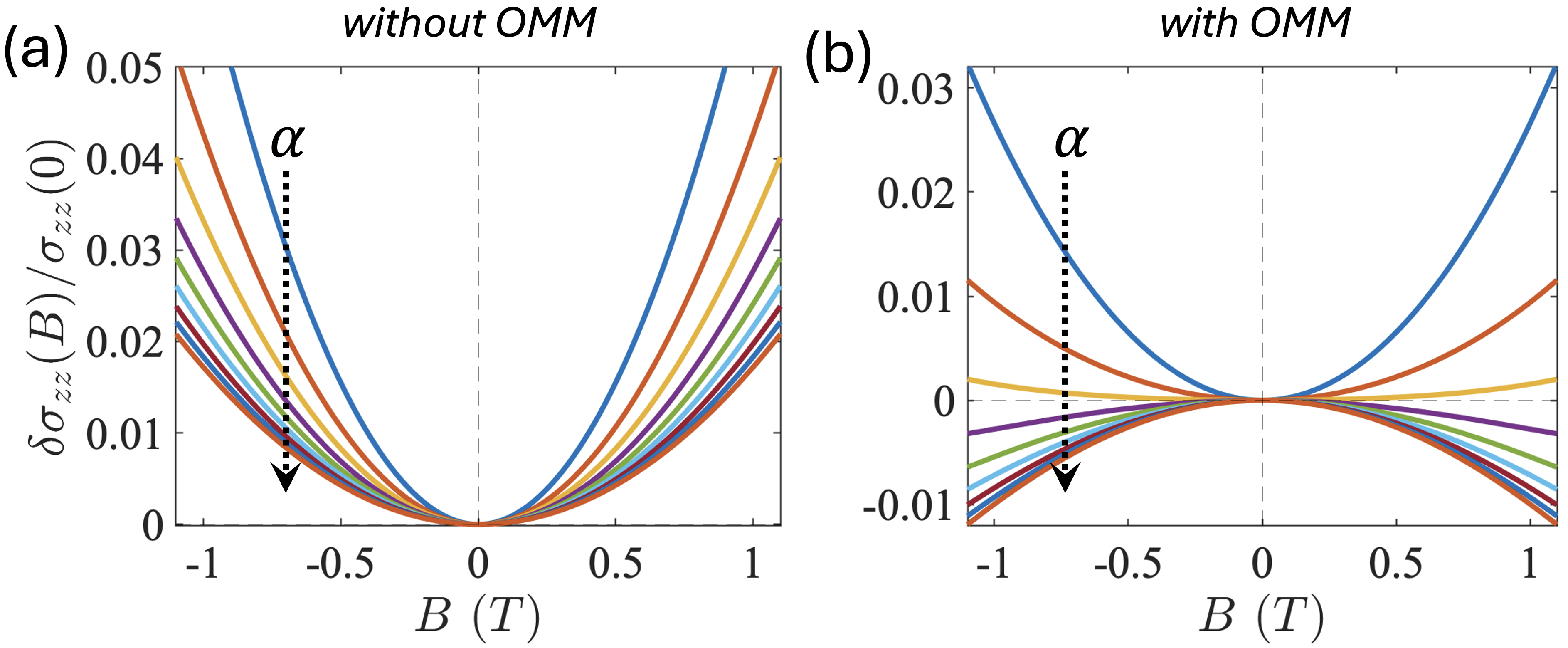}
    \caption{Longitudinal magnetoconductivity in a pseudospin-1 semimetal with and without including the effect of orbital magnetic moment. Increasing the relative magnitude of the internode scattering strength ($\alpha$) results in a reversal of the sign of LMC from positive to negative.}
    \label{fig:lmc:spin1}
\end{figure}
\section{Results and Discussion} We study the longitudinal magnetoconductance of the pseudospin-1 semimetal subjected to parallel electric and magnetic fields ($\mathbf{E}\parallel\mathbf{B}$) in the semiclassical regime. This specific field arrangement allows us to restrict the current flow along the $z$-direction, making it easy to make analytical progress. For parallel electric and magnetic fields, the Lorentz force ($\mathbf{F}\propto \mathbf{v}\times\mathbf{B}$) is zero, which makes no conventional transverse current possible in this orientation. LMC is defined as
\begin{align}
\delta\sigma_{zz}(B) = \sigma_{zz} (B) - \sigma_{zz} (0). 
\end{align} 
In Fig.~\ref{fig:lmc:spin1} we plot LMC (normalized with respect to the zero-field conductivity) as a function of the magnetic field. The conductivity is quadratic in the magnetic field. When $\alpha<\alpha_c^{(1)}$, the conductivity is positive, and switches sign when $\alpha>\alpha_c^{(1)}$, where $\alpha_c^{(1)}$ is the critical relative internode scattering strength. When the effects of orbital magnetic moment are turned off, we only obtain positive LMC irrespective of $\alpha$. Increasing $\alpha$ only reduces the magnitude of the conductivity but does not change the sign. Before we discuss further, it is useful to compare and contrast our results with a conventional spin-$1/2$ Weyl semimetal.   
Son \& Spivak~\cite{son2013chiral} first predicted positive and quadratic LMC in Weyl semimetals driven by chirality-breaking internode scattering. However, several recent studies later proposed that intranode scattering, by itself, could lead to positive longitudinal magnetoconductivity by including a force term $\mathbf{E}\cdot\mathbf{B}$ in equations of motion~\cite{kim2014boltzmann,lundgren2014thermoelectric,cortijo2016linear,sharma2016nernst,zyuzin2017magnetotransport,das2019berry,kundu2020magnetotransport}. However, these results do not account for global charge conservation, orbital magnetic moment, and assume a constant relaxation time, both of which are crucial to obtain the correct magnetoconductivity results. Going beyond the standard approximations reveals that LMC in Weyl fermions is negative for zero internode scattering~\cite{sharma2023decoupling}, positive for infinitesimal internode scattering, and negative if the internode scattering is large enough~\cite{knoll2020negative,sharma2023decoupling}. The results obtained here for pseudospin-1 fermions also paint a similar picture, but a closer comparison sheds further light on their universal features.
\begin{figure}
    \centering
    \includegraphics[width= 0.98 \columnwidth]{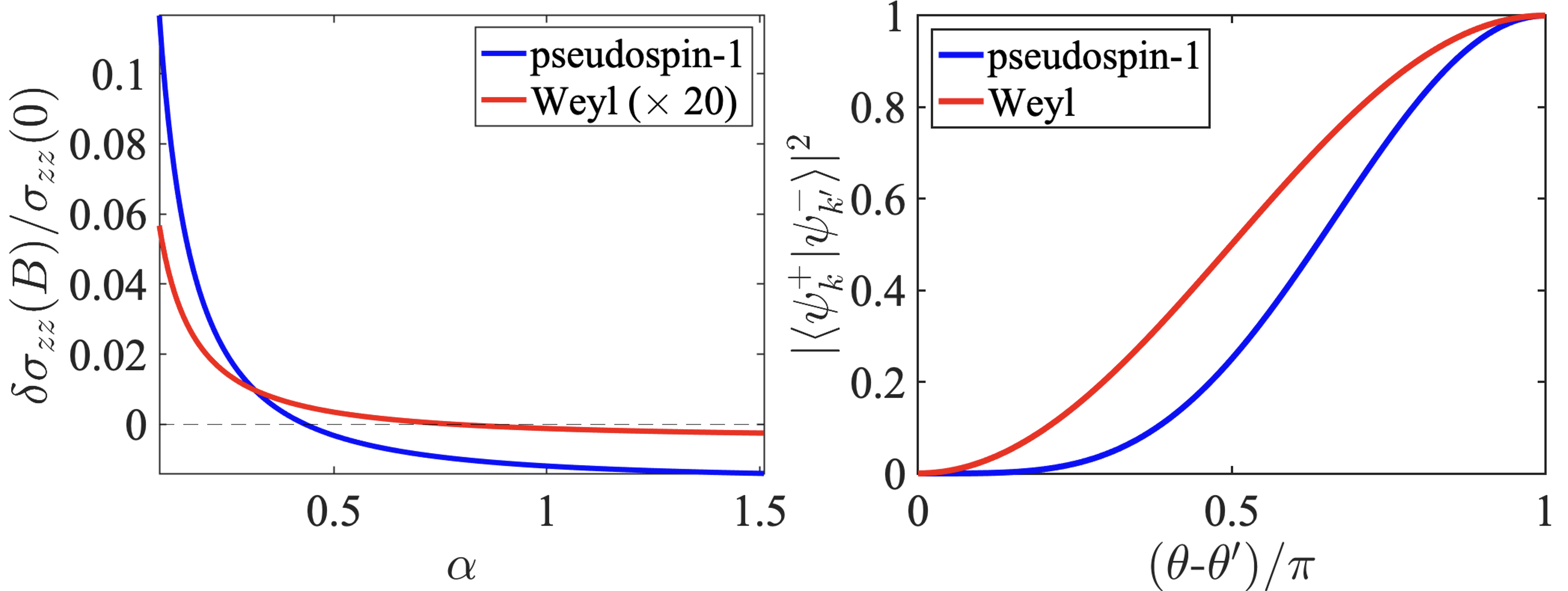}
    \caption{(a) Longitudinal magnetoconductance for a fixed magnetic field as a function of the intervalley scattering strength $\alpha$. The switch from positive to negative LMC happens at a higher value for Weyl fermions compared to pseudospin-1. (b) Overlap between fermions at different valleys $|\langle \psi_\mathbf{k}^+|\psi_\mathbf{k}^- \rangle|^2$ for a fixed azimuthal angle $\phi$ as a function of the difference of incoming ($\theta$) and outgoing ($\theta'$) polar angle.}
    \label{fig:wsm-spin1-compare}
\end{figure}

A comparison of LMC between Weyl fermions and pseudospin-1 fermions is made in Fig.~\ref{fig:wsm-spin1-compare} (a). Note that magnetoconductivity has a larger magnitude for pseudospin-1 fermions and switches its sign from positive to negative for a much smaller value of internode scattering strength. We intuitively understand this by first examining the physics at zero magnetic field. We 
compare the overlap of fermions in opposite valleys ($|\langle \psi_\mathbf{k}^+|\psi_\mathbf{k}^- \rangle|^2$), which is responsible for chirality breaking transport (chiral anomaly). This is shown in Fig.~\ref{fig:wsm-spin1-compare} (b). Clearly, pseudospin-1 fermions are more likely to get backscattered while flipping their chirality compared to Weyl fermions. The exact likelihood ($l$) is calculated to be 
\begin{align}
l=\frac{(8 + 3 \pi)}{(6 + 3 \pi)}\sim 1.12.
\end{align}
This increased likelihood of backscattering results in a quicker conductivity decrease due to internode scattering. We further analytically evaluate the zero-field conductivity using the Boltzmann formalism to 
\begin{align}
\sigma_{zz}^{s=1} = \frac{\mathrm{e}^2 v_{\mathrm{F}}^2}{ V \pi^2 \left(3 \alpha + 1\right)}, \nonumber\\
\sigma_{zz}^{s=1/2} = \frac{\mathrm{e}^2 v_{\mathrm{F}}^2}{16 V \pi^2 \left(2 \alpha + 1\right)},
\label{Eq:LMC_B0}
\end{align}
where $V= U^2\mathcal{V}/\hbar$ ($U$ being the strength of the impurity potential). The conductivity for pseudospin-1 fermions is greater in magnitude than Weyl fermions for the same set of parameters, and 
again, we note that conductivity depends more strongly on internode scattering in pseudospin-1 fermions than in Weyl fermions. In the absence of intervalley scattering, it is straightforward to evaluate that for fermions with higher pseudospin ($s>1$), the zero-field conductivity also increases with $s$. We conjecture that the conductivity in the presence of intervalley scattering ($\alpha$) should also drop more dramatically with increasing $s$ (although this must be confirmed by explicit calculations). This suggests that we may need more diagnostic tools other than relying on negative magnetoresistance studies to confirm CA in systems with higher pseudospin-$s$ fermions.

\section{Outlook} Condensed matter systems provide a unique platform for studying emergent fermions that otherwise have no analogs in high-energy physics. Pseudospin-1 fermions form one such example that can emerge in candidate materials with space group symmetries 
{199 (tetragonal), 214 (cubic), and 220 (orthorhombic)}. Similarly, higher pseudospin excitations are possible as well~\cite{bradlyn2016beyond}. 
Investigating chiral anomaly and its manifestation in transport experiments can reveal fascinating properties of (pseudo)relativistic fermions beyond the standard Dirac fermions and shed light on the universality of (pseudo)relativistic fermions.  We conjecture that magnetoconductivity in the presence of intervalley scattering should quickly become negative with increasing pseudospin index, which suggests that we may need more diagnostic tools other than relying on negative magnetoresistance studies to confirm CA in systems with higher pseudospin-$s$ fermions. This will help design upcoming experiments on candidate materials where such excitations can emerge.
\chapter{\label{chap6}Conclusion}
Weyl materials not only host quasiparticle excitations that mimic the Weyl Hamiltonian—originally introduced a century ago in high-energy physics—but also serve as a bridge between geometry, topology, high-energy physics, and condensed matter. This makes their study highly significant from multiple perspectives and has garnered considerable interest. This fascination arises from the intricate interplay between the geometric and topological properties of Weyl semimetals (WSMs) and high-energy physics phenomena, leading to exotic, anomalous, and topological effects that are absent in conventional metals. Analyzing chiral fermions on a lattice reveals that: (i) energy bands can feature degenerate points that remain stable against arbitrary perturbations in three dimensions; (ii) these degenerate points always appear in pairs, ensuring that their total topological index sums to zero; and (iii) this system suggests a condensed matter counterpart to the Adler-Bell-Jackiw (ABJ) anomaly observed in high-energy physics. The dynamics of charge carriers in the classical regime are influenced by the presence of a non-trivial Berry phase and anomalous orbital motion. Although these materials fall under the category of quantum materials, their transport properties can still be effectively described using the semiclassical equations of motion, provided the semiclassical approximation remains valid. In this thesis, encompassing the above three points, we explore the transport behavior of Weyl materials under various physical conditions.

\textit{\textbf{Ambiguity in the sign of magnetoconductance in Weyl semimetals: }}The linear response formalism for conductivity~\cite{mahan20089} assumes a characteristic timescale, $\tau_{\phi}$, representing inelastic energy exchange. Ideally, $\tau_{\phi}$ is the longest relevant timescale. In weakly disordered Weyl semimetals under strong magnetic fields, chiral anomaly ({CA}) manifests as a positive longitudinal magnetoconductivity ({LMC}), given by $\mathbf{j} \propto B(\mathbf{E} \cdot \mathbf{B})$. Current is constrained by internode scattering time ($\tau_\mathrm{inter}$), which must exceed $\tau_{\phi}$ for linear response validity. If intranode scattering dominates, chiral charge conservation gains significance, requiring reconsideration of calculations. Despite chiral charge non-conservation, global charge remains conserved. Son and Spivak~\cite{son2013chiral} predicted positive longitudinal magnetoconductivity (LMC) in Weyl semimetals (WSMs) due to internode scattering using the Boltzmann approach. Later studies suggested that intranode scattering alone could generate positive LMC via the $\mathbf{E}\cdot\mathbf{B}$ term. However, they did not distinguish between two regimes: $\tau_\mathrm{intra}\ll\tau_\phi\ll\tau_\mathrm{inter}$ (chiral charge conservation) and $\tau_\mathrm{intra}\ll\tau_\mathrm{inter}\ll\tau_\phi$ (global charge conservation). Sharma \textit{et al.}~\cite{sharma2023decoupling} resolved this by correctly computing LMC across different parameter values and demonstrating that constant relaxation time approximations violate charge conservation. A proper study of magnetotransport necessitates moving beyond the constant relaxation-time approximation. By applying the first Born approximation (Fermi's golden rule), we make $\tau$ dependent on $\theta, \phi$, which greatly enhances the understanding of the conductivity sign in Weyl materials.

\textbf{Chapter~\ref{chap2}} establishes a detailed framework for analyzing the electronic transport properties of Weyl fermions using the semiclassical Boltzmann formalism. By incorporating a smooth lattice cutoff, it overcomes the limitations of the constant relaxation time approximation and captures the gradual transition of band dispersion. The study reveals the role of geometric properties such as Berry curvature and the orbital magnetic moment in magnetotransport. The results emphasize the importance of considering band nonlinearity and ultraviolet cutoffs to obtain physically meaningful predictions for longitudinal magnetoconductance (LMC) and the planar Hall effect (PHE) in Weyl semimetals. 

\textbf{Chapter~\ref{chap3}} investigates the effects of strain-induced chiral gauge fields on magnetotransport in Weyl semimetals. It demonstrates that strain acts as an axial magnetic field, altering impurity-driven transport. A key finding is the emergence of 'strong sign-reversal' in LMC due to strain, leading to a flipped magnetoconductance parabola. Additionally, the interplay between strain, external magnetic fields, and intervalley scattering introduces both `strong and weak sign-reversals', enriching the LMC phase diagram. The study also explores the impact of strain on planar Hall conductance, predicting experimentally observable features that can distinguish strain-induced transport effects in Weyl systems. Having explored the linear regime conductivity, we moved to the nonlinear part and predicted Chiral anomaly-induced nonlinear hall effects. 

\textbf{Chapter~\ref{chap4}} formulates a comprehensive theory of the chiral anomaly-induced nonlinear Hall effect (CNLHE) in three-dimensional chiral quasiparticles. It highlights the influence of momentum-dependent scattering and global charge conservation in Weyl semimetals and spin-orbit-coupled non-centrosymmetric metals (SOC-NCMs). The findings reveal nonmonotonic behavior and strong sign-reversal of nonlinear Hall conductivity in WSMs, while in SOC-NCMs, the orbital magnetic moment dominates CNLHE with a quadratic magnetic field dependence. Furthermore, spin Zeeman coupling effectively enhances the nonlinear Hall current by mimicking an effective tilt. These results provide a foundation for experimental investigations into chiral transport phenomena.

\textbf{Chapter~\ref{chap5}} extends the analysis of chiral anomaly to pseudospin-1 fermions, offering new insights into their magnetotransport properties. By incorporating momentum-dependent relaxation times, the orbital magnetic moment, and global charge conservation, it provides a refined framework for understanding multifold fermions. The study finds that LMC remains positive under weak internode scattering but turns negative beyond a lower critical threshold compared to Weyl fermions. Additionally, internode scattering significantly impacts zero-field conductivity, distinguishing pseudospin-1 systems from Weyl semimetals. These results guide experimental efforts in identifying candidate materials and interpreting magnetotransport data in systems with specific space group symmetries.

\section{\textit{Future scope}} More precise, quick, and effective computing is required due to growing technology, innovation, and dependence of intelligence on it.  There is now a race underway to develop a new method of computation using quantum bits (\textit{qbits}), which, in theory, will be so powerful that a single quantum computer can complete more processing tasks than all of the computers ever created.  Addressing the theoretical difficulties engineers face is essential to turning dreams become reality. The Weyl materials are significant in this context due to their non-trivial properties arising from band structure.  Examples include, but are not limited to, the existence of topologically protected surface states that exhibit immunity to local perturbations.  This robustness may enhance the stability of qubits, thereby increasing the reliability of quantum computations. The `masslessness' of Weyl fermions contributes to high electron mobility and low energy dissipation, mitigating decoherence, a significant obstacle in quantum computing. The chiral anomaly in Weyl materials, the primary focus of this thesis, facilitates unique charge transport properties that could be utilized for the development of novel quantum gates or the improvement of information transfer in quantum circuits. Additionally, the coupling of Weyl materials with superconductors offers a platform for the realization of Majorana fermions, which are potential candidates for topological quantum computing.  Majorana modes offer fault-tolerant qubits as a result of their non-Abelian statistics ~\cite{das2023search,kitaev2001unpaired,karoliya2025majorana,marra2022majorana}.  In addition to promising applications, there are notable challenges, such as preserving Weyl properties at low temperatures and effectively integrating Weyl semimetals with current quantum computing architectures.  
This thesis establishes a basis for comprehending the complex transport phenomena in Weyl and pseudospin-1 fermions, highlighting the significant impact of momentum-dependent scattering, Berry curvature, orbital magnetic moments, and global charge conservation. Nevertheless, the realm of uncharted territories persists, inviting additional investigation in areas like non-equilibrium dynamics, time-resolved transport studies, thermoelectric effects, Floquet engineering, Anderson localization, electron-electron interactions, and so on in Weyl semimetals.

\include{Conclusion}
\chapter*{List of publications}

\section*{Publications in refereed journals}	
\begin{enumerate}[label=(\roman*), noitemsep]

    \item Longitudinal magnetoconductance and the planar Hall effect in a lattice model of tilted Weyl fermions
    \item[] \textbf{Azaz Ahmad} and Gargee Sharma, \textit{Phys. Rev. B} \textbf{103}, 115146 (2021).

    \item Longitudinal magnetoconductance and the planar Hall conductance
in inhomogeneous Weyl semimetals
    \item[] \textbf{Azaz Ahmad}, Karthik V. Raman, Sumanta Tewari, and Gargee Sharma, \textit{Phys. Rev. B} \textbf{107}, 144206 (2023).
    
   \item Chiral anomaly induced nonlinear Hall effect in three-dimensional chiral fermions
    \item[] \textbf{Azaz Ahmad}, Gautham Varma K, and Gargee sharma, \textit{Phys. Rev. B} \textbf{111}, 035138 (2025).

    \item Chiral anomaly and longitudinal magnetoconductance in pseudospin-1 fermions
    \item[] \textbf{Azaz Ahmad} and Gargee sharma, \textit{arXiv, cond-mat.mes-hall }\textbf{2412.10500 } (2024). \\ Under review.

    \item Geometry, anomaly, topology, and transport in Weyl fermions (invited review)
    \item[] \textbf{Azaz Ahmad}, Gautham Varma K, and Gargee sharma, \textit{Journal of Physics: Condensed Matter} \textbf{37}, 043001 (2024).

    \item Magnetotransport in spin-orbit coupled noncentrosymmetric and Weyl metals
    \item[] Gautham Varma K,\textbf{Azaz Ahmad}, Sumanta Tewari, and Gargee sharma, \textit{Physical Review B} \textbf{109}, 165114 (2024).
   
\end{enumerate}
\newpage
\section*{Presentations in conferences and workshops}
\begin{enumerate}[label=(\roman*)] 

    \item Oral presentations (Virtual) at the \href{https://www.aps.org/}{\textit{March Meeting}} 2023 and 2024, organized by \textbf{\href{https://www.aps.org/}{\textit{American Physical Society}}}. 

    \item Presented a poster at the \textit{IPS Meeting 2023}, organized by the \textbf{\href{https://ipssingapore.org/}{\textit{Institute of Physics, National University of Singapore (NUS)}}}, held from 27th to 29th September 2023.

    \item Presented a poster at the \textit{\href{https://sites.google.com/acads.iiserpune.ac.in/yimqcmt-2024/home}{YIMQCMT-2024}}, organized by the \textbf{\href{https://www.iiserpune.ac.in/}{\textit{IISER Pune}}}, held from 16-18 December 2024.

    \item Participated in and presented a poster at the conference \textbf{\href{https://sites.google.com/view/psces2023}{\textit{Physics of Strongly Correlated Electron Systems (PSCES-2023 and PSCES-2024)}}} held at \href{https://www.iiserpune.ac.in/}{\textit{IISER Pune}} and \href{https://www.iiserb.ac.in/}{\textit{IISER Bhopal}}, respectively.

    \item The abstract has been accepted for an oral presentation at the conference \textit{Emerging Phenomena in Quantum Materials}, held from December 11–15, 2023, in Bharatpur, India. The event was jointly organized by \textbf{\href{https://www.uu.se/en}{\textit{Uppsala University, Sweden}}} and \textbf{\href{http://www.unipune.ac.in/}{\textit{Savitribai Phule Pune University, India}}}.  

    \item Attended and did a poster presentation at the \textit{Science Day Celebration} at \textbf{\href{https://www.iitmandi.ac.in/}{ \textit{IIT Mandi}}}.
    
    \item Attended International Conference \href{https://sites.google.com/view/ifimac-icmm-joint-seminars/home}{\textbf{ \textit{ICMM+IFIMAC} joint Seminar Series on Condensed Matter Physics}}.
    
    \item Attended International Conference \textbf{\textit{Quantum-2020}} Organized by IOP publishing House.
    
    \item Attended National Short Term Course on \textbf{\textit{Current Trends in Condensed Matter Physics}} Organized by NIT JALANDHAR-INDIA.
    
    \item Attended 28th National Conference \textbf{\textit{CONDENSED MATTER DAYS 2020}} Organized by NIT SILCHAR-INDIA.		
\end{enumerate}
\begin{appendices}
\chapter{\label{appendix_ch2}}

\section{Lattice Weyl fermion}
The Hamiltonian of a Weyl node with smooth lattice cutoff can be expressed as 
\begin{align}
 H^{\chi}=\chi E_{0}\sin({a\mathbf{k}\cdot\boldsymbol{\sigma}}),\
 \label{ch2_Hk_lattice_no_tilt}
\end{align}
where $k$ is measured from the nodal point, $\chi$ is the chirality index, $E_0$ is an energy parameter, and $a$ is constant with dimensions of length. Using the relations, $\sin{\theta}=({\exp^{i\theta}-\exp^{-i \theta}})/{2i}$, and $    \exp{\{i a (\mathbf{\sigma \cdot k})\}}={{I}} \cos{\theta}+i(\mathbf{\sigma \cdot \hat{k}})\sin{ ak}$, one can  rewrite down the Hamiltonian in the following form, (with $\theta$ and $\phi$ as polar and azimuthal angles respectively )  
\begin{align}
  H^{\chi}=\chi E_{0}\sin(ak)\begin{pmatrix}
  \cos{\theta} & \sin{\theta} e^{-i\phi} \\
  \sin{\theta} e^{i\phi} & -\cos{\theta} \\
  \label{ch2_Hk_lattice_latrix_form}
  \end{pmatrix}   
\end{align}
Here we are going to use the property of matrices that for a matrix $M=[...]_{N \times N}$ having eigenvalues,$\lambda_1,\lambda_2,\lambda_3,......,\lambda_N,$ and eigenfunctions $\mathbf{\alpha_1,\alpha_2,\alpha_3,.......,\alpha_N,}$ respectively, then for matrix $C M$, the same will be $C \lambda_1,C \lambda_2,C.\lambda_3,......,C \lambda_N,$ \&
 $\alpha_1,\alpha_2,\alpha_3,.......,\alpha_N$, (where $C$ is constant ). Thus the eigenvalues of the Hamiltonian are 
\begin{align}
     \epsilon(\mathbf{k})=\pm{E_{0}\sin{(ak)}},
     \label{ch2_Ek_lattice_notilt}
\end{align}
and eigenfunctions for positive band with different chirality are
\begin{align}
    \ket{u^+{(k)}}=\begin{pmatrix}
     e^{-i \phi} \cos{\frac{\theta}{2}}\\
      \sin{\frac{\theta}{2}}
    \end{pmatrix}
    \end{align}
\begin{align}
        \ket{u^-{(k)}}=\begin{pmatrix} -e^{-i \phi}
        \sin{\frac{\theta}{2}}\\
        \cos{\frac{\theta}{2}}
  \end{pmatrix}
\end{align}
The expressions for Berry curvature and orbital magnetic moments(OMM) are given by
\begin{align}
    \Omega^{\chi}_{\mathbf{k}}&=i\mathbf{\nabla}_{\mathbf{k}}\times(\bra{u^{\chi}({\mathbf{k}})}\mathbf{\nabla}_{\mathbf{k}}\ket{u^{\chi}{\mathbf{(k)}}})\nonumber\\
    m^{\chi}_{\mathbf{k}}&=\frac{-i e}{2\hbar}\bra{\nabla_{\mathbf{k}} u^{\chi}({\mathbf{k}})}\mathbf{\times}[H^{\chi}\mathbf{(k)}-\epsilon(\mathbf{k})]\ket{\nabla_\mathbf{k} u^{\chi}{\mathbf{(k)}}},
    \label{ch2_OMM_and_BC_formula}
\end{align}
from which one can easily find the expressions for Berry curvature and OMM
\begin{align}
   \boldsymbol{\Omega}^{\chi}_k&=\frac{-\chi\mathbf{k}}{2 k^3}\nonumber\\
   \mathbf{m}^{\chi}_\mathbf{k}&=\frac{-e \chi E_{0}\sin{(ka)} \mathbf{k}}{2\hbar k^3}
   \label{ch2_OMM_and_BC_for_latt_WSM}
\end{align}

\begin{figure*}
\includegraphics[width=1\columnwidth]{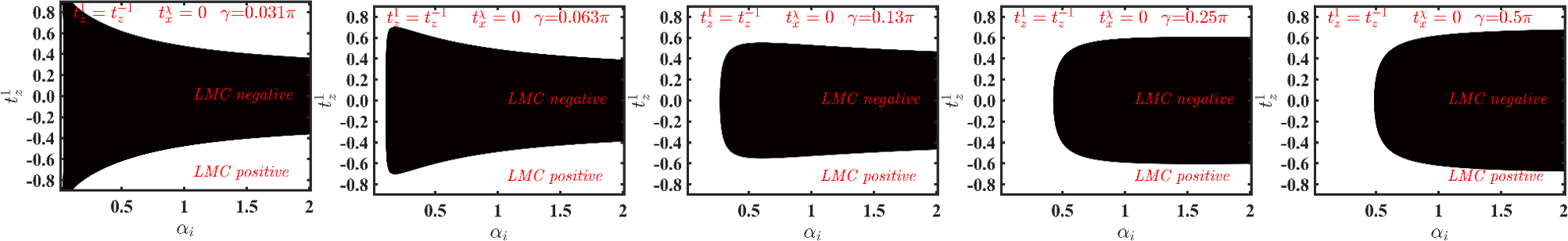}
	\caption{Evolution of the phase diagram in the $t^1_z-\alpha^i$ parameter space as a function of the angle of the magnetic field $\gamma$, when the Weyl cones are tilted along the same direction.}
	\label{Fig_phase-evolution-tz-same}
\end{figure*}
\begin{figure*}
    \centering
    \includegraphics[width=1\columnwidth]{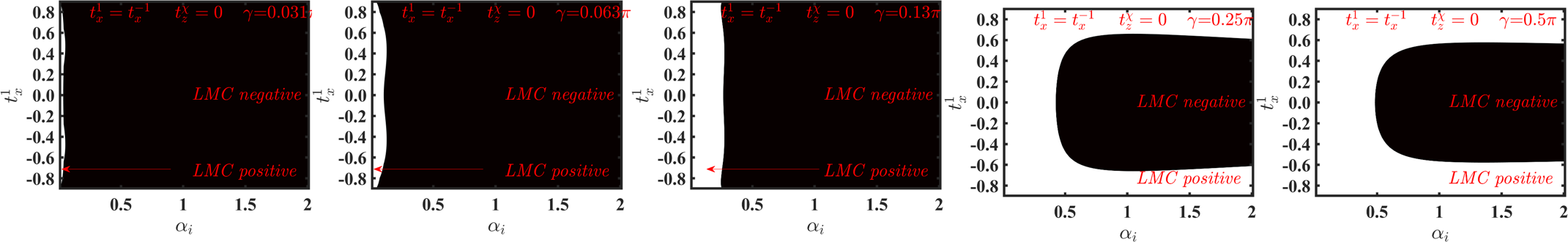}
    \caption{Evolution of the phase diagram in the $t^1_x-\alpha^i$ parameter space as a function of the angle of the magnetic field $\gamma$, when the Weyl cones are tilted along the same direction.}
    \label{Fig_phase-evolution-tx-same}
\end{figure*}
\begin{figure*}
    \centering
    \includegraphics[width=1\columnwidth]{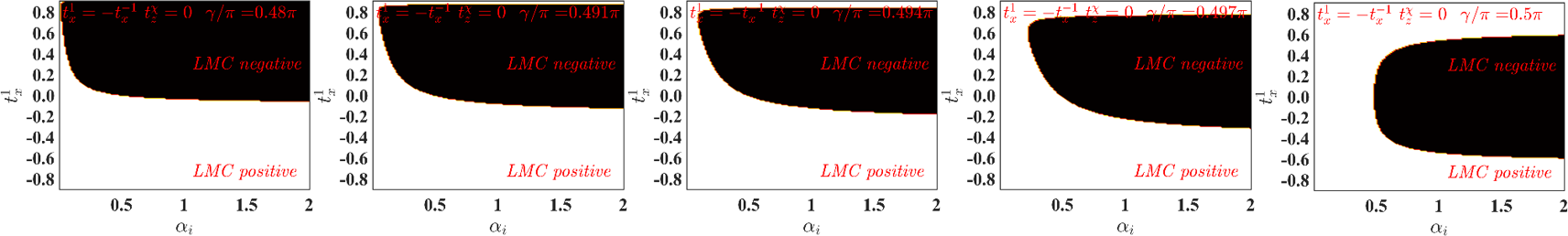}
    \caption{Evolution of the phase diagram in the $t^1_x-\alpha^i$ parameter space as a function of the angle of the magnetic field $\gamma$, when the Weyl cones are tilted opposite to each other.}
    \label{Fig_phase-evolution-tx-opp}
\end{figure*}
\begin{figure*}
	\includegraphics[width=.49\columnwidth]{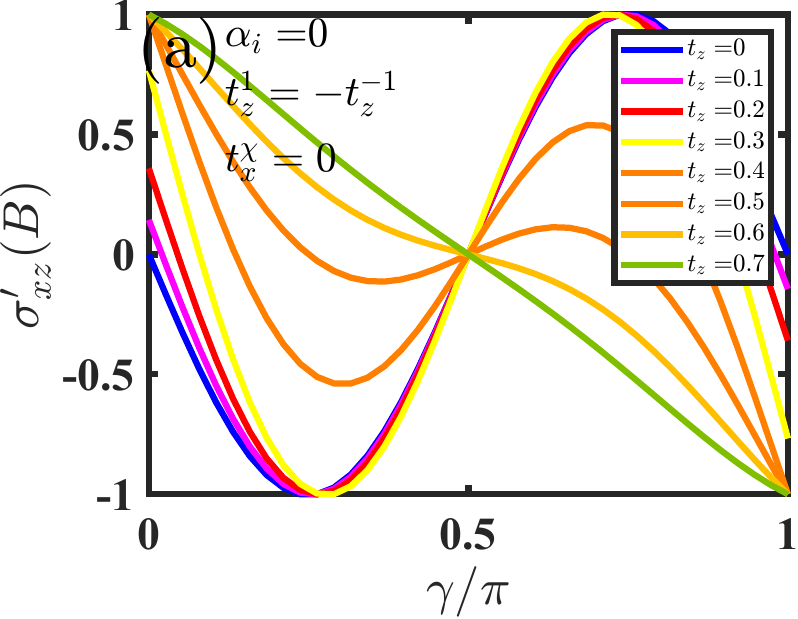}
	\includegraphics[width=.49\columnwidth]{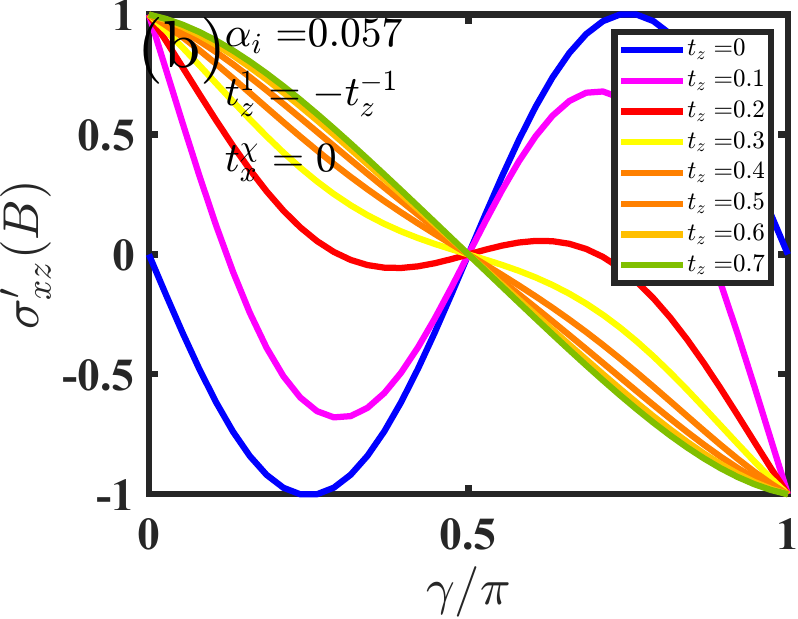}
	\includegraphics[width=.49\columnwidth]{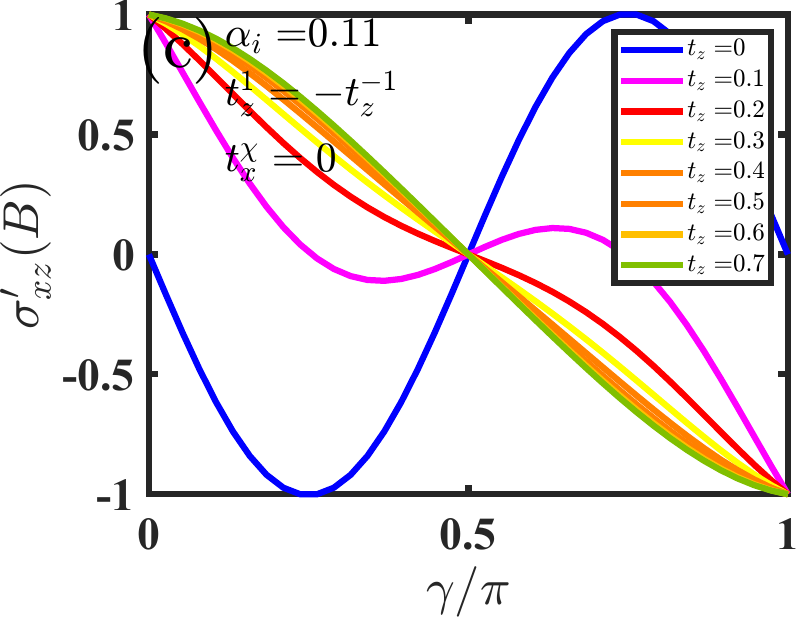}
	\includegraphics[width=.49\columnwidth]{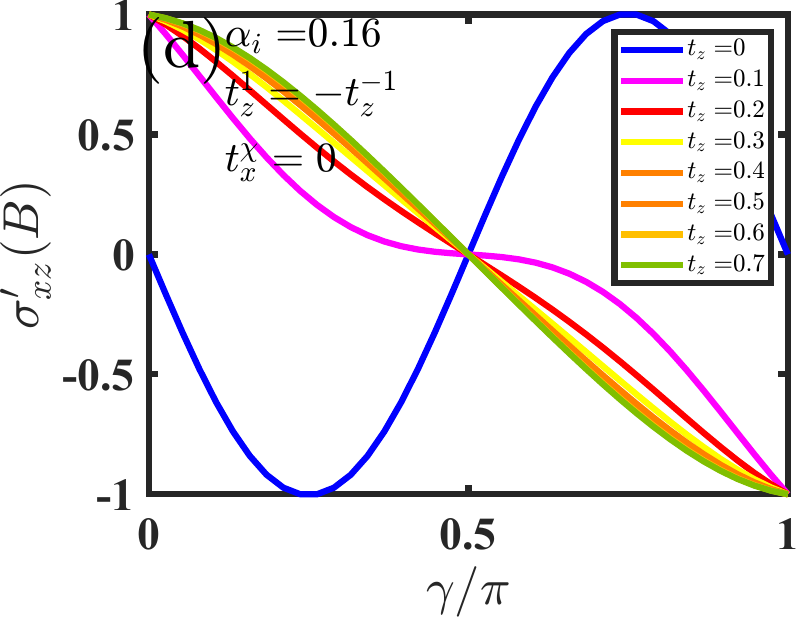}
	\caption{Normalized planar Hall conductivity ($\sigma_{xz}'$) as a function of the angle $\gamma$ for several values of tilt parameter $t_z$ for oppositely tilted Weyl cones. In the absence of tilt the behavior follows the trend $\sin(2\gamma)$, while in the presence of tilt, a $\cos\gamma$ component is added. Beyond a critical $t_z^c$, the $\cos\gamma$ term dominates and $\sigma_{xz}'(\pi/2 + \epsilon)$ changes from positive to negative, where $\epsilon$ is a small positive angle.
		A finite intervalley scattering further enhances the $\cos\gamma$ trend (however only in the presence of a finite tilt). It's effect is to lower the critical tilt $t_z^c$ where the sign change occurs.}
	\label{Fig_sxz_vs_gamma_tiltz_opp_omm_on}
\end{figure*}
\begin{figure*}
	\centering
	\includegraphics[width=0.49\columnwidth]{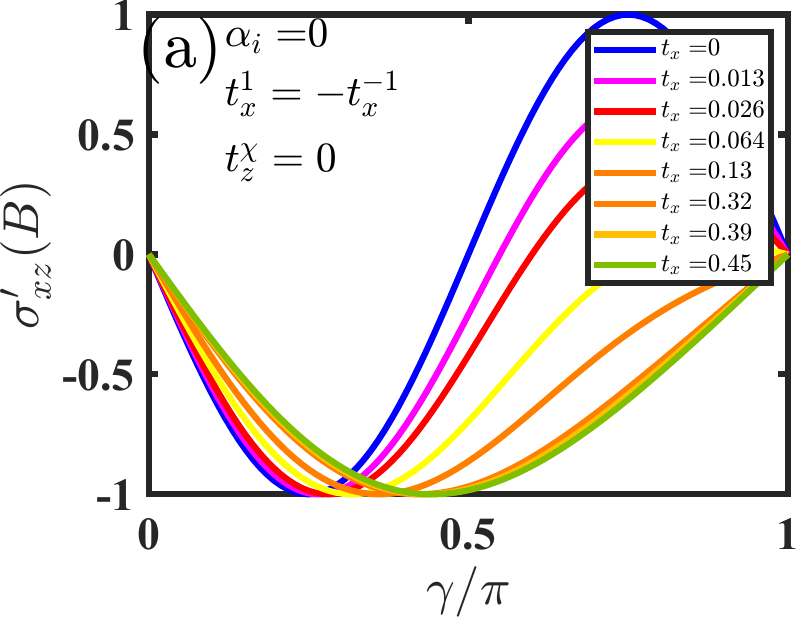}
	\includegraphics[width=0.49\columnwidth]{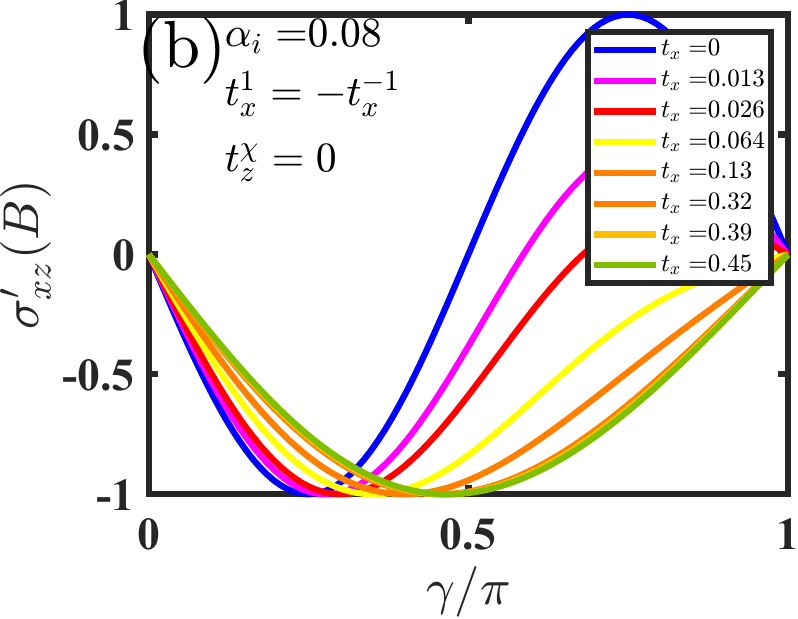}
	\includegraphics[width=0.49\columnwidth]{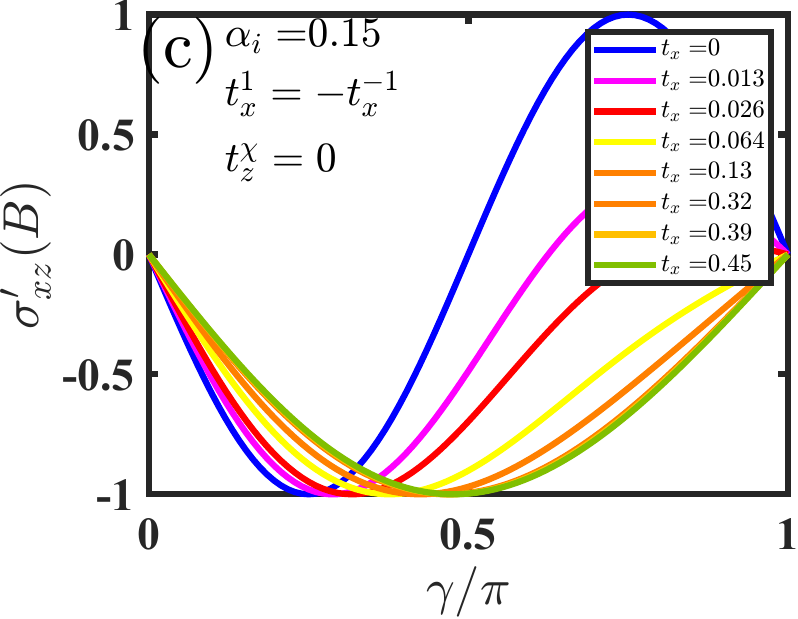}
	\includegraphics[width=0.49\columnwidth]{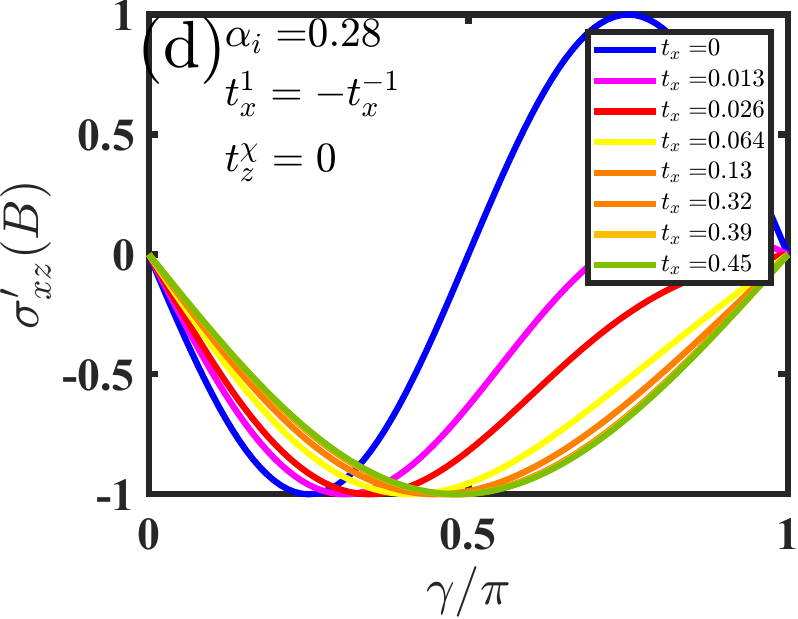}
	\caption{Normalized planar Hall conductivity ($\sigma_{xz}'$) as a function of the angle $\gamma$ for several values of tilt parameter $t_x$ for oppositely tilted Weyl cones. In the absence of tilt the behavior follows the trend $\sin(2\gamma)$, while in the presence of tilt, a $\sin\gamma$ component is added. Beyond a critical $t_x^c$, the $\sin\gamma$ term dominates and $\sigma_{xz}'(\pi/2 + \epsilon)$ changes from positive to negative, where $\epsilon$ is a small positive angle.
		A finite intervalley scattering further enhances the $\sin\gamma$ trend (however only in the presence of a finite tilt). It's effect is to lower the critical tilt $t_x^c$ where the sign change occurs.}
	\label{Fig_sxz_vs_gamma_tiltx_opp_omm_on}
\end{figure*}
\section{Boltzmann transport equation}
The Boltzmann equation is reduced to the following form 
\begin{align}
    \mathbb{Z} = \mathbb{A}\mathbb{Z} - \mathbb{Y}, 
\end{align}
where 
\begin{align}
\mathbb{Z} = \begin{pmatrix}
\lambda^+\\  a^+\\ b^+\\ c^+\\
\lambda^-\\ a^-\\ b^-\\ c^-\\
\end{pmatrix}
\end{align}
\begin{align}
\mathbb{A}=\begin{pmatrix}
\alpha^{++}F^+&\alpha^{++}G^+&\alpha^{++}I^+&\alpha^{++}J^+&\alpha^{+-}F^-&\alpha^{+-}G^-&\alpha^{+-}I^-&\alpha^{+-}J^-\\
\alpha^{++}G^+&\alpha^{++}O^+&\alpha^{++}P^+&\alpha^{++}Q^+&\alpha^{+-}G^-&\alpha^{+-}O^-&\alpha^{+-}P^-&\alpha^{+-}Q^-\\
\alpha^{++}I^+&\alpha^{++}P^+&\alpha^{++}S^+&\alpha^{++}U^+&\alpha^{+-}I^-&\alpha^{+-}P^-&\alpha^{+-}S^-&\alpha^{+-}U^-\\
\alpha^{++}J^+&\alpha^{++}Q^+&\alpha^{++}U^+&\alpha^{++}V^+&\alpha^{+-}J^-&\alpha^{+-}Q^-&\alpha^{+-}U^-&\alpha^{+-}V^-\\
\alpha^{-+}F^+&\alpha^{-+}G^+&\alpha^{-+}I^+&\alpha^{-+}J^+&\alpha^{--}F^-&\alpha^{--}G^-&\alpha^{--}I^-&\alpha^{--}J^-\\
\alpha^{-+}G^+&\alpha^{-+}O^+&\alpha^{-+}P^+&\alpha^{-+}Q^+&\alpha^{--}G^-&\alpha^{--}O^-&\alpha^{--}P^-&\alpha^{--}Q^-\\
\alpha^{-+}I^+&\alpha^{-+}P^+&\alpha^{-+}S^+&\alpha^{-+}U^+&\alpha^{--}I^-&\alpha^{--}P^-&\alpha^{--}S^-&\alpha^{--}U^-\\
\alpha^{-+}J^+&\alpha^{-+}Q^+&\alpha^{-+}U^+&\alpha^{-+}V^+&\alpha^{--}J^-&\alpha^{--}Q^-&\alpha^{--}U^-&\alpha^{--}V^-\\
\end{pmatrix}
\end{align}

\begin{align}
\mathbb{Y}=
\begin{pmatrix}
\alpha^{++}H^++\alpha^{+-}H^-\\
\alpha^{++}N^+-\alpha^{+-}N^-\\
\alpha^{++}L^+-\alpha^{+-}L^-\\
\alpha^{-+}M^+-\alpha^{--}M^-\\
\alpha^{--}H^++\alpha^{-+}H^-\\
\alpha^{--}N^--\alpha^{-+}N^+\\
\alpha^{--}L^--\alpha^{-+}L^+\\
\alpha^{--}M^--\alpha^{-+}M^+\\
\end{pmatrix}
\end{align}

The relevant integrals involved in the above matrices are:
\begin{align}
    \iint d\theta' d\phi'f^{\chi'}(\theta',\phi')&=F^{\chi'}\nonumber\\
    \iint d\theta' d\phi'f^{\chi'}(\theta',\phi') h^{\chi'}&=H^{\chi'}
\end{align}
\begin{align}
    \iint d\theta' d\phi'f^{\chi'}(\theta',\phi')\cos{\theta'}&=G^{\chi'}\nonumber\\
    \iint d\theta' d\phi'f^{\chi'}(\theta',\phi')\sin{\theta'}\cos{\phi'}&=I^{\chi'}
\end{align}
\begin{align}
    \iint d\theta' d\phi'f^{\chi'}(\theta',\phi')\sin{\theta'}\sin{\phi'}&=J^{\chi'}\nonumber\\
    \iint d\theta' d\phi'f^{\chi'}(\theta',\phi')\sin^2{\theta'}\cos^2{\phi'}&=S^{\chi'}
\end{align}
\begin{align}
    \iint d\theta' d\phi'f^{\chi'}(\theta',\phi')h^{\chi'}(\theta',\phi')\cos{\theta'}&=N^{\chi'}\nonumber\\
    \iint d\theta' d\phi'f^{\chi'}(\theta',\phi')h^{\chi'}(\theta',\phi')\sin{\theta'}\cos{\phi'}&=L^{\chi'}
\end{align}
\begin{align}
    \iint d\theta' d\phi'f^{\chi'}(\theta',\phi')h^{\chi'}(\theta',\phi')\sin{\theta'}\sin{\phi'}&=M^{\chi'}\nonumber\\
    \iint d\theta' d\phi'f^{\chi'}(\theta',\phi')\cos^2{\theta'}&=O^{\chi'}
\end{align}
\begin{align}
    \iint d\theta' d\phi'f^{\chi'}(\theta',\phi')\sin{\theta'}\cos{\theta'}\cos{\phi'}&=P^{\chi'}\nonumber\\
    \iint d\theta' d\phi'f^{\chi'}(\theta',\phi')\sin{\theta'}\cos{\theta'}\sin{\phi'}&=Q^{\chi'}
\end{align}
\begin{align}
    \iint d\theta' d\phi'f^{\chi'}(\theta',\phi')\sin^2{\theta'}\cos{\phi'}\sin{\phi'}&=U^{\chi'}\nonumber\\
    \iint d\theta' d\phi'f^{\chi'}(\theta',\phi')\sin^2{\theta'}\sin^2{\phi'}&=V^{\chi'}
\end{align}
\section{Evolution of the LMC phase diagram with magnetic field}
In Fig.~\ref{Fig_phase-evolution-tz-same} we plot the evolution of the contour shape in the $t^1_z-\alpha^i$ parameter space as a function of the angle of the magnetic field $\gamma$. When $\gamma$ is directed away from $\pi/2$ the shape of the zero LMC contour looks like a curved trapezoid instead of $U$. The critical value $\alpha_i^c$ where the sign change first occurs is seen to reduce and elongate its region from $|t_z|\approx 0.5$ when $\gamma=\pi/2$ to $|t_z|\approx 1$ as $\gamma\rightarrow 0$. In FIg.~\ref{Fig_phase-evolution-tx-same} we plot the evolution of the contour shape in the $t^1_x-\alpha^i$ parameter space as a function of the angle of the magnetic field $\gamma$. It is noted that region of negative LMC expands in the parameter space along with the reduction of the critical intervalley strength $\alpha_i^c$ where the sign change first
occurs. The reduction of the critical intervalley strength can again be understood as a combination of the two factors  (i) a finite tilt and intervalley scattering
(when $\gamma=\pi/2$) drives the system to change the LMC sign from positive to negative, and
secondly directing the magnetic field away from the $z$-axis in the presence of intervalley scattering (in the absence of tilt) drives the system to change LMC sign from positive to
negative much below the critical intervalley strength. The different shape of the contour (negative LMC filling out the parameter space instead of a curved trapezoid) is essentially because the cones are now tilted along the $x$-direction and
the magnetic field has an $x$-component to it, which is
qualitatively different from the tilt occurring in the $z$-direction. Finally in Fig.~\ref{Fig_phase-evolution-tx-opp} we plot the evolution of the phase diagram when the Weyl cones are tilted along the $x$ direction but oriented opposite to each other. Directing the magnetic field even slightly away from the $z$-axis changes the qualitative behavior since a linear-in-$B$ component is added in the LMC response. This is because the magnetic field now has a finite component along the tilt direction, and the tilts are oppositely oriented to each
other. 
\section{Angular dependence on PHC}
In Fig.~\ref{Fig_sxz_vs_gamma_tiltz_opp_omm_on} we plot the normalized planar Hall conductivity ($\sigma_{xz}'$) as a function of the angle $\gamma$ for several values of tilt parameter $t_z$ for oppositely tilted Weyl cones. In the absence of tilt the behavior follows the trend $\sin(2\gamma)$, while in the presence of tilt, a $\cos\gamma$ component is added. Beyond a critical $t_z^c$, the $\cos\gamma$ term dominates and $\sigma_{xz}'(\pi/2 + \epsilon)$ changes from positive to negative, where $\epsilon$ is a small positive angle. A finite intervalley scattering further enhances the $\cos\gamma$ trend (however only in the presence of a finite tilt). Its effect is to lower the critical tilt $t_z^c$ where the sign change occurs.

In Fig.~\ref{Fig_sxz_vs_gamma_tiltx_opp_omm_on} we plot the normalized planar Hall conductivity as a function of the angle $\gamma$ for several values of tilt parameter $t_x$ for oppositely tilted Weyl cones. In the absence of tilt the behavior follows the expected trend of $\sin(2\gamma)$, while in the presence of tilt, a $\sin\gamma$ component is added.  Beyond a critical value of the tilt ($t_x^c$), the $\sin\gamma$ term dominates the behavior $\sigma_{xz}'$ never changes sign as a function of the parameter $\gamma$. A finite intervalley scattering further enhances the $\sin\gamma$ trend (however only in the presence of a finite tilt).  Its effect is to lower the critical value of the tilt $t_x^c$.

\section{Boltzmann transport for a system with multiple nodes}
For a system with multiple Weyl nodes, the distribution function at each node can be represented by $f_\mathbf{k}^m$. Generalizing the formalism presented in the main text, the collision integral must take into account scattering between multiple Weyl cones that may or may not be of the same chirality or tilt. Thus $\mathcal{I}_{{col}}[f^m_\mathbf{k}]$ can be expressed as 
\begin{align}
\mathcal{I}_{{col}}[f^m_\mathbf{k}] = \sum\limits_{p}\sum\limits_{\mathbf{k}'} W^{mp}_{\mathbf{k},\mathbf{k}'} (f^{p}_{\mathbf{k}'} - f^m_\mathbf{k}),
\end{align}
where the sum $p$ runs over nodes, and scattering rate $W^{mp}_{\mathbf{k},\mathbf{k}'}$ in the first Born approximation is given by
\begin{align}
W^{mp}_{\mathbf{k},\mathbf{k}'} = \frac{2\pi}{\hbar} \frac{n}{\mathcal{V}} |\langle \psi^{p}_{\mathbf{k}'}|U^{mp}_{\mathbf{k}\mathbf{k}'}|\psi^m_\mathbf{k}\rangle|^2 \delta(\epsilon^{p}_{\mathbf{k}'}-\epsilon_F)
\end{align}
The scattering potential profile $U^{mp}_{\mathbf{k}\mathbf{k}'}$ can be chosen such that scattering between all or some of the nodes (internode) as well as within each node (intranode) is considered. Proceeding as before, we define the valley scattering time ${\tau^m_\mathbf{k}}$ as

\begin{align}
\frac{1}{\tau^m_\mu(\theta,\phi)} = \mathcal{V} \sum\limits_{p} \iint{\frac{\beta^{mp}(k')^3}{|\mathbf{v}^p_{\mathbf{k}'}\cdot \mathbf{k}'^p|}\sin\theta'\mathcal{G}^{mp}(\mathcal{D}^p_{\mathbf{k}'})^{-1} d\theta'd\phi'},
\label{Eq_tau2m}
\end{align}
and the Boltzmann equation becomes
\begin{align}
&h^m_\mu(\theta,\phi) + \frac{\Lambda^m_\mu(\theta,\phi)}{\tau^m_\mu(\theta,\phi)} =\nonumber\\ &\mathcal{V}\sum_p \iint {\frac{\beta^{mp}(k')^3}{|\mathbf{v}^p_{\mathbf{k}'}\cdot \mathbf{k}'^p|} \sin\theta'\mathcal{G}^{mp}(\mathcal{D}^p_{\mathbf{k}'})^{-1}\Lambda^p_{\mu}(\theta',\phi') d\theta'd\phi'}.
\label{Eq_boltz4m}
\end{align}
Making the ansatz $\Lambda^m_\mu(\theta,\phi) = (\lambda^m - h^m_\mu(\theta,\phi) + a^m \cos\theta +b^m \sin\theta\cos\phi + c^m \sin\theta\sin\phi)\tau^m_\mu(\theta,\phi)$, and using the particle number conservation constraint, the Boltzmann equation is reduced to a system of $4N$ equations to be solved for $4N$ unknowns ($N$ being the number of nodes).
\end{appendices}

\setcounter{equation}{0}
\setcounter{table}{0}
\setcounter{figure}{0}

\begin{appendices}
\chapter{\label{appendix_ch3}}

\section{Boltzmann formalism for magnetotransport}
Using the quasiclassical Boltzmann theory, we  study transport in Weyl semimetals in the limit of weak electric and magnetic fields. Since quasiclassical Boltzmann theory is valid away from the nodal point such that $\mu^2\gg \hbar v_F^2 e B$, therefore without any loss of generality we will assume that the chemical potential lies in the conduction band. The phenomenological Boltzmann equation for the non-equilibrium distribution function $f^\chi_\mathbf{k}$ can be expressed as~\cite{bruus2004many} 
\begin{align}
\left(\frac{\partial}{\partial t} + \dot{\mathbf{r}}^\chi\cdot \nabla_\mathbf{r}+\dot{\mathbf{k}}^\chi\cdot \nabla_\mathbf{k}\right)f^\chi_\mathbf{k} = \mathcal{I}_{\mathrm{coll}}[f^\chi_\mathbf{k}],
\label{Eq_boltz1}
\end{align}
where the collision term on the right-hand side of the equation incorporates the effects of scattering due to impurities.
In the presence of electric ($\mathbf{E}$) and magnetic ($\mathbf{B}$) fields, the semiclassical dynamics of the Bloch electrons is~\cite{son2012berry} 
\begin{align}
\dot{\mathbf{r}}^\chi &= \mathcal{D}^\chi \left( \frac{e}{\hbar}(\mathbf{E}\times \boldsymbol{\Omega}^\chi + \frac{e}{\hbar}(\mathbf{v}^\chi\cdot \boldsymbol{\Omega}^\chi) \mathbf{B} + \mathbf{v}_\mathbf{k}^\chi)\right) \nonumber\\
\dot{\mathbf{p}}^\chi &= -e \mathcal{D}^\chi \left( \mathbf{E} + \mathbf{v}_\mathbf{k}^\chi \times \mathbf{B} + \frac{e}{\hbar} (\mathbf{E}\cdot\mathbf{B}) \boldsymbol{\Omega}^\chi \right),
\end{align}
where $\mathbf{v}_\mathbf{k}^\chi$ is the band velocity, $\boldsymbol{\Omega}^\chi = -\chi \mathbf{k} /2k^3$ is the Berry curvature, and $\mathcal{D}^\chi = (1+e\mathbf{B}\cdot\boldsymbol{\Omega}^\chi/\hbar)^{-1}$. The self-rotation of Bloch wavepacket also gives rise to an orbital magnetic moment (OMM)~\cite{xiao2010berry} $\mathbf{m}^\chi_\mathbf{k}$. In the presence of magnetic field, the OMM shifts the energy dispersion as $\epsilon^{\chi}_{\mathbf{k}}\rightarrow \epsilon^{\chi}_{\mathbf{k}} - \mathbf{m}^\chi_\mathbf{k}\cdot \mathbf{B}$. Interestingly, the Berry curvature and the orbital magnetic moment turn out to be independent of the tilting of the Weyl cones.

The collision integral must take into account scattering between the two Weyl nodes (internode, $\chi\Longleftrightarrow\chi'$), as well as scattering withing a Weyl node (intranode, $\chi\Longleftrightarrow\chi$), and thus $\mathcal{I}_{\mathrm{coll}}[f^\chi_\mathbf{k}]$ can be expressed as 
\begin{align}
\mathcal{I}_{\mathrm{coll}}[f^\chi_\mathbf{k}] = \sum\limits_{\chi'}\sum\limits_{\mathbf{k}'} W^{\chi\chi'}_{\mathbf{k},\mathbf{k}'} (f^{\chi'}_{\mathbf{k}'} - f^\chi_\mathbf{k}),
\end{align}
where the scattering rate $W^{\chi\chi'}_{\mathbf{k},\mathbf{k}'}$ is given by~\cite{bruus2004many} 
\begin{align}
W^{\chi\chi'}_{\mathbf{k},\mathbf{k}'} = \frac{2\pi}{\hbar} \frac{n}{\mathcal{V}} |\langle \psi^{\chi'}_{\mathbf{k}'}|U^{\chi\chi'}_{\mathbf{k}\mathbf{k}'}|\psi^\chi_\mathbf{k}\rangle|^2 \delta(\epsilon^{\chi'}_{\mathbf{k}'}-\epsilon_F)
\label{Eq_W_1}
\end{align}
In the above expression $n$ is the impurity concentration, $\mathcal{V}$ is the system volume, $|\psi^\chi_\mathbf{k}\rangle$ is the Weyl spinor wavefunction (which is obtained by diagonalizing the low-energy Weyl Hamiltonian given in the main text), $U^{\chi\chi'}_{\mathbf{k}\mathbf{k}'}$ is the scattering potential, and $\epsilon_F$ is the Fermi energy. The scattering potential profile $U^{\chi\chi'}_{\mathbf{k}\mathbf{k}'}$ is determined by the nature of impurities. Here we restrict ourselves to only non-magnetic point-like impurity, but distinguish between intervalley and intravalley scattering.  This can be controlled independently in our formalism. Thus, the scattering matrix is momentum-independent but has a dependence on the chirality, i.e.,  $U^{\chi\chi'}_{\mathbf{k}\mathbf{k}'} = U^{\chi\chi'}\mathbb{I}$.

The distribution function is assumed to take the form $f^\chi_\mathbf{k} = f_0^\chi + g^\chi_\mathbf{k}$, where $f_0^\chi$ is the equilibrium Fermi-Dirac distribution function and $g^\chi_\mathbf{k}$ indicates the deviation from equilibrium. 
In the steady state, the Boltzmann equation (Eq.~\ref{Eq_boltz1}) takes the following form 
\begin{align}
&\left[\left(\frac{\partial f_0^\chi}{\partial \epsilon^\chi_\mathbf{k}}\right) \mathbf{E}\cdot \left(\mathbf{v}^\chi_\mathbf{k} + \frac{e\mathbf{B}}{\hbar} (\boldsymbol{\Omega}^\chi\cdot \mathbf{v}^\chi_\mathbf{k}) \right)\right]\nonumber\\
 &= -\frac{1}{e \mathcal{D}^\chi}\sum\limits_{\chi'}\sum\limits_{\mathbf{k}'} W^{\chi\chi'}_{\mathbf{k}\mathbf{k}'} (g^\chi_{\mathbf{k}'} - g^\chi_\mathbf{k})
 \label{Eq_boltz2}
\end{align}
The deviation $g^\chi_\mathbf{k}$ is assumed to be linearly proportional to the applied electric field 
\begin{align}
g^\chi_\mathbf{k} = e \left(-\frac{\partial f_0^\chi}{\partial \epsilon^\chi_\mathbf{k}}\right) \mathbf{E}\cdot \boldsymbol{\Lambda}^\chi_\mathbf{k}
\end{align}
We fix the direction of the applied external electric field to be along $+\hat{z}$, i.e., $\mathbf{E} = E\hat{z}$. Therefore only ${\Lambda}^{\chi z}_\mathbf{k}\equiv {\Lambda}^{\chi}_\mathbf{k}$, is relevant. Further, we rotate the magnetic field along the $xz$-plane such that it makes an angle $\gamma$ with respect to the $\hat{x}-$axis, i.e., $\mathbf{B} = B(\cos\gamma,0,\sin\gamma)$. When $\gamma=\pi/2$, the electric and magnetic fields are parallel to each other. Similarly, the strain induced chiral gauge field is rotated in the $xz$-plane, i.e,. $\mathbf{B_5}^\chi = \chi B_5(\cos\gamma_5,0,\sin\gamma_5)$. 
When $\gamma_5\neq \pi/2$, the electric and gauge field are non-collinear and this geometry will be useful in analyzing the strain induced planar Hall effect. Thus the net magnetic field at each valley becomes $\mathbf{B}^\chi\longrightarrow \mathbf{B}+\chi\mathbf{B_5}$.

Keeping terms only up to linear order in the electric field, Eq.~\ref{Eq_boltz2} takes the following form 
\begin{align}
\mathcal{D}^\chi \left[v^{\chi z}_{\mathbf{k}} + \frac{e B}{\hbar} \sin \gamma (\boldsymbol{\Omega}^\chi\cdot \mathbf{v}^\chi_\mathbf{k})\right] = \sum\limits_{\eta}\sum\limits_{\mathbf{k}'} W^{\eta\chi}_{\mathbf{k}\mathbf{k}'} (\Lambda^{\eta}_{\mathbf{k}'} - \Lambda^\chi_\mathbf{k})
\label{Eq_boltz3}
\end{align} 
In order to solve the above equation, we first define the valley scattering rate as follows
\begin{align}
\frac{1}{\tau^\chi_\mathbf{k}} = \mathcal{V} \sum\limits_{\eta} \int{\frac{d^3 \mathbf{k}'}{(2\pi)^3} (\mathcal{D}^\eta_{\mathbf{k}'})^{-1} W^{\eta\chi}_{\mathbf{k}\mathbf{k}'}}
\label{Eq_tau11}
\end{align}
Due to the tilting of the Weyl cones the azimuthal symmetry is destroyed even when the electric and magnetic fields are parallel to each other, and therefore all the integrations are performed over both $\theta$ and $\phi$. The radial integration is simplified due to the delta-function in Eq.~\ref{Eq_W_1}.

Substituting the scattering rate from Eq.~\ref{Eq_W_1} in the above equation, we have 
\begin{align}
\frac{1}{\tau^\chi_\mathbf{k}} = \frac{\mathcal{V}N}{8\pi^2 \hbar} \sum\limits_{\eta} |U^{\chi\eta}|^2 \iiint{(k')^2 \sin \theta' \mathcal{G}^{\chi\eta}(\theta,\phi,\theta',\phi') \delta(\epsilon^{\eta}_{\mathbf{k}'}-\epsilon_F)(\mathcal{D}^\eta_{\mathbf{k}'})^{-1}dk'd\theta'd\phi'},
\label{Eq_tau1}
\end{align}
where $N$ now indicates the total number of impurities, and $ \mathcal{G}^{\chi\eta}(\theta,\phi,\theta',\phi') = (1+\chi\eta(\cos\theta \cos\theta' + \sin\theta\sin\theta' \cos(\phi-\phi')))$ is the Weyl chirality factor defined by the overlap of the wavefunctions. The Fermi wavevector contour $k^\chi$ is evaluated by equating the energy expression with the Fermi energy.
The three-dimensional integral in Eq.~\ref{Eq_tau1} is  reduced to just integration in $\phi'$ and $\theta'$. The scattering time ${\tau^\chi_\mathbf{k}}$ depends on the chemical potential ($\mu$), and is a function of the angular variables $\theta$ and $\phi$. 

\begin{align}
\frac{1}{\tau^\chi_\mu(\theta,\phi)} = \mathcal{V} \sum\limits_{\eta} \iint{\frac{\beta^{\chi\eta}(k')^3}{|\mathbf{v}^\eta_{\mathbf{k}'}\cdot \mathbf{k}'^\eta|}\sin\theta'\mathcal{G}^{\chi\eta}(\mathcal{D}^\eta_{\mathbf{k}'})^{-1} d\theta'd\phi'},
\label{Eq_tau2}
\end{align}
where $\beta^{\chi\eta} = N|U^{\chi\eta}|^2 / 4\pi^2 \hbar^2$. The Boltzmann equation (Eq~\ref{Eq_boltz3}) assumes the form  
\begin{align}
&h^\chi_\mu(\theta,\phi) + \frac{\Lambda^\chi_\mu(\theta,\phi)}{\tau^\chi_\mu(\theta,\phi)} =\nonumber\\ &\mathcal{V}\sum_\eta \iint {\frac{\beta^{\chi\eta}(k')^3}{|\mathbf{v}^\eta_{\mathbf{k}'}\cdot \mathbf{k}'^\eta|} \sin\theta'\mathcal{G}^{\chi\eta}(\mathcal{D}^\eta_{\mathbf{k}'})^{-1}\Lambda^\eta_{\mu}(\theta',\phi') d\theta'd\phi'}
\label{Eq_boltz4}
\end{align}
We make the following ansatz for $\Lambda^\chi_\mu(\theta,\phi)$
\begin{align}
\Lambda^\chi_\mu(\theta,\phi) &= (\lambda^\chi - h^\chi_\mu(\theta,\phi) + a^\chi \cos\theta +\nonumber\\
&b^\chi \sin\theta\cos\phi + c^\chi \sin\theta\sin\phi)\tau^\chi_\mu(\theta,\phi),
\label{Eq_Lambda_1}
\end{align}
where we solve for the eight unknowns ($\lambda^{\pm 1}, a^{\pm 1}, b^{\pm 1}, c^{\pm 1}$). The L.H.S in Eq.~\ref{Eq_boltz4} simplifies to $\lambda^\chi + a^\chi \cos\theta + b^\chi \sin\theta\cos\phi + c^\chi \sin\theta\sin\phi$. The R.H.S of Eq.~\ref{Eq_boltz4} simplifies to
\begin{align}
\mathcal{V}\sum_\eta \beta^{\chi\eta} \iint &f^{\eta} (\theta',\phi') \mathcal{G}^{\chi\eta} (\lambda^\eta - h^\eta_\mu(\theta',\phi') + a^\eta \cos\theta' +\nonumber\\
	&b^\eta \sin\theta'\cos\phi' + c^\eta \sin\theta'\sin\phi')d\theta'd\phi',
	\label{Eq_boltz5_rhs}
\end{align}
where the function
\begin{align}
f^{\eta} (\theta',\phi') = \frac{(k')^3}{|\mathbf{v}^\eta_{\mathbf{k}'}\cdot \mathbf{k}'^\eta|} \sin\theta' (\mathcal{D}^\eta_{\mathbf{k}'})^{-1} \tau^\chi_\mu(\theta',\phi')
\label{Eq_f_eta}
\end{align}
The above equations, when written down explicitly take the form of seven simultaneous equations to be solved for eight variables. The final constraint comes from the particle number conservation 
\begin{align}
\sum\limits_{\chi}\sum\limits_{\mathbf{k}} g^\chi_\mathbf{k} = 0
\label{Eq_sumgk}
\end{align}
Eq.~\ref{Eq_Lambda_1}, Eq.~\ref{Eq_boltz5_rhs}, Eq.~\ref{Eq_f_eta} and Eq.~\ref{Eq_sumgk} are solved together with Eq~\ref{Eq_tau2}, simultaneously for the eight unknowns ($\lambda^{\pm 1}, a^{\pm 1}, b^{\pm 1}, c^{\pm 1}$). Due to the complicated nature of the equations, all the two dimensional integrals w.r.t \{$\theta'$, $\phi'$\}, and the solution of the simultaneous equations are performed numerically. 

For the inversion asymmetric WSM with four Weyl nodes, the distribution function at each node can be represented by $f_\mathbf{k}^m$. Generalizing the formalism presented above, the collision integral must take into account scattering between multiple Weyl cones. Thus $\mathcal{I}_{\mathrm{coll}}[f^m_\mathbf{k}]$ can be expressed as 
\begin{align}
\mathcal{I}_{\mathrm{coll}}[f^m_\mathbf{k}] = \sum\limits_{p}\sum\limits_{\mathbf{k}'} W^{mp}_{\mathbf{k},\mathbf{k}'} (f^{p}_{\mathbf{k}'} - f^m_\mathbf{k}),
\end{align}
where $p$ runs over all the nodes, and scattering rate $W^{mp}_{\mathbf{k},\mathbf{k}'}$ is given by
\begin{align}
W^{mp}_{\mathbf{k},\mathbf{k}'} = \frac{2\pi}{\hbar} \frac{n}{\mathcal{V}} |\langle \psi^{p}_{\mathbf{k}'}|U^{mp}_{\mathbf{k}\mathbf{k}'}|\psi^m_\mathbf{k}\rangle|^2 \delta(\epsilon^{p}_{\mathbf{k}'}-\epsilon_F)
\end{align}
The scattering potential profile $U^{mp}_{\mathbf{k}\mathbf{k}'}$ can be chosen such that scattering between the nodes (internode) as well as within each node (intranode) is considered. Proceeding as before, we define ${\tau^m_\mathbf{k}}$ as

\begin{align}
\frac{1}{\tau^m_\mu(\theta,\phi)} = \mathcal{V} \sum\limits_{p} \iint{\frac{\beta^{mp}(k')^3}{|\mathbf{v}^p_{\mathbf{k}'}\cdot \mathbf{k}'^p|}\sin\theta'\mathcal{G}^{mp}(\mathcal{D}^p_{\mathbf{k}'})^{-1} d\theta'd\phi'},
\label{Eq_tau2m_ch3}
\end{align}
and the Boltzmann equation becomes
\begin{align}
&h^m_\mu(\theta,\phi) + \frac{\Lambda^m_\mu(\theta,\phi)}{\tau^m_\mu(\theta,\phi)} =\nonumber\\ &\mathcal{V}\sum_p \iint {\frac{\beta^{mp}(k')^3}{|\mathbf{v}^p_{\mathbf{k}'}\cdot \mathbf{k}'^p|} \sin\theta'\mathcal{G}^{mp}(\mathcal{D}^p_{\mathbf{k}'})^{-1}\Lambda^p_{\mu}(\theta',\phi') d\theta'd\phi'}.
\label{Eq_boltz4m_ch3}
\end{align}
Making the ansatz $\Lambda^m_\mu(\theta,\phi) = (\lambda^m - h^m_\mu(\theta,\phi) + a^m \cos\theta +b^m \sin\theta\cos\phi + c^m \sin\theta\sin\phi)\tau^m_\mu(\theta,\phi)$, and using the constraint for particle number conservation, the Boltzmann equation is reduced to a system of sixteen equations to be solved for sixteen unknowns.

\end{appendices}

\setcounter{equation}{0}
\setcounter{table}{0}
\setcounter{figure}{0}

\begin{appendices}
\chapter{\label{appendix_ch5}}

\section{Substitution and expression for system of linear equations}
\label{Appendix:Substitution and expression}
The system of linear equations of the unknown variables involved in the $\Lambda^{\chi}$ is written as follows:
\begin{align}
\begin{bmatrix}
      a^+\\
      b^+\\
      c^+\\
      a^-\\
      b^-\\
      c^-\\
\end{bmatrix}
&=
\begin{bmatrix}
\Pi^{++}T^+ & \Pi^{++}I^+ & \Pi^{++}J^+ & \Pi^{+-}I^- & \Pi^{+-}M^- & \Pi^{+-}N^-\\
\Pi^{++}I^+ & \Pi^{++}M^+ & \Pi^{++}N^+ & \Pi^{+-}T^- & \Pi^{+-}I^- & \Pi^{+-}J^-\\
\Pi^{++}J^+ & \Pi^{++}N^+ & \Pi^{++}Q^+ & \Pi^{+-}J^- & \Pi^{+-}N^- & \Pi^{+-}Q^-\\
\Pi^{-+}I^+ & \Pi^{-+}M^+ & \Pi^{-+}N^+ & \Pi^{--}T^- & \Pi^{--}I^- & \Pi^{--}J^-\\
\Pi^{-+}T^+ & \Pi^{-+}I^+ & \Pi^{-+}J^+ & \Pi^{--}I^- & \Pi^{--}M^- & \Pi^{--}N^-\\
\Pi^{-+}J^+ & \Pi^{-+}N^+ & \Pi^{-+}Q^+ & \Pi^{--}J^- & \Pi^{--}N^- & \Pi^{--}Q^-\\
\end{bmatrix}
\cdot
\begin{bmatrix}
      a^+\\
      b^+\\
      c^+\\
      a^-\\
      b^-\\
      c^-\\
\end{bmatrix} \nonumber\\
&-
\begin{bmatrix}
\Pi^{++}G^+ + \Pi^{+-}L^-\\
\Pi^{++}L^+ + \Pi^{+-}G^-\\
\Pi^{++}P^+/4 + \Pi^{+-}P^-/4\\
\Pi^{-+}L^+ + \Pi^{--}G^-\\
\Pi^{-+}G^+ + \Pi^{--}L^-\\
\Pi^{-+}P^+/4 + \Pi^{--}P^-/4\\
\end{bmatrix}
\label{Eq:System of linear equation}
\end{align}

Substitution in the above matrix equation is written as:
\begin{align*}
\int d\theta' f^{\chi}(\theta') \cos^4(\theta'/2) &= R^{\chi}, \quad 
\int d\theta' f^{\chi}(\theta') \sin^4(\theta'/2) = S^{\chi}, \\
\int d\theta' f^{\chi}(\theta') \sin^8(\theta'/2) &= M^{\chi}, \quad 
\int d\theta' f^{\chi}(\theta') h^{\chi}(\theta') = U^{\chi}, \\
\int d\theta' f^{\chi}(\theta') \cos^8(\theta'/2) &= T^{\chi}, \quad 
\int d\theta' f^{\chi}(\theta') \sin^2(\theta') = O^{\chi}, \\
\int d\theta' f^{\chi}(\theta') h^{\chi}(\theta')\sin^2(\theta') &= P^{\chi}, \quad 
\int d\theta' f^{\chi}(\theta') h^{\chi}(\theta') \sin^4(\theta'/2) = L^{\chi}, \\
\int d\theta' f^{\chi}(\theta') h^{\chi}(\theta') \cos^4(\theta'/2) &= G^{\chi},  \\
\int d\theta' f^{\chi}(\theta') h^{\chi}(\theta') \cos^4(\theta'/2) \sin^2(\theta') &= J^{\chi}, \\
\int d\theta' f^{\chi}(\theta') h^{\chi}(\theta') \sin^4(\theta'/2) \sin^2(\theta') &= N^{\chi},\\ \quad 
\int d\theta' f^{\chi}(\theta') h^{\chi}(\theta') \sin^4(\theta'/2) \cos^4(\theta'/2) &= I^{\chi},\\
\int d\theta' f^{\chi}(\theta') \sin^4(\theta') &= Q^{\chi}.
\end{align*}
\section{Zero-field conductivity}\label{Sec:Zero magnetic field coductivity}
We calculate the zero-field conductivity and derive its analytical expressions for two systems, i.e., spin-1/2 and pseudospin-1 using the formalism mentioned in the main text. For $\mathbf{B} = 0 $, the the coupled Eq.~\ref{Couplled_equation} acquire the following form:
\begin{align}
\dot{\mathbf{r}}^\chi &= \frac{e}{\hbar}(\mathbf{E}\times \boldsymbol{\Omega}^\chi) + \mathbf{v}_\mathbf{k}^\chi \nonumber\\ 
\dot{\mathbf{k}}^\chi &= -\frac{e}{\hbar} ~\mathbf{E}  .
\label{Eq:Couplled_equation_B0}
\end{align}
Using this, the Boltzmann equation is simplified to:
\begin{align}
v^{\chi}_{z,\mathbf{k}} + \frac{\Lambda^{\chi}(\theta)}{\tau^{\chi}_{\mu}}=\sum_{\chi'}\int\frac{d^3\mathbf{k}'}{(2\pi)^3} \mathbf{W}^{\chi \chi'}_{\mathbf{k k'}}\Lambda^{\chi'}(\theta').
\label{Eq:MB_in_term_Wkk'_B0}
\end{align}
We solve this analytically using the ansatz defined in the main text and using the ansatz defined in Ref.~\cite{ahmad2021longitudinal,ahmad2024nonlinear} for Weyl (spin-1/2). The net scattering rate at two valleys ($\chi=\pm1$) for $B=0$ is calculated to be:
\begin{align}
\frac{1}{\tau^{\chi,s=1}_{\mu}} = \frac{4 \pi V \epsilon_{\mathrm{F}}^2 (\alpha + 1)}{3 \hbar^3 v_{\mathrm{F}}^3} \equiv \frac{8}{\tau^{\chi,s=1/2}_{\mu}}.
\label{Eq:tauinv_pm1_B0}
\end{align}
Returning back to Eq.~\ref{Eq:MB_in_term_Wkk'_B0}, it is written explicitly in terms of unknown variables involved in ansatz for spin-1 WSMs as: 
\begin{multline}
\lambda^{+} - \frac{\sin^2(\theta) \left( 10\lambda^{+} + 3a^{+} + 3b^{+} - 12c^{+} + 10\lambda^{-}\alpha + 3\alpha a^{-} + 3\alpha b^{-} - 20\alpha c^{+} + 8\alpha c^{-} \right)}{20(\alpha + 1)} \\
- \frac{\cos^4\left(\frac{\theta}{2}\right) \left( 10\lambda^{+} - 4a^{+} + b^{+} + 6c^{+} - 5v_{\mathrm{F}} + 10\lambda^{-}\alpha - 10\alpha a^{+} + \alpha a^{-} + 6\alpha b^{-} + 6\alpha c^{-} + 5\alpha v_{\mathrm{F}} \right)}{10(\alpha + 1)}\\
- \frac{\sin^4\left(\frac{\theta}{2}\right) \left( 10\lambda^{+} + a^{+} - 4b^{+} + 6c^{+} + 5v_{\mathrm{F}} + 10\lambda^{-}\alpha + 6\alpha a^{-} - 10\alpha b^{+} + \alpha b^{-} + 6\alpha c^{-} - 5\alpha v_{\mathrm{F}} \right)}{10(\alpha + 1)}=0
\end{multline}
\begin{multline}
\lambda^{-} - \frac{\sin^2(\theta) \big( 10\lambda^{-} + 3a^{-} + 3b^{-} - 12c^{-} + 10\lambda^{+}\alpha + 3\alpha a^{+} + 3\alpha b^{+} + 8\alpha c^{+} - 20\alpha c^{-} \big)}{20(\alpha + 1)} \\
- \frac{\cos^4\big(\frac{\theta}{2}\big) \big( 10\lambda^{-} - 4a^{-} + b^{-} + 6c^{-} - 5v_{\mathrm{F}} + 10\lambda^{+}\alpha 
+ \alpha a^{+} - 10\alpha a^{-} + 6\alpha b^{+} + 6\alpha c^{+} + 5\alpha v_{\mathrm{F}} \big)}{10(\alpha + 1)}\\
- \frac{\sin^4\big(\frac{\theta}{2}\big) \big( 10\lambda^{-} + a^{-} - 4b^{-} + 6c^{-} + 5v_{\mathrm{F}} + 10\lambda^{+}\alpha + 6\alpha a^{+} + \alpha b^{+} - 10\alpha b^{-} + 6\alpha c^{+} - 5\alpha v_{\mathrm{F}} \big)}{10(\alpha + 1)}=0.\\
\label{Eq:MBT_in_terms_unknown_variables_B0}
\end{multline}
Equating the coefficient of  $\cos^4\big(\frac{\theta}{2}\big),~\sin^4\big(\frac{\theta}{2}\big)$ and $\sin^2(\theta)$ in these two equations results in a system of eight coupled equations to be solved for eight unknown variables ($\lambda^{\pm}, a^\pm, b^\pm, c^\pm$). The solution obtained is as follows: $\lambda^{\pm} = 0, a^{\pm} = \frac{v_{\mathrm{F}} (\alpha-1)}{3\alpha+1}, 
b^{\pm} = a^{\pm}, c^{\pm} = 0$. 
Using Eq.~\ref{Eq:g1}, Eq.~\ref{Eq:J_formula} and Eq.~\ref{Eq:tauinv_pm1_B0} LMC is evaluated to be:
\begin{align}
\sigma_{zz}^{s=1} = \frac{\mathrm{e}^2 v_{\mathrm{F}}^2}{ V \pi^2 \left(3 \alpha + 1\right)}, \nonumber\\
\sigma_{zz}^{s=1/2} = \frac{\mathrm{e}^2 v_{\mathrm{F}}^2}{16 V \pi^2 \left(2 \alpha + 1\right)}.
\label{Eq:LMC_B0_ch5}
\end{align}
\end{appendices}

\setcounter{equation}{0}
\setcounter{table}{0}
\setcounter{figure}{0}

\bibliography{reference}
\end{document}